\pdfoutput=1

\documentclass{sty/thesis}
\usepackage{sty/mythesis}
\usepackage{amsmath, textcomp, gensymb, pifont}
\usepackage{url}
\usepackage{tikz}
\usepackage{setspace}
\usepackage{graphicx}
\usepackage{flexisym}

\usepackage{wrapfig}
\usepackage{hyperref}
\usepackage{xspace}
\usepackage{xfrac}

\newcommand{\ignore}[1]{}   

\newcommand\vdd{$V_{\mathit{DD}}$\xspace}

\let\oldcelsius\celsius
\renewcommand{\celsius}{~\oldcelsius\xspace}

\newlength{\realbaselineskip}
\newboolean{publicversion}
\setboolean{publicversion}{true}

\ifthenelse{\boolean{publicversion}}{
  \newcommand{\grumbler}[2]{}
}{
  \newcommand{\grumbler}[2]{\textcolor{red}{\bf #1: #2}}
}

\usepackage{ifthen}
\usepackage{calc}
\usepackage{pifont}
\usepackage{color}
\usepackage{fancyhdr}
\usepackage{tikz}
\usepackage{anyfontsize}

\newcounter{hours}
\newcounter{minutes}

\newcommand{\bl}[0]{{\small{$BL$}}\xspace}
\newcommand{\blbar}[0]{{\small{$\overline{BL}$}}\xspace}
\newcommand{\vdddelta}[0]{{\small{$V_{DD}/2+\Delta$}}\xspace}
\newcommand{\vddhalf}[0]{{\small{$V_{DD}/2$}}\xspace}

\newcommand{\halfvdd}[0]{{$V_{DD}/2$}\xspace}

\newcommand{\zerov}[0]{{$0$V}\xspace}
\newcommand{\newt}[0]{}
\newcommand{\new}[0]{}
\newcommand{\newthree}[0]{}
\newcommand{\dhlfix}[0]{}

\newcommand{\reftiny}[0]{}

\newcommand{\fix}[1]{#1}

\newcommand{\fixII}[1]{#1}
\newcommand{\fixIII}[1]{#1}
\newcommand{\fixIV}[1]{#1}
\newcommand{\fixV}[1]{#1}
\newcommand{\fixVI}[1]{#1}

\newdimen\origiwspc%
\origiwspc=\fontdimen2\font

\newcommand{\myitem}[1]{{\em (#1)}\xspace}

\newcommand{\varr}[0]{{{$V_{array}$}}\xspace}
\newcommand{\vperi}[0]{{{$V_{peri}$}}\xspace}
\newcommand{\vmin}[0]{$V_{min}$\xspace}

\newcommand{\voltron}[0]{{Voltron}\xspace}
\newcommand{\memdvfs}[0]{{MemDVFS}\xspace}

\newboolean{timeofmake}
\setboolean{timeofmake}{false}

\newcommand{\todo}[1]{{\color{red}TODO: \textbf{#1}}\xspace}

\newcommand{\changes}[0]{}

\newcommand{\cc}[1]{#1}

\newcommand{\refsec}[1]{\S\ref{sec:#1}}

\newcommand{\villa}[0]{{VILLA-DRAM}\xspace}

\newcommand{\acro}[0]{{LISA}\xspace}
\newcommand{\lisa}[0]{{LISA}\xspace}
\newcommand{\lisarc}[0]{{LISA-RISC}\xspace}
\newcommand{\lisarcnol}[0]{{RISC}\xspace}

\newcommand{\lisapre}[0]{{LISA-LIP}\xspace}
\newcommand{\lisaprenol}[0]{{LIP}\xspace}
\newcommand{\lisavilla}[0]{{LISA-VILLA}\xspace}
\newcommand{\lisavillanol}[0]{{VILLA}\xspace}
\newcommand{\lisarcvilla}[0]{{LISA-(RISC+VILLA)}\xspace}
\newcommand{\baseline}[0]{\texttt{memcpy}\xspace}
\newcommand{\baselinenorm}[0]{memcpy\xspace}
\newcommand{\xferfull}[0]{{row buffer movement}\xspace}
\newcommand{\xfernorm}[0]{{{RBM}}\xspace}
\newcommand{\xfer}[0]{{RBM}\xspace}

\newcommand{\rb}[1]{\texttt{RB#1}\xspace}

\newcommand{\rc}[0]{{RowClone}\xspace}
\newcommand{\iso}[0]{{link}\xspace}
\newcommand{\isos}[0]{{links}\xspace}
\newcommand{\capfullname}[0]{{Low-Cost Inter-Linked Subarrays}\xspace}
\newcommand{\fullname}[0]{{low-cost inter-linked subarrays}\xspace}

\newcommand{\lisafullname}[0]{{\underline{L}ow-cost \underline{I}nter-linked
\underline{S}ub\underline{A}rrays}\xspace}

\newcommand{\refab}[0]{{\small{$REF_{ab}$}}\xspace}
\newcommand{\refpb}[0]{{\small{$REF_{pb}$}}\xspace}
\newcommand{\trefi}[0]{{\small{$tREFI_{ab}$}}\xspace}
\newcommand{\trefipb}[0]{{\small{$tREFI_{pb}$}}\xspace}
\newcommand{\trfc}[0]{{\small{$tRFC_{ab}$}}\xspace}
\newcommand{\trfcpb}[0]{{\small{$tRFC_{pb}$}}\xspace}

\newcommand{\tras}[0]{{\small{tRAS}}\xspace}

\newcommand{\trp}[0]{{\small{tRP}}\xspace}
\newcommand{\trpe}[0]{{\small{tRP}}=}
\newcommand{\trrd}[0]{{\small{tRRD}}\xspace}
\newcommand{\tfaw}[0]{{\small{tFAW}}\xspace}
\newcommand{\fgr}[0]{{\small{FGR}}\xspace}
\newcommand{\trcd}[0]{{\small{tRCD}}\xspace}
\newcommand{\trcde}[0]{{\small{tRCD}}=}
\newcommand{\tcl}[0]{{\small{tCL}}\xspace}
\newcommand{\tcle}[0]{{\small{tCL}}=}

\newcommand{\tbl}[0]{{\small{tBL}}\xspace}
\newcommand{\tble}[0]{{\small{tBL}}=}

\newcommand{\src}[0]{\texttt{\small{src}}\xspace}
\newcommand{\dst}[0]{\texttt{\small{dst}}\xspace}
\newcommand{\dsttwo}[0]{\texttt{\small{dst2}}\xspace}

\newcommand{\trcdmin}[0]{$tRCD_{min}$\xspace}
\newcommand{\trpmin}[0]{$tRP_{min}$\xspace}

\newcommand{\act}[0]{\textsc{{activate}}\xspace}
\newcommand{\acts}[0]{\textsc{{activates}}\xspace}
\newcommand{\crd}[0]{\textsc{{read}}\xspace}
\newcommand{\cwr}[0]{\textsc{{write}}\xspace}
\newcommand{\cpre}[0]{\textsc{{precharge}}\xspace}
\newcommand{\pre}[0]{\textsc{{precharge}}\xspace}

\newcommand{\intersub}[0]{{Subarray Access Refresh Parallelization}\xspace}

\newcommand{\is}[0]{{\text{SARP}}\xspace}
\newcommand{\sarp}[0]{{\text{SARP}}\xspace}
\newcommand{\isab}[0]{{$\text{SARP}_{\text{ab}}$}\xspace}
\newcommand{\ispb}[0]{{$\text{SARP}_{\text{pb}}$}\xspace}
\newcommand{\warplongCap}[0]{{\mbox{Write-refresh} Parallelization}\xspace}
\newcommand{\warplong}[0]{{\mbox{write-refresh} parallelization}\xspace}
\newcommand{\Warplong}[0]{{\mbox{Write-refresh} parallelization}\xspace}
\newcommand{\darp}[0]{{DARP}\xspace}
\newcommand{\darplong}[0]{{Dynamic Access Refresh Parallelization}\xspace}
\newcommand{\Ooolong}[0]{{\mbox{Out-of-order} per-bank refresh}\xspace}
\newcommand{\ooolong}[0]{{\mbox{out-of-order} per-bank refresh}\xspace}
\newcommand{\ooolongCap}[0]{{\mbox{Out-of-order} Per-bank Refresh}\xspace}
\newcommand{\ib}[0]{{\darp}\xspace}

\newcommand{\combo}[0]{{DSARP}\xspace}

\newcommand{\romnum}[1]{{\em (#1)}\xspace}

\newcommand{\dimm}[2]{$\texttt{D}^{#1}_{#2}$\xspace}

\newcommand{\response}[1]{#1}
\newcommand{\vddbl}[0]{\emph{full voltage}\xspace}
\newcommand{\vddblhalf}[0]{\emph{half voltage}\xspace}
\newcommand{\vddzero}[0]{\emph{zero voltage}\xspace}
\newcommand{\patt}[1]{\texttt{#1}\xspace}

\algnewcommand\algorithmicforeach{\textbf{for each}}
\algdef{S}[FOR]{ForEach}[1]{\algorithmicforeach\ #1\ \algorithmicdo}

\newfloat{afloat}{t!}{log}
\newcommand\afloatname{Algorithm}
\floatname{afloat}{\afloatname}

\newcounter{observation}

\newcommand{\obs}[1]{
    \refstepcounter{observation}%
    \textbf{Observation \theobservation:}~\textit{#1}}

\newenvironment{observation}[0]
{
    \par
    \refstepcounter{observation}%
    \textbf{Observation \theobservation:}
    \begin{itshape}%
    }%
    {
    \end{itshape}%
}

\newcommand{\mech}[0]{\mbox{FLY-DRAM}\xspace}
\newcommand{\mechlong}[0]{{Flexible-LatencY DRAM}\xspace}

\newcommand{\caprefab}[0]{\small{REF\textsubscript{ab}}\xspace}
\newcommand{\caprefpb}[0]{\small{REF\textsubscript{pb}}\xspace}
\newcommand{\captrfc}[0]{\small{tRFC\textsubscript{ab}}\xspace}

\newcommand{\rciasa}[0]{{RC-IntraSA}\xspace}
\newcommand{\rcirsa}[0]{{RC-InterSA}\xspace}

\newcommand*\circled[1]{\tikz[baseline=(char.base)]{
            \node[shape=circle,draw,inner sep=0.8pt,fill=white,text=black] (char) {#1};}}
\newcommand*\darkcircled[1]{\tikz[baseline=(char.base)]{
            \node[shape=circle,draw,inner sep=0.8pt,fill=black,text=white] (char) {#1};}}

\newcommand{\figputWS}[3]{
\begin{figure*}[t]
\begin{minipage}{\linewidth}
\begin{center}
\includegraphics[scale=#2]{plots/#1}
\end{center}
\vspace{-0.15in}
\caption{#3 \label{fig:#1}}
\end{minipage}
\end{figure*}
}

\newcommand{\figputGTW}[2]{
\begin{figure}[!t]
\begin{minipage}{\linewidth}
\begin{center}
\includegraphics[width=1.0\linewidth]{plots/gnuplots/#1}
\end{center}
\vspace{-0.15in}
\caption{#2 \label{fig:#1}}
\end{minipage}
\end{figure}
}

\newcommand{\figputHS}[3]{
\begin{figure}[h]
\begin{minipage}{\linewidth}
\begin{center}
\includegraphics[scale=#2]{plots/#1}
\end{center}
\vspace{-0.1in}
\caption{#3 \label{fig:#1}}
\end{minipage}
\end{figure}
}

\newcommand{\figputGHS}[3]{
\begin{figure}[h]
\begin{minipage}{\linewidth}
\begin{center}
\includegraphics[scale=#2]{plots/gnuplots/#1}
\end{center}
\vspace{-0.1in}
\caption{#3 \label{fig:#1}}
\end{minipage}
\end{figure}
}
\newcommand{\figputGTS}[3]{
\begin{figure}[t]
\begin{minipage}{\linewidth}
\begin{center}
\includegraphics[scale=#2]{plots/gnuplots/#1}
\end{center}
\vspace{-0.1in}
\caption{#3 \label{fig:#1}}
\end{minipage}
\end{figure}
}

\newcommand{\figputHSL}[4]{
\begin{figure}[h]
\begin{minipage}{\linewidth}
\begin{center}
\includegraphics[scale=#2]{plots/#1}
\end{center}
\vspace{-0.1in}
\caption{#3 \label{fig:#4}}
\end{minipage}
\end{figure}
}

\newcommand{\figref}[1]{Figure~\ref{fig:#1}}
\newcommand{\tabref}[1]{Table~\ref{tab:#1}}

\newcommand{\symsref}[1]{\S\ref{sec:#1}}
\newcommand{\secref}[1]{Section~\ref{sec:#1}}
\newcommand{\chapref}[1]{Chapter~\ref{chap:#1}}
\newcommand{\ssecref}[1]{Section~\ref{ssec:#1}}

\newcommand{\paratitle}[1]{\textbf{#1.}\xspace}

\newcommand{\module}[3]{{{\scriptsize\textit #1}$_{\mathrm{#2}}^{\mathrm{#3}}$}\xspace}

\usepackage[numbers,sort]{natbib}
\usepackage{subcaption}
\usepackage{afterpage}
\usepackage{tcolorbox}
\usepackage{afterpage}

\usepackage{appendix}
\usepackage{chngcntr}
\AtBeginEnvironment{subappendices}{%
  \chapter*{Appendix}
  \addcontentsline{toc}{chapter}{Appendices for Chapter 7}
  \counterwithin{figure}{section}
  \counterwithin{table}{section}
}

\def\secondpage{\clearpage\null\vfill
  \pagestyle{empty}
  \begin{minipage}[b]{0.9\textwidth}
    \footnotesize\raggedright
    \begin{center}
    \setlength{\parskip}{0.5\baselineskip}
    Copyright \copyright \the\year\ Kevin K. Chang
    \end{center}
  \end{minipage}
  \vspace*{2\baselineskip}
  \cleardoublepage
  \cfoot{\thepage}
}

\newcommand{\squishlist}{
	\begin{list}{$\bullet$}
		{ \setlength{\itemsep}{0pt}      \setlength{\parsep}{3pt}
			\setlength{\topsep}{3pt}       \setlength{\partopsep}{0pt}
			\setlength{\leftmargin}{1.5em} \setlength{\labelwidth}{1em}
			\setlength{\labelsep}{0.5em} } }
\newcommand{\squishend}{
    \end{list}  }

\title{{\bf {\Large Understanding and Improving\\ the Latency of DRAM-Based Memory
    Systems}}}

\begin{document}

\maketitle
\pagenumbering{roman}
\chapter*{Abstract}

Over the past two decades, the storage capacity and access bandwidth of main
memory have improved tremendously, by 128x and 20x, respectively. These
improvements are mainly due to the continuous technology scaling of \emph{DRAM
  (dynamic random-access memory)}, which has been used as the physical substrate
for main memory. In stark contrast with capacity and bandwidth, DRAM
\emph{latency} has remained almost constant, reducing by only 1.3x in the same
time frame. Therefore, long DRAM latency continues to be a critical performance
bottleneck in modern systems. Increasing core counts, and the emergence of
increasingly more \mbox{data-intensive} and \mbox{latency-critical} applications further
stress the importance of providing low-latency memory access.

In this dissertation, we identify three main problems that contribute
significantly to long latency of DRAM accesses. To address these problems, we present
a series of new techniques. Our new techniques significantly improve both system
performance and energy efficiency. We also examine the critical relationship
between supply voltage and latency in modern DRAM chips and develop new
mechanisms that exploit this voltage-latency trade-off to improve energy
efficiency.

First, while bulk data movement is a key operation in many applications and
operating systems, contemporary systems perform this movement inefficiently, by
transferring data from DRAM to the processor, and then back to DRAM, across a
narrow off-chip channel. The use of this narrow channel for bulk data movement
results in high latency and high energy consumption. This dissertation
introduces a new DRAM design, \mbox{Low-cost} \mbox{Inter-linked} SubArrays
(\lisa), which provides fast and energy-efficient bulk data movement across
subarrays in a DRAM chip. We show that the \lisa substrate is very powerful and
versatile by demonstrating that it efficiently enables several new architectural
mechanisms, including low-latency data copying, reduced DRAM access latency for
frequently-accessed data, and reduced preparation latency for subsequent
accesses to a DRAM bank.

Second, DRAM needs to be periodically refreshed to prevent data loss due to
leakage. Unfortunately, while DRAM is being refreshed, a part of it becomes
unavailable to serve memory requests, which degrades system performance. To
address this \emph{refresh interference} problem, we propose two access-refresh
parallelization techniques that enable more overlapping of accesses with
refreshes inside DRAM, at the cost of very modest changes to the memory
controllers and DRAM chips. These two techniques together achieve performance
close to an idealized system that does not require refresh.

Third, we find, for the first time, that there is significant latency variation
in accessing different cells of a single DRAM chip due to the irregularity in
the DRAM manufacturing process. As a result, some DRAM cells are inherently faster to
access, while others are inherently slower. Unfortunately, existing systems do
not exploit this variation and use a fixed latency value based on the slowest
cell across all DRAM chips. To exploit latency variation within the DRAM chip,
we experimentally characterize and understand the behavior of the variation that
exists in real commodity DRAM chips. Based on our characterization, we propose
\mechlong (\mech), a mechanism to reduce DRAM latency by categorizing the DRAM
cells into fast and slow regions, and accessing the fast regions with a reduced
latency, thereby improving system performance significantly. Our extensive
experimental characterization and analysis of latency variation in DRAM chips
can also enable the development of other new techniques to improve performance or
reliability.

Fourth, this dissertation, for the first time, develops an understanding of the
latency behavior due to another important factor -- \emph{supply voltage}, which
significantly impacts DRAM performance, energy consumption, and reliability. We
take an experimental approach to understanding and exploiting the behavior of
modern DRAM chips under different supply voltage values. Our detailed
characterization of real commodity DRAM chips demonstrates that memory access
latency can be reliably reduced by increasing the DRAM array supply voltage.
Based on our characterization, we propose Voltron, a new mechanism that improves
system energy efficiency by dynamically adjusting the DRAM supply voltage using
a new performance model. Our extensive experimental data on the relationship
between DRAM supply voltage, latency, and reliability can further enable
developments of other new mechanisms that improve latency, energy efficiency, or
reliability.

The key conclusion of this dissertation is that augmenting DRAM architecture
with simple and \mbox{low-cost} features, and developing a better understanding
of manufactured DRAM chips together lead to significant memory latency reduction
as well as energy efficiency improvement. We hope and believe that the proposed
architectural techniques and the detailed experimental data and observations on
real commodity DRAM chips presented in this dissertation will enable development
of other new mechanisms to improve the performance, energy efficiency, or
reliability of future memory systems.

%
%

\chapter*{Acknowledgments}
The pursuit of Ph.D. has been a period of fruitful learning experience for me,
not only in the academic arena, but also on a personal level. I would like to
reflect on the many people who have supported and helped me to become who I am
today. First and foremost, I would like to thank my advisor, Prof. Onur Mutlu,
who has taught me how to think critically, speak clearly, and write thoroughly.
Onur generously provided the resources and the open environment that enabled me
to carry out my research. I am also very thankful to Onur for giving me the
opportunities to collaborate with students and researchers from other
institutions. They have broadened my knowledge and improved my
research.

I am grateful to the members of my thesis committee: Prof. James Hoe, Prof.
Kayvon Fatahalian, Prof. Moinuddin Qureshi, and Prof. Steve Keckler for serving
on my defense. They provided me valuable comments and helped make the final
stretch of my Ph.D. very smooth. I would like to especially thank Prof.
James Hoe for introducing me to computer architecture and providing me
numerous pieces of advice throughout my education.

I would like to thank my internship mentors, who provided the guidance to make
my work successful for both sides: Gabriel Loh, Mithuna Thottethodi, Yasuko
Eckert, Mike O'Connor, Srilatha Manne, Lisa Hsu, Zeshan Chishti, Alaa
Alameldeen, Chris Wilkerson, Shih-Lien Lu, and Manu Awasthi. I thank the Intel
Corporation and Semiconductor Research Corporation (SRC) for their generous
financial support.

During graduate school, I have met many wonderful fellow graduate students and
friends whom I am grateful to. Members in SAFARI research group have been both
great friends and colleagues to me. Rachata Ausavarungnirun was my great cubic
mate who supported and tolerated me for many years. Donghyuk Lee was our DRAM
guru who introduced expert DRAM knowledge to many of us. Saugata Ghose was my
collaborator and mentor who assisted me in many ways. Hongyi Xin has been a good
friend who taught me a great deal about bioinformatics and amused me with his
great sense of humor. Yoongu Kim taught me how to conduct research and think
about problems from different perspectives. Samira Khan provided me insightful
academic and life advice. Gennady Pekhimenko was always helpful when I am in
need. Vivek Seshadri is someone I aspire to be because of his creative and
methodical approach to problem solving and thinking. Lavanya Subramanian was
always warm and welcoming when I approached her with ideas and problems. Chris
Fallin's critical feedback on research during the early years of my research was
extremely helpful. Kevin Hsieh was always willing to listen to my problems in
school and life, and provided me with the right advice. Nandita Vijaykumar was a
strong-willed person who gave me unwavering advice. I am grateful to other
SAFARI members for their companionship: Justin Meza, Jamie, Ben, HanBin Yoon,
Yang Li, Minesh Patel, Jeremie Kim, Amirali Boroumand, and Damla Senol. I also
thank graduate interns and visitors who have assisted me in my research: Hasan
Hassan, Abhijith Kashyap, and Abdullah Giray Yaglikci.

Graduate school is a long and lonely journey that sometimes hits you the hardest
when working at midnight. I feel very grateful for the many friends who were
there to help me grind through it. I want to thank Richard Wang, Yvonne Yu,
Eugene Wang, and Hongyi Xin for their friendship and bringing joy during my
low time.

I would like to thank my family for their enormous support and sacrifices that
they made. My mother, Lili, has been a pillar of support throughout my life. She
is the epitome of love, strength, and sacrifice. I am grateful to her strong
belief in education. My sister, Nadine, has always been there kindly supporting
me with unwavering love. I owe all of my success to these two strong women in my
life. I would like to thank my late father, Pei-Min, for his love and full
support. My childhood was full of joy because of him. I am sorry that he has not
lived to see me finish my Ph.D. I also thank my step father, Stephen, for his
encouragement and support. Lastly, I thank my girlfriend Sherry Huang, for all
her love and understanding.

Finally, I am grateful to the Semiconductor Research Corporation and Intel for
providing me a fellowship. I would like to acknowledge the support of Google,
Intel, NVIDIA, Samsung, VMware, and the United States Department of Energy. This
dissertation was supported in part by the Intel Science and Technology Center
for Cloud Computing, Semiconductor Research Corporation, and National Science
Foundation (grants 1212962 and 1320531).

\newpage
\tableofcontents
\listoffigures
\listoftables
\newpage

\doublespacing

\setcounter{page}{1}
\pagenumbering{arabic}
\pagestyle{fancy}

\chapter{Introduction}
\label{chap:intro}

\section{Problem} Since the inception of general-purpose electronic computers
from more than half a century ago, the computer technology has seen tremendous
improvements in system performance, main memory, and disk storage. Main memory,
a major system component, has served the essential role of storing data and
instructions for computer systems to operate. For decades, \emph{semiconductor
  DRAM (dynamic random-access memory)} has been the building foundation of main
memory.

DRAM-based main memory has made rapid progress on capacity and bandwidth,
improving by 128x and 20x, respectively, over the past two decades
~\cite{micronSDR_128Mb,jedec-ddr2,jedec-ddr3, jedec-ddr4, lee-hpca2013,
  son-isca2013, lee-hpca2015, chang-sigmetrics2016}, as shown in
\figref{dram_scaling_improvement}, which illustrates the historical scaling
trends of a DRAM chip from 1999 to 2017. These capacity and bandwidth
improvements mainly follow Moore's Law~\cite{moore-law} and Dennard
scaling~\cite{dennard-scaling}, which enable more and faster transistors along
with more pins. On the contrary, DRAM latency has improved (i.e., reduced) by
only 1.3x, which is a drastic underperformer compared to capacity and bandwidth.
As a result, long DRAM latency remains as a significant system performance
bottleneck for many modern applications~\cite{superfri,mutlu-imw2013}, \cc{such
  as in-memory databases~\cite{ailamaki-vldb1999, clapp-2015, mao-2013,
    boncz-1999, xi-2015}, data analytics (e.g.,
  Spark)~\cite{clapp-2015,awan-bdcloud2015,yasin-iiswc2014,awan-bdcloud2016},
  graph traversals~\cite{xu-iiswc2014, umuroglu-fpl2015, ahn-isca2015}, pointer
  chasing workloads~\cite{hsieh-iccd2016}, Google's datacenter
  workloads~\cite{kanev-isca2015}, and buffers for network packets in routers or
  network
  processors~\cite{hermsmeyer-bell2009,appenzeller-sigcomm2004,toal-ahs2007,zabinski-csm2013,kleveland-ieeemicro2013,vishwanath-conext2011}.
  For example, a recent study by Google reported that memory latency is more
  important than memory bandwidth for the applications running in Google's
  datacenters~\cite{kanev-isca2015}. Another example is that, to achieve 100
  Gb/s Ethernet, network processors require low DRAM latency to access and
  process network packets buffered in the DRAM~\cite{hermsmeyer-bell2009}.}

\figputHS{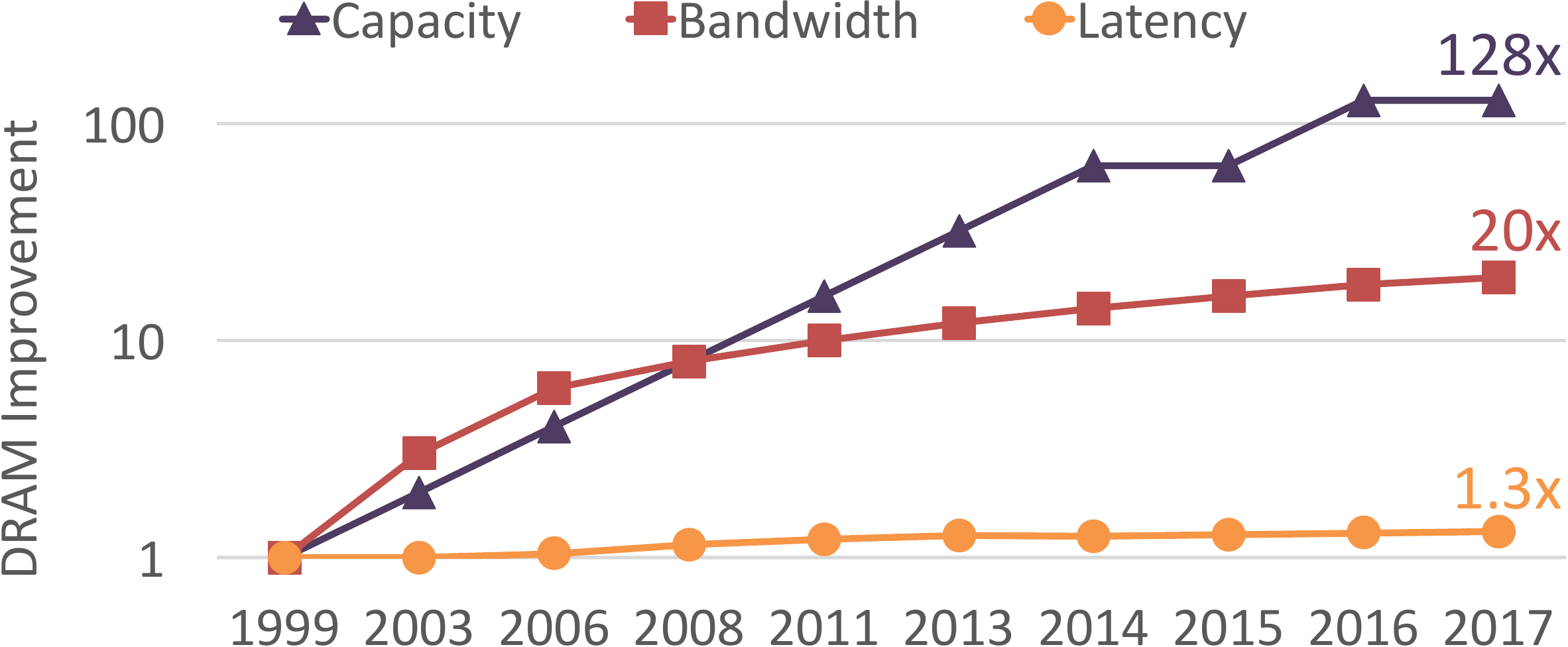}{0.48}{DRAM scaling trends over
  time~\cite{micronSDR_128Mb,jedec-ddr2,jedec-ddr3, jedec-ddr4, lee-hpca2013,
    son-isca2013, lee-hpca2015, chang-sigmetrics2016}.}

\vspace{-0.15in} \cc{To provide low DRAM access latency, DRAM manufacturers
design specialized \mbox{low-latency} DRAM chips (e.g.,
RLDRAM~\cite{micron-rldram3} and FCRAM~\cite{sato-vlsic1998}) at the cost of
higher price and lower density than the commonly-used DDRx DRAM (e.g.,
DDR3~\cite{jedec-ddr3}, DDR3L~\cite{jedec-ddr3l}, DDR4~\cite{jedec-ddr4},
LPDDR4~\cite{jedec-lpddr4}) chips. \figref{dram_cost} compares RLDRAM2/3
(low-latency) to DDR3L/4 DRAM (high-density) chips in terms of cost (i.e., price
per bit) and access latency. We obtain the pricing information (for buying a
bulk of 1000 DRAM chips) from a major electronic component
distributor~\cite{digikey}. Although the RLDRAMx chip attains 4x lower latency
than the DDRx DRAM chip, its cost for each bit is significantly higher, 39x. We
provide further discussion on how the RLDRAMx chip achieves low latency at a
high cost in \secref{special-dram}. One main reason for the high increase in the
price is the high area overhead incurred by the architectural designs in RLDRAMx
chips. In contrast to the density of a DDRx chip, which ranges from 2Gb to 8Gb,
an RLDRAMx chip typically has a low density, ranging from 256Mb to 1.125Gb.
Therefore, this dissertation focuses on understanding, characterizing, and
addressing the long latency problem of DRAM-based memory systems at \emph{low
  cost} (i.e., low DRAM chip area overhead) \emph{without} intrusive changes to
DRAM chips and/or memory controllers}.

\figputWS{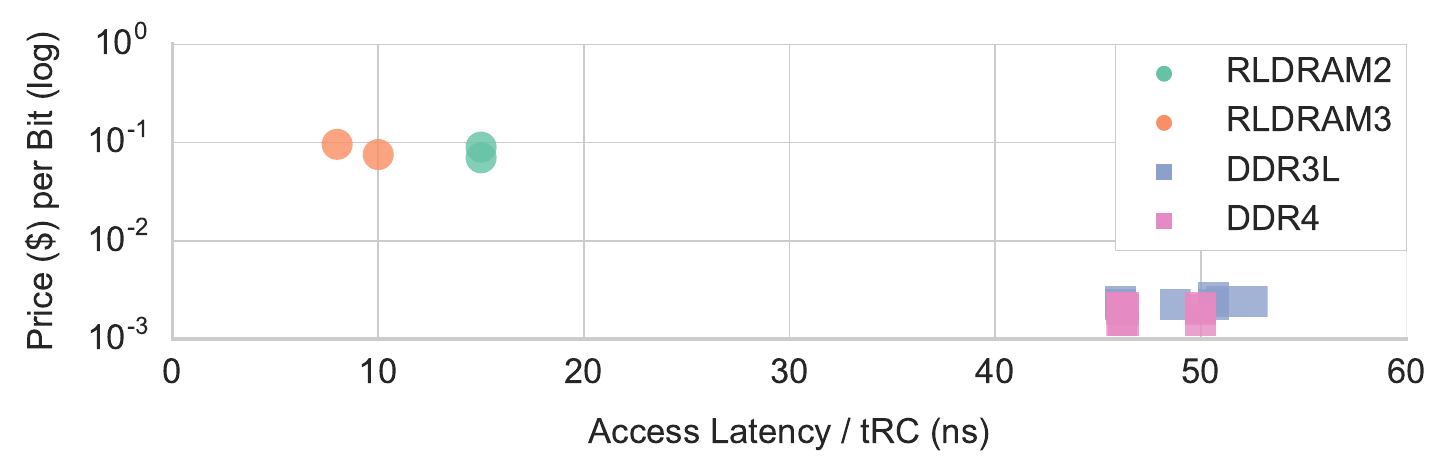}{1}{Cost and latency comparison between RLDRAMx and DDRx
  DRAM chips.}


We first identify three specific problems that cause, incur, or affect long
memory latency. First, bulk data movement, the movement of thousands or millions
of bytes between two \emph{memory locations}, is a common operation performed by
an increasing number of real-world applications (e.g.,~\cite{kanev-isca2015,
  lee-hpca2013, ousterhout-usenix1990, rosenblum-sosp1995, seshadri-cal2015,
  seshadri-micro2013, son-isca2013, sudan-asplos2010,zhao-iccd2005,
  seshadri-arxiv2016}). In current systems, since memory is designed as a simple
data repository that supplies data, performing a bulk data movement operation
between two locations in memory requires the data to go through the processor
\emph{even though both the source and destination are within the memory}. To
perform the movement, the data is first read out one cache line at a time from
the source location in memory into the processor caches, over a pin-limited
off-chip channel (typically 64-bit wide in current
systems~\cite{chang-sigmetrics2016}). Then, the data is written back to memory,
again one cache line at a time over the pin-limited channel, into the
destination location. By going through the processor, this data movement across
memory incurs a significant penalty in terms of both latency and energy
consumption (as well as consumed memory bandwidth).

Second, due to the increasing difficulty of efficiently manufacturing smaller
DRAM cells with smaller technology nodes, DRAM cells are becoming slower and
faultier than they were in the past~\cite{superfri, mutlu-imw2013, kim-iedm2005,
  khan-dsn2016, kang14, meza-dsn2015, kim-isca2014}. At smaller technology
nodes, DRAM cells are more susceptible to imperfect manufacturing process, which
causes the characteristics (e.g., latency) of the cells to deviate from the DRAM
design specification. As a result, \emph{latency variation} -- the phenomenon
that cells within the same DRAM chip or across different DRAM chips require
different access latencies -- becomes a problem in commodity DRAM chips. In
order to preserve chip production yield, DRAM manufacturers choose to tolerate
latency variation across cells within a chip or from different chips by
conservatively setting the standard DRAM latency to be determined by the
worst-case latency of any cell in any acceptable chip~\cite{lee-hpca2015,
  chang-sigmetrics2016}. This very high worst-case latency is applied uniformly
across \emph{all} DRAM cells in \emph{all} DRAM chips. As a result, even though
some fraction of a DRAM chip and some DRAM chips can inherently be accessed with
a latency that is shorter than the standard specification, the standard latency,
which is pessimistically set to a very conservative value, prevents systems from
attaining higher performance.

Third, since a DRAM cell stores data in a capacitor, which leaks charge over
time, DRAM needs to be periodically refreshed to prevent data loss due to
leakage. While DRAM is being refreshed, a part of it becomes \emph{unavailable}
to serve memory requests~\cite{liu-isca2012, chang-hpca2014}, which prolongs the
already-long memory latency by delaying the demand requests of processors and
accelerators.
This problem will become more prevalent as DRAM density
increases~\cite{liu-isca2012, chang-hpca2014}, leading to more DRAM cells to be
refreshed within the same refresh interval.

These three problems cause or exacerbate the long memory latency, which is
already a critical bottleneck in system performance. The trend of increasing
memory latency penalty is expected to continue to grow due to increasing core
and accelerator counts (and, hence, increasing memory interference) and the
emergence of increasingly more data-intensive and latency-critical applications.
Thus, low-latency memory accesses are now even more important than the past on
improving overall system performance and energy efficiency.

In addition, there is a critical trade-off between DRAM latency and supply
voltage, which greatly affects both the performance and energy efficiency of
DRAM chips. There is little experimental understanding of this trade-off and
hence almost no mechanisms taking advantage of it in existing systems, which
apply a fixed supply voltage value during the runtime.
If this voltage-latency trade-off is well understood, one can devise mechanisms
that can improve energy efficiency, latency, or both, by achieving a good
trade-off depending on system design goals.

\section{Thesis Statement and Overview}

The goal of this thesis is to enable low-latency DRAM memory systems, based on a
solid understanding of the causes of and trade-offs related to long DRAM
latency. Towards this end, we explore the causes of the three latency problems
that we described in the previous section, by \emph{(i)} examining the internal
DRAM chip architecture and memory controller designs, and \emph{(ii)}
experimentally characterizing commodity DRAM chips under various
conditions. With the understanding of the causes of long latency, our
thesis statement is that


\begin{tcolorbox}[notitle,boxrule=0pt,colback=white!20,colframe=white!20]
  \emph{\textbf{memory latency can be significantly reduced with a multitude of
    low-cost architectural techniques that aim to reduce different causes of
    long latency.}} \end{tcolorbox}


To this end, we \emph{(i)} propose a series of mechanisms that augment the DRAM
chip architecture with simple and low-cost features that better utilize the
existing DRAM designs, \emph{(ii)} develop a better understanding of latency
behavior and trade-offs by conducting extensive experiments on real commodity
DRAM chips, and \emph{(iii)} propose techniques to enhance memory controllers to
take advantage of the inherent, heterogeneous latency and voltage
characteristics of individual DRAM chips employed in the systems rather than
treating all the chips as having the same latency. We give a brief overview of
our mechanisms and experimental characterizations in the rest of this section.

%

\subsection{Low-Cost Inter-Linked Subarrays: Enabling Fast Data Movement}

To enable fast and efficient data movement across a wide range of memory at low
cost, we propose a new DRAM substrate, \capfullname (\lisa). To achieve this,
\lisa adds low-cost connections \emph{between} adjacent subarrays--the smallest
building block in today's DRAM chips. By using these connections to link the
existing internal wires (\emph{bitlines}) of adjacent subarrays, \lisa enables
wide-bandwidth data transfer across multiple subarrays with only 0.8\% DRAM area
overhead. As a DRAM substrate, \lisa is versatile, enabling an array of new
applications that reduce various latency components. We describe and evaluate
three such applications in detail: (1)~fast inter-subarray bulk data copy,
(2)~in-DRAM caching using a DRAM architecture whose rows have heterogeneous
access latencies, and (3)~accelerated bitline precharging (an operation that
prepares DRAM for subsequent accesses) by linking multiple precharge units
together. Our extensive evaluations show that combining LISA's three
applications attains
1.9x system performance improvement and 2x DRAM energy reduction on average
across a variety of workloads running on a quad-core system. To our knowledge,
LISA is the first DRAM substrate that supports fast inter-subarray data
movement, which enables a wide variety of performance enhancement mechanisms for
DRAM systems.

\subsection{Refresh Parallelization with Memory Accesses}

To mitigate the negative performance impact of DRAM refresh, we propose two
complementary mechanisms, \ib (\darplong) and \is (\intersub). The goal is to
address the drawbacks of per-bank refresh by building more efficient techniques
to parallelize refreshes and accesses within DRAM. Per-bank refresh is a
state-of-the-art DRAM refresh mechanism that refreshes only a single bank (a
bank is a collection of subarrays, and multiple banks are organized into a DRAM
chip) at a time. Although per-bank refresh enables a bank to be accessed while
another bank is being refreshed, it suffers from two shortcomings that limit the
ability of DRAM to serve demand requests while refresh operations are being
performed.

First, today's memory controllers issue per-bank refreshes in a strict
round-robin order, which can unnecessarily delay a bank's demand requests when
there are idle banks. To avoid refreshing a bank with pending demand requests,
\ib issues per-bank refreshes to idle banks in an out-of-order manner.
Furthermore, DRAM writes are not latency-critical because processors do not
stall to wait for them. Taking advantage of this observation, \ib proactively
schedules refreshes during intervals when a batch of writes are draining to
DRAM. Second, \is exploits the existence of mostly-independent {\em subarrays}
within a bank. With the cost of only 0.7\% DRAM area overhead, it allows a bank
to serve memory accesses to an idle subarray while another subarray is being
refreshed. Our extensive evaluations on a wide variety of workloads and systems
show that our mechanisms improve system performance by 3.3\%/7.2\%/15.2\% on
average (and up to 7.1\%/14.5\%/27.0\%) across 100 workloads over per-bank
refresh for 8/16/32Gb DRAM chips.
To our knowledge, these two techniques are the first mechanisms to \emph{(i)}
enhance refresh scheduling policy of per-bank refresh and \emph{(ii)} achieve
parallelization of refresh and memory accesses within a refreshing bank.

\subsection{Understanding and Exploiting Latency Variation Within a DRAM Chip}

To understand the characteristics of latency variation in modern DRAM chips, we
comprehensively characterize 240 DRAM chips from three major vendors and make
several new observations about latency variation within DRAM. We find that {\em
  (i)}~there is large latency variation across the DRAM cells, and {\em
  (ii)}~variation characteristics exhibit significant spatial locality: slower
cells are clustered in certain regions of a DRAM chip
Based on our observations, we propose \mechlong (\mech), a mechanism that
exploits latency variation across DRAM cells within a DRAM chip to improve
system performance. The key idea of \mech is to enable the memory controller to
exploit the spatial locality of slower cells within DRAM and access the faster
DRAM regions with reduced access latency. \mech requires modest modification in
the memory controller without introducing any changes to the DRAM chips. Our
evaluations show that FLY-DRAM improves the performance of a wide range of
applications by 13.3\%, 17.6\%, and 19.5\%, on average, for each of the three
different vendors' real DRAM chips, in a simulated 8-core system. To our
knowledge, this is the first work to \emph{(i)} provide a detailed experimental
characterization and analysis of latency variation  across different cells
within a DRAM chip, \emph{(ii)} show that access latency variation exhibits
spatial locality, and \emph{(iii)} propose mechanisms that take advantage of
variation within a DRAM chip to improve system performance.

\subsection{Understanding and Exploiting Trade-off Between Latency and Voltage
Within a DRAM Chip}

To understand the critical relationship and trade-off between DRAM latency and
supply voltage, which greatly affects both DRAM performance, energy efficiency,
and reliability, we perform an experimental study on 124 real DDR3L
(low-voltage) DRAM chips manufactured recently by three major DRAM vendors. We
find that reducing the supply voltage below a certain point introduces bit
errors in the data, and we comprehensively characterize the behavior of these
errors. We discover that these errors can be avoided by increasing the access
latency. This key finding demonstrates that there exists a trade-off between
access latency and supply voltage, i.e., increasing supply voltage enables lower
access latency (or vice versa). Based on this trade-off, we propose a new
mechanism, Voltron, which aims to improve energy efficiency of DRAM. The key
idea of Voltron is to use a performance model to determine how much we can
reduce the supply voltage without introducing errors and without exceeding a
user-specified threshold for performance loss. Our evaluations show that Voltron
reduces the average system energy consumption by 7.3\%, with a small system
performance loss of 1.8\% on average, for a variety of memory-intensive
quad-core workloads.

\section{Contributions}

The overarching contribution of this dissertation is the three new mechanisms
that reduce DRAM access latency and experimental characterizations for
understanding latency behavior in DRAM chips. More specifically, this
dissertation makes the following main contributions.

\begin{enumerate}

  \item We propose a new DRAM substrate, \emph{\capfullname} (\emph{\lisa}),
    which provides high-bandwidth connectivity between subarrays within the
    same bank to support bulk data movement at low latency, energy, and cost.
    Using the LISA substrate, we propose and evaluate three new applications:
    (1)~Rapid Inter-Subarray Copy (\emph{\lisarcnol}), which copies data across
    subarrays at low latency and low DRAM energy; (2)~Variable Latency
    (\emph{\lisavillanol}) DRAM, which reduces the access latency of
    frequently-accessed data
    by caching it in fast subarrays; and (3)~Linked Precharge
    (\emph{\lisaprenol}), which reduces the precharge latency for a subarray by
    linking its precharge units with neighboring idle precharge units.
    Chapter~\ref{chap:lisa} describes LISA and its applications in detail.

  \item We propose two new refresh mechanisms: (1) \darp (\darplong), a new
    per-bank refresh scheduling policy, which proactively schedules refreshes to
    banks that are idle or that are draining writes and (2) \is (\intersub), a new
    refresh architecture, that enables a bank to serve memory requests in idle
    subarrays while other subarrays are being refreshed. Chapter~\ref{chap:parref}
    describes these two refresh techniques in detail.

  \item We experimentally demonstrate and characterize the significant variation
    in DRAM access latency
    across different cells within a DRAM chip. Our experimental
    characterization on modern DRAM chips yields six new fundamental
    observations about latency variation. Based on this experimentally-driven
    characterization and understanding, we propose a new mechanism, \mech, which
    exploits the lower latencies of DRAM regions with faster cells by
    introducing heterogeneous timing parameters into the memory controller.
    Chapter~\ref{chap:latvar} describes our experiments, analysis, and
    optimization in detail.

  \item We perform a detailed experimental characterization of the effect of
    varying supply voltage on DRAM latency, reliability, and data
    retention on real DRAM chips. Our comprehensive experimental
    characterization provides four major observations on how DRAM latency and
    reliability is affected by supply voltage. These observations allow us to
    develop a deep understanding of the critical relationship and trade-off
    between DRAM latency and supply voltage. Based on this trade-off, we propose
    a new low-cost DRAM energy optimization mechanism called Voltron, which
    improves system energy efficiency by dynamically adjusting the voltage based
    on a performance model.
    Chapter~\ref{chap:voltron} describes our experiments, analysis, and
    optimization in detail.

\end{enumerate}

\section{Outline}

This thesis is organized into \ref{chap:conclusions} chapters. Chapter 2
describes necessary background on DRAM organization, operations, and latency.
Chapter 3 discusses related prior work on providing low-latency DRAM systems.
\chapref{lisa} presents the design \lisa and the three new
architectural mechanisms enabled by it. \chapref{parref} presents the two new
refresh mechanisms (\darp or \is) that address the refresh interference problem.
\chapref{latvar} presents our experimental study on DRAM latency variation and
our mechanism (FLY-DRAM) that exploits it to reduce latency. \chapref{voltron}
presents our experimental study on the trade-off between latency and voltage in
DRAM and our mechanism (Voltron) that exploits it to improve energy efficiency.
Finally,~\chapref{conclusions} presents conclusions and future research
directions that are enabled by this dissertation.

\chapter{Background}
\label{chap:background}

In this chapter, we provide necessary background on DRAM organization and
operations used to access data in DRAM. Each operation requires a certain
latency, which contributes to the overall DRAM access latency. Understanding of
these fundamental operations and their associated latencies provides the core
basics required for understanding later chapters in this dissertation.

\section{High-Level DRAM System Organization}
\label{sec:dram_sys_back}

\begin{figure}[h]
    \centering
		\captionsetup[subfigure]{justification=centering}
    \subcaptionbox{DRAM System\label{dram-organization}}[0.48\linewidth] {
        \includegraphics[scale=1.3]{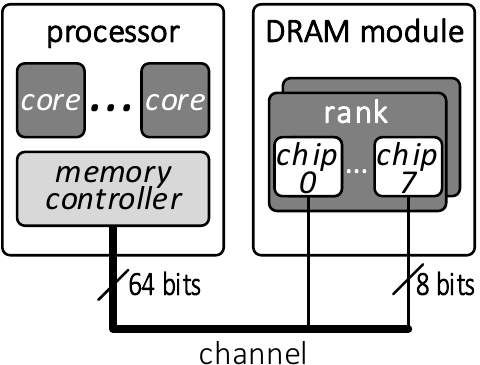}
		}
    \subcaptionbox{DRAM Bank\label{bank-organization}}[0.48\linewidth] {
        \includegraphics[scale=1.3]{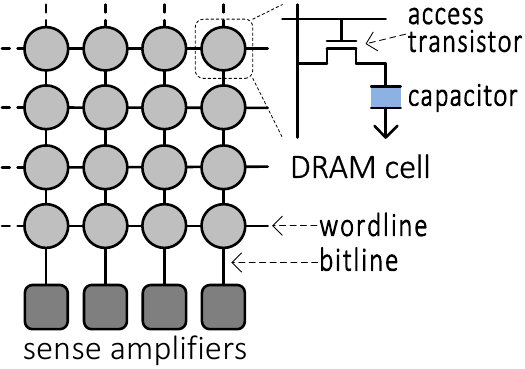}
		}%
        \caption{DRAM \new{system organization}.}
    \label{memory-organization}
\end{figure}

A modern DRAM system consists of a hierarchy of channels, modules, ranks, and
chips, as shown in Figure~\ref{dram-organization}. Each \emph{memory channel}
drives DRAM commands, addresses, and data between a memory controller in the
processor and one or more DRAM modules. Each \emph{module} contains multiple
DRAM chips that are organized into one or more ranks. A \emph{rank} refers to a
group of chips that operate in lock step to provide a wide data bus (usually
64 bits), as a single DRAM chip is designed to have a narrow data bus width
(usually 8 bits) to minimize chip cost. Each of the eight chips in the rank
shown in Figure~\ref{dram-organization} transfers 8 bits simultaneously to
supply 64 bits of data.

\section{Internal DRAM Logical Organization}

Within a DRAM chip, there are multiple banks (e.g., eight in a typical DRAM
chip~\cite{jedec-ddr3,kim-isca2012}) that can process DRAM commands
independently from each other to increase parallelism. A \emph{bank} consists of
a 2D-array of DRAM cells that are organized into rows and columns, as shown in
Figure~\ref{bank-organization}\footnote{Note that the figure shows a logical
  representation of the bank to ease the understanding of the DRAM operations
  required to access data and their associated latency. After we explain the
  DRAM operations in the next section, we will show the detailed physical
  organization of a bank in \secref{background}.}. A row typically consists of
8K cells. The number of rows varies depending on the chip density. Each DRAM
cell has {\em (i)} a \emph{capacitor} that stores binary data in the form of
electrical charge (e.g., fully charged and discharged states represent
\texttt{1} and \texttt{0}, respectively), and {\em (ii)} an \emph{access
  transistor} that serves as a switch to connect the capacitor to the
\emph{bitline}. Each column of cells share a bitline, which connects them to a
\emph{sense amplifier}.  The sense amplifier senses the charge stored in a cell,
converts the charge to digital binary data, and buffers it. Each row of cells
share a wire called the \emph{wordline}, which controls the cells' access
transistors. When a row's wordline is enabled, the entire row of cells gets
connected to the row of sense amplifiers through the bitlines, enabling the
sense amplifiers to sense and latch that row's data. The row of sense amplifiers
is also called the \emph{row buffer}.

\section{Accessing DRAM}
\label{sec:dram_access}

\begin{figure}[h]
    \centering
    \captionsetup[figure]{justification=centering}
    \includegraphics[scale=1.0]{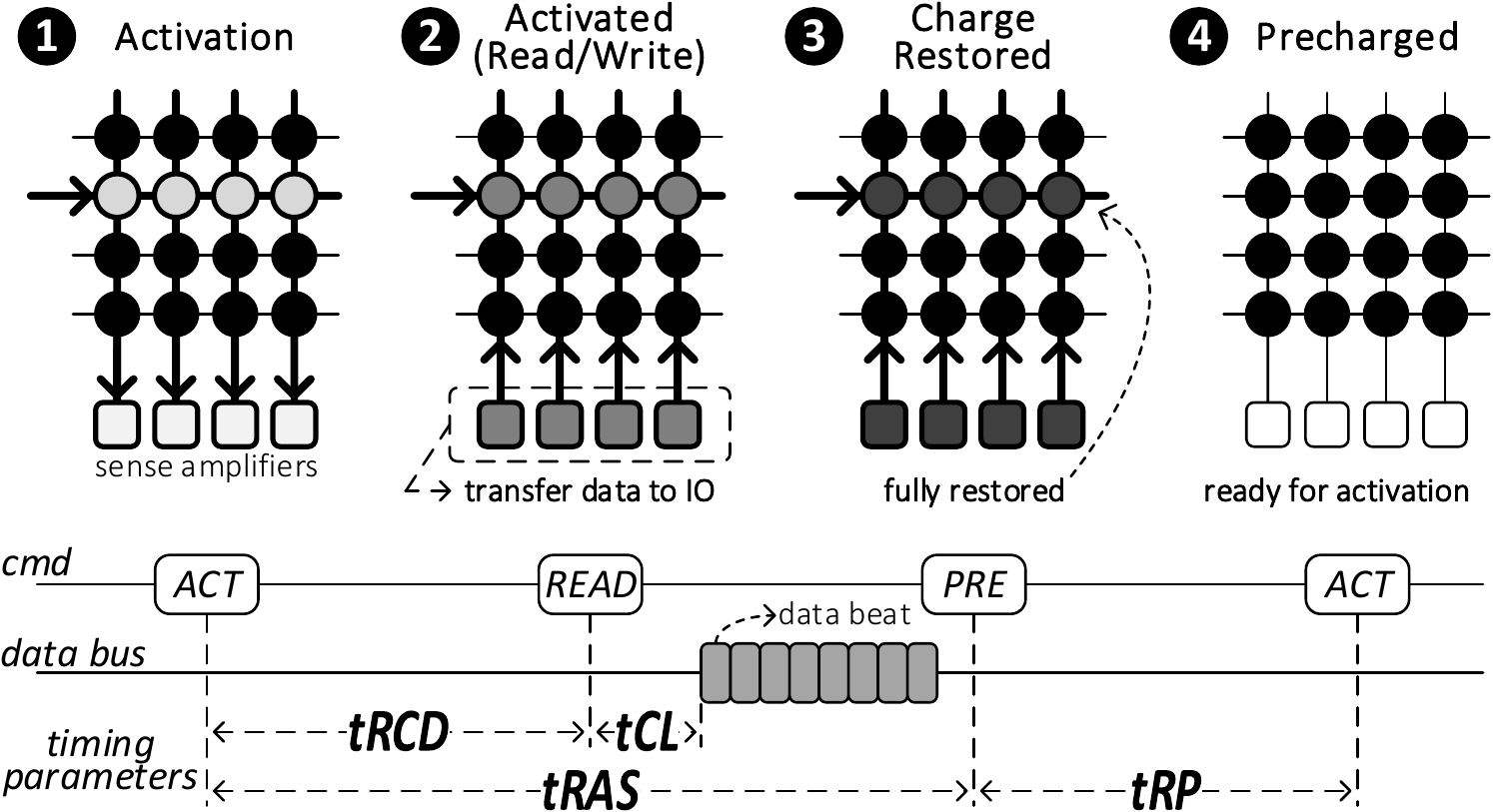}
    \caption{Internal DRAM phases, DRAM command/data timelines, and timing
        parameters to read a cache line.}
    \label{fig:timing_standard}
\end{figure}

Accessing (i.e., reading from or writing to) a bank consists of three steps:
{\em (i)} {\bf Row Activation \& Sense Amplification}: opening a row to transfer
its data to the row buffer, {\em (ii)} {\bf Read/Write}: accessing the target
\new{column} in the row buffer, and {\em (iii)} {\bf Precharge}: closing the row
and the row buffer. We use Figure~\ref{fig:timing_standard} to explain these
three steps in detail. The top part of the figure shows the phase of the cells
within the row that is being accessed. The bottom part shows both the DRAM
command and data bus timelines, and demonstrates the associated timing
parameters.


\noindent{\bf Initial State.} Initially, the bank is in the
\emph{precharged} state (\darkcircled{4} in Figure~\ref{fig:timing_standard}), where
all of the components are ready for activation. All cells are fully charged,
represented with the black color (a darker cell color indicates more
charge). Second, the bitlines are charged to \halfvdd, represented as a thin
line (a thin bitline indicates the initial voltage state of \halfvdd; a thick
bitline means the \new{bitline} is being driven). Third, the wordline is disabled
with \zerov (a thin wordline indicates \zerov; a thick wordline indicates \vdd).
Fourth, the sense amplifier is {\em off} without any data latched in it
(indicated by light color in the sense amplifier).\\\vspace{-2mm}

\noindent{\bf Row Activation \& Sense Amplification Phases.} To open a row, the
memory controller sends an \act command to raise the wordline of the
corresponding row, which connects the row to the bitlines (\darkcircled{1}).
This triggers an \emph{activation}, where charge starts to flow from the cell to
the bitline (or the other way around, depending on the initial charge level in
the cell) via a process called \emph{charge sharing}. This process perturbs the
voltage level on the corresponding bitline by a small amount. If the cell is
initially charged (which we assume for the rest of this explanation, without
loss of generality), the bitline voltage is perturbed upwards.
Note that this causes the cell itself to discharge, losing its data temporarily
(hence \new{the lighter color of the accessed row}), but this charge will be
restored as we will describe below. After the activation phase, the sense
amplifier \emph{senses} the voltage perturbation on the bitline, and turns
\emph{on} to further \emph{amplify} the voltage level on the bitline by
injecting more charge into the bitline and the cell (\new{making the activated
    row's cells darker in \darkcircled{2}}). When the bitline is amplified to a certain
    voltage level (e.g., $0.8V_{DD}$), the sense amplifier latches in the cell's
    data, which transforms it into binary data (\darkcircled{2}). At this point in
    time, the data can be read from the sense amplifier. The latency of these
    two phases (activation and sense amplification) is called the
    \emph{activation latency}, and is defined as \textbf{\small tRCD} in the
    standard DDR interface~\cite{jedec-ddr3,jedec-ddr4}. This activation latency
    specifies the latency from the time an \act command is issued to the time
    the data is ready to be accessed in the sense amplifier.\\ \vspace{-2mm}

\noindent{\bf Read/Write \& Restoration Phases.} Once the sense amplifier (row
buffer) latches in the data, the memory controller can send a \crd or \cwr
command to access the corresponding column of data within the row buffer (called
a \emph{column access}). The column access time to read the cache line data is
called \textbf{\small tCL} (\textbf{\small tCWL} for writes). These parameters
define the time between the column command and the appearance of the \emph{first
beat of data} on the data bus, shown at the bottom of
Figure~\ref{fig:timing_standard}. A \emph{data beat} is a 64-bit data transfer
from the DRAM to the processor.
In a typical DRAM~\cite{jedec-ddr3}, a column \crd command reads out 8 data
beats (also called an 8-beat burst), thus reading a complete 64-byte cache line.

After the bank becomes activated and the sense amplifier latches in the binary
data of a cell, it starts to \emph{restore} the connected cell's charge back to
its original fully-charged state (\darkcircled{3}). This phase is known as
\emph{restoration}, and can happen in parallel with column accesses. The
restoration latency (from issuing an \act command to fully restoring a row of
cells) is defined as \textbf{\small tRAS} in the standard DDR
interface~\cite{jedec-ddr3,jedec-ddr4,kim-isca2012,lee-hpca2013,lee-hpca2015},
as shown in Figure~\ref{fig:timing_standard}.\\ \vspace{-2mm}

\noindent{\bf Precharge Phase.} In order to access data from a different row,
the bank needs to be re-initialized back to the precharged state
(\darkcircled{4}). To achieve this, the memory controller sends a \cpre command,
which {\em (i)} disables the wordline of the corresponding row, disconnecting
the row from the sense amplifiers, and {\em (ii)} resets the voltage level on
the bitline back to the initial state, \halfvdd, so that the sense amplifier can
sense the charge from the new row that is to be opened (i.e., acitviated). The
latency of a precharge operation is defined as \textbf{\small tRP} in the
standard DDR
interface~\cite{jedec-ddr3,jedec-ddr4,kim-isca2012,lee-hpca2013,lee-hpca2015},
which is the latency between a \cpre and a subsequent \act within the same bank.

\section{DRAM Refresh}
\label{sec:refresh_background}

Since the capacitor in a DRAM cell leaks charge over time, to retain data in all
cells, DRAM needs to be refreshed periodically~\cite{liu-isca2012,
  liu-isca2013}. There are two state-of-the-art methods for DRAM refresh: 1)
all-bank refresh (\refab) and 2) per-bank refresh (\refpb).

\subsection{All-Bank Refresh (\refab)} 
\begin{figure}[h]
  \centering
  \subcaptionbox{All-bank refresh (\refab) frequency and
    granularity.\label{fig:ab_timeline}}[\linewidth]{
    \includegraphics[scale=0.7]{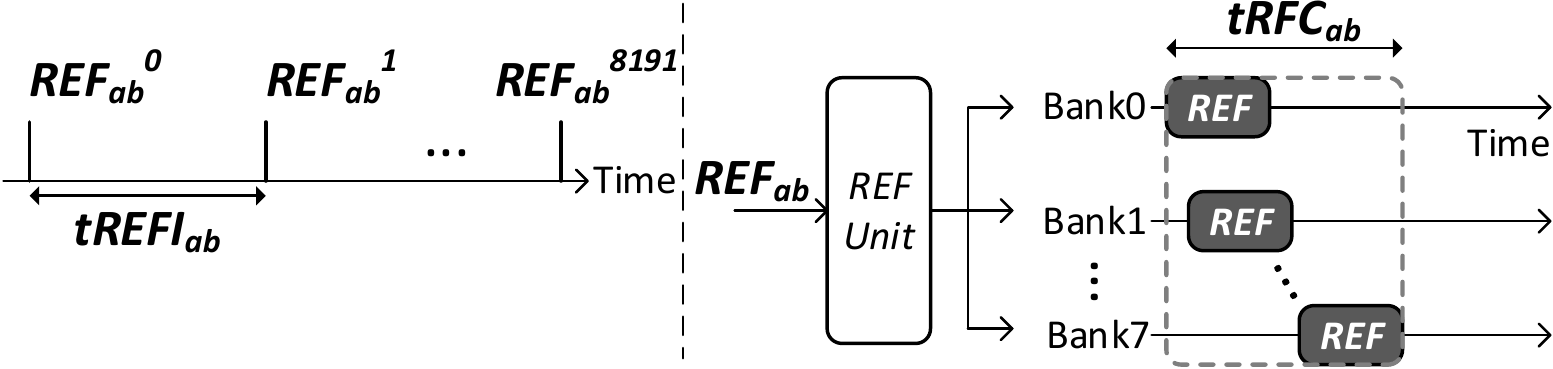}
  }
  \vspace{0.4in}

  \subcaptionbox{Per-bank refresh (\refpb) frequency and granularity.
    \label{fig:pb_timeline}}[\linewidth]{
    \includegraphics[scale=0.7]{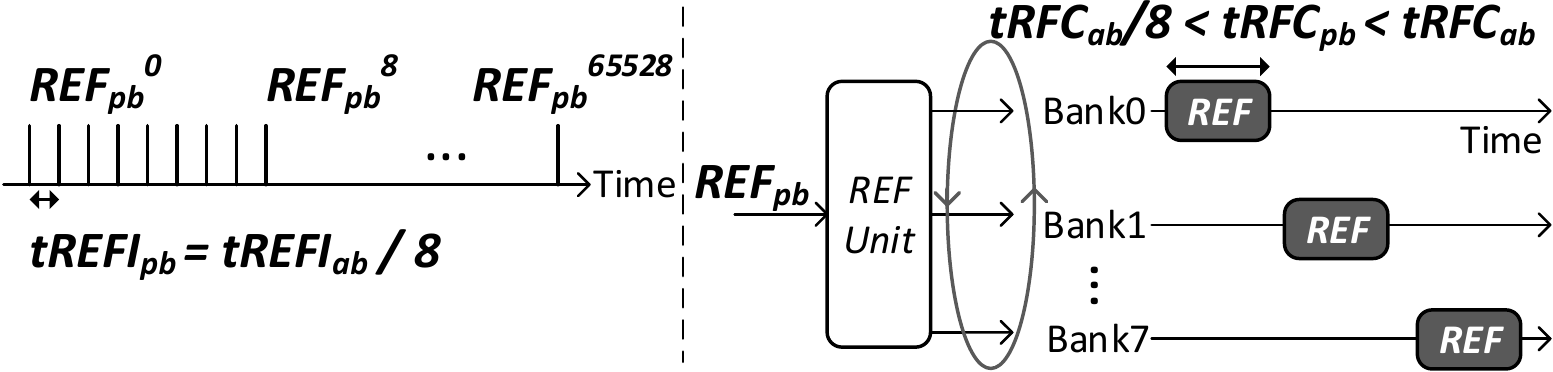}
  }
\caption{Refresh command service timelines.}
\label{fig:inter-bank-service-timeline}
\end{figure}

The minimum time interval during which
any cell can retain its electrical charge without being refreshed is called the
\emph{minimum retention time}, which depends on the operating temperature and
DRAM type. Because there are tens of thousands of rows in DRAM, refreshing all
of them in bulk incurs high latency. Instead, memory controllers send a number
of refresh commands that are evenly distributed throughout the retention time to
trigger refresh operations, as shown in \figref{ab_timeline}. Because a typical
refresh command in a commodity DDR DRAM chip operates at an entire rank level,
it is also called an \emph{all-bank refresh} or \emph{\refab} for
short~\cite{jedec-ddr3,jedec-lpddr3,micronLPDDR2_2Gb}. The timeline shows that the time
between two \refab commands is specified by \trefi (e.g., 7.8$\mu$s for 64ms
retention time). Therefore, refreshing a rank requires $\sfrac{64ms}{7.8\mu
s}\approx8192$ refreshes and each operation refreshes exactly $\sfrac{1}{8192}$
of the rank's rows.

When a rank receives a refresh command, it sends the command to a DRAM-internal
refresh unit that selects which specific rows or banks to refresh. A \refab
command triggers the refresh unit to refresh a number of rows in every bank for
a period of time called \trfc (\figref{ab_timeline}). During \trfc, banks are
not refreshed simultaneously. Instead, refresh operations are staggered
(pipelined) across banks~\cite{mukundan-isca2013,chang-hpca2014}. The main
reason is that refreshing every bank simultaneously would draw more current than
what the power delivery network can sustain, leading to potentially incorrect
DRAM operation~\cite{mukundan-isca2013, shevgoor-micro2013}. Because a \refab
command triggers refreshes on all the banks within a rank, the entire rank
cannot process any memory requests during \trfc, The length of \trfc is a
function of the number of rows to be refreshed.

\subsection{Per-Bank Refresh (\refpb)} To allow partial access to DRAM during
refresh, LPDDR DRAM (which is designed for mobile platforms), supports an
additional, finer-granularity refresh scheme, called \emph{per-bank refresh}
(\refpb for short)~\cite{jedec-lpddr3,micronLPDDR2_2Gb,chang-hpca2014}. This
refresh scheme splits up a \refab operation into eight separate operations
scattered across eight banks (\figref{pb_timeline}).  Therefore, a \refpb
command is issued eight times more frequently than a \refab command (i.e.,
\trefipb = \trefi / 8).

Similar to issuing a \refab, a controller simply sends a \refpb command to DRAM
every \trefipb without specifying which particular bank to refresh. Instead,
when a rank's internal refresh unit receives a \refpb command, it refreshes only
one bank for each command following a \emph{sequential round-robin order} as
shown in \figref{pb_timeline}. The refresh unit inside the DRAM chip uses an
internal counter to keep track of which bank to refresh next. The round-robin
order is known to the memory controller, so the memory controller \emph{knows}
which bank is being refreshed at any point in time.

By scattering refresh operations from \refab into multiple and
non-overlapping per-bank refresh operations, the refresh latency of
\refpb (\trfcpb) becomes shorter than \trfc. Disallowing \refpb
operations from overlapping with each other is a design decision made
by the LPDDR3 DRAM standard committee~\cite{jedec-lpddr3}. The reason
is simplicity: to avoid the need to introduce new timing constraints,
such as the timing between two overlapped refresh
operations.\footnote{At slightly increased complexity, one can
  potentially propose a modified standard that allows overlapped
  refresh of a subset of banks within a rank.}

\begin{figure}[h]
\begin{center}
\includegraphics[scale=0.7]{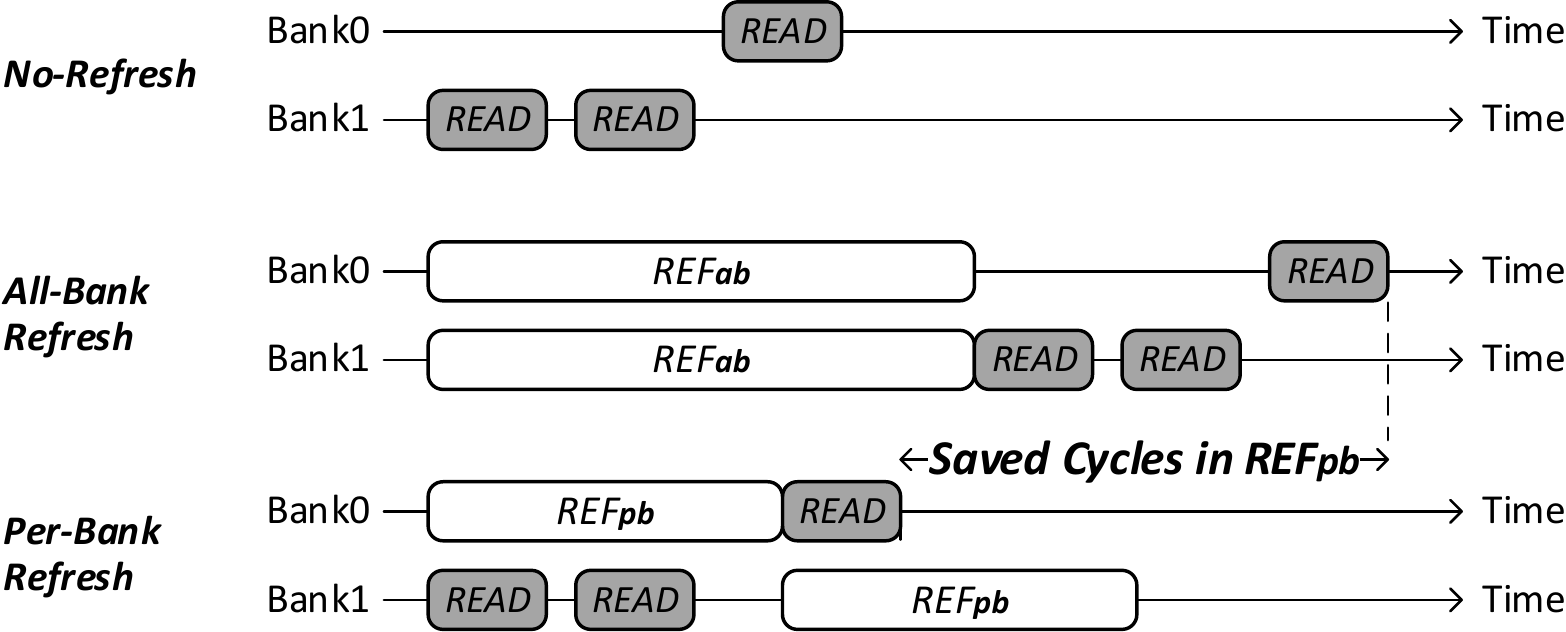}
\end{center}
\caption{Service timelines of all-bank and per-bank refresh.}
\label{fig:per-bank-refresh-timeline}
\end{figure}

With the support of \refpb, LPDDR DRAM can serve memory requests to
non-refreshing banks in parallel with a refresh operation in a single bank.
\figref{per-bank-refresh-timeline} shows pictorially how \refpb provides
performance benefits over \refab by enabling the parallelization of refreshes
and reads.  \refpb reduces refresh interference on reads by issuing reads to
Bank 1 while Bank 0 is being refreshed. Subsequently, it refreshes Bank 1 while
allowing Bank 0 to serve a read at the same time. As a result, \refpb alleviates
part of the performance loss due to refreshes by enabling parallelization of
refreshes and accesses across banks.

\section{Physical Organization of a DRAM Bank: DRAM Subarrays and Open-Bitline Architecture}
\label{sec:background:dram}
\label{sec:background}
\label{sec:openbitline}


In this section, we delve deeper into the physical organization of a bank. This
knowledge is required for understanding our proposals described in
\chapref{lisa} and \chapref{parref}. However, such knowledge is not required for
our other two proposals in \chapref{latvar} and \chapref{voltron}.

Typically, a bank is subdivided into multiple
\emph{subarrays}~\cite{chang-hpca2014, kim-isca2012, seshadri-micro2013,
  vogelsang-micro2010}, as shown in \figref{background_bank_sa_detail}.
Each subarray consists of a 2D-array of DRAM cells that are connected to sense
amplifiers through \emph{bitlines}. Because the size of a sense amplifier is
more than 100x the size of a cell~\cite{lee-hpca2013}, modern DRAM designs fit
in only enough sense amplifiers in a row to sense \emph{half a row of cells}. To
sense the \emph{entire} row of cells, each subarray has bitlines that connect to
\emph{two rows} of sense amplifiers --- one above and one below the cell array
(\circled{1} and \circled{2} in \figref{background_bank_sa_detail}, for Subarray
1). This DRAM design is known as the \emph{open bitline architecture}, and is
commonly used to achieve high density in modern DRAM chips~\cite{lim-isscc2012,
  takahashi-jsscc2001}. A single row of sense amplifiers, which holds the data
from \emph{half} a row of activated cells, is also referred as a row buffer.

\figputHS{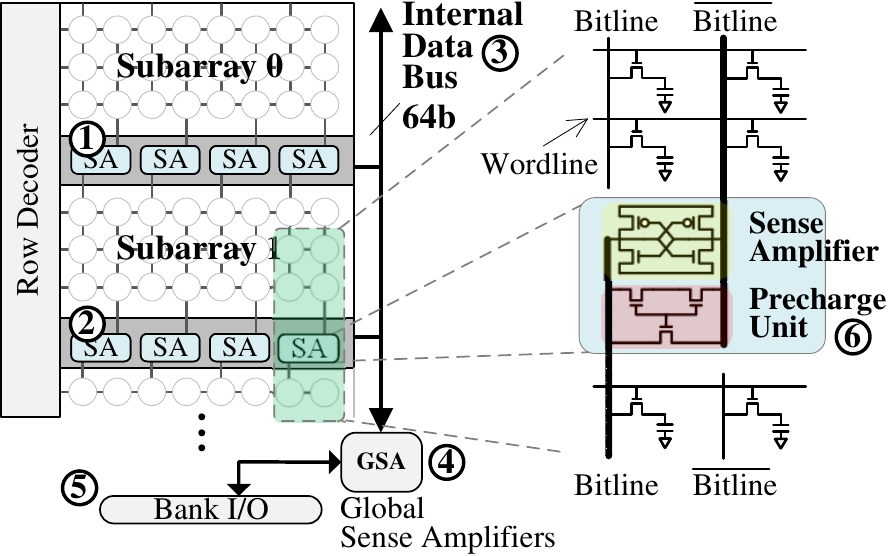}{1.1}{Bank and subarray organization in a
DRAM chip.}

\subsection{DRAM Subarray Operation}
\label{sec:background:access}

In \secref{dram_access}, we describe the details of major DRAM operations to
access data in a bank. In this section, we describe the same set of operations
to understand how they work at the \emph{subarray-level} within a bank. Accessing data
in a subarray requires two steps. The DRAM row (typically 8KB across a rank of
eight x8 chips) must first be \emph{activated}. Only after activation completes,
a column command (i.e., a \crd/\cwr) can operate on a piece of data (typically
64B across a rank; the size of a single cache line) from that row.

When an \act command with a row address is issued, the data stored within a row
in a subarray is read by \emph{two} row buffers (i.e., the row buffer at the
top of the subarray \circled{1} and the one at the bottom \circled{2}). First,
a wordline corresponding to the row address is selected by the subarray's row
decoder. Then, the top row buffer and the bottom row buffer
each sense the charge stored in half of the row's cells through the bitlines,
and amplify the charge to full digital logic values (0 or 1) to latch in the
cells' data.

After an \act finishes latching a row of cells into the row buffers, a \crd or a
\cwr can be issued. Because a typical read/write memory request is made at the
granularity of a single cache line, only a subset of bits are selected from a
subarray's row buffer by the column decoder.
On a \crd, the selected column bits are sent to the global sense amplifiers
through the \emph{internal data bus} (also known as the global data lines)
\circled{3}, which has a narrow width of 64B across a rank of eight chips (64
bits within a chip). The global sense amplifiers \circled{4} then drive the data
to the bank I/O logic \circled{5}, which sends the data out of the DRAM chip to
the memory controller.

While the row is activated, a consecutive column command to the same row can
access the data from the row buffer without performing an additional \act. This
is called a \emph{row buffer hit}. In order to access a different row in the
same bank, a \cpre command is required to reinitialize the bitlines' values for
another \act. This re-initialization process is completed by a set of
\emph{precharge units} \circled{6} in the row buffer.



\chapter{Related Work}

Many prior works propose mechanisms to reduce or mitigate DRAM latency. In
this chapter, we describe the closely relevant works by dividing them into
different categories based on their high-level approach.

\section{Specialized Low-Latency DRAM Architecture}
\label{sec:special-dram}

RLDRAM~\cite{micron-rldram3} and FCRAM~\cite{sato-vlsic1998} enable lower DRAM
timing parameters by reducing the length of bitlines (i.e., with a fewer number
of cells attached to each bitline). Because the bitline parasitic capacitance
reduces with bitline length, shorter bitlines enable faster charge sharing
between the cells and the sense amplifiers, thus reducing the latency of DRAM
operations~\cite{lee-thesis2016, lee-hpca2013}. The main drawback of this simple approach is that it leads to lower
chip density due to a significant amount of area overhead (30-40\% for FCRAM,
40-80\% for RLDRAM) induced by the additional peripheral logic (e.g., row
decoders) required to support shorter bitlines~\cite{kim-isca2012,lee-hpca2013}.
In contrast, our proposals do not require as significant and intrusive changes
to a DRAM chip.

\section{Cached DRAM}

Several prior works (e.g., \cite{ hart-compcon1994, hidaka-ieeemicro90,
hsu-isca1993,kedem-1997}) propose to add a small SRAM cache to a DRAM chip to
lower the access latency for data that is kept in the SRAM cache (e.g.,
frequently or recently used data). There are two main disadvantages of these
works. First, adding an SRAM cache into a DRAM chip is very intrusive: it incurs
a high area overhead (38.8\% for 64KB in a 2Gb DRAM chip) and significant design
complexity~\cite{lee-hpca2013,kim-isca2012}. Second, transferring data from DRAM
to SRAM uses a narrow global data bus, internal to the DRAM chip, which is
typically 64-bit wide. Thus, installing data into the DRAM cache incurs high
latency, especially if the SRAM cache stores data at the row granularity.
Compared to these works, our proposals in this dissertation reduce DRAM latency
without significant area overhead or complexity.

\section{Heterogeneous-Latency DRAM}

Prior works propose DRAM architectures that provide heterogeneous latency
either \emph{spatially} (dependent on \emph{where} in the memory an access
targets) or \emph{temporally} (dependent on \emph{when} an access occurs).

\subsection{Spatial Heterogeneity}

Prior work introduces spatial heterogeneity into DRAM, where one region has a
fast access latency but fewer DRAM rows, while the other has a slower access
latency but many more rows~\cite{lee-hpca2013, son-isca2013}. The fast region is
mainly utilized as a caching area, for the frequently or recently accessed data.
We briefly describe two state-of-the-art works that offer different
heterogeneous-latency DRAM designs.

CHARM~\cite{son-isca2013} introduces heterogeneity \emph{within a rank} by
designing a few fast banks with (1)~shorter bitlines for faster data sensing,
and (2)~closer placement to the chip I/O for faster data transfers. To exploit
these low-latency banks, CHARM uses an OS-managed mechanism to \emph{statically}
map hot data to these banks, based on profiled information from the compiler or
programmers. Unfortunately, this approach \emph{cannot adapt} to program phase
changes, limiting its performance gains. If it were to adopt dynamic hot data
management, CHARM would incur high migration costs over the narrow 64-bit bus
that internally connects the fast and slow banks.

Tiered-Latency DRAM (TL-DRAM)~\cite{lee-hpca2013} provides heterogeneity \emph{within a subarray} by
dividing the subarray into fast (near) and slow (far) segments that have short and long
bitlines, respectively, using isolation transistors. The fast segment can be
managed as a software-transparent hardware cache. The main disadvantage is that it
needs to cache each hot row in \emph{two near segments} as each subarray uses
two row buffers on \emph{opposite ends} to sense data in the open-bitline
architecture (as we discussed in Section~\ref{sec:openbitline}). This prevents
TL-DRAM from using the full near segment capacity. As we can see, neither CHARM
nor TL-DRAM strike a good design balance for heterogeneous-latency DRAM. In this
dissertation, we propose a new heterogeneous DRAM design that offers fast data
movement with a low-cost and easy-to-implement design.

Several prior works~\cite{luo-dsn2014,chatterjee-micro2012,phadke-date2011}
propose to employ different types of DRAM modules to provide heterogeneous
latency at the memory module level. These works are orthogonal to the proposals
in this dissertation because we focus on reducing latency at the chip level.




\subsection{Temporal Heterogeneity}

Prior work observes that DRAM latency can vary depending on \emph{when} an
access occurs. The key observation is that a \emph{recently-accessed or
a recently-refreshed} row has nearly full electrical charge in the cells, and thus the
following access to the same row can be performed faster~\cite{hassan-hpca2016,
  hassan-hpca2017, shin-hpca2014}. We briefly describe two state-of-the-art
works that focus on providing heterogeneous latency temporally.

ChargeCache~\cite{hassan-hpca2016} enables faster access to
\emph{recently-accessed} rows in DRAM by tracking the addresses of
recently-accessed rows in the memory controller. NUAT~\cite{shin-hpca2014}
enables accesses to recently-refreshed rows at low latency because these rows
are already highly-charged. The main issue with these works is that the proposed
effect of highly-charged cells can be accessed with lower latency, is slightly
observable only when very long refresh intervals are used on existing DRAM
chips, as demonstrated by a recent DRAM characterization work~\cite{hassan-hpca2017}.
However, within the duration of the standard 64ms refresh interval, no latency
benefits can be directly observed on existing DRAM chips. As a result, these
ideas likely require changes to the DRAM chips to provide benefits as suggested
by a prior work~\cite{hassan-hpca2017}. In contrast, our work in this
dissertation does not require data to be recently-accessed or recently-refreshed to
benefit from reduced latency, but it focuses on providing low latency by
exploiting spatial heterogeneity. Hence, our techniques are independent of
access or refresh patterns.

\section{Bulk Data Transfer Mechanisms}

Prior works~\cite{gschwind-cf2006, gummaraju-pact2007,
  kahle-ibmjrd2005,carter-hpca1999,zhang-ieee2001} propose to add
scratchpad memories to reduce CPU pressure during bulk data transfers,
which can also enable sophisticated data movement (e.g., scatter-gather), but
they still require data to first be moved on-chip. A patent~\cite{seo-patent}
proposes a DRAM design that can copy a page across memory blocks, but lacks concrete
analysis and evaluation of the underlying copy operations. Intel I/O
Acceleration Technology~\cite{intelioat} allows for memory-to-memory DMA
transfers \emph{across a network}, but cannot transfer data \emph{within the
  main memory}.

Zhao et al.~\cite{zhao-iccd2005} propose to add a bulk data movement engine
inside the memory controller to speed up bulk-copy operations. Jiang et
al.~\cite{jiang-pact2009} design a different copy engine, placed within the
cache controller, to alleviate pipeline and cache stalls that occur when these
transfers occur. However, these works do not directly address the problem
of data movement \emph{across} the narrow memory channel as they still require
the data to move between the main memory and the processor.

Seshadri et al.~\cite{seshadri-micro2013} propose RowClone to perform data
movement within a DRAM chip, avoiding costly data transfers over the pin-limited
channels. However, its effectiveness is limited because RowClone enables very
fast data movement only when the source and destination rows are within the
\emph{same DRAM subarray}. The reason is that while two DRAM rows in the same
subarray are connected by row-wide bitlines (e.g., 8K bits), rows in different
subarrays are connected through a narrow 64-bit data bus (albeit an internal
DRAM bus). Therefore, even for an in-DRAM data movement mechanism such as
RowClone, \emph{inter-subarray} bulk data movement incurs long latency even
though data does not move out of the DRAM chip. In contrast, one of our
proposals, LISA (\chapref{lisa}), enables fast and energy-efficient bulk data
movement \emph{across} subarrays. We provide more detailed qualitative and quantitative
comparisons between LISA and RowClone in \secref{lisaclone}.

Lu et al.~\cite{lu-micro2015} propose a heterogeneous DRAM design called
\mbox{DAS-DRAM} that consists of fast and slow subarrays. It introduces a row of
\emph{migration cells} into each subarray to move rows across different
subarrays. Unfortunately, the latency of DAS-DRAM is not scalable with movement
distance, because DAS-DRAM requires writing the migrating row into each
intermediate subarray's migration cells before the row reaches its destination,
which prolongs the data transfer latency. In contrast, LISA (\chapref{lisa})
provides a \emph{direct path} to transfer data \emph{between row buffers} of
different subarrays without requiring intermediate data writes into the
subarray.

\section{DRAM Refresh Latency Mitigation}

Prior works (e.g.,~\cite{liu-isca2012, venkatesan-hpca2006,bhati-isca2015,
  lin-iccd2012, agrawal-memsys2016, nair-isca2013,
  kim-asic2001,baek-tc2014,agrawal-hpca2014,ohsawa-islped1998,qureshi-dsn2015,
  patel-isca2017}) propose mechanisms to reduce unnecessary refresh operations
by taking advantage of the fact that different DRAM cells have widely different
retention times~\cite{liu-isca2013, kim-edl2009}. These works assume that the
retention time of DRAM cells can be \emph{accurately} profiled and they depend
on having this accurate profile to guarantee data integrity~\cite{liu-isca2013}.
However, as shown in Liu et al.~\cite{liu-isca2013} and later analyzed in detail
by several other works~\cite{khan-sigmetrics2014, khan-dsn2016, khan-cal2016,
  patel-isca2017}, accurately determining the retention time profile of DRAM is
an outstanding research problem due to the Variable Retention Time (VRT) and
Data Pattern Dependence (DPD) phenomena, which can cause the retention time of a
cell to fluctuate over time. As such, retention-aware refresh techniques need to
overcome the profiling challenges to be viable. A recent work,
AVATAR~\cite{qureshi-dsn2015}, proposes a retention-aware refresh mechanism that
addresses VRT by using ECC chips, which introduces extra cost. In contrast, our
refresh mitigation techniques (\chapref{parref}) enable parallelization of
refreshes and accesses \emph{without} relying on cell data retention profiles or
ECC, thus reducing the performance overhead of refresh at high reliability and
low cost.

Several other works propose different refresh mechanisms. Nair et
al.~\cite{nair-hpca2013} propose Refresh Pausing, which pauses a refresh
operation to serve pending memory requests when the refresh causes conflicts
with the requests. Although our work already significantly reduces conflicts
between refreshes and memory requests by enabling parallelization, it can be
combined with Refresh Pausing to address rare conflicts. Tavva et
al.~\cite{tavva-taco2014} propose EFGR, which exposes non-refreshing banks
during an all-bank refresh operation so that a few accesses can be scheduled to
those non-refreshing banks during the refresh operation. However, such a
mechanism does not provide additional performance and energy benefits over
state-of-the-art per-bank refresh, which we use to build our mechanism in this
dissertation. Isen and John~\cite{isen-micro2009} propose ESKIMO, which modifies
the ISA to enable memory allocation libraries to skip refreshes on memory
regions that do not affect programs' execution. ESKIMO is orthogonal to our
mechanism, and it requires high system-level complexity by requiring
system software libraries to make refresh decisions.

Another technique to address refresh latency is through refresh scheduling
(e.g.,~\cite{stuecheli-micro2010,mukundan-isca2013,agrawal-hpca2013,
  ishii-msc2012, bhati-ispled2013}). Stuecheli et al.~\cite{stuecheli-micro2010}
propose elastic refresh, which postpones refreshes by a time delay that varies
based on the number of postponed refreshes and the predicted rank idle time, to
avoid interfering with demand requests. Elastic refresh has two shortcomings.
First, it becomes less effective when the average rank idle period is shorter
than the refresh latency as the refresh latency cannot be fully hidden in that
period. This occurs especially with 1) more memory-intensive workloads that
inherently have less idleness and 2) higher density DRAM chips that have higher
refresh latencies. Second, elastic refresh incurs higher refresh latency when it
incorrectly predicts that a period is idle without pending memory requests in
the memory controller. In contrast, our mechanisms parallelize refresh
operations with accesses even if there is no idle period and they therefore
outperform elastic refresh. We quantitatively demonstrate the benefits of our
mechanisms over elastic refresh~\cite{stuecheli-micro2010} in
\secref{parref_evaluation}.

Mukundan et al.~\cite{mukundan-isca2013} propose scheduling techniques to
address the problem of command queue seizure, whereby a command queue gets
filled up with commands to a refreshing rank, blocking commands to another
non-refreshing rank. In our dissertation, we use a different memory controller
design that does not have command queues, similarly to another prior
work~\cite{herrero-tc2013,subramanian-iccd2014,subramanian-tpds2016,
  subramanian-thesis2015}. Our
controller generates a command for a scheduled request right before the request
is sent to DRAM instead of pre-generating the commands and queueing them up.
Thus, our baseline refresh design does not suffer from the problem of command
queue seizure.


\section{Exploiting DRAM Latency Variation}

Adaptive-Latency DRAM (AL-DRAM)~\cite{lee-hpca2015} also characterizes and
exploits DRAM latency variation, but does so at a much coarser granularity. This
work experimentally characterizes latency variation across different DRAM chips
under different operating temperatures. AL-DRAM sets a uniform operation latency
for the \emph{entire} DIMM and does not exploit heterogeneity at the chip-level
or within a chip. Chandrasekar et al.\ study the potential of reducing some DRAM
timing parameters~\cite{chandrasekar-date2014}. Similar to AL-DRAM, our
dissertation observes and characterizes latency variation \emph{across} DIMMs.
Different from prior works, this dissertation also characterizes latency
variation \emph{within a chip}, at the granularity of individual DRAM cells and
exploits the latency variation that exists within a DRAM chip. Our proposal can
be combined with AL-DRAM to improve performance further.

A recent work by Lee et al.~\cite{lee-arxiv2016, lee-sigmetrics2017} also
observes latency variation within DRAM chips. The work analyzes the variation
that is due to the circuit design of DRAM components, which it calls
\emph{design-induced variation}. Furthermore, it proposes a new profiling
technique to identify the lowest DRAM latency without introducing errors. In
this dissertation, we provide the \emph{first} detailed experimental
characterization and analysis of the general latency variation phenomenon within
real DRAM chips. Our analysis is broad and is not limited to design-induced
variation. Our proposal of exploiting latency variation, FLY-DRAM
(\chapref{latvar}), can employ Lee et al.'s new profiling
mechanism~\cite{lee-arxiv2016, lee-sigmetrics2017} to identify additional
latency variation regions for reducing access latency.

\section{In-Memory Computation}

Modern execution models rely on transferring data from the memory to the
processor to perform computation. Since a large number of modern applications
consume a large amount of data, this model incurs high latency, bandwidth, and
energy due to the excessive use of the narrow memory channel that is typically
as wide as only 64 bits. To avoid the memory channel bottleneck, many prior
works
(e.g.,~\cite{ahn-isca2015,ahn-isca2015-2,7056040,7429299,guo-wondp14,592312,
  seshadri-cal2015,mai-isca2000,draper-ics2002,
  seshadri-micro2015,seshadri-arxiv2016,hsieh-iccd2016,hsieh-isca2016,
  amirali-cal2016, stone-1970, fraguela-2003,375174,808425,
  4115697,694774,sura-2015,zhang-2014,akin-isca2015,
  babarinsa-2015,7446059,6844483,pattnaik-pact2016, seshadri-thesis2016})
propose different frameworks and mechanisms to enable processing-in-memory (PIM)
to accelerate parts of the applications. However, these works do \emph{not}
fundamentally reduce the \emph{raw} memory access latency within a DRAM chip.
Therefore, our dissertation is complementary to these mechanisms. Furthermore,
one of our proposals, LISA (\chapref{lisa}) is also complementary to a
previously proposed in-memory bulk processing mechanism that can perform bulk
bitwise AND, OR~\cite{seshadri-cal2015,seshadri-arxiv2016}. LISA can enhance the
speed and range of such operations as these operations require copying data
between rows.

\section{Mitigating Memory Latency via Memory Scheduling}

Since memory has limited bandwidth and parallelism to serve memory requests
concurrently, contention for memory bandwidth across different applications can
cause significant performance slowdown for individual applications as well as
the entire system. Many prior works propose to address bandwidth contention by
using more intelligent memory scheduling policies. A number of prior works focus
on improving DRAM throughput without being aware of the characteristics of the
running applications in the system
(e.g.,~\cite{rixner-isca2000,zuravleff-patent, shao-hpca2007,hur-micro2004,
  lee-micro2009}). Many other works observe that application-unaware memory
scheduling provides low performance, unfairness, and cases that lead to denial
of memory service~\cite{moscibroda-usenix2007}. As a result, these prior works
(e.g.,~\cite{lee-micro2009,kim-hpca2010,kim-micro2010,mutlu-isca2008,mutlu-micro2007,nesbit-micro2006,rafique-pact2007,
  ipek-isca2008,
  ausavarungnirun-isca2012,usui-taco2016,ebrahimi-micro2011,subramanian-iccd2014,subramanian-hpca2013,subramanian-micro2015,subramanian-thesis2015,
  moscibroda-usenix2007, lee-micro2008, muralidhara-micro2011,
  subramanian-tpds2016, moscibroda-podc2008, zhao-micro2014, das-hpca2013,
  jog-sigmetrics2016}) propose scheduling policies that take into account
individual applications' characteristics to perform better memory request
scheduling to improve overall system performance and fairness. While these works
reduce the queueing latency experienced by the applications and the system, they
do \emph{not} fundamentally reduce the DRAM access latency of memory requests.
The various proposals in this dissertation do, and thus they are complementary
to memory scheduling mechanisms.



\section{Improving Parallelism in DRAM to Hide Memory Latency}

A number of prior works propose new DRAM architectures to increase parallelism
within DRAM and thus overlap memory latency of different DRAM operations. Kim et
al.~\cite{kim-isca2012} propose subarray-level parallelism (SALP) to take
advantage of the existing subarray architecture to overlap multiple memory
requests going to different subarrays within the same bank. O et
al.~\cite{o-isca2014} propose to add isolation transistors in each subarray to
separate the bitlines from the sense amplifiers, so that the bitlines can be
precharged while the row buffer is still activated. Lee et
al.~\cite{lee-pact2015} propose to add a data channel dedicated for I/O to serve
accesses from both the CPU and the I/O subsystem in parallel. Several
works~\cite{zheng-micro2008,ware-iccd2006,ahn-cal2009,ahn-taco2012} propose to
divide a DRAM rank into multiple smaller ranks (i.e., sub-ranks) to serve memory
requests independently from each sub-rank at the cost of higher read or write
latency. All these prior works do \emph{not} fundamentally reduce the access
latency of DRAM operations. Their benefits decrease when more memory accesses
interfere with each other at a single subarray, bank, or rank. Our proposals
in this dissertation reduce the DRAM access latency directly. These prior works are
complementary to our proposals, and combined together with our techniques can
provide further system performance improvement.

%


%

\section{Other Prior Works on Mitigating High Memory Latency}

\subsection{Data Prefetching} Many prior works propose data prefetching
techniques to load data speculatively from memory into the cache (before the
data is accessed), to hide the memory latency with computation
(e.g.,~\cite{lee-micro2008,seshadri-taco2015,lee-micro2009,nesbit-pact2004,srinath-hpca2007,lai-isca2001,alameldeen-hpca2007,baer-1995,cao-sigmetrics1995,dahlgren-1995,1669154,jouppi-isca90,hur-micro2006,joseph-isca1997,cooksey-asplos2002,ebrahimi-hpca2009,ebrahimi-isca2011,mutlu-isca2005,
  mutlu-hpca2003, mutlu-micro2005,
  hashemi-isca2016,lee-tc2011,hashemi-micro2016}). However, prefetching does
\emph{not} reduce the fundamental DRAM latency required to fetch data, and
prefetch requests can cause interference with demand requests, thereby
introducing performance
overhead~\cite{ebrahimi-isca2011,srinath-hpca2007,1669154}. On the other hand,
our proposals can reduce the DRAM access latency for all types of memory
requests, without causing interference to other requests.

\subsection{Multithreading} To hide memory latency, prior
works~\cite{smith-mimd, thornton-1964, smith-spie1981,
  lindholm-2008,kongetira-ieeemicro2005,tullsen-isca1995} propose to use
multithreading to overlap the DRAM latency of one thread with computation by
another thread. While multithreading can tolerates the latency experienced by
the applications or threads, the technique does \emph{not} reduce the memory
access latency. In fact, multithreading can cause additional delays due to the
contention that arises between threads on shared resource accesses. For example,
on a GPU system that runs a large number of threads, memory latency can still be
a performance limiter when threads stalling on memory requests delay other
threads from being issued~\cite{owl-asplos13, caba, osp-isca13, largewarps,
  medic}. Exploiting the potential of multithreading provided by the hardware
also requires non-trivial effort from programmers to write bug-free
programs~\cite{lee-2006}. Furthermore, multithreading does not improve
single-thread performance, which is still important for many modern
applications, e.g., mobile applications~\cite{halper-hpca2016}. Critical threads
that are delayed on a memory access can be bottlenecks that degrade the
performance of an entire multi-threaded application by delaying other
threads~\cite{ebrahimi-micro2011,suleman-asplos2009,joao-asplos2012,joao-isca2013,suleman-isca2010,dubois-isca2013}.
Our proposals in this dissertation reduce the memory access latency directly. As
a result, these proposals not only improve single-thread performance but
also the performance of multithreading processors by reducing the amount of
memory stall time of critical threads that can stall other threads.




%

\subsection{Processor Architecture Design to Tolerate Memory Latency}

A single processor core can employ various techniques to tolerate memory latency
by generating multiple DRAM accesses that can potentially be served concurrently
by the DRAM system (e.g., out-of-order execution~\cite{tomasulo-ibmj67,patt-micro1985},
non-blocking caches~\cite{kroft-isca81}, and runahead
execution~\cite{mutlu-isca2005, mutlu-hpca2003,
  mutlu-micro2005,hashemi-micro2016,mutlu-ieeemicro2003,mutlu-ieeemicro2006}). The effectiveness of
these latency tolerance techniques highly depends on whether DRAM can serve the
generated memory accesses in parallel as these techniques do not directly reduce
the latency of individual accesses.

Other prior works
(e.g.~\cite{mutlu-ieeetc2006,lipasti-asplos1996,lipasti-micro1996,sazeides-micro1997,yazdanbakhsh-taco2016,thwaites-pact2014,mutlu-micro2005,wang-micro1997})
propose to use value prediction to avoid pipeline stalls due to memory by
predicting the requested data value. However, incorrect value prediction incurs
high cost due to pipeline flushes and re-executions. Although this cost can be
mitigated with approximate value
prediction~\cite{yazdanbakhsh-taco2016,thwaites-pact2014}, approximation is not
applicable to all applications as some require precise correctness for
execution.

Our proposals in this
dissertation directly reduce DRAM access latency even if the accesses cannot be
served in parallel. Our proposals are also complementary to these processor
architectural techniques as we introduce low-cost modifications to DRAM chips
and memory controllers.

\subsection{System Software to Mitigate Application Interference} Prior works
(e.g.,~\cite{kim-rts2016,liu2014going,kim-rtas2014,
  ebrahimi-asplos2010,liu-pact2012,lin-hpca2008,zhuravlev-asplos2010}) propose
system software techniques to manage inter-application interference in the
memory to reduce interference-induced memory latency. These works do not reduce
the access latency to memory. However, their techniques are complementary to our
proposals.

\subsection{Reducing Latency of On-Chip Interconnects}

Prior works
(e.g.~\cite{xiang-ics2017,chang-sbacpad2012,das-hpca2013,grot-micro2009,das-isca2010,sharifi-micro2012,das-micro2009,lee-pact2010,fallin-nocs2012,grot-isca2011,grot-hpca2009,moscibroda-isca2009,fallin-hpca2011})
propose mechanisms to reduce the latency of memory requests when they are
traversing the on-chip interconnects. These works are complementary to the
proposals presented in this dissertation since our works reduce the fundamental
memory device access latency.

\subsection{Reducing Latency of Non-Volatile Memory}

In this dissertation, we focus on the DRAM technology, which is the predominant
physical substrate for main memory in today's systems. On the other hand, a new
class of \emph{non-volatile memory (NVM)} technology is becoming a potential
substrate to replace DRAM or co-exist with DRAM in future
systems~\cite{meza-weed2013,ku-ispass2013,meza-cal2012,yoon-iccd2012,qureshi-isca2009,qureshi-micro2009,lee-isca2009,lee-ieeemicro2010,lee-cacm2010}.
Since NVM has substantially longer latency than DRAM, prior works
(e.g.,~\cite{nair-hpca2015, yoon-taco2014, meza-weed2013,
  ku-ispass2013,hoseinzadeh-taco2015,li2016maxpb,qureshi-isca2010,zhang-micro2009,jiang-hpca2012,lee-isca2009,lee-ieeemicro2010,lee-cacm2010})
propose various techniques to reduce the access latency of different types of
NVM (e.g., PCM and STT-RAM). However, these techniques are not directly
applicable to DRAM devices because each NVM technology has a fundamentally
different way of accessing its memory cells (i.e., devices) from DRAM.

\section{Experimental Studies of Memory Chips}

In this dissertation, we provide extensive detailed experimental
characterization and analysis of latency behavior in modern commodity DRAM
chips. There have been other experimental studies of DRAM
chips~\cite{hassan-hpca2017,khan-cal2016,chandrasekar-date2014,jung-memsys2016,jung-patmos2016,lee-hpca2015,kim-isca2014,liu-isca2013,
  khan-sigmetrics2014,
  khan-dsn2016,patel-isca2017,lee-sigmetrics2017,lee-arxiv2016,lee-thesis,kim-thesis}
that study various issues including data retention, read disturbance, latency,
address mapping, and power. There have also been field studies of the
characteristics of DRAM memories employed in large-scale
systems~\cite{el-sayed-sigmetrics2012,hwang-asplos2012,schroeder-sigmetrics2009,
  meza-dsn2015, li-usenixatc2010, sridharan-sc2012, sridharan-asplos2015}. Both
of these types works are complementary to the works presented in this
dissertation.

Similarly, there have been experimental studies of other types of memories,
especially NAND flash memory~\cite{cai.date13, parnell.globecom14,
  luo.jsac16,cai.iccd13, cai.sigmetrics14,cai.iccd12,cai.date12,cai-itj2013,
  cai.hpca15,cai.dsn15,cai.hpca17,cai-ieee2017,aya-2017}. These studies develop a similar FPGA-based
infrastructure~\cite{cai-fccm2011,aya-2017} used in this dissertation and examine various
issues including data retention, read disturbance, latency, P/E cycling errors,
programming errors, and cell-to-cell program interference. There have also been
field studies of the characteristics of flash memories employed in large-scale
systems~\cite{meza-sigmetrics2015,schroeder2016flash,ouyang-asplos2014,narayanan.systor16}.
These works are also complementary to the experimental works presented in this
dissertation.

Furthermore, there have been experimental studies of other memory and storage
technologies, such as hard
disks~\cite{schroeder-fast07,pinheiro2007failure,bairavasundaram2007analysis,bairavasundaram2008analysis},
SRAM~\cite{maiz-iedm2003,autran-tns2009,toh-vlsic2010,radaelli2005investigation,tipton2006multiple,tosaka2004comprehensive},
and PCM~\cite{6378664,1369204}. All of these works are also complementary to the
experimental works presented in this dissertation.

\chapter{Low-Cost Inter-Linked Subarrays (LISA)}
\label{chap:lisa}

Bulk data movement, the movement of thousands or millions of bytes between two
memory locations, is a common operation performed by an increasing number of
real-world applications (e.g.,~\cite{kanev-isca2015, lee-hpca2013,
    ousterhout-usenix1990, rosenblum-sosp1995, seshadri-cal2015,
    seshadri-micro2013, son-isca2013, sudan-asplos2010,zhao-iccd2005}).
    Therefore, it has been the target of several architectural optimizations
    (e.g.,~\cite{blagodurov-usenix2011, jiang-pact2009, seshadri-micro2013,
    wong-fpt2006, zhao-iccd2005}). In fact, bulk data movement is important
    enough that modern commercial processors are adding specialized support to
    improve its performance, such as the ERMSB instruction recently added to the
    x86 ISA~\cite{intel-optmanual2012}.

In today's systems, to perform a bulk data movement between two locations in
memory, the data needs to go through the processor \emph{even though both the
  source and destination are within memory}. To perform the movement, the data
is first read out one cache line at a time from the source location in memory
into the processor caches, over a pin-limited off-chip channel (typically
64~bits wide). Then, the data is written back to memory, again one cache line at
a time over the pin-limited channel, into the destination location. By going
through the processor, this data movement incurs a significant penalty in terms
of latency and energy consumption. In this chapter, we introduce a new DRAM
substrate, \capfullname (\lisa), whose goal is to enable fast and efficient data
movement across a large range of memory at low cost. We show that, as a DRAM
substrate, \lisa is versatile, enabling an array of new applications that reduce
the fundamental access latency of DRAM.

\section{Motivation: Low Subarray Connectivity Inside DRAM}

\begin{figure}[h]
  \centering
  \begin{minipage}[t][3.5cm]{\linewidth/2-0.08in}
    \centering
    \includegraphics[scale=1.2]{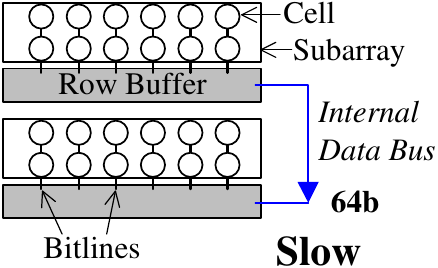}
    \subcaption{RowClone~\cite{seshadri-micro2013}} \label{fig:intro_baseline}
  \end{minipage}
  \hfill
  \begin{minipage}[t][3.5cm]{\linewidth/2-0.22in}
    \centering
    \includegraphics[scale=1.2]{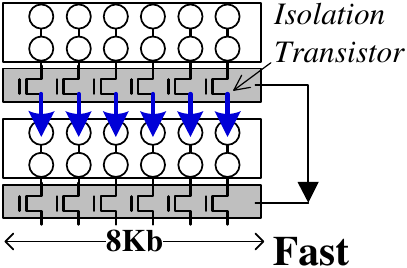}
    \subcaption{\acro} \label{fig:intro_lisa}
  \end{minipage}
  \caption{Transferring data between subarrays using the internal data bus takes a long time in
    state-of-the-art DRAM design, \rc~\cite{seshadri-micro2013} (a). Our work, LISA, enables fast
    inter-subarray data movement with a low-cost substrate (b).}
  \label{fig:intro}
\end{figure}

To address the inefficiencies of traversing the pin-limited channel, a number of
mechanisms have been proposed to accelerate bulk data movement
(e.g.,~\cite{jiang-pact2009,lu-micro2015,seshadri-micro2013,zhao-iccd2005}). The
state-of-the-art mechanism, \rc~\cite{seshadri-micro2013}, performs data
movement \emph{completely within a DRAM chip}, avoiding costly data transfers
over the pin-limited memory channel. However, its effectiveness is limited
because \rc can enable \emph{fast} data movement \emph{only} when the source and
destination are within the same DRAM \emph{subarray}. A DRAM chip is divided
into multiple \emph{banks} (typically 8), each of which is further split into
many \emph{subarrays} (16 to 64)~\cite{kim-isca2012}, shown in
\figref{intro_baseline}, to ensure reasonable read and write latencies at high
density~\cite{ chang-hpca2014,jedec-ddr3, jedec-ddr4, kim-isca2012,
udipi-isca2010}. Each subarray is a two-dimensional array with hundreds of rows
of DRAM cells, and contains only a few megabytes of data (e.g., 4MB in a rank of
eight 1Gb~DDR3 DRAM chips with 32 subarrays per bank). While two DRAM rows in the
\emph{same} subarray are connected via a wide (e.g., 8K~bits) bitline interface,
rows in \emph{different} subarrays are connected via only a \emph{narrow 64-bit
data bus} within the DRAM chip (\figref{intro_baseline}). Therefore, even for
previously-proposed in-DRAM data movement mechanisms such as
\rc~\cite{seshadri-micro2013}, \emph{inter-subarray} bulk data movement incurs
long latency and high memory energy consumption even though data does not move
out of the DRAM chip.


While it is clear that fast \emph{inter-subarray} data movement can have several
applications that improve system performance and memory energy
efficiency~\cite{kanev-isca2015, ousterhout-usenix1990,
  rosenblum-sosp1995,seshadri-cal2015,seshadri-micro2013,zhao-iccd2005}, there
is currently no mechanism that performs such data movement quickly and
efficiently.  This is because \emph{no wide datapath exists today between
  subarrays} within the same bank (i.e., the connectivity of subarrays is low in
modern DRAM). \textbf{Our goal} is to design a low-cost DRAM substrate that
enables fast and energy-efficient data movement \emph{across subarrays}.

\section{Design Overview and Applications of LISA}

We make two key observations that allow us to improve the connectivity of
subarrays within each bank in modern DRAM. First, accessing data in DRAM causes
the transfer of an entire row of DRAM cells to a buffer (i.e., the \emph{row
buffer}, where the row data temporarily resides while it is read or written)
via the subarray's \emph{bitlines}. Each bitline connects a column of cells
to the row buffer, interconnecting every row within the same subarray
(\figref{intro_baseline}). Therefore, the bitlines essentially serve as a very
wide bus that transfers a \emph{row's worth of data} (e.g., 8K~bits) at once.
Second, subarrays within the same bank are placed in close proximity to each
other. Thus, the bitlines of a subarray are very close to (but are not currently
connected to) the bitlines of neighboring subarrays (as shown in
\figref{intro_baseline}).

\vspace{5pt}
\noindent\textbf{Key Idea.} Based on these two observations, we introduce a new
DRAM substrate, called \emph{\lisafullname} (\emph{\lisa}). \lisa enables
\emph{low-latency, high-bandwidth inter-subarray connectivity} by linking
neighboring subarrays' bitlines together with \emph{isolation transistors}, as
illustrated in \figref{intro_lisa}. We use the new inter-subarray connection in
\lisa to develop a new DRAM operation, \emph{\xferfull (\xfer)}, which moves
data that is latched in an activated row buffer in one subarray into an inactive
row buffer in another subarray, without having to send data through the narrow
internal data bus in DRAM. \xfer exploits the fact that
the activated row buffer has enough drive strength to induce charge perturbation
within the idle (i.e., \emph{precharged}) bitlines of neighboring subarrays,
allowing the destination row buffer to sense and latch this data when the
isolation transistors are enabled.

By using a rigorous DRAM circuit model that conforms to the JEDEC
standards~\cite{jedec-ddr3} and ITRS
specifications~\cite{itrs-fep-2013,itrs-interconnect-2013}, we show that \xfer
performs inter-subarray data movement at 26x the bandwidth of a modern 64-bit
DDR4-2400 memory channel (500~GB/s vs.\ 19.2~GB/s; see~\refsec{trans_lat}), even
after we conservatively add a large (60\%) timing margin to account for process
and temperature variation.

\vspace{5pt}
\noindent\textbf{Applications of \lisa.} We exploit LISA's \emph{fast
inter-subarray movement} to enable many applications that can improve system
performance and energy efficiency. We implement and evaluate the following three
applications of \lisa:

\begin{itemize}

    \item \textbf{Bulk data copying.} Fast inter-subarray data movement can eliminate
     long data movement latencies for copies between two locations in the same
     DRAM chip. Prior work showed that such copy operations are widely used in
     today's operating systems~\cite{ousterhout-usenix1990,rosenblum-sosp1995}
     and datacenters~\cite{kanev-isca2015}. We propose \underline{R}apid
     \underline{I}nter-\underline{S}ubarray \underline{C}opy
     (\emph{\lisarcnol}), a new bulk data copying mechanism based on LISA's \xfer
     operation, to reduce the latency and DRAM energy of an inter-subarray copy
     by 9.2x and 48.1x, respectively, over the best previous mechanism,
     \rc~\cite{seshadri-micro2013} (\refsec{lisaclone}).

    \item \textbf{Enabling access latency heterogeneity within DRAM.} Prior
        works~\cite{lee-hpca2013, son-isca2013} introduced non-uniform access
        latencies within DRAM, and harnessed this heterogeneity to provide a
        data caching mechanism within DRAM for hot (i.e., frequently-accessed)
        pages. However, these works do not achieve either one of the following
        goals: (1)~low area overhead, and (2)~fast data movement from the slow
        portion of DRAM to the fast portion. By exploiting the \lisa substrate,
        we propose a new DRAM design,
        \underline{V}ar\underline{I}ab\underline{L}e \underline{LA}tency
        (\emph{\lisavillanol}) DRAM, with asymmetric subarrays that reduce the
        access latency to hot rows by up to 63\%, delivering high system
        performance and achieving both goals of low overhead and fast data
        movement (\refsec{villa}).

    \item \textbf{Reducing precharge latency.} Precharge is the process of
        preparing the subarray for the next memory access~\cite{jedec-ddr3,
        kim-isca2012, lee-hpca2015, lee-hpca2013}. It incurs latency that is on
        the critical path of a bank-conflict memory access. The precharge
        latency of a subarray is limited by the drive strength of the precharge
        unit attached to its row buffer. We demonstrate that \lisa enables a new
        mechanism, \underline{LI}nked \underline{P}recharge
        (\emph{\lisaprenol}), which connects a subarray's precharge unit with
        the idle precharge units in the neighboring subarrays, thereby
        accelerating precharge and reducing its latency by 2.6x (\refsec{lisapre}).

\end{itemize}
These three mechanisms are complementary to each other, and we show that when
combined, they provide additive system performance and energy efficiency
improvements (\refsec{everything_results}). LISA is a versatile DRAM substrate, capable of supporting several
other applications beyond these three, such as performing efficient data
remapping to avoid conflicts in systems that support subarray-level
parallelism~\cite{kim-isca2012}, and improving the efficiency of bulk bitwise
operations in DRAM~\cite{seshadri-cal2015} (see \refsec{other_apps}).

\section{Mechanism}

First, we discuss the low-cost design changes to DRAM to enable high-bandwidth
connectivity across neighboring subarrays (Section~\ref{sec:lisa:design}).  We
then introduce a new DRAM command that uses this new connectivity to perform
bulk data movement (Section~\ref{sec:bulk_lisa_transfer}).  Finally, we conduct
circuit-level studies to determine the latency of this command
(Sections~\ref{sec:trans_lat} and~\ref{sec:lisa:variation}).

\subsection{\lisa Design in DRAM}
\label{sec:lisa:design}

\lisa is built upon two key characteristics of DRAM. First, large data bandwidth
\emph{within} a subarray is already available in today's DRAM chips. A row activation
transfers an entire DRAM row (e.g., 8KB across all chips in a rank) into the row buffer via the bitlines of
the subarray. These bitlines essentially serve as a wide bus that transfers an entire
row of data in parallel to the respective subarray's row buffer. Second, every subarray
has its own set of bitlines, and subarrays within the same bank are placed in
close proximity to each other. Therefore, a subarray's bitlines are very close
to its neighboring subarrays' bitlines, although these bitlines are \emph{not} directly
connected together.\footnote{Note that matching the bitline pitch across
    subarrays is important for a high-yield DRAM
process~\cite{lim-isscc2012,takahashi-jsscc2001}.}

By leveraging these two characteristics, we propose to \emph{build a wide
connection path between subarrays} within the same bank at low cost, to
overcome the problem of a narrow connection path between subarrays in commodity
DRAM chips (i.e., the internal data bus \circled{3} in
\figref{background_bank_sa_detail}). \figref{lisa_links_hpca} shows the
subarray structures in \lisa. To form a new, low-cost inter-subarray datapath
with the same wide bandwidth that already exists inside a subarray, we join
neighboring subarrays' bitlines together using \emph{isolation transistors}. We
call each of these isolation transistors a \emph{\iso}. A link connects the
bitlines for the same column of two adjacent subarrays.

\figputHS{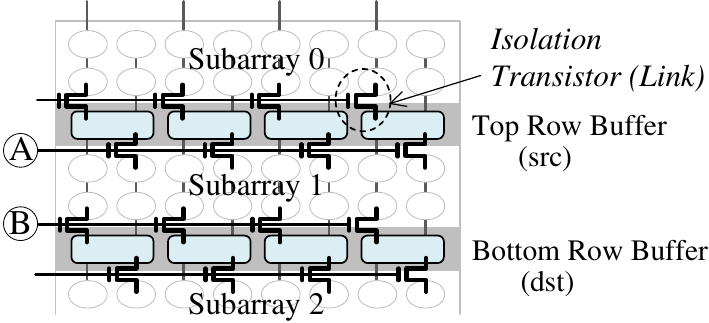}{1.2}{Inter-linked subarrays in \lisa.}

When the isolation transistor is turned on (i.e., the link is enabled), the
bitlines of two adjacent subarrays are connected. Thus, the sense amplifier of a
subarray that has already driven its bitlines (due to an \act) can also drive
its neighboring subarray's precharged bitlines through the enabled link. This
causes the neighboring sense amplifiers to sense the charge difference, and
simultaneously help drive both sets of bitlines.
When the isolation transistor is turned off (i.e., the link is disabled),
the neighboring subarrays are disconnected from each other and thus operate as
in conventional DRAM.

\ignore{
As a result, an enabled sense amplifier that has already fully driven its bitlines
(to $V_{DD}$ or 0V) can also drive its neighbor subarray's precharged bitline
through an enabled \iso. This causes the neighbor sense amplifiers to sense the
voltage difference, and simultaneously help drive the bitlines.
}

\subsection{Row Buffer Movement (\xfernorm) Through \lisa}
\label{sec:bulk_lisa_transfer}

Now that we have inserted physical \isos to provide high-bandwidth connections
across subarrays, we must provide a way for the memory controller to make use of
these new connections. Therefore, we introduce a new DRAM command, \xfer, which
triggers an operation to move data from one row buffer (half a row of data) to
another row buffer within the same bank through these \isos. This operation serves
as the building block for our architectural optimizations.

To help explain the \xfer process between two row buffers, we assume that the
\emph{top row buffer} and the \emph{bottom row buffer} in
\figref{lisa_links_hpca} are the source (\src) and destination (\dst) of an
example \xfer operation, respectively, and that \src is activated with the
content of a row from Subarray~0. To perform this \xfer, the memory controller
enables the \isos (\circled{A} and \circled{B}) between \src and \dst, thereby
connecting the two row buffers' bitlines together ($bitline$ of \src to
$bitline$ of \dst, and $\overline{bitline}$ of \src to $\overline{bitline}$ of
\dst).

\figref{lisa_transfer_states_longer_filled} illustrates how \xfer drives the
data from \src to \dst. For clarity, we show only one column from each row
buffer. State \circled{1} shows the initial values of the bitlines (\bl and
\blbar) attached to the row buffers --- \src is activated and has fully driven
its bitlines (indicated by thick bitlines), and \dst is in the precharged state
(thin bitlines indicating a voltage state of \vddhalf). In state \circled{2},
the links between \src and \dst are turned \emph{on}. The charge of the \src
bitline (\bl) flows to the connected bitline (\bl) of \dst, raising the voltage
level of \dst's \bl to \vdddelta. The other bitlines (\blbar) have the opposite
charge flow direction, where the charge flows from the \blbar of \dst to the
\blbar of \src. This phase of charge flowing between the bitlines is known as
\emph{charge sharing}. It triggers \dst's row buffer to sense the increase of
differential voltage between \bl and \blbar, and amplify the voltage difference
further. As a result, \emph{both} \src and \dst start driving the bitlines with
the same values. This \emph{double sense amplification} process pushes both
sets of bitlines to reach the final \emph{fully sensed} state (\circled{3}),
thus completing the \xfer from \src to \dst.

\figputHS{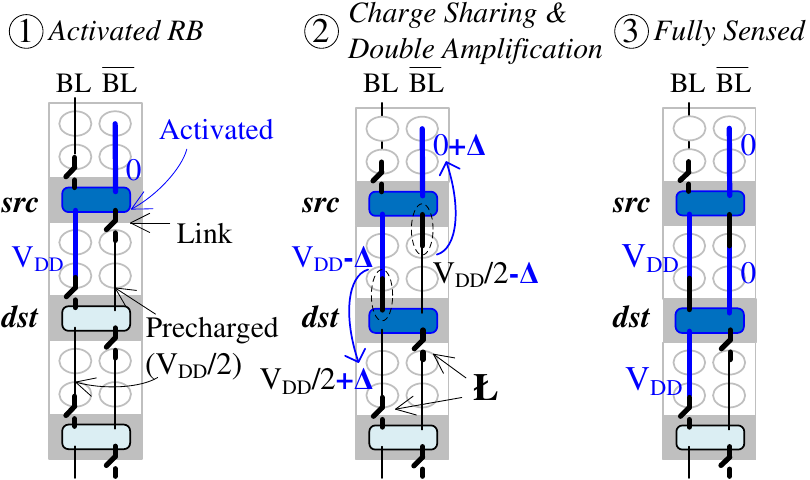}{1.2}{Row buffer movement process using \lisa.}

Extending this process, \xfer can move data between two row buffers that are
\emph{not} adjacent to each other as well.  For example, \xfer can move data
from the \src row buffer (in \figref{lisa_links_hpca}) to a row buffer,
\dsttwo, that is \emph{two} subarrays away (i.e., the bottom row buffer of
Subarray~2, not shown in \figref{lisa_links_hpca}). This operation is similar
to the movement shown in \figref{lisa_transfer_states_longer_filled}, except
that the \xfer command turns on \emph{two extra links} ($\L$ \circled{2} in
\figref{lisa_transfer_states_longer_filled}), which connect the bitlines of
\dst to the bitlines of \dsttwo, in state \circled{2}. By enabling \xfer to
perform row buffer movement across non-adjacent subarrays via a single command,
instead of requiring multiple commands, the movement latency and command
bandwidth are reduced.


\subsection{Row Buffer Movement (\xfernorm) Latency}
\label{sec:trans_lat}

To validate the \xfer process over \lisa links and evaluate its latency, we
build a model of \lisa using the Spectre Circuit Simulator~\cite{spectre}, with
the NCSU FreePDK 45nm library~\cite{ncsu-freepdk45}. We configure the DRAM
using the JEDEC DDR3-1600 timings~\cite{jedec-ddr3}, and attach each bitline to
512 DRAM cells~\cite{lee-hpca2013,son-isca2013}. We conservatively perform our
evaluations using \emph{worst-case cells}, with the resistance and capacitance
parameters specified in the ITRS
reports~\cite{itrs-fep-2013,itrs-interconnect-2013} for the metal lanes.
Furthermore, we conservatively model the worst RC drop (and hence latency) by
evaluating cells located at the edges of subarrays.

\newcommand{\samp}[1]{\texttt{SAmp#1}\xspace}


We now analyze the process of using one \xfer operation to move data between
\emph{two non-adjacent row buffers} that are two subarrays apart. To help the
explanation, we use an example that performs \xfer from \rb{0} to \rb{2}, as
shown on the left side of \figref{lisa_clone_spice_hpca}. The right side of the
figure shows the voltage of a single bitline $BL$ from each subarray during the
\xfer process over time. The voltage of the $\overline{BL}$ bitlines show the
same behavior, but have inverted values. We now explain this \xfer process step
by step.

\figputHS{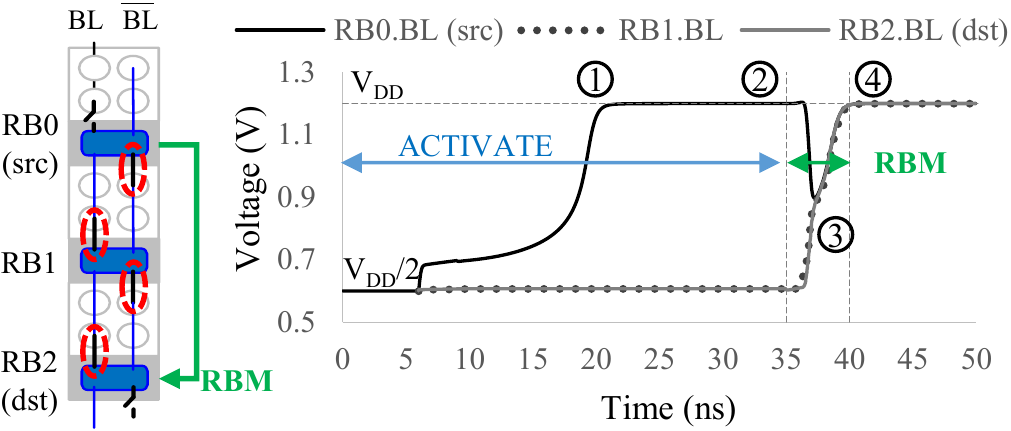}{1.1}{SPICE simulation results for transferring
data across two subarrays with \lisa.}

First, before the \xfer command is issued, an \act command is sent to \rb{0} at
time~0. After roughly 21ns (\circled{1}), the bitline reaches $V_{DD}$, which
indicates the cells have been fully restored (\tras). Note that, in our
simulation, restoration happens more quickly than the standard-specified \tras
value of 35ns, as the standard includes a guardband on top of the typical cell
restoration time to account for process and temperature
variation~\cite{chandrasekar-date2014, lee-hpca2015}. This amount of margin is
on par with values experimentally observed in commodity DRAMs at
55{\textdegree}C~\cite{lee-hpca2015}.

Second, at 35ns (\circled{2}), the memory controller sends the \xfer command to
move data from \rb{0} to \rb{2}. \xfer simultaneously turns on the \emph{four}
links (circled on the left in \figref{lisa_clone_spice_hpca}) that connect the
subarrays' bitlines.

Third, after a small amount of time (\circled{3}), the voltage of \rb{0}'s
bitline drops to about 0.9V, as the fully-driven bitlines of \rb{0} are now
charge sharing with the precharged bitlines attached to \rb{1} and \rb{2}. This
causes both \rb{1} and \rb{2} to sense the charge difference and start
amplifying the bitline values. Finally, after amplifying the bitlines for a few
nanoseconds (\circled{4} at 40ns), all three bitlines become fully driven with
the value that is originally stored in \rb{0}.

We thus demonstrate that \xfer moves data from one row buffer to a row buffer
two subarrays away at very low latency. Our SPICE simulation shows that the
\xfer latency across two \lisa links is approximately 5ns (\circled{2}
$\rightarrow$ \circled{4}). To be conservative, we do \emph{not} allow data
movement across more than two subarrays with a single \xfer command.\footnote{In
    other words, \xfer has two variants, one that moves data between
    immediately adjacent subarrays
    (\figref{lisa_transfer_states_longer_filled}) and one that moves data
between subarrays that are one subarray apart from each other
(\figref{lisa_clone_spice_hpca}).}



\subsection{Handling Process and Temperature Variation}
\label{sec:lisa:variation}

On top of using worst-case cells in our SPICE model, we add in a latency
guardband to the \xfer latency to account for process and temperature variation,
as DRAM manufacturers commonly do~\cite{chandrasekar-date2014, lee-hpca2015}.
For instance, the \act timing (\trcd) has been observed to have margins of
13.3\%~\cite{chandrasekar-date2014} and 17.3\%~\cite{lee-hpca2015} for different
types of commodity DRAMs. To conservatively account for process and temperature
variation in \lisa, we add a large timing margin, of \emph{60\%}, to the \xfer
    latency. Even then, \xfer latency is 8ns and \xfer provides a 500~GB/s data
    transfer bandwidth across two subarrays that are one subarray apart from
    each other, which is 26x the bandwidth of a DDR4-2400 DRAM
    channel (19.2~GB/s)~\cite{jedec-ddr4}.


\section{Application 1: Rapid Inter-Subarray Bulk Data Copying (\lisarc)}
\label{sec:lisaclone}

Due to the narrow memory channel width, bulk copy operations used by
applications and operating systems are performance limiters in today's
systems~\cite{jiang-pact2009, kanev-isca2015,
seshadri-micro2013,zhao-iccd2005}. These operations are commonly performed due
to the \texttt{memcpy} and \texttt{memmov}. Recent work reported
that these two operations consume 4-5\% of \emph{all of} Google's datacenter
cycles, making them an important target for lightweight hardware
acceleration~\cite{kanev-isca2015}. As we show in
Section~\ref{sec:lisaclone:shortcomings}, the \mbox{state-of-the-art} solution,
RowClone~\cite{seshadri-micro2013}, has poor performance for such operations
when they are performed \emph{across} subarrays in the same bank.

Our goal is to provide an architectural mechanism to accelerate these
inter-subarray copy operations in DRAM. We propose \lisarc, which uses the
\xfer operation in \lisa to perform rapid data copying.  We describe
the high-level operation of \lisarc (Section~\ref{sec:lisaclone:overview}), and
then provide a detailed look at the memory controller command sequence required
to implement \lisarc (Section~\ref{sec:lisaclone:commands}).

\subsection{Shortcomings of the State-of-the-Art}
\label{sec:lisaclone:shortcomings}

Previously, we have described the state-of-the-art work,
\rc~\cite{seshadri-micro2013}, which addresses the problem of costly data
movement over memory channels by coping data completely in DRAM.
However, \rc does \emph{not} provide fast data copy between subarrays.
The main latency benefit of RowClone comes from intra-subarray copy
(\emph{\rciasa} for short) as it copies data at the row granularity. In
contrast, inter-subarray RowClone (\emph{\rcirsa}) requires transferring data at
the cache line granularity (64B) through the internal data bus in DRAM.
Consequently, \rcirsa incurs \emph{16x longer latency} than \rciasa.
Furthermore, \rcirsa is a long \emph{blocking operation} that prevents
reading from or writing to the other banks in the same rank, reducing
bank-level parallelism~\cite{lee-micro2009,mutlu-isca2008}.


To demonstrate the ineffectiveness of \rcirsa, we compare it to today's
currently-used copy mechanism, \baseline, which moves data via the memory
channel. In contrast to \mbox{\rcirsa}, which copies data in DRAM, \baseline
copies data by sequentially reading out source data from the memory and then
writing it to the destination data in the on-chip caches. \figref{rc_vs_base}
compares the average system performance and queuing latency of \rcirsa and
\baseline, on a quad-core system across 50 workloads that contain bulk (8KB)
data copies (see Section~\ref{sec:meth} for our methodology). \rcirsa actually
\emph{degrades} system performance by 24\% relative to \baseline, mainly
because \rcirsa increases the overall memory queuing latency by 2.88x, as it
blocks other memory requests from being serviced by the memory controller
performing the \rcirsa copy. In contrast, \baseline is \emph{not} a long or
blocking DRAM command, but rather a long sequence of memory requests that can
be interrupted by other critical memory requests, as the memory scheduler can
issue memory requests out of order~\cite{kim-hpca2010,
kim-micro2010,mutlu-micro2007, mutlu-isca2008, rixner-isca2000,
subramanian-iccd2014, usui-taco2016, zuravleff-patent}.


\figputGHS{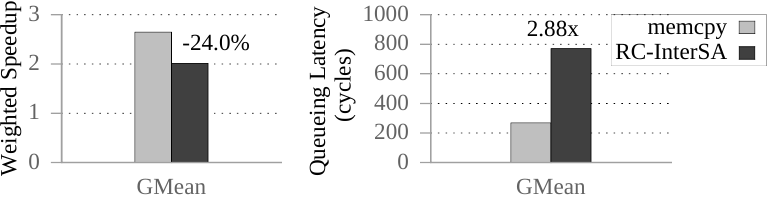}{1.3}{Comparison of \rc to \baselinenorm over the memory
    channel, on workloads that perform bulk data copy across subarrays on a
    4-core system.}

\begin{figure*}[!hb]
\begin{minipage}{\linewidth}
\vspace{6pt}
\begin{center}
\includegraphics[width=\linewidth]{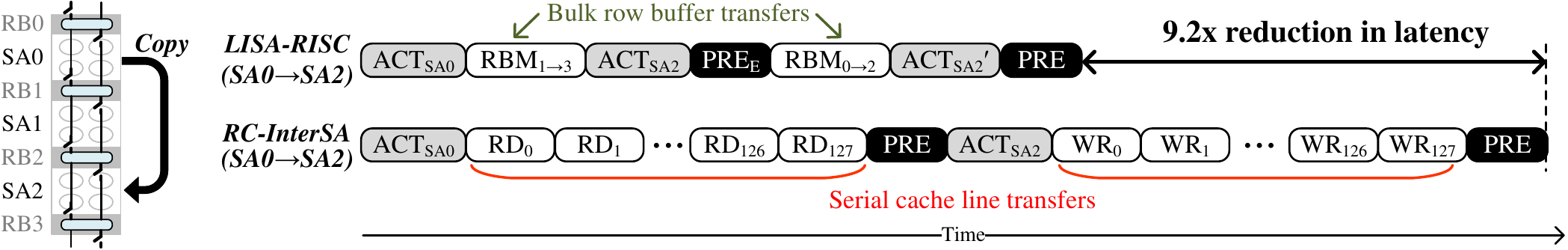}
\end{center}
\vspace{-0.1in}
\caption{Command service timelines of a row copy for \lisarc and \rcirsa
(command latencies not drawn to scale). \label{fig:lisa_timelines_irsa}}
\end{minipage}
\end{figure*}

On the other hand, \rcirsa offers energy savings of 5.1\% on average over
\baseline by \emph{not} transferring the data over the memory channel. Overall,
these results show that neither of the existing mechanisms (\baseline or \rc)
offers \emph{fast and energy-efficient} bulk data copy across subarrays.

\ignore{ Second, the fixed row-size copy granularity of \rciasa can also become
a disadvantage as it cannot control how much data to copy between two rows. The
typical page size in modern operating systems is 4KB, but commodity DDR3/4
memory technology only support 8KB row size~\cite{jedec-ddr3,jedec-ddr4},
creating a mismatch between the copy granularity of \rciasa and page-size data
copies. This can result in potential data loss as \rciasa copies an extra OS
page into another physical page that may be currently used by the operating
system. \todo{? Therefore, the RowClone work assumes a row size of 4KB} Third,
while \rcirsa is copying data, it prevents other banks from reading or writing
data as it uses the internal data bus shared across all banks to transfer data.
\todo{Add this? Fourth, \rciasa restricts its copy range to a smaller physical
address space, that is within a subarray.} }

\subsection{In-DRAM Rapid Inter-Subarray Copy (RISC)}
\label{sec:lisaclone:overview}

Our goal is to design a new mechanism that enables
\emph{low-latency} and \emph{energy-efficient} memory copy between rows \emph{in
different subarrays} within the same bank. To this end, we propose a new in-DRAM
copy mechanism that uses \lisa to exploit the high-bandwidth links between
subarrays. The key idea, step by step, is to: (1)~activate a source row in a
subarray; (2)~rapidly transfer the data in the activated source row buffers to
the destination subarray's row buffers, through LISA's wide inter-subarray
links, without using the narrow internal data bus; and (3)~activate
the destination row, which enables the contents of the destination row buffers
to be latched into the destination row. We call this inter-subarray row-to-row
copy mechanism \emph{\lisa-\underline{R}apid
\underline{I}nter-\underline{S}ubarray \underline{C}opy} (\lisarc).

As \lisarc uses the full row bandwidth provided by \lisa, it reduces the copy
latency by 9.2x compared to \rcirsa (see Section~\ref{sec:comp-copy}). An
additional benefit of using \lisarc is that its inter-subarray copy operations
are performed \emph{completely inside a bank}. As the internal DRAM data bus is
untouched, \emph{other} banks can concurrently serve memory requests,
exploiting bank-level parallelism. This new mechanism is complementary to \rc,
which performs fast \emph{intra-subarray} copies.  Together, our mechanism and
\rc can enable a complete set of fast in-DRAM copy techniques in future
systems. We now explain the step-by-step operation of how \lisarc copies data
across subarrays.

\ignore{Our goal is to design a new in-DRAM cloning mechanism that addresses some of the
downsides of \rc with the characteristics of low copy latency and flexible copy
granularity across subarrays within individual banks. This new mechanism will
also be complementary to \rc so that together our mechanism and \rc can enable
a more complete set of in-DRAM copy techniques to accelerate copying or moving a
large amount of data in future systems.

\lisarc offers several benefits over \rc. First, \lisarc enables fast data
transfer within the entire bank, as each transfer between two subarrays utilizes
the full row bandwidth. The copy latency only increases by a small amount (8ns,
Section~\ref{sec:trans_lat}) with every additional two subarrays between the
source and destination subarray. Second, we can enable a finer granularity
data copy between two rows, by providing independent
enable signals for each of the \isos in the same row (e.g., enabling the \isos
for only half of a row buffer).
Third, all inter-subarray copy operations for \lisarc are fully performed
inside the bank, allowing other banks to concurrently serve memory
requests. In addition, \lisarc can also be applied to intra-subarray copying by
simply writing the row buffer content first to the neighbor row buffer, and
then writing it back to its subarray. \lisarc still provides some benefit in
this case, offering finer-grained control over the amount of data being copied.
}

\subsection{Detailed Operation of \lisarc}
\label{sec:lisaclone:commands}

\newcommand{\sa}[1]{\texttt{SA#1}\xspace}
\newcommand{\tr}[2]{\texttt{RBM\textsubscript{#1$\rightarrow$#2}}\xspace}
\newcommand{\actcopy}[1]{\texttt{ACT\textsubscript{SA#1}}\xspace}

\figref{lisa_timelines_irsa} shows the command service timelines for both
\lisarc and \rcirsa, for copying a single row of data across two subarrays, as
we show on the left.  Data is copied from subarray \texttt{SA0} to \texttt{SA2}.
We illustrate four row buffers (\texttt{RB0}--\rb{3}): recall from
Section~\ref{sec:background:dram} that in order to activate one row, a subarray
must use \emph{two} row buffers (at the top and bottom), as each row buffer
contains only half a row of data.  As a result, \lisarc must copy half a row at
a time, first moving the contents of \texttt{RB1} into \texttt{RB3}, and then
the contents of \texttt{RB0} into \texttt{RB2}, using two \xfer commands.


First, the \lisarc memory controller activates the source row (\actcopy{0}) to
latch its data into two row buffers (\rb{0} and \rb{1}). Second, \lisarc
invokes the first \xfer operation (\tr{1}{3}) to move data from the bottom
source row buffer (\rb{1}) to the respective destination row buffer (\rb{3}),
thereby linking \rb{1} to both \rb{2} and \rb{3}, which activates both \rb{2}
and \rb{3}. After this step, \lisarc \emph{cannot} immediately invoke another
\xfer to transfer the remaining half of the source row in \rb{0} into \rb{2},
as a row buffer (\rb{2}) needs to be in the \emph{precharged state} in order to
receive data from an activated row buffer (\rb{0}). Therefore, \lisarc
completes copying the first half of the source data into the destination row
before invoking the second \xfer, by writing the row buffer (\rb{3}) into the
cells through an activation (\actcopy{2}). This activation enables the contents
of the sense amplifiers (\rb{3}) to be driven into the destination row. To
address the issue that modern DRAM chips do not allow a second \act to an
already-activated bank, we use the \emph{back-to-back} \act command that is
used to support \rc~\cite{seshadri-micro2013}.



Third, to move data from \rb{0} to \rb{2} to complete the copy transaction, we
need to precharge both \rb{1} and \rb{2}. The challenge here is to precharge
all row buffers \emph{except} \rb{0}. This cannot be accomplished in today's
DRAM because a precharge is applied at the bank level to \emph{all} row
buffers. Therefore, we propose to add a new \emph{precharge-exception} command,
which prevents a row buffer from being precharged and keeps it activated. This
bank-wide exception signal is supplied to all row buffers, and when raised for
a particular row buffer, the selected row buffer retains its state while the
other row buffers are precharged. After the precharge-exception
(\texttt{PRE\textsubscript{E}}) is complete, we then invoke the second \xfer
(\tr{0}{2}) to copy \rb{0} to \rb{2}, which is followed by an activation
(\texttt{ACT\textsubscript{SA2}{\small\textprime}}) to write \rb{2} into
\sa{2}. Finally, \lisarc finishes the copy by issuing a \cpre command
(\texttt{PRE} in \figref{lisa_timelines_irsa}) to the bank.

In comparison, the command service timeline of \rcirsa is much longer, as \rc
can copy only \emph{one cache line} of data at a time (as opposed to half a row
buffer).  This requires 128 \emph{serial cache line transfers} to read the data
from \texttt{RB0} and \texttt{RB1} into a temporary row in another bank,
followed by another 128 serial cache line transfers to write the data into
\texttt{RB2} and \texttt{RB3}.  \lisarc, by moving half a row using a single
\xfer command, achieves 9.2x lower latency than \rcirsa.

\subsection{Data Coherence}
\label{sec:copy:coherence}

When a copy is performed in DRAM, one potential disadvantage is that the data
stored in the DRAM may not be the most recent version, as the processor may
have dirty cache lines that belong to the section of memory being copied. Prior
works on in-DRAM migration have proposed techniques to accommodate data
coherence~\cite{seshadri-cal2015,seshadri-micro2013}. Alternatively, we can
accelerate coherence operations by using structures like the \mbox{Dirty-Block}
Index~\cite{seshadri-isca2014}.

\ignore{, which we can also use for \lisarc.  When the memory controller
initiates a copy in DRAM, it also performs an \emph{in-cache} copy of the dirty
lines belonging to the section of memory being copied.  These in-cache copies
are marked as dirty.  This approach avoids forcing the in-DRAM copy
operation to wait for dirty data to be flushed back to memory,
because the cache has the most up-to-date copy for dirty lines that xx cache
before the copy operation has initiated
}

\subsection{Comparison of Copy Techniques}
\label{sec:comp-copy}

\figref{copy_latency_energy} shows the DRAM latency and DRAM energy consumption
of different \emph{copy commands} for copying a row of data (8KB). The exact
latency and energy numbers are listed in \tabref{copy_latency}.\footnote{Our
    reported numbers differ from prior work~\cite{seshadri-micro2013} because:
(1) we use faster DRAM timing parameters (1600-11-11-11 vs 1066-8-8-8), and (2)
we use the 8KB row size of most commercial DRAM instead of
4KB~\cite{seshadri-micro2013}.} We derive the copy latency of each command
sequence using equations based on the DDR3-1600 timings~\cite{jedec-ddr3}
(available in our technical report~\cite{chang-tr2016}), and the DRAM
energy using the Micron power calculator~\cite{micron-tr}. For \lisarc, we
define a \emph{hop} as the number of subarrays that \lisarc needs to copy data
\emph{across} to move the data from the source subarray to the destination
subarray. For example, if the source and destination subarrays are adjacent to
each other, the number of hops is 1. The DRAM chips that we evaluate have 16
subarrays per bank, so the maximum number of hops is 15.

\figputGTS{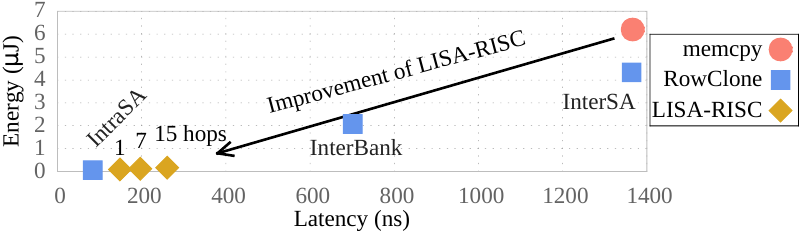}{1.5}{Latency and DRAM energy of 8KB copy.}

%

\begin{table}[t]
    \fontsize{8.5}{10.2}\selectfont
    \newdimen\origwspace
    \origwspace=\fontdimen2\font
    \fontdimen2\font=0.22em
    \setlength{\tabcolsep}{.2em}

    \centering
    \begin{tabular}{lll}
        \toprule
        \textbf{Copy Commands} (8KB) & \textbf{Latency} (ns) & \textbf{Energy} ($\micro$J)\\
        \cmidrule(rl){1-3}

        \baseline (via mem. channel) & 1366.25 & 6.2 \\
        RC-InterSA / Bank / IntraSA & 1363.75 / 701.25 / 83.75 & 4.33 / 2.08 / 0.06\\
        \lisarc (1 / 7 / 15 hops) & 148.5 / 196.5 / 260.5 & 0.09 / 0.12 / 0.17 \\
        \bottomrule
    \end{tabular}
    \caption{Copy latency and DRAM energy.}
    \label{tab:copy_latency}%
    \fontdimen2\font=\origwspace
\end{table}

We make two observations from these numbers. First, although \rcirsa incurs
similar latencies as \baseline, it consumes 29.6\% less energy, as it does
\emph{not} transfer data over the channel and DRAM I/O for each copy operation.
However, as we showed in \secref{lisaclone:shortcomings}, \rcirsa incurs a higher system performance
penalty because it is a long-latency \emph{blocking} memory command. Second,
copying between subarrays using \lisa achieves significantly lower
latency and energy compared to \rc, even though the total latency of
\lisarc grows linearly with the hop count.

By exploiting the \lisa substrate, we thus provide a more complete set of
in-DRAM copy mechanisms. Our workload evaluation results show that \lisarc
outperforms \rcirsa and \baseline: its average performance improvement and energy
reduction over the best performing inter-subarray copy mechanism (i.e.,
\baseline) are 66.2\% and 55.4\%, respectively, on a quad-core system, across 50
workloads that perform bulk copies (see Section~\ref{sec:eval_lisaclone}).

%


\section{Application 2: In-DRAM Caching Using Heterogeneous Subarrays
(\lisavilla)}
\label{sec:villa}

Our second application aims to reduce the DRAM access latency for
frequently-accessed (hot) data. Prior work introduces heterogeneity into DRAM,
where one region has a fast access latency but small capacity (fewer DRAM rows),
while the other has a slow access latency but high capacity (many more
rows)~\cite{lee-hpca2013, son-isca2013}. To yield the highest performance
benefits, the fast region is used as a dynamic cache that stores the hot rows.
There are two design constraints that must be considered: (1)~\emph{ease of
implementation}, as the fast caching structure needs to be low-cost and
non-intrusive; and (2)~\emph{data movement cost}, as the caching mechanism
should adapt to dynamic program phase changes, which can lead to changes in the
set of hot DRAM rows.  As we show in Section~\ref{sec:villa:shortcomings}, prior
work has not balanced the trade-off between these two constraints.


Our goal is to design a heterogeneous DRAM that offers fast data movement with
a low-cost and easy-to-implement design. To this end, we propose
\emph{\lisavilla (\underline{V}ar\underline{I}ab\underline{L}e
\underline{LA}tency)}, a mechanism that uses \lisa to provide fast row movement
into the cache when the set of hot DRAM rows changes. \mbox{\lisavilla} is also
easy to implement, as discussed in Section~\ref{sec:villa:dram}. We describe
our hot row caching policy in Section~\ref{sec:villa:caching}.

\subsection{Shortcomings of the State-of-the-Art}
\label{sec:villa:shortcomings}

We observe that two state-of-the-art techniques for heterogeneity
within a DRAM chip are not effective at providing \emph{both}
ease of implementation and low movement cost.

\begin{figure}[h]
  \centering
  \subcaptionbox{CHARM~\cite{son-isca2013}\label{fig:charm}} {
  \includegraphics[scale=1.3]{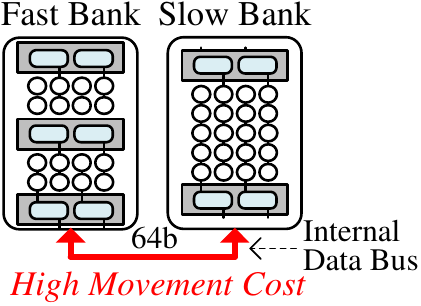}
  }
  \qquad
  \subcaptionbox{TL-DRAM~\cite{lee-hpca2013}\label{fig:tldram}} {
    \includegraphics[scale=1.3]{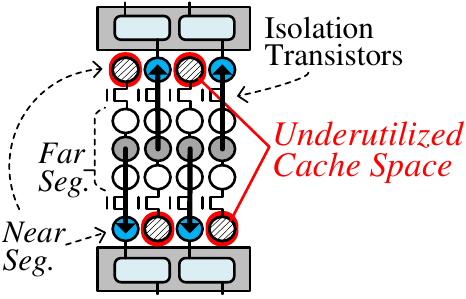}
  }
  \caption{Drawbacks of existing heterogeneous DRAMs.}
  \label{fig:villa_compare}
\end{figure}

CHARM~\cite{son-isca2013} introduces heterogeneity \emph{within a rank} by
designing a few fast banks with (1)~shorter bitlines for faster data sensing,
and (2)~closer placement to the chip I/O for faster data transfers. To exploit
these low-latency banks, CHARM uses an OS-managed mechanism to statically
allocate hot data to them based on program profile information. Unfortunately,
this approach cannot adapt to program phase changes, limiting its performance
gains. If it were to adopt dynamic hot data management, CHARM would incur high
movement cost over the narrow 64-bit internal data bus in DRAM, as illustrated
in \figref{charm}, since it does not provide high-bandwidth connectivity
between banks.

TL-DRAM~\cite{lee-hpca2013} provides heterogeneity \emph{within a subarray} by
dividing it into fast (near) and slow (far) segments that have short and long
bitlines, respectively, using isolation transistors. To manage the fast segment
as an OS-transparent hardware cache, TL-DRAM proposes a \emph{fast
intra-subarray movement} scheme similar to RowClone~\cite{seshadri-micro2013}.
The main disadvantage is that TL-DRAM needs to cache each hot row in \emph{two
near segments}, as shown in \figref{tldram}, as each subarray uses two row
buffers on \emph{opposite ends} to sense data in the open-bitline architecture.
This prevents TL-DRAM from using the \emph{full} near segment capacity.
TL-DRAM's area overhead is also sizable (3.15\%) in an open-bitline
architecture. As we can see, neither CHARM nor TL-DRAM strike a good trade-off
between the two design constraints.

\subsection{Variable Latency (VILLA) DRAM}
\label{sec:villa:dram}

We propose to introduce heterogeneity \emph{within a bank} by designing
\emph{heterogeneous-latency subarrays}. We call this heterogeneous DRAM design
\emph{\underline{V}ar\underline{I}ab\underline{L}e \underline{LA}tency DRAM}
(\villa). To design a low-cost fast subarray, we take an approach similar to
prior work, attaching fewer cells to each bitline to reduce the parasitic
capacitance and resistance. This reduces the sensing (\trcd), restoration
(\tras), and precharge (\trp) time of the fast subarrays~\cite{lee-hpca2013,
    micron-rldram3, son-isca2013}. In this chapter, we focus on managing the fast
    subarrays in hardware, as it offers better adaptivity to dynamic changes in
    the hot data set.

In order to take advantage of \villa, we rely on \lisarc to rapidly copy rows
across subarrays, which significantly reduces the caching latency. We call this
synergistic design, which builds \villa using the \lisa substrate,
\emph{\lisavilla}. Nonetheless, the cost of transferring data to a fast
subarray is still non-negligible, especially if the fast subarray is far from
the subarray where the data to be cached resides. Therefore, an intelligent
cost-aware mechanism is required to make astute decisions on which data to
cache and when.

\subsection{Caching Policy for \lisavilla}
\label{sec:villa:caching}

We design a simple epoch-based caching policy to evaluate the benefits of
caching a row in \lisavilla. Every epoch, we track the number of accesses to
rows by using a set of 1024 saturating counters for each bank.\footnote{The
    hardware cost of these counters is low, requiring only 6KB of storage in
the memory controller (see \secref{hwarea}).} The counter values are halved
every epoch to prevent staleness. At the end of an epoch, we mark the 16 most
frequently-accessed rows as \emph{hot}, and cache them when they are accessed
the next time. For our cache replacement policy, we use the \emph{benefit-based
caching} policy proposed by Lee et al.~\cite{lee-hpca2013}. Specifically, it
uses a benefit counter for each row cached in the fast subarray: whenever a
cached row is accessed, its counter is incremented. The row with the least
benefit is replaced when a new row needs to be inserted. Note that a large body
of work proposed various caching policies (e.g., \cite{hart-compcon1994,
hidaka-ieeemicro90,hsu-isca1993,jiang-hpca2010,kedem-1997,meza-cal2012,qureshi-isca2007,
seshadri-pact2012, yoon-iccd2012}), each of which can potentially be used with
\lisavilla.

Our evaluation shows that \lisavilla improves system performance by 5.1\% on
average, and up to 16.1\%, for a range of 4-core workloads
(see Section~\ref{sec:eval_lisavilla}).


\section{Application 3: Fast Precharge Using Linked Precharge Units
(\lisapre)}
\label{sec:lisapre}

Our third application aims to accelerate the process of precharge. The
precharge time for a subarray is determined by the drive strength of the
precharge unit. We observe that in modern DRAM, while a subarray is being
precharged, the precharge units (PUs) of \emph{other} subarrays remain idle.

We propose to exploit these idle PUs to accelerate a precharge operation by
connecting them to the subarray that is being precharged. Our mechanism,
\emph{\lisa-\underline{LI}nked \underline{P}recharge} (\lisapre), precharges a
subarray using \emph{two} sets of PUs: one from the row buffer that is being
precharged, and a second set from a neighboring subarray's row buffer (which is
already in the precharged state), by enabling the links between the two
subarrays.


\figref{lisa_pre_states_longer_filled} shows the process of \emph{linked
precharging} using \lisa. Initially, only one subarray (top) is fully activated
(state \circled{1}) while the neighboring (bottom) subarray is in the precharged
state. The neighboring subarray is in the precharged state, as only one subarray
in a bank can be activated at a time, while the other subarrays remain
precharged.
In state \circled{2}, we begin the precharge operation by disabling the sense
amplifier in the top row buffer and enabling its PU. After we enable the links
between the top and bottom subarrays, the bitlines start sharing charge with
each other, and both PUs \emph{simultaneously} reinitialize the bitlines,
eventually fully pulling the bitlines to $V_{DD}/2$ (state \circled{3}).
Note that we are using \emph{two} PUs to pull down \emph{only one} set of
activated bitlines, which is why the precharge process is shorter.

\figputHS{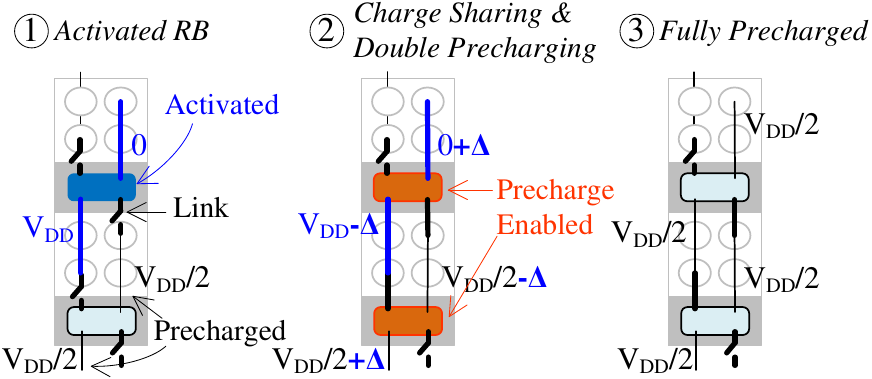}{1.2}{Linked precharging through \lisa.}

To evaluate the accelerated precharge process, we use the same methodology
described in Section~\ref{sec:trans_lat} and simulate the linked precharge
operation in SPICE. \figref{lisa_pre_spice_hpca} shows the resulting timing
diagram. During the first 2ns, the wordline is lowered to disconnect the cells
from the bitlines \circled{1}. Then, we enable the \isos to begin precharging
the bitlines \circled{2}. The result shows that the precharge latency reduces
significantly due to having two PUs to perform the precharge. \lisa enables a
shorter precharge latency of approximately 3.5ns \circled{3} versus the
baseline precharge latency of 13.1ns \circled{4}.

\ignore{ This is due not only to having two PUs to perform the precharge, but
also because of charge sharing between the connected bitlines. As we recall from
\figref{lisa_clone_spice_hpca} (\refsec{trans_lat}), when the activated bitline
connects with the precharged bitlines (state \circled{2}), the charge starts to
flow from the activated bitline to the other bitlines, thus helping to bring the
voltage closer to $V_{DD}/2$. Problem: I need to show the other BL's voltage to
prove this, which I do not have. }

\figputHS{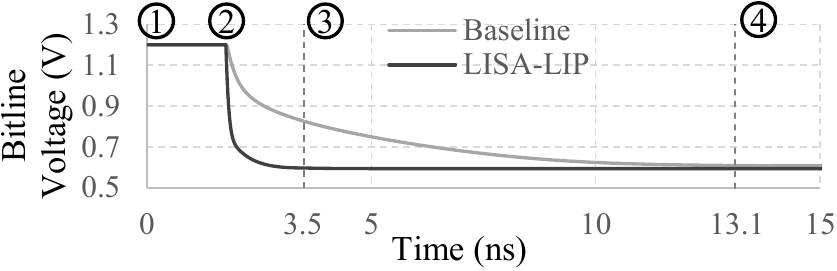}{1.1}{SPICE simulation of precharge operation.}

To account for \textbf{process and temperature variation}, we add a guardband to
the SPICE-reported latency, increasing it to 5ns (i.e., by \emph{42.9\%}), which
    still achieves 2.6x lower precharge latency than the baseline. Our
    evaluation shows that \mbox{\lisapre} improves performance by 10.3\% on average,
    across 50 \mbox{four-core} workloads (see Section~\ref{sec:eval_lisapre}).


\section{Hardware Cost}


\subsection{Die Area Overhead}
\label{sec:hwarea}
 To evaluate the area overhead of adding isolation
transistors, we use area values from prior work, which adds isolation
transistors to disconnect bitlines from sense amplifiers~\cite{o-isca2014}. That
work shows that adding an isolation transistor to every bitline incurs a total of
0.8\% die area overhead in a 28nm DRAM process technology. Similar to prior work
that adds isolation transistors to DRAM~\cite{lee-hpca2013,o-isca2014}, our
\lisa substrate also requires additional control logic outside the DRAM banks to
control the isolation transistors, which incurs a small amount of area and is
non-intrusive to the cell arrays. For \lisavilla, we use 1024 six-bit saturating
counters to track the access frequency of rows in every bank; this
requires an additional 6KB storage within a memory controller connected to one rank.


\subsection{Handling Repaired Rows} To improve yield, DRAM manufacturers often
employ post-manufacturing repair techniques that can remap faulty rows to
spare rows provisioned in every subarray~\cite{keeth-dram-tutorial}.
Therefore, consecutive row addresses as observed by the memory controller may
physically reside in \emph{different} subarrays. To handle this issue for
techniques that require the controller to know the subarray a row resides in
(e.g., RowClone~\cite{seshadri-micro2013}, \lisarc), a simple approach can be
used to expose the repaired row information to the memory controller. Since
DRAM already stores faulty rows' remapping information inside the chip, this
information can be exposed to the controller through the serial presence
detect (SPD)~\cite{jedec-spd}, which is an EEPROM that stores DRAM information
such as timing parameters. The memory controller can read this stored
information at system boot time so that it can correctly determine a repaired
row's location in DRAM. Note that similar techniques may be necessary for
other mechanisms that require information about physical location of rows in
DRAM (e.g.,~\cite{chang-hpca2014, kang14, kim-isca2014, kim-isca2012,
lee-hpca2013, liu-isca2013}).

\section{Methodology} \label{sec:meth}

We evaluate our system using a variant of Ramulator~\cite{kim-cal2015},
an open-source cycle-accurate DRAM simulator, driven by traces generated from
Pin~\cite{luk-pldi2005}. We will make our simulator publicly
available~\cite{safari-github}. We use a row buffer policy that closes a row
only when there are no more outstanding requests in the memory controller to
the same row~\cite{rixner-isca2000}. Unless stated otherwise, our simulator
uses the parameters listed in \tabref{sys-config}.

\begin{table}[h]
    \fontsize{8.5}{10.2}\selectfont
    \renewcommand{\arraystretch}{0.8}
    \newdimen\origwspace
    \origwspace=\fontdimen2\font

    \centering
    \setlength{\tabcolsep}{.3em}
    \begin{tabular}{ll}
        \toprule
        \textbf{Processor} & 1--4 OoO cores, 4GHz, 3-wide issue \\
        \midrule

        \textbf{Cache} & L1: 64KB, L2: 512KB per core, L3: 4MB, 64B lines \\
        \midrule

        \textbf{Mem. Controller} & 64/64-entry read/write queue,
        FR-FCFS~\cite{rixner-isca2000,zuravleff-patent} \\

        \midrule

        \multirow{2}{*}{\textbf{DRAM}} & DDR3-1600~\cite{micronDDR3_4Gb}, 1--2 channels, 1
        rank/channel,\\
        & 8 banks/rank, 16 subarrays/bank \\

        \bottomrule
    \end{tabular}
    \fontdimen2\font=\origwspace
    \caption{Evaluated system configuration.}
    \label{tab:sys-config}%
\end{table}

\cc{ To evaluate the benefits of different data copy mechanisms in isolation, we
  use a \mbox{copy-aware} page mapping policy that allocates destination pages
  to the same DRAM structures (i.e., subarrays, banks) where the source pages
  are allocated. As a result, our evaluation of different data copy mechanisms
  is a \emph{limit study} as only the specified copy mechanism (e.g.,
  \lisarcnol) is used for copy operations. For example, when evaluating
  \lisarcnol, the page mapper allocates both the source and destination pages within the
  same bank to evaluate the benefits of RISC's fast data movement between
  subarrays.
}

\noindent\textbf{Benchmarks and Workloads.} We primarily use benchmarks from
TPC(-C/-H)~\cite{tpc}, DynoGraph (BFS, Page\-Rank)~\cite{dynograph}, SPEC
CPU2006~\cite{spec2006}, and STREAM~\cite{stream}, along with a random-access
microbenchmark similar to HPCC RandomAccess~\cite{randombench}. Because these
benchmarks predominantly stress the CPU and memory while rarely invoking
\baseline, we use the following benchmarks to evaluate different copy
mechanisms: (1)~\texttt{bootup}, (2)~\texttt{forkbench}, and
(3)~\texttt{Unix}~\texttt{shell}. These were shared by the authors of
RowClone~\cite{seshadri-micro2013}. The \texttt{bootup} benchmark consists of a
trace collected while a Debian operating system was booting up. The
\texttt{forkbench} kernel forks a child process that copies 1K pages from the
parent process by randomly accessing them from a 64MB array. The \texttt{Unix}
\texttt{shell} is a script that runs \texttt{find} in a directory along with
\texttt{ls} on each subdirectory. More information on these is in
\cite{seshadri-micro2013}.

To construct multi-core workloads for evaluating the benefits of data copy
mechanisms, we randomly assemble 50 workloads, each comprising 50\%
copy-intensive benchmarks and 50\% non-copy-intensive benchmarks. To evaluate
the benefits of in-DRAM caching and reduced precharge time, we restrict our
workloads to randomly-selected memory-intensive ($\ge$ 5 misses per thousand
instructions) non-copy-intensive benchmarks. Due to the large number of
workloads, we present detailed results for only five workload mixes
(\tabref{wkld-mixes}), along with the average results across all 50 workloads.


\begin{table}[h]
    \small
    \centering
    \renewcommand{\arraystretch}{0.8}
    \begin{tabular}{ll}
        \toprule
        \textbf{Mix 1} & \texttt{tpcc64, forkbench, libquantum, bootup} \\
        \textbf{Mix 2} & \texttt{bootup, xalancbmk, pagerank, forkbench} \\
        \textbf{Mix 3} & \texttt{libquantum, pagerank, forkbench, bootup} \\
        \textbf{Mix 4} & \texttt{mcf, forkbench, random, forkbench} \\
        \textbf{Mix 5} & \texttt{bfs, bootup, tpch2, bootup} \\
        \bottomrule
    \end{tabular}
    \caption{A subset of copy workloads with detailed results.}
    \label{tab:wkld-mixes}%
\end{table}

\noindent\textbf{Performance Metrics.} We measure single-core and
\mbox{multi-core}
performance using IPC and Weighted Speedup (WS)~\cite{snavely-asplos2000},
respectively. Prior work showed that WS is a measure of system
throughput~\cite{eyerman-ieeemicro2008}. To report DRAM energy consumption, we
use the Micron power calculator~\cite{micron-tr}. We run all workloads for 100
million instructions, as done in many recent
works~\cite{kim-micro2010,kim-isca2012,lee-hpca2013,lee-taco2016,mutlu-micro2007}.

\vspace{3pt} \noindent\textbf{\villa Configuration.} For our simulated \villa,
each fast subarray consists of 32 rows to achieve low latency on sensing,
precharge, and restoration (a typical subarray has 512 rows). Our SPICE
simulation reports the following new timing parameters for a 32-row subarray:
{\small{tRCD}}=7.5ns, {\small{tRP}}=8.5ns, and {\small{tRAS}}=13ns, which are
reduced from the original timings by respectively, 45.5\%, 38.2\%, and 62.9\%.
For each bank, we allocate 4 fast subarrays in addition to the 16 512-row
subarrays, incurring a 1.6\% area overhead. We set the epoch length for our
caching policy to 10,000~cycles.


\section{Evaluation} \label{sec:eval}

We quantitatively evaluate our proposed applications of \lisa: (1)~rapid bulk
copying (\lisarc), (2)~in-DRAM caching with heterogeneous subarrays
(\lisavilla), and (3)~reduced precharge time (\lisapre).

\subsection{Bulk Memory Copy}
\label{sec:eval_lisaclone}

\subsubsection{Single-Core Workloads} 
\begin{figure}[t]
    \centering
    \subcaptionbox{IPC\label{fig:1c_ipc}}[\linewidth] {
        \includegraphics[scale=1.5, trim=0 8pt 0 0, clip]{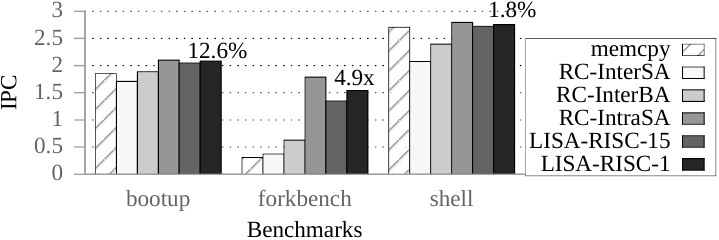}
    }

    \subcaptionbox{Energy\label{fig:1c_e}}[\linewidth]  {
        \includegraphics[scale=1.5, trim=0 8pt 0 0, clip]{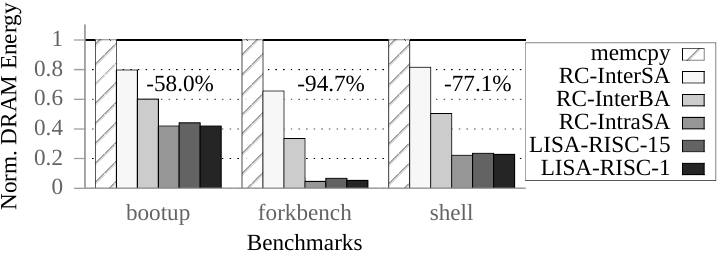}
    }

    \vspace{0.2in}
    \subcaptionbox{LISA's performance improvement over \baseline as LLC size
      varies \label{fig:1c_cache_sweep}}[\linewidth] {
        \includegraphics[scale=1.5, trim=0 6pt 0 0, clip]{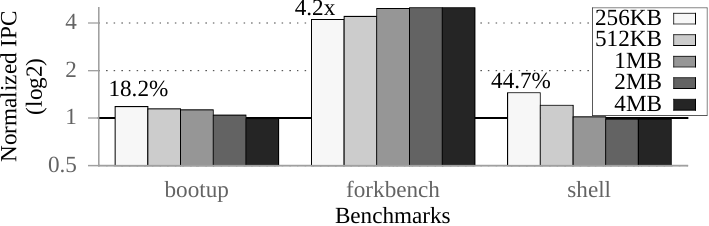}
    }

    \caption{Comparison of copy mechanisms in a single-core system. Value (\%)
    on top indicates the improvement of \lisarc-1 over \baselinenorm.}
\label{fig:1c_perf}
\end{figure}

\figref{1c_perf} shows the performance of
three copy benchmarks on a single-core system with one memory channel and 1MB
of last-level cache (LLC). We evaluate the following bulk copy mechanisms:
(1)~\baseline, which copies data over the memory channel; (2)
\rc~\cite{seshadri-micro2013}; and (3) \lisarc. We use two different hop counts
between the source and destination subarray for \mbox{\lisarc}: 15 (longest)
and 1 (shortest). They are labeled as \lisarc-15 and \lisarc-1, respectively,
in the figure. We make four major observations.

First, \lisarc achieves significant improvement over \mbox{\rcirsa} for all three
benchmarks in terms of both IPC and memory energy consumption, shown in
\figref{1c_ipc} and \figref{1c_e}, respectively. This shows that the LISA
substrate is effective at performing fast inter-subarray copies.

Second, both \lisarc-1 and \lisarc-15 significantly reduce the memory energy
consumption over \baseline. This is due to (1)~reduced memory traffic over the
channel by keeping the data within DRAM, and (2)~higher performance.


Third, \lisarc-1/-15 provides 12.6\%/10.6\%, 4.9x/4.3x, and 1.8\%/0.7\% speedup
for \texttt{bootup}, \texttt{forkbench}, and \texttt{shell}, respectively, over
\baseline. The performance gains are smaller for \texttt{bootup} and
\texttt{shell}. Both of these benchmarks invoke fewer copy operations (i.e.,
2171 and 2682, respectively) than \texttt{forkbench}, which invokes a large
number (40952) of copies. As a result, \texttt{forkbench} is more sensitive to
the memory latency of copy commands. Furthermore, the large LLC capacity (1MB)
helps absorb the majority of memory writes resulting from \baseline for
\texttt{bootup} and \texttt{shell}, thereby reducing the effective latency of
\baseline.

Fourth, \rcirsa performs \emph{worse} than \baseline for \texttt{bootup} and
\texttt{shell} due to its long \emph{blocking} copy operations. Although, it
attains a 19.4\% improvement on \texttt{forkbench} because \baseline causes
severe \emph{cache pollution} by installing a large amount of copied data into
the LLC. Compared to the 20\% cache hit rate for \baseline, \rcirsa has a much
higher hit rate of 67.2\% for \texttt{forkbench}. The copy performance of
\baseline is strongly correlated with the LLC management policy and size.

To understand performance sensitivity to LLC size, \figref{1c_cache_sweep} shows
the speedup of \lisarc-1 over \baseline for different LLC capacities. We make
two observations, which are also similar for \lisarc-15 (not shown). First, for
\texttt{bootup} and \texttt{shell}, the speedup of LISA over \baseline reduces
as the LLC size increases because the destination locations of \baseline
operations are more likely to hit in the larger cache.

Second, for \texttt{forkbench}, LISA-RISC's performance gain over \baseline
decreases as cache size reduces from 1MB to 256KB. This is because the LLC hit
rate reduces much more significantly for \lisarc, from 67\% (1MB) to 10\%
(256KB), than for \baseline (from 20\% at 1MB, to 19\% at 256KB). When
\texttt{forkbench} uses \lisarc for copying data, its working set mainly
consists of non-copy data, which has good locality. As the LLC size reduces by
4x, the working set no longer fits in the smaller cache, thus causing a
significant hit rate reduction. On the other hand, when \baseline is used as the
copy mechanism, the working set of \texttt{forkbench} is mainly from bulk copy
data, and is less susceptible to cache size reduction. Nonetheless,
\mbox{\lisarc} still provides an improvement of 4.2x even with a 256KB cache.

We conclude that \lisarc significantly improves performance and memory energy
efficiency in single-core workloads that invoke bulk copies.


\subsubsection{Multi-Core Workloads}

\begin{figure}[h]
    \centering
    \subcaptionbox{Weighted speedup normalized to \baseline
      \label{fig:lisa_4c_ws}}{
        \includegraphics[scale=1.5]{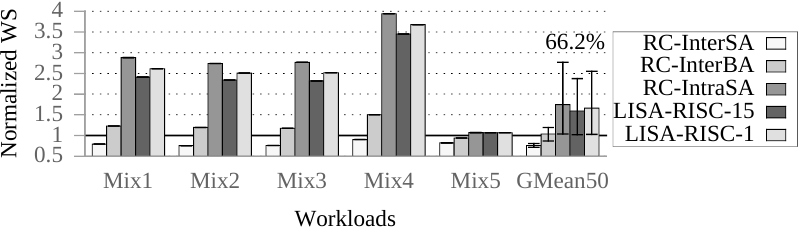}
    }

    \subcaptionbox{Memory energy efficiency normalized to \baseline
      \label{fig:lisa_4c_energy}}{
        \includegraphics[scale=1.5]{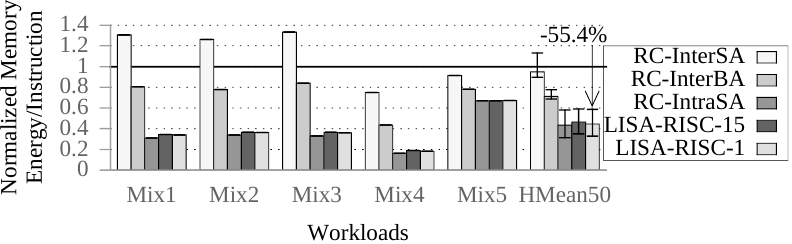}
    }
    \caption{Four-core system evaluation: (a) weighted speedup and (b)
    memory energy per instruction.}
    \label{fig:4c_copy_perf}
\end{figure}

\figref{4c_copy_perf} shows the system performance and energy efficiency (i.e.,
memory energy per instruction) of different copy mechanisms across 50 workloads,
on a quad-core system with two channels and 4MB of LLC. The error bars in this
figure (and other figures) indicate the 25th and 75th percentile values across
all 50 workloads. Similar to the performance trends seen in the single-core
system, \lisarc consistently outperforms other mechanisms at copying data
between subarrays. \lisarc-1 attains a high average system performance
improvement of 66.2\% and 2.2x over \baseline and \rcirsa, respectively.
Although Mix~5 has the smallest number of copy operations out of the five
presented workload mixes, \lisarc still improves its performance by 6.7\% over
\baseline. By moving copied data only within DRAM, \lisarc significantly reduces
memory energy consumption (55.4\% on average) over \baseline. In summary,
\lisarc provides both high performance and high memory energy efficiency for
bulk data copying for a wide variety of single- and multi-core workloads.

\subsection{In-DRAM Caching with \lisavilla}
\label{sec:eval_lisavilla}

\figref{villa_multic_errbar} shows the system performance improvement of
\lisavilla over a baseline without any fast subarrays in a four-core system. It
also shows the hit rate in \villa, i.e., the fraction of accesses that hit in
the fast subarrays. We make two main observations. First, by exploiting \lisarc
to quickly cache data in \villa, \lisavilla improves system performance for a
wide variety of workloads --- by up to 16.1\%, with a geometric mean of 5.1\%.
This is mainly due to reduced DRAM latency of accesses that hit in the fast
subarrays (which comprise 16MB of total storage across two memory channels).
The performance improvement heavily correlates with the VILLA cache hit rate.
Our work does not focus on optimizing the caching scheme, but the hit rate may
be increased by an enhanced caching policy
(e.g.,~\cite{qureshi-isca2007,seshadri-pact2012}), which can further improve
system performance.

\figputGHS{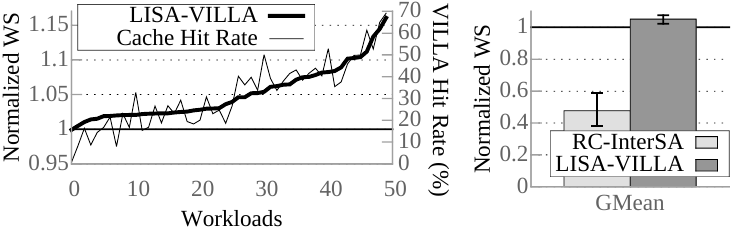}{1.5}{Performance improvement and hit rate with \lisavilla, and
performance comparison to using \rcirsa with \villa.}

Second, the \villa design, which consists of heterogeneous subarrays, is not
practical without \lisa. \figref{villa_multic_errbar} shows that using
\rcirsa to move data into the cache \emph{reduces} performance by 52.3\% due to
slow data movement, which overshadows the benefits of caching. The results indicate
that \lisa is an important substrate to enable not only fast bulk data
copy, but also a fast in-DRAM caching scheme.



\subsection{Accelerated Precharge with \lisapre}
\label{sec:eval_lisapre}

\figref{pre_norm_ws_4c_multic} shows the system performance improvement of
\lisapre over a baseline that uses the standard DRAM precharge latency, as well
as \lisapre's row-buffer hit rate, on a four-core system across 50
workloads. \lisapre attains a maximum gain of 13.2\%, with a mean improvement of
8.1\%. The performance gain becomes higher as the row-buffer hit rate decreases,
which leads to more precharge commands. These results show that \lisa is a
versatile substrate that effectively reduces precharge latency in addition to
accelerating data movement.

\figputGHS{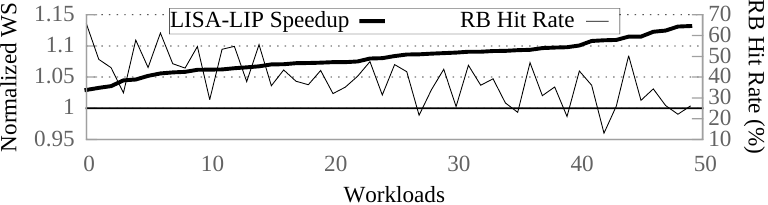}{1.5}{Speedup and row buffer (RB) hit rate of
\lisapre.}

We also evaluate the effectiveness of combining \lisavilla and \lisapre (not
shown, but available in our technical report~\cite{chang-tr2016}). The combined
mechanism, which is transparent to software, improves system performance by
12.2\% on average and up to 23.8\% across the same set of 50 workloads without
bulk copies. Thus, LISA is an effective substrate that can enable mechanisms to
fundamentally reduce memory latency.


\subsection{Putting Everything Together}
\label{sec:everything_results}

As all of the three proposed applications are complementary to each other, we
evaluate the effect of putting them together on a four-core system.
\figref{all_norm_ws_4c} shows the system performance improvement of adding
\lisavilla to \lisarc (15 hops), as well as combining all three optimizations,
compared to our baseline using \baseline and standard DDR3-1600 memory. We draw
several key conclusions. First, the performance benefits from each scheme are
additive. On average, adding \lisavilla improves performance by 16.5\% over
\lisarc alone, and adding \lisapre further provides an 8.8\% gain over
\lisarcvilla. Second, although \lisarc alone provides a majority of the
performance improvement over the baseline (59.6\% on average), the use of
both \mbox{\lisavilla} and \mbox{\lisapre} further improves performance,
resulting in an average performance gain of 94.8\% and memory energy reduction
(not plotted) of 49.0\%. Taken together, these results indicate that \lisa is
an effective substrate that enables a wide range of high-performance and
energy-efficient applications in the DRAM system.

\figputGHS{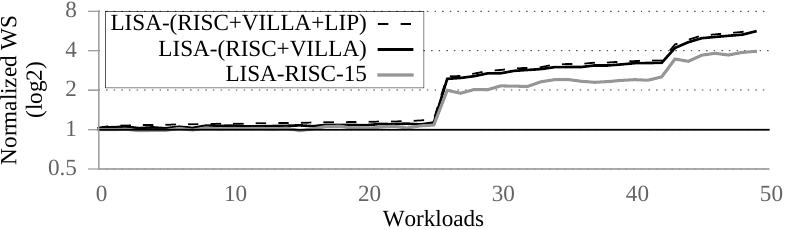}{1.5}{Combined WS improvement of \lisa applications.}
\vspace{-2pt}

\subsection{Sensitivity to System Configuration}

\figref{4c_sensitivity_errbar} shows the weighted speedup for \baseline and
\mbox{\lisa-All} (i.e., all three applications) on a 4-core system using varying
memory channel counts and LLC sizes. The results show that performance
improvement increases with fewer memory channels, as memory contention
increases.  On the other hand, adding more memory channels increases
memory-level parallelism, allowing more of the copy latency to be hidden.
Similar trends are observed with the LLC capacity. As LLC size decreases,
the working set becomes less likely to fit with \baseline,
worsening its performance. \lisa-All provides significant performance
benefits for all configurations.

\figputGHS{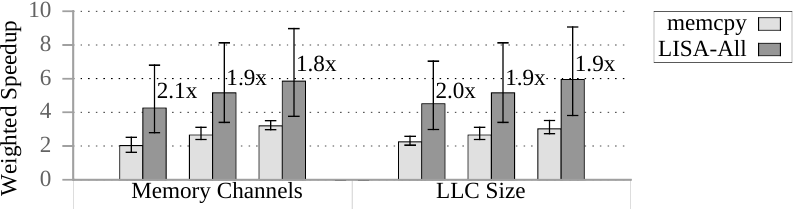}{1.5}{Performance sensitivity to channels and LLC size.}
\vspace{-2pt}

\subsection{Effect of Copy Distance on \lisarc}

\tabref{4c_lisa_sweep} shows that the performance gain and memory energy savings
of \lisarc over \baseline increases as the copy distance reduces. This is
because with fewer subarrays between the source and destination subarrays, the
number of \xfer commands invoked by \lisarc reduces accordingly, which decreases
the latency and memory energy consumption of bulk data copy.

\begin{table}[h]
\fontsize{8.5}{10.2}\selectfont
\centering
    \setlength{\tabcolsep}{.33em}
    \renewcommand{\arraystretch}{1}
    \begin{tabular}{lrrrrrr}
      \toprule
      \textbf{Copy Distance} (hops) & 1 & 3 & 7 & 15 & 31 & 63 \\
      \midrule
      \textbf{RISC Copy Latency} (ns) & 148.5 & 164.5 & 196.5 &
          260.5 & 388.5 & 644.5 \\
      \textbf{WS Improvement} (\%) & 66.2 & 65.3 & 63.3 & 59.6 & 53.0 & 42.4 \\
      \textbf{DRAM Energy Savings} (\%) & 55.4 & 55.2 & 54.6 & 53.6 & 51.9 & 48.9 \\
      \bottomrule
    \end{tabular}
\caption{Effect of copy distance on \lisarc.}
\label{tab:4c_lisa_sweep}
\vspace{-2pt}
\end{table}


\section{Other Applications Enabled by \lisa}
\label{sec:other_apps}

We describe two additional applications that can potentially benefit from
\lisa. We describe them at a high level, and defer evaluations to future work.

\vspace{3pt} \noindent\textbf{Reducing Subarray Conflicts via Remapping.} When
two memory requests access two different rows in the same bank, they have to be
served serially, even if they are to different subarrays. To mitigate such
\emph{bank conflicts}, Kim et al.~\cite{kim-isca2012} propose
\emph{subarray-level parallelism (SALP)}, which enables multiple subarrays to
remain activated at the same time. However, if two accesses are to the same
subarray, they still have to be served serially. This problem is exacerbated
when \mbox{frequently-accessed} rows reside in the same subarray. To help
alleviate such \emph{subarray conflicts}, \lisa can enable a simple mechanism
that efficiently remaps or moves the conflicting rows to different subarrays by
exploiting fast \xfer operations.

\vspace{3pt} \noindent\textbf{Extending the Range of In-DRAM Bulk Operations.}
To accelerate bitwise operations, Seshadri et al.~\cite{seshadri-cal2015}
propose a new mechanism that performs bulk bitwise AND and OR operations in
DRAM. Their mechanism is restricted to applying bitwise operations
only on rows within the \emph{same subarray} as it requires the copying of
source rows before performing the bitwise operation. The high cost of
\emph{inter-subarray} copies makes the benefit of this mechanism inapplicable to
data residing in rows in different subarrays. LISA can enable efficient
inter-subarray bitwise operations by using \lisarc to copy rows to the same
subarray at low latency and low energy.


\ignore{
\noindent\textbf{Micro-Pages.} To mitigate bank conflicts, Sudan et
al.~\cite{sudan-asplos2010} observe that only a small portion of each OS page is
accessed in a time period. Therefore, they propose to co-locate these small
chunks (micro-pages) from each page into a DRAM row to increase row buffer hit
rate. This technique requires them to reduce the OS page size to 1KB to co-locate
different pages. In addition, it requires a heavy page movement scheme.

\lisa can efficiently implement their proposal. Instead of shrinking the page
size, which is intrusive to the operating system, we can use \lisavilla to cache
the frequently-accessed chunks from every rows inside a bank and modify the
transfer granularity of \lisa to the chunk size (i.e., 1KB, which is a quarter
of each row buffer). This would result in a less heavy-overhead implementation
for micro-pages.
}


\section{Summary} \label{sec:conclusion}

We present a new DRAM substrate, \emph{\fullname (\lisa)}, that expedites bulk
data movement across subarrays in DRAM. \lisa achieves this by creating a new
high-bandwidth datapath at low cost between subarrays, via the insertion of a
small number of isolation transistors. We describe and evaluate three
applications that are enabled by \lisa. First, \lisa significantly reduces the
latency and memory energy consumption of bulk copy operations between subarrays
over two state-of-the-art mechanisms~\cite{seshadri-micro2013}. Second, \lisa
enables an effective in-DRAM caching scheme on a new heterogeneous DRAM
organization, which uses fast subarrays for caching hot data in every bank.
Third, we reduce precharge latency by connecting two precharge units of
adjacent subarrays together using \lisa. We experimentally show that the three
applications of LISA greatly improve system performance and memory energy
efficiency when used individually or together, across a variety of workloads
and system configurations.

We conclude that \lisa is an effective substrate that enables several effective
applications. We believe that this substrate, which enables low-cost
interconnections between DRAM subarrays, can pave the way for other applications
that can further improve system performance and energy efficiency through fast
data movement in DRAM.

\chapter{Mitigating Refresh Latency by Parallelizing
Accesses with Refreshes}
\label{chap:parref}

In the previous chapter, we describe LISA, a new DRAM substrate, that
significantly reduces inter-subarray movement latency to enable several
low-latency optimizations. While LISA primarily targets the latency incurred on
demand requests issued from the applications, it does not address the latency
problem due to a DRAM maintenance operation, refresh, which is issued
periodically by the memory controllers to recharge the cells’ data.

Each DRAM cell must be refreshed periodically every {\em refresh interval} as
specified by the DRAM standards~\cite{jedec-ddr3,jedec-lpddr3}. The exact
refresh interval time depends on the DRAM type (e.g., DDR or LPDDR) and the
operating temperature.  While DRAM is being refreshed, it becomes unavailable to
serve memory requests. As a result, refresh latency significantly degrades
system performance~\cite{liu-isca2012, mukundan-isca2013,
  nair-hpca2013,stuecheli-micro2010} by delaying in-flight memory requests. This
problem will become more prevalent as DRAM density increases, leading to more
DRAM rows to be refreshed within the same refresh interval.  DRAM chip density
is expected to increase from 8Gb to 32Gb by 2020 as it doubles every two to
three years~\cite{itrs-dram}. Our evaluations show that DRAM refresh, as it is
performed today, causes an average performance degradation of 8.2\% and 19.9\%
for 8Gb and 32Gb DRAM chips, respectively, on a variety of memory-intensive
workloads running on an 8-core system. Hence, it is important to develop
practical mechanisms to mitigate the performance penalty of DRAM refresh. In
this chapter, we propose two complementary mechanisms to mitigate the negative
performance impact of refresh: DARP (Dynamic Access Refresh Parallelization) and
SARP (Subarray Access Refresh Parallelization) The goal is to address the draw-
backs of per-bank refresh by building more efficient techniques to parallelize
refreshes and accesses within DRAM.

\section{Motivation}
\label{sec:motivation}


In this section, we first describe the scaling trend of commonly used
all-bank refresh in both LPDDR and DDR DRAM as chip density increases
in the future. We then provide a quantitative analysis of all-bank
refresh to show its performance impact on multi-core systems followed
by performance comparisons to per-bank refresh that is only supported
in LPDDR.

\subsection{Increasing Performance Impact of Refresh}
\label{sec:refresh-challenge}


During the \trfc time period, the entire memory rank is locked up, preventing
the memory controller from sending any memory request. As a result, refresh
operations degrade system performance by increasing the latency of memory
accesses. The negative impact on system performance is expected to be
exacerbated as \trfc increases with higher DRAM density. The value of \trfc is
currently 350ns for an 8Gb memory device~\cite{jedec-ddr3}. \figref{rfc_trend}
shows our estimated trend of \trfc for future DRAM generations using linear
extrapolation on the currently available and previous DRAM devices. The same
methodology is used in prior works~\cite{liu-isca2012,stuecheli-micro2010}.
\emph{Projection 1} is an extrapolation based on 1, 2, and 4Gb devices;
\emph{Projection 2} is based on 4 and 8Gb devices. We use the more optimistic
\emph{Projection 2} for our evaluations. As it shows, \trfc may reach up to
1.6$\mu$s for future 64Gb DRAM devices. This long period of unavailability to
process memory accesses is detrimental to system performance.

\figputGHS{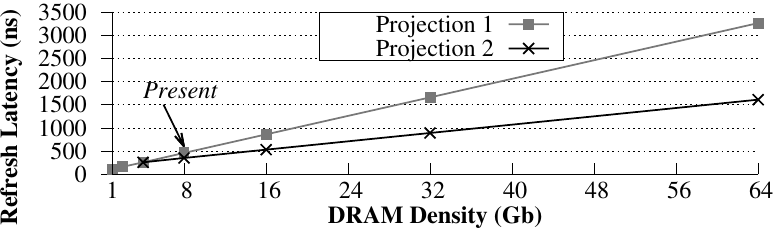}{1.3}{Refresh latency (\captrfc) trend.}

To demonstrate the negative system performance impact of DRAM refresh,
we evaluate 100 randomly mixed workloads categorized to five different
groups based on memory intensity on an 8-core system using various
DRAM densities.\footnote{Detailed methodology is described in
  Section~\ref{methodology}, including workloads, simulation
  methodology, and performance metrics.\label{fnote}} We use up to
32Gb DRAM density that the ITRS predicts to be manufactured by
2020~\cite{itrs-dram}. \figref{perf_loss_mpki} shows the average
performance loss due to all-bank refresh compared to an ideal baseline
without any refreshes for each memory-intensity category. The
performance degradation due to refresh becomes more severe as either
DRAM chip density (i.e., \trfc) or workload memory intensity increases
(both of which are trends in systems), demonstrating that it is
increasingly important to address the problem of DRAM refresh.

\figputGHS{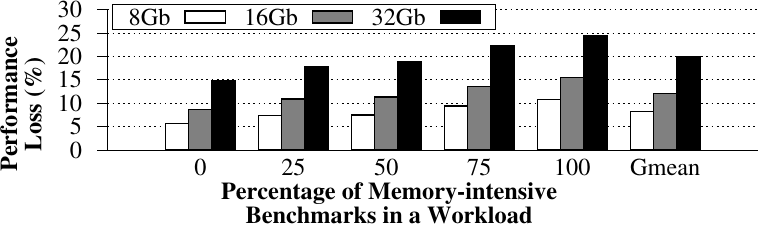}{1.3}{Performance degradation due to refresh.}

Even though the current DDR3 standard does not support \refpb, we believe that
it is important to evaluate the performance impact of \refpb on DDR3 DRAM
because DDR3 DRAM chips are widely deployed in desktops and servers.
Furthermore, adding per-bank refresh support to a DDR3 DRAM chip should be
non-intrusive because it does not change the internal bank organization. We
estimate the refresh latency of \refpb in a DDR3 chip based on the values used
in an LPDDR2 chip. In a 2Gb LPDDR2 chip, the per-bank refresh latency (\trfcpb)
is 90ns and the all-bank refresh latency (\trfc) is 210ns, which takes
\emph{2.3x} longer than \trfcpb~\cite{micronLPDDR2_2Gb}.\footnote{LPDDR2 has a
shorter \trfc than DDR3 because LPDDR2 1) has a retention time of 32ms instead
of 64ms in DDR3 under normal operating temperature and 2) each operation
refreshes fewer rows.} We apply this multiplicative factor to \trfc to calculate
\trfcpb.

Based on the estimated \trfcpb values, we evaluate the performance impact of
\refpb on the same 8-core system and workloads.\textsuperscript{\ref{fnote}}
\figref{perf_loss_all} shows the average performance degradation of \refab and
\refpb compared to an ideal baseline without any refreshes. Even though \refpb
provides performance gains over \refab by allowing DRAM accesses to
non-refreshing banks, its performance degradation becomes exacerbated as \trfcpb
increases with higher DRAM density. With 32Gb DRAM chips using \refpb, the
performance loss due to DRAM refresh is still a significant 16.6\% on average,
which motivates us to address issues related to \refpb.

\figputGHS{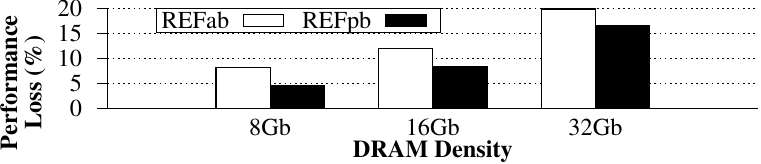}{1.3}{Performance loss due to \caprefab and \caprefpb.}

\subsection{Our Goal} We identify two main problems that \refpb faces. First,
\refpb commands are scheduled in a very restrictive manner in today's systems.
Memory controllers have to send \refpb commands in a sequential round-robin
order without any flexibility. Therefore, the current implementation does not
exploit the full benefit from overlapping refreshes with accesses across banks.
Second, \refpb cannot serve accesses to a refreshing bank until the refresh of
that bank is complete.  Our goal is to provide practical mechanisms to address
these two problems so that we can minimize the performance overhead of DRAM
refresh. 




%

\section{Mechanisms}
\label{sec:mechanism}

\subsection{Overview}


We propose two mechanisms, \emph{\darplong (\darp)} and
\emph{\intersub (\is)}, that hide refresh latency by parallelizing
refreshes with memory accesses across \emph{banks} and
\emph{subarrays}, respectively. \darp is a new refresh scheduling
policy that consists of two components. The first component is
\emph{\ooolong} that enables the memory controller to specify a
particular (idle) bank to be refreshed as opposed to the standard
per-bank refresh policy that refreshes banks in a strict round-robin
order. With out-of-order refresh scheduling, \darp can avoid
refreshing (non-idle) banks with pending memory requests, thereby
avoiding the refresh latency for those requests. The second component
is \emph{\warplong} that proactively issues per-bank refresh to a bank
while DRAM is draining write batches to other banks, thereby
overlapping refresh latency with write latency. The second mechanism,
\is, allows a bank to serve memory accesses in idle subarrays while
other subarrays within the same bank are being refreshed. \is exploits
the fact that refreshing a row is contained within a subarray, without
affecting the other subarrays' components and the I/O bus used for
transferring data. We now describe each mechanism in detail.





\subsection{\darplong}


\subsubsection{\ooolongCap} The limitation of the current per-bank refresh
mechanism is that it disallows a memory controller from specifying
which bank to refresh. Instead, a DRAM chip has internal logic that
strictly refreshes banks in a \emph{sequential round-robin order}.
Because DRAM lacks visibility into a memory controller's state (e.g.,
request queues' occupancy), simply using an in-order \refpb policy can
unnecessarily refresh a bank that has multiple pending memory requests
to be served when other banks may be free to serve a refresh
command. To address this problem, we propose the first component of
\darp, \emph{\ooolong}. The idea is to remove the bank selection logic
from DRAM and make it the memory controller's responsibility to
determine which bank to refresh. As a result, the memory controller
can refresh an idle bank to enhance parallelization of refreshes and
accesses, avoiding refreshing a bank that has pending memory requests
as much as possible.

Due to \refpb reordering, the memory controller needs to guarantee
that deviating from the original in-order schedule still preserves
data integrity. To achieve this, we take advantage of the fact that
the contemporary DDR JEDEC standard~\cite{jedec-ddr3,jedec-ddr4}
actually provides some refresh scheduling flexibility. The standard
allows up to \emph{eight} all-bank refresh commands to be issued late
(postponed) or early (pulled-in). This implies that each bank can
tolerate up to eight \refpb to be postponed or pulled-in. Therefore,
the memory controller ensures that reordering \refpb preserves data
integrity by limiting the number of postponed or pulled-in commands.

\figref{darp_flow_chart} shows the algorithm of our mechanism. The \ooolong
scheduler makes a refresh decision every DRAM cycle. There are three key steps.
First, when the memory controller hits a per-bank refresh schedule time (every
\trefipb), it postpones the scheduled \refpb if the to-be-refreshed bank
(\texttt{\emph{R}}) has pending demand requests (read or write) {\em and} it has
postponed fewer refreshes than the limit of eight (\circled{1}). The hardware
counter that is used to keep track of whether or not a refresh can be postponed
for each bank is called the \emph{refresh credit (ref\_credit)}. The counter
decrements on a postponed refresh and increments on a pulled-in refresh for each
bank. Therefore, a \refpb command can be postponed if the bank's ref\_credit
stays between values of 0 and 8 ($0\le ref\_credit \le8$). Otherwise the memory
controller is required to send a \refpb command when more than eight \refpb
commands have been postponed (i.e., $8 < ref\_credit$) to comply with the
standard. Each \refpb resets the bank's ref\_credit value bank to 0. Second, the
memory controller prioritizes issuing commands for a demand request if a refresh
is not sent at any given time (\circled{2}). Third, if the memory controller
cannot issue any commands for demand requests due to the timing constraints, it
instead randomly selects one bank (\texttt{\emph{B}}) from a list of banks that
have no pending demand requests to refresh. Such a refresh command is either a
previously postponed \refpb or a new pulled-in \refpb (\circled{3}).



\figputHS{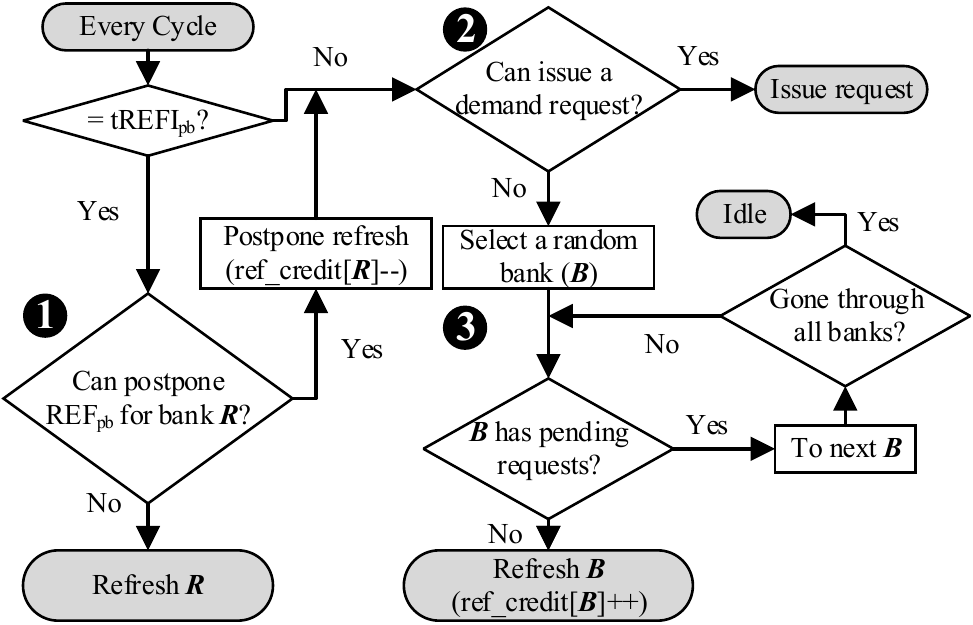}{1.2}{Algorithm of \ooolong.}

\subsubsection{\warplongCap} The key idea of the second component of \darp is to
actively avoid refresh interference on read requests and instead enable more
parallelization of refreshes with \emph{write requests}. We make two
observations that lead to our idea. First, {\em write batching} in DRAM creates
an opportunity to overlap a refresh operation with a sequence of writes, without
interfering with reads. A modern memory controller typically buffers DRAM writes
and drains them to DRAM in a batch to amortize the \emph{bus turnaround
latency}, also called \emph{tWTR} or
\emph{tRTW}~\cite{jedec-ddr3,kim-isca2012,lee-tech2010}, which is the additional
latency incurred from switching between serving writes to reads because DRAM I/O
bus is half-duplex. Typical systems start draining writes when the write buffer
occupancy exceeds a certain threshold until the buffer reaches a low watermark.
This draining time period is called the \emph{writeback mode}, during which no
rank within the draining channel can serve read
requests~\cite{chatterjee-hpca2012,lee-tech2010,stuecheli-isca2010}. Second,
DRAM writes are not latency-critical because processors do not stall to wait for
them: DRAM writes are due to dirty cache line evictions from the last-level
cache~\cite{lee-tech2010,stuecheli-isca2010}.

Given that writes are not latency-critical and are drained in a batch
for some time interval, we propose the second component of \darp,
\emph{\warplong}, that attempts to maximize parallelization of
refreshes and writes. \Warplong selects the bank with the minimum
number of pending demand requests (both read and write) and preempts
the bank's writes with a per-bank refresh. As a result, the bank's
refresh operation is hidden by the writes in other banks.


The reasons why we select the bank with the lowest number of demand
requests as a refresh candidate during writeback mode are
two-fold. First, the goal of the writeback mode is to drain writes as
fast as possible to reach a low watermark that determines the end of
the writeback
mode~\cite{chatterjee-hpca2012,lee-tech2010,stuecheli-isca2010}. Extra
time delay on writes can potentially elongate the writeback mode by
increasing queueing delay and reducing the number of writes served in
parallel across banks. Refreshing the bank with the lowest write
request count (zero or more) has the smallest impact on the writeback
mode length because other banks can continue serving their writes to
reach to the low watermark. Second, if the refresh scheduled to a bank
during the writeback mode happens to extend beyond writeback mode, it
is likely that the refresh 1) does not delay immediate reads within
the same bank because the selected bank has no reads or 2) delays
reads in a bank that has less contention. Note that we only preempt
one bank for refresh because the JEDEC standard~\cite{jedec-lpddr3}
disallows overlapping per-bank refresh operations across banks within
a rank.





\begin{figure}[t]
\centering
\subcaptionbox{Scenario 1: Parallelize postponed refresh with writes.
  \label{fig:inter-bank-postpone}}[\linewidth]{
    \includegraphics[scale=0.8]{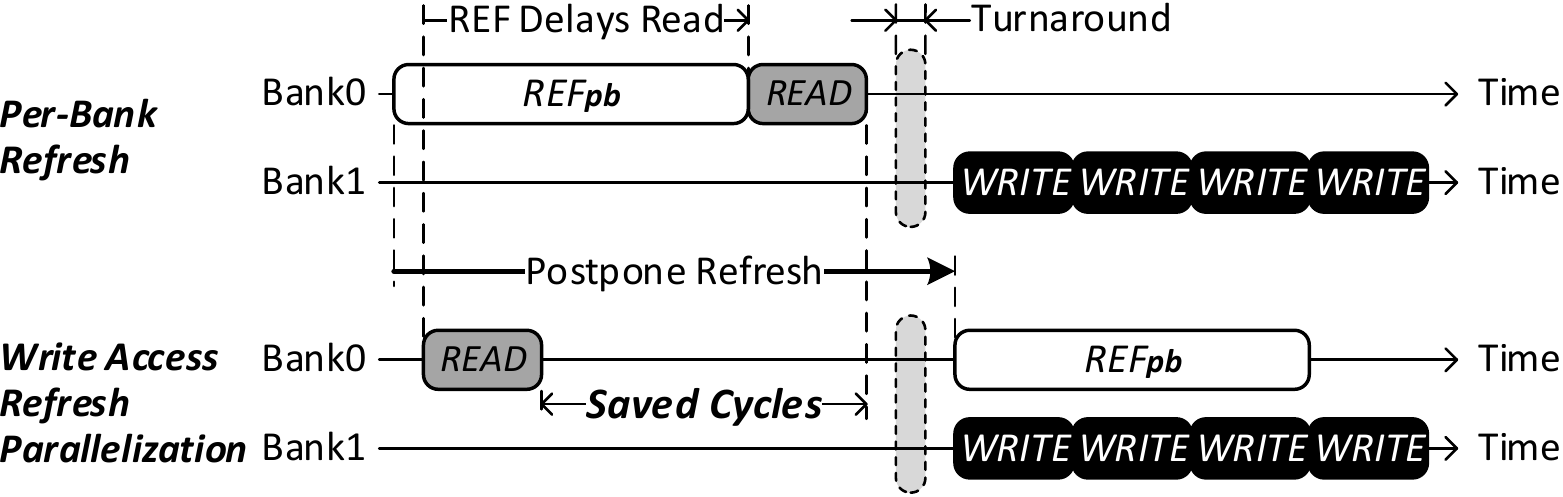}
}

\subcaptionbox{Scenario 2: Parallelize pulled-in refresh with writes.
  \label{fig:inter-bank-pull}}[\linewidth]{
    \includegraphics[scale=0.8]{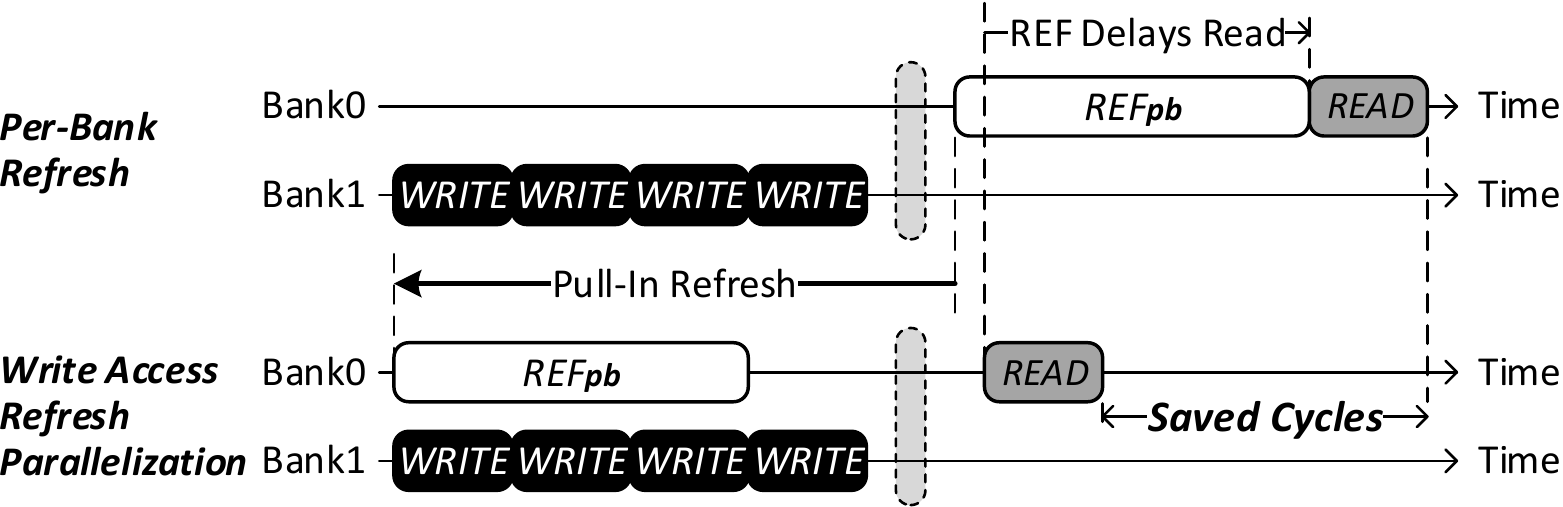}
}
\caption{Service timeline of a per-bank refresh operation along with read and
write requests using different scheduling policies.}
\label{fig:inter-bank-service-timeline}
\end{figure}

\figref{inter-bank-service-timeline} shows the service timeline and
benefits of \warplong. There are \textbf{two scenarios} when the
scheduling policy parallelizes refreshes with writes to increase
DRAM's availability to serve read
requests. \figref{inter-bank-postpone} shows the first scenario when
the scheduler \emph{postpones} issuing a \refpb command to avoid
delaying a read request in Bank 0 and instead serves the refresh in
parallel with writes from Bank 1, effectively hiding the refresh
latency in the writeback mode. Even though the refresh can potentially
delay individual write requests during writeback mode, the delay does
not impact performance as long as the length of writeback mode remains
the same as in the baseline due to longer prioritized write request
streams in other banks. In the second scenario shown in
\figref{inter-bank-pull}, the scheduler proactively \emph{pulls in} a
\refpb command early in Bank 0 to fully hide the refresh latency from
the later read request while Bank 1 is draining writes during the
writeback mode (note that the read request cannot be scheduled during
the writeback mode).

The crucial observation is that \warplong improves performance because
it avoids stalling the read requests due to refreshes by postponing or
pulling in refreshes in parallel with writes without extending the
writeback period.

\ignore{\figref{inter-bank-service-timeline} shows the service timeline and benefits of
\ib. There are \textbf{two scenarios} when the scheduling policy parallelizes
refresh operations with writes to increase DRAM's availability to serve read
requests. The first scenario is when the scheduler \emph{postpones} issuing a
\refpb to allow a DRAM bank to serve read requests. Then when the DRAM enters
writeback mode issuing write requests to other banks, the scheduler can send
the postponed refresh to the bank. As a result, the refresh operation is
parallelized with writes, effectively hiding the refresh latency in the
writeback mode to reduce the negative impact of refresh on read requests. Even
though refresh can potentially delay individual write requests during
writeback mode, the delay may not impact performance as long as the length
of writeback mode remains approximately the same due to longer prioritized
write request streams in other banks. \figref{inter-bank-postpone} shows the
performance benefit of postponing a refresh and overlapping it with writes.

In the second scenario, the scheduler takes preventive action by issuing a \refpb
command early when there is an opportunity to fully hide the refresh latency
during the writeback mode. As shown in \figref{inter-bank-pull}, the scheduler
speculatively \emph{pulls in} a \refpb command during the writeback mode before
it reaches its refresh schedule in order to free up an upcoming refresh
interval to serve a read request.}

Algorithm~\ref{algo:warp-algo} shows the operation of \warplong. When
the memory controller enters the writeback mode, the scheduler selects
a bank candidate for refresh when there is no pending refresh. A bank
is selected for refresh under the following criteria: 1) the bank has
the lowest number of demand requests among all banks and 2) its
refresh credit has not exceeded the maximum \emph{pulled-in} refresh
threshold. After a bank is selected for refresh, its credit increments
by one to allow an additional refresh postponement.

\begin{algorithm}
\caption{\small Write-refresh parallelization}
\label{algo:warp-algo}
\footnotesize{
\textbf{\underline{Every \trfcpb in Writeback Mode:}}
\begin{algorithmic}
\If{\textit{refresh\_queue[0:N-1].isEmpty()}}
    \State \textit{b = find\_bank\_with\_lowest\_request\_queue\_count AND $ref\_credit<8$}
        \State \textit{refreshBank($b$)}
        \State \textit{ref\_credit[$b$]} += 1
\EndIf
\end{algorithmic}
}
\end{algorithm}




\subsubsection{Implementation}
\darp incurs a small overhead in the memory controller and DRAM
without affecting the DRAM cell array organization. There are five
main modifications. First, each refresh credit is implemented with a
hardware integer counter that either increments or decrements by up to
eight when a refresh command is pulled-in or postponed,
respectively. Thus, the storage overhead is very modest with 4 bits
per bank (32 bits per rank). Second, \darp requires logic to monitor
the status of various existing queues and schedule refreshes as
described. Despite reordering refresh commands, all DRAM timing
constraints are followed, notably \trrd and \trfcpb that limit when
\refpb can be issued to DRAM.  Third, the DRAM command decoder needs
modification to decode the bank ID that is sent on the address bus
with the \refpb command.  Fourth, the refresh logic that is located
outside of the banks and arrays needs to be modified to take in the
specified bank ID. Fifth, each bank requires a separate row counter to
keep track of which rows to refresh as the number of postponed or
pulled-in refresh commands differs across banks.  Our proposal limits
the modification to the least invasive part of the DRAM without
changing the structure of the dense arrays that consume the majority
of the chip area.

\subsection{\intersub} Even though \darp allows refreshes and accesses to occur
in parallel across different banks, \darp cannot deal with their collision
\emph{within a bank}. To tackle this problem, we propose \emph{\sarp
(\intersub)} that exploits the existence of subarrays within a bank. The key
observation leading to our second mechanism is that refresh occupies only a few
\emph{subarrays} within a bank whereas the other \emph{subarrays} and the
\emph{I/O bus} remain idle during the process of refreshing. The reasons for
this are two-fold.  First, refreshing a row requires only its subarray's sense
amplifiers that restore the charge in the row without transferring any data
through the I/O bus.  Second, each subarray has its own set of \emph{sense
amplifiers} that are not shared with other subarrays.

Based on this observation, SARP's key idea is to allow memory accesses
to an \emph{idle} subarray while another subarray is refreshing.
\figref{subarray-service-timeline} shows the service timeline and the
performance benefit of our mechanism. As shown, \is reduces the read
latency by performing the read operation to Subarray 1 in parallel
with the refresh in Subarray 0. Compared to \ib, \is provides the
following advantages: 1) \is is applicable to both all-bank and
per-bank refresh, 2) \is enables memory accesses to a refreshing bank,
which cannot be achieved with \ib, and 3) \is also utilizes bank-level
parallelism by serving memory requests from multiple banks while the
entire rank is under refresh. \is requires modifications to 1) the
DRAM architecture because two distinct wordlines in different
subarrays need to be raised simultaneously, which cannot be done in
today's DRAM due to the shared peripheral logic among subarrays, 2)
the memory controller such that it can keep track of which subarray is
under refresh in order to send the appropriate memory request to an
idle subarray.


\figputHS{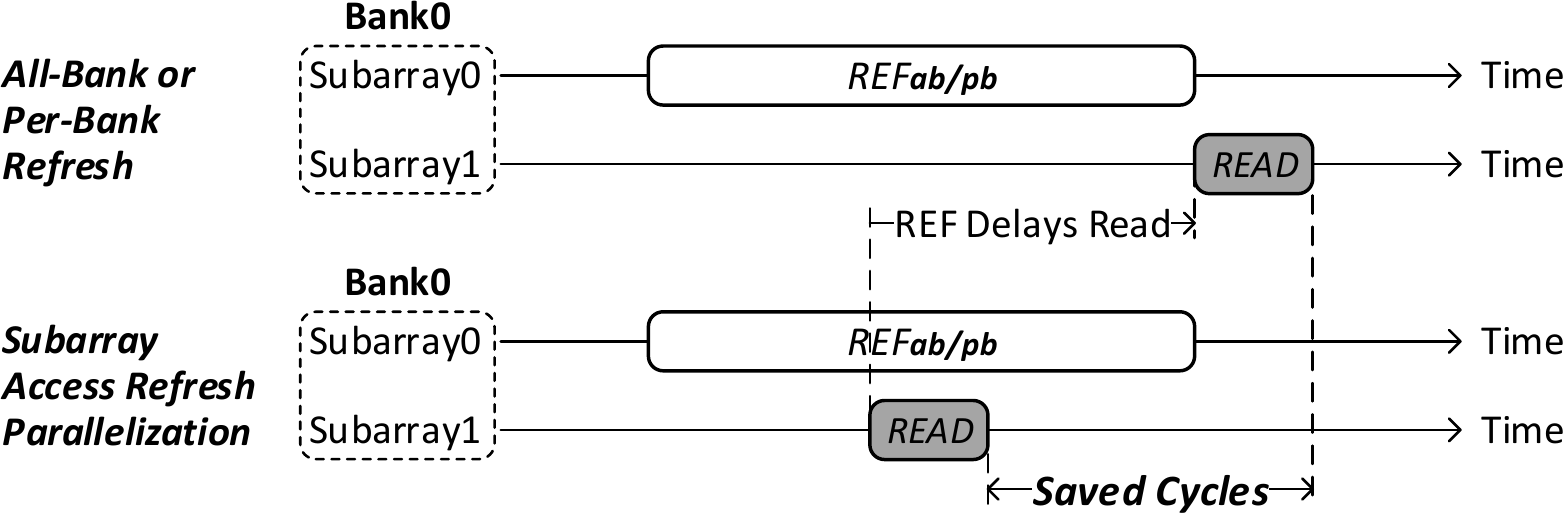}{0.9}{Service timeline of a refresh and a read request to two
different subarrays within the same bank.}


\subsubsection{DRAM Bank Implementation for \is} \label{sec:sarp_implementation}
As opposed to \ib, \is requires modifications to DRAM to support accessing
subarrays individually. While subarrays are equipped with dedicated local
peripheral logic, what prevents the subarrays from being operated independently
is the global peripheral logic that is shared by all subarrays within a bank.

\begin{figure}[t]
\centering
\subcaptionbox{Existing organization without SARP.
  \label{fig:detail-bank-organization}}[0.49\linewidth]{
    \includegraphics[scale=0.7]{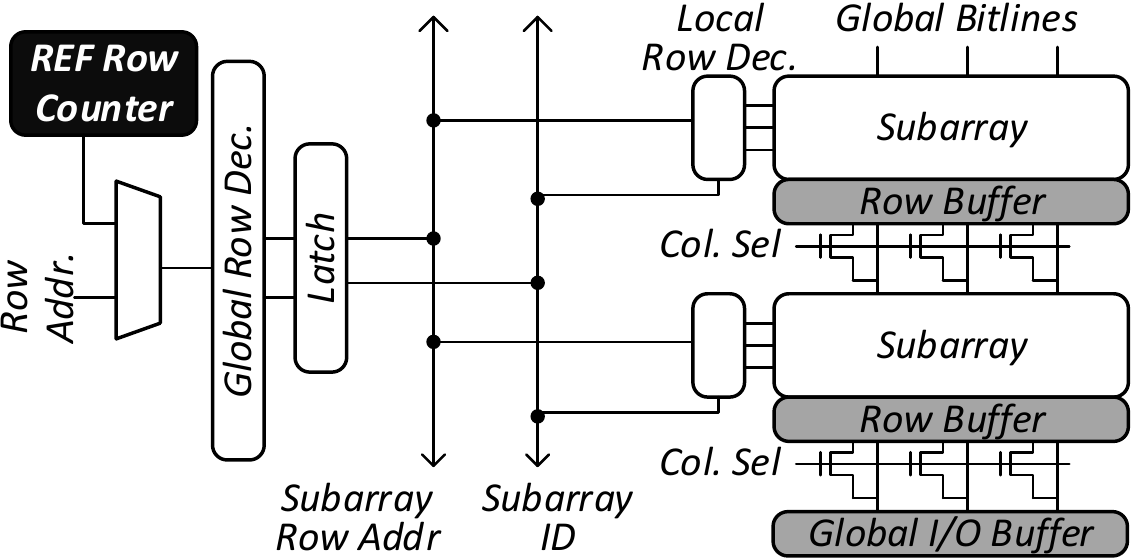}
}
\subcaptionbox{New organization with SARP.
  \label{fig:inter-sub-implementation}}[0.49\linewidth]{
    \includegraphics[scale=0.7]{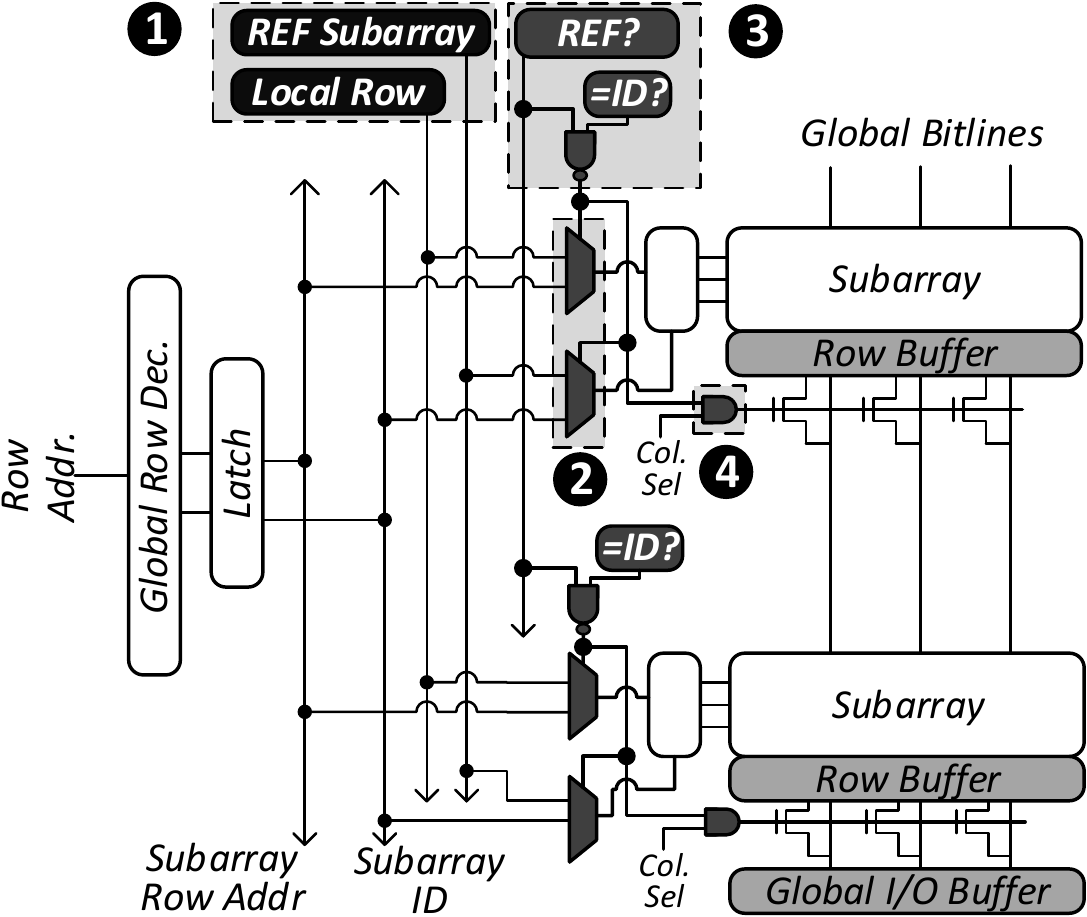}
}
\caption{DRAM bank without and with SARP.}
\end{figure}

\figref{detail-bank-organization} shows a detailed view of an existing DRAM
bank's organization. There are two major shared peripheral components within a
bank that prevent modern DRAM chips to refresh at subarray level. First, each
bank has a \emph{global row decoder} that decodes the incoming row's addresses.
To read or write a row, memory controllers first issue an \act command with the
row's address.
Upon receiving this command, the bank feeds the row address to the \emph{global
row decoder} that broadcasts the partially decoded address to all subarrays
within the bank. After further decoding, the row's subarray then raises its
wordline to begin transferring the row's cells' content to the row
buffer.\footnote{The detailed step-to-step explanation of the activation process
can be found in prior
works~\cite{kim-isca2012,lee-hpca2013,seshadri-micro2013}.} During the transfer,
the row buffer also restores the charge in the row. Similar to an \act,
refreshing a row requires the refresh unit to \act the row to restore its
electrical charge (only the refresh row counter is shown for clarity in
\figref{detail-bank-organization}).
Because a bank has only one global row decoder and one pair of address wires
(for subarray row address and ID), it cannot simultaneously activate two
different rows (one for a memory access and the other for a refresh).

Second, when the memory controller sends a read or write command, the required
column from the activated row is routed through the \emph{global bitlines} into
the \emph{global I/O buffer} (both of which are shared across all subarrays' row
buffers) and is transferred to the I/O bus. This is done by asserting a
\emph{column select} signal that is routed globally to {\em all} subarrays,
which enables {\em all} subarrays' row buffers to be concurrently connected to
the global bitlines. Since this signal connects all subarrays' row buffers to
the global bitlines at the same time, if more than one activated row buffer
(i.e., activated subarray) exists in the bank, an electrical short-circuit
occurs, leading to incorrect operation. As a result, two subarrays cannot be
kept activated when one is being read or written to, which prevents a refresh to
one subarray from happening concurrently with an access in a different subarray
in today's DRAM.

The key idea of \is is to allow the concurrent activation of multiple subarrays,
but to only connect the accessed subarray's row buffer to the global bitlines
while another subarray is refreshing.
\figref{inter-sub-implementation} shows
our proposed changes to the DRAM microarchitecture. There are two major enablers
of \is.


The first enabler of \is allows both refresh and access commands to
simultaneously select their designated rows and subarrays with three
new components. The first component (\circled{1}) provides the subarray and
row addresses for refreshes without relying on the global row
decoder. To achieve this, it decouples the refresh row counter into a
\emph{refresh-subarray} counter and a \emph{local-row} counter that
keep track of the currently refreshing subarray and the row address
within that subarray, respectively. The second component (\circled{2})
allows each subarray to activate a row for either a refresh or an
access through two muxes. One mux is a row-address selector and the
other one is a subarray selector. The third component (\circled{3})
serves as a control unit that chooses a subarray for refresh. The
\texttt{\small{REF?}} block indicates if the bank is currently under
refresh and the \texttt{\small{=ID?}} comparator determines if the
corresponding subarray's ID matches with the refreshing subarray
counter for refresh. These three components form a new address path
for the refresh unit to supply refresh addresses in parallel with
addresses for memory accesses.

The second enabler of \is allows accesses to one activated subarray while
another subarray is kept activated for refreshes. We add an \emph{AND} gate
(\circled{4}) to each subarray that ensures the refreshing subarray's row buffer
is {\em not} connected to the global bitlines when the {\em column select}
signal is asserted on an access. At any instance, there is at most one activated
subarray among all non-refreshing subarrays because the global row decoder
activates only one subarray at a time. With the two proposed enablers, \is
allows one activated subarray for refreshes in parallel with another activated
subarray that serves data to the global bitlines.

\subsubsection{Detecting Subarray Conflicts in the Memory Controller} To avoid
accessing a refreshing subarray, which is determined internally by the DRAM chip
in our current mechanism, the memory controller needs to know the current
refreshing subarray and the number of subarrays. We create shadow copies of the
\emph{refresh-subarray} and \emph{local-row} counters in the memory controller
to keep track of the currently-refreshing subarray.
We store the number of subarrays in an EEPROM called the \emph{serial presence
detect (SPD)}~\cite{jedec-spd}, which stores various timing and DRAM
organization information in existing DRAM modules. The memory controller reads
this information at system boot time so that it can issue commands
correctly.\footnote{Note that it is possible to extend our mechanisms such that
the memory controller specifies the subarray to be refreshed instead of the DRAM
chip. This requires changes to the DRAM interface.}




\subsubsection{Power Integrity} Because an \act draws a lot of current, DRAM
standards define two timing parameters to constrain the activity rate
of DRAM so that \acts do not over-stress the power delivery
network~\cite{jedec-ddr3,shevgoor-micro2013}.  The first parameter is
the \emph{row-to-row activation delay} (\trrd) that specifies the
minimum waiting time between two subsequent \act commands within a
DRAM device. The second is called the \emph{four activate window}
(\tfaw) that defines the length of a rolling window during which a
maximum of four \acts can be in progress. Because a refresh operation
requires activating rows to restore charge in DRAM cells, \is consumes
additional power by allowing accesses during refresh. To limit the
power consumption due to \acts, we further constrain the activity rate
by increasing both \tfaw and \trrd, as shown below. This results in
fewer \act commands issued during refresh.




{
\small
\vspace{-0.15in}
\begin{align}
\label{eq:1}
 &PowerOverhead_{FAW} = \frac{4*I_{ACT} + I_{REF}}{4*I_{ACT}} \\
\label{eq:2}
 &t_{FAW\_SARP} = t_{FAW} * PowerOverhead_{FAW} \\
\label{eq:3}
 &t_{RRD\_SARP} = t_{RRD} * PowerOverhead_{FAW}
\end{align}
\vspace{-0.2in}
}


{\small $I_{ACT}$} and {\small $I_{REF}$} represent the current values
of an \act and a refresh, respectively, based on the Micron
Power Calculator~\cite{micron-tr}. We calculate the power overhead of
parallelizing a refresh over a {\em four activate window} using
\eqref{eq:1}. Then we apply this power overhead to both \tfaw
\eqref{eq:2} and \trrd \eqref{eq:3}, which are enforced during refresh
operations. Based on the $IDD$ values in the Micron 8Gb
DRAM~\cite{micronDDR3_8Gb} data sheet, \is increases \tfaw and \trrd
by 2.1x during all-bank refresh operations. Each per-bank refresh consumes 8x
lower current than an all-bank refresh, thus increasing \tfaw and \trrd by only
13.8\%.


\subsubsection{Die Area Overhead} In our evaluations, we use 8 subarrays per
bank and 8 banks per DRAM chip. Based on this configuration, we calculate the
area overhead of \is using parameters from a Rambus DRAM model at 55$nm$
technology~\cite{rambus_powermodel}, the best publicly available model that we
know of, and find it to be 0.71\% in a 2Gb DDR3 DRAM chip with a die area of
73.5$mm^2$. The power overhead of the additional components is negligible
compared to the entire DRAM chip.

\ignore{
\subsection{Summary} We  proposed two mechanisms to hide
refresh latency by parallelizing refresh operations with memory
accesses at different granularities within the DRAM architecture. Our
first mechanism, \darp, exploits the refresh scheduling flexibility in
JEDEC standards to enable the memory controller to make more
intelligent decisions on when and which specific bank to refresh. The
second mechanism, \is, pushes the boundaries of today's DRAMs to
enable a DRAM to serve memory requests in idle subarrays while another
subarray is under refresh. As a result, it requires modifications to
the existing DRAM microarchitecture to achieve refresh-access
parallelization without impacting the array structures.  \is is
complementary to \darp and the combination of both, called \combo,
offers additive performance benefits with the implementation cost
incurred from both mechanisms.}

\section{Methodology}
\label{methodology}


To evaluate our mechanisms, we use an in-house cycle-level x86
multi-core simulator with a front end driven by
Pin~\cite{luk-pldi2005} and an in-house cycle-accurate DRAM timing
model validated against DRAMSim2~\cite{rosenfeld-cal2011}. Unless
stated otherwise, our system configuration is as shown in
Table~\ref{table:sys-config}.


%
%
%
%

\begin{table}[h]
\begin{footnotesize}
  \centering
    \setlength{\tabcolsep}{.55em}
    \begin{tabular}{ll}
        \toprule
\multirow{2}{*}{Processor} &
8 cores, 4GHz, 3-wide issue, 8 MSHRs/core,\\
& 128-entry instruction window
\\

        \cmidrule(rl){1-2}
\multirow{2}{*}{\begin{minipage}{0.5in}Last-level Cache\end{minipage}} &
64B cache-line, 16-way associative,\\
& 512KB private cache-slice per core \\

        \cmidrule(rl){1-2}
\multirow{3}{*}{\begin{minipage}{0.5in}Memory Controller\end{minipage}} &
64/64-entry read/write request queue, FR-FCFS~\cite{rixner-isca2000}, \\
& writes are scheduled in batches~\cite{chatterjee-hpca2012, lee-tech2010, stuecheli-isca2010}
with \\
& low watermark = 32, closed-row policy~\cite{chatterjee-hpca2012, kim-micro2010, rixner-isca2000} \\

        \cmidrule(rl){1-2}
        \multirow{2}{*}{DRAM} & DDR3-1333~\cite{micronDDR3_8Gb}, 2 channels, 2 ranks per
        channel,\\
            & 8 banks/rank, 8 subarrays/bank, 64K rows/bank, 8KB rows \\

        \cmidrule(rl){1-2}

\multirow{2}{*}{\begin{minipage}{0.5in}Refresh Settings\end{minipage}}
        & $tRFC_{ab}$ = 350/530/890ns for 8/16/32Gb
        DRAM chips,\\
        & $tREFI_{ab}$ = 3.9\micro s, $tRFC_{ab}$-to-$tRFC_{pb}$ ratio = 2.3  \\

        \bottomrule
    \end{tabular}
  \caption{Evaluated system configuration.}
  \label{table:sys-config}%
\end{footnotesize}
\end{table}



In addition to 8Gb DRAM, we also evaluate systems using 16Gb and 32Gb
near-future DRAM chips~\cite{itrs-dram}. Because commodity DDR DRAM does not
have support for \refpb, we estimate the \trfcpb values for DDR3 based on the
ratio of \trfc to \trfcpb in LPDDR2~\cite{micronLPDDR2_2Gb} as described in
Section~\ref{sec:refresh-challenge}. We evaluate our systems with 32ms retention
time, which is a typical setting for a server environment and LPDDR DRAM, as
also evaluated in previous work~\cite{nair-hpca2013,stuecheli-micro2010}.

We use benchmarks from \emph{SPEC CPU2006~\cite{spec2006}, STREAM~\cite{stream},
TPC~\cite{tpc}}, and a microbenchmark with random-access behavior similar to
HPCC RandomAccess~\cite{randombench}. We classify each benchmark as either
memory intensive (MPKI $\ge$ 10) or memory non-intensive (MPKI $<$ 10). We then
form five intensity categories based on the fraction of memory intensive
benchmarks within a workload: 0\%, 25\%, 50\%, 75\%, and 100\%. Each category
contains 20 randomly mixed workloads, totaling to 100 workloads for our main
evaluations. For sensitivity studies in Sections~\ref{sec:sense_core},
\ref{sec:sense_tfaw}, \ref{sec:sense_sa}, and \ref{sec:sense_ref}, we run 16
randomly selected memory-intensive workloads using 32Gb DRAM to observe the
performance trend.

We measure system performance with the commonly-used \emph{weighted
  speedup (WS)}~\cite{eyerman-ieeemicro2008,snavely-asplos2000}
metric. To report the DRAM system power, we use the methodology from
the \emph{Micron power calculator}~\cite{micron-tr}. The DRAM device
parameters are obtained from~\cite{micronDDR3_8Gb}. Every workload
runs for 256 million cycles to ensure the same number of refreshes. We
report DRAM system power as \emph{energy per memory access serviced}
to fairly compare across different workloads.

\section{Evaluation}
\label{sec:parref_evaluation}

In this section, we evaluate the performance of the following
mechanisms: 1) the \emph{all-bank} refresh scheme (\refab), 2) the
\emph{per-bank} refresh scheme (\refpb), 3) elastic
refresh~\cite{stuecheli-micro2010}, 4) our first mechanism, \ib, 5)
our second mechanism, \is, that is applied to either \refab (\isab) or
\refpb (\ispb), 6) the combination of \ib and \ispb, called \combo,
and 7) an ideal scheme that eliminates refresh. Elastic
refresh~\cite{stuecheli-micro2010} takes advantage of the refresh
scheduling flexibility in the DDR standard: it postpones a refresh if
the refresh is predicted to interfere with a demand request, based on
a prediction of how long a rank will be idle, i.e., without any demand
request.

\subsection{Multi-Core Results}
\label{sec:multi-core-results}

\figref{scurves} plots the system performance improvement of \refpb, \ib, \ispb,
and \combo over the all-bank refresh baseline (\refab) using various densities
across 100 workloads (sorted based on the performance improvement due to \ib).
The x-axis shows the sorted workload numbers as categorized into five
memory-intensive groups with 0 to 19 starting in the least memory-intensive
group and 80 to 99 in the most memory-intensive one.
Table~\ref{table:scurve_summary} shows the maximum and geometric mean of system
performance improvement due to our mechanisms over \refpb and \refab for
different DRAM densities. We draw five key conclusions from these results.

\figputGTW{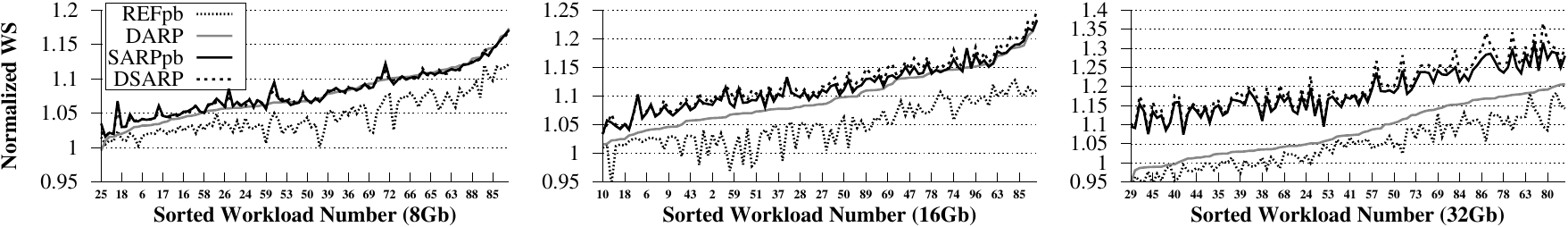}{Multi-core system performance improvement over
\caprefab across 100 workloads.}

First, \ib provides system performance gains over both \refpb and
\refab schemes: 2.8\%/4.9\%/3.8\% and 7.4\%/9.8\%/8.3\% on average in
8/16/32Gb DRAMs, respectively. The reason is that \ib hides refresh
latency with writes and issues refresh commands in out-of-order
fashion to reduce refresh interference on reads. Second, \ispb
provides significant system performance improvement over \ib and
refresh baselines for all the evaluated DRAM densities as \ispb
enables accesses to idle subarrays in the refreshing banks.  \ispb's
average system performance improvement over \refpb and \refab is
3.3\%/6.7\%/13.7\% and 7.9\%/11.7\%/18.6\% in 8/16/32Gb DRAMs,
respectively.  Third, as density increases, the performance
benefit of \ispb over \ib gets larger. This is because the longer
refresh latency becomes more difficult to hide behind writes or idle
banks for \ib. This is also the reason why the performance improvement
due to \ib drops slightly at 32Gb compared to 16Gb.  On the other
hand, \ispb is able to allow a long-refreshing bank to serve some
memory requests in its subarrays.

Fourth, combining both \ispb and \darp (\combo) provides additive system
performance improvement by allowing even more parallelization of
refreshes and memory accesses. As DRAM density (refresh latency)
increases, the benefit becomes more apparent, resulting in improvement
up to 27.0\% and 36.6\% over \refpb and \refab in 32Gb DRAM,
respectively.

Fifth, \refpb performs worse than \refab for some workloads (the
curves of \refpb dropping below one) and the problem is exacerbated
with longer refresh latency. Because \refpb commands cannot overlap
with each other~\cite{jedec-lpddr3}, their latencies are
serialized. In contrast, \refab operates on every bank in parallel, which is
triggered by a single command that partially overlaps refreshes across
different banks~\cite{mukundan-isca2013}. Therefore, in a pathological
case, the \refpb latency for refreshing every bank (eight in most
DRAMs) in a rank is
$8 \times tRFC_{pb} = 8 \times
  \frac{tRFC_{ab}}{2.3} \approx 3.5 \times tRFC_{ab}$, whereas all-bank
refresh takes \trfc (see Section~\ref{sec:refresh-challenge}). If a
workload cannot
effectively utilize multiple banks during a per-bank refresh operation, \refpb
may potentially degrade system performance compared to \refab.

\begin{table}[h]
\small
\centering
\renewcommand{\arraystretch}{0.7}
    \begin{tabular}{llrrrr}
      \toprule
      \multirow{2}{*}{\textbf{Density}} &
      \multirow{2}{*}{\textbf{Mechanism}} &
      \multicolumn{2}{c}{\textbf{Max (\%)}} &
      \multicolumn{2}{c}{\textbf{Gmean (\%)}} \\
      & & \refpb & \refab & \refpb & \refab \\
      \midrule

      \multirow{3}{*}{8Gb} & {\ib} & 6.5 & 17.1 & 2.8 & 7.4\\
      &{\ispb} & 7.4 & 17.3 & 3.3 & 7.9\\
      &{\combo} & 7.1 & 16.7 & 3.3 & 7.9\\
      \midrule

      \multirow{3}{*}{16Gb} & {\ib} & 11.0 & 23.1& 4.9& 9.8\\
      &{\ispb} & 11.0 & 23.3 & 6.7 & 11.7\\
      &{\combo} & 14.5 & 24.8 & 7.2 & 12.3\\
      \midrule

      \multirow{3}{*}{32Gb} & {\ib} & 10.7 & 20.5 & 3.8 & 8.3\\
      &{\ispb} & 21.5 & 28.0 & 13.7 & 18.6\\
      &{\combo} & 27.0 & 36.6 & 15.2 & 20.2\\

      \bottomrule
    \end{tabular}
\caption{Maximum and average WS improvement due to our mechanisms
over \caprefpb and \caprefab.}
\label{table:scurve_summary}%
\end{table}

\subsubsection{All Mechanisms' Results}
Figure~\ref{fig:multi-core-perf-density-sweep} shows the average
performance improvement due to all the evaluated refresh mechanisms
over \refab.  The weighted speedup value for \refab is 5.5/5.3/4.8
using 8/16/32Gb DRAM density.  We draw three major conclusions. First,
using \is on all-bank refresh (\isab) also significantly improves
system performance. This is because \is allows a rank to continue
serving memory requests while it is refreshing. Second, elastic
refresh does not substantially improve performance, with an average of
1.8\% over all-bank refresh.  This is because elastic refresh does not
attempt to pull in refresh opportunistically, nor does it try to
overlap refresh latency with other memory accesses. The observation is
consistent with prior work~\cite{nair-hpca2013}. Third, \combo
captures most of the benefit of the ideal baseline ("No REF"),
performing within 0.9\%, 1.2\%, and 3.7\% of the ideal for 8, 16, and
32Gb DRAM, respectively.

\figputGHS{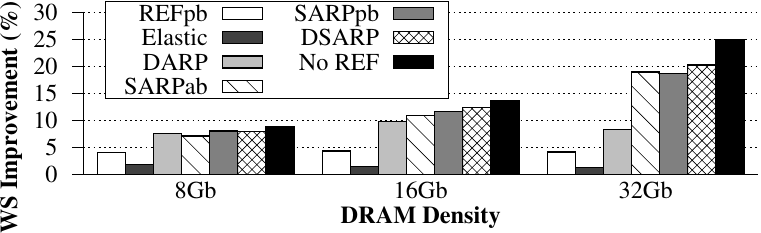}{1.3}{Average system performance
improvement over \caprefab.}

\subsubsection{Performance Breakdown of \ib} To understand the observed
performance gain in more detail, we evaluate the performance of \darp's two
components separately. \emph{\Ooolong} improves performance by 3.2\%/3.9\%/3.0\%
on average and up to 16.8\%/21.3\%/20.2\% compared to \refab in 8/16/32Gb DRAMs.
Adding \emph{\warplong} to \emph{\ooolong} (\ib) provides additional performance
gains of 4.3\%/5.8\%/5.2\% on average by hiding refresh latency with write
accesses.





\subsubsection{Energy}

Our techniques reduce energy per memory access compared to existing
policies, as shown in \figref{energy_reduction}. The main reason is
that the performance improvement reduces average static energy for
each memory access. Note that these results conservatively assume the
same power parameters for 8, 16, and 32 Gb chips, so the savings in
energy would likely be more significant if realistic power parameters
are used for the more power-hungry 16 and 32 Gb nodes.

\figputGHS{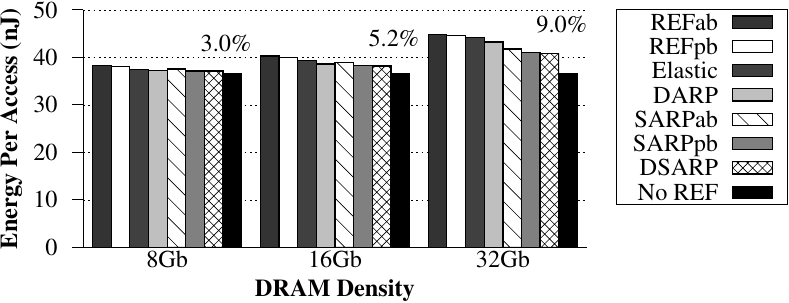}{1.3}{Energy consumption. Value on top indicates
percentage reduction of \combo compared to
\caprefab.}

\subsubsection{Effect of Memory Intensity}
Figure~\ref{fig:mpki_ws} shows the performance improvement of \combo
compared to \refab and \refpb on workloads categorized by memory
intensity (\% of memory-intensive benchmarks in a workload),
respectively. We observe that \combo outperforms \refab and \refpb
consistently. Although the performance improvement of \combo over
\refab increases with higher memory intensity, the gain over \refpb
begins to plateau when the memory intensity grows beyond 25\%. This is
because \refpb's benefit over \refab also increases with memory
intensity as \refpb enables more accesses to be be parallelized with
refreshes. Nonetheless, our mechanism provides the highest system
performance compared to prior refresh policies.

\figputGHS{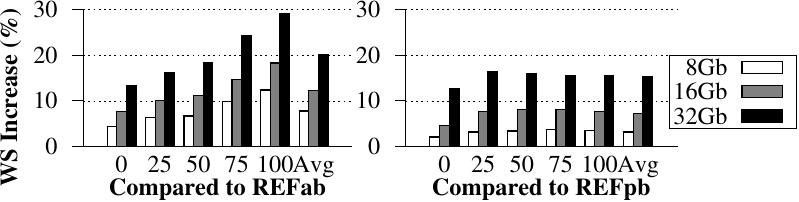}{1.3}{WS improvement of {\small \combo}
over {\small \caprefab} and {\small \caprefpb} as memory intensity and DRAM density vary.}


\subsubsection{Effect of Core Count}
\label{sec:sense_core}

Table~\ref{table:core_sweep} shows the weighted speedup, harmonic speedup,
fairness, and energy-per-access improvement due to \combo compared to \refab for
systems with 2, 4, and 8 cores. For all three systems, \combo consistently
outperforms the baseline without unfairly penalizing any specific application.
We conclude that \combo is an effective mechanism to improve performance,
fairness and energy of multi-core systems employing high-density DRAM.

\begin{table}[h]
\small
\centering
    \setlength{\tabcolsep}{.35em}
    \begin{tabular}{lrrr}
      \toprule
      \textbf{Number of Cores} & 2 & 4 & 8 \\
      \midrule
      \textbf{Weighted Speedup Improvement (\%)} & 16.0 & 20.0 & 27.2 \\
      \textbf{Harmonic Speedup Improvement~\cite{luo-ispass2001} (\%)} & 16.1 & 20.7 & 27.9 \\
      \textbf{Maximum Slowdown Reduction~\cite{das-micro2009, kim-hpca2010,
      kim-micro2010} (\%)} & 14.9 & 19.4 & 24.1 \\
      \textbf{Energy-Per-Access Reduction (\%)} & 10.2 & 8.1 & 8.5 \\
      \bottomrule
    \end{tabular}
\caption{Effect of \combo on multi-core system metrics.}
\label{table:core_sweep}
\end{table}


\subsection{Effect of \tfaw}
\label{sec:sense_tfaw}
Table~\ref{table:faw_sweep} shows the performance improvement of \ispb
over \refpb when we vary \tfaw in DRAM cycles (20 cycles for the
baseline as specified by the data sheet) and when \trrd scales
proportionally with \tfaw.\footnote{We evaluate only \ispb because it
  is sensitive to \tfaw and \trrd as it extends these parameters
  during parallelization of refreshes and accesses to compensate for
  the power overhead.} As \tfaw reduces, the performance benefit of
\ispb increases over \refpb. This is because reduced \tfaw enables
more accesses/refreshes to happen in parallel, which our mechanism
takes advantage of.

\begin{table}[h]
\small
\centering
    \setlength{\tabcolsep}{.33em}
    \begin{tabular}{lrrrrrr}
      \toprule
      \textbf{\tfaw/\trrd} (DRAM cycles) & 5/1 & 10/2 & 15/3 & \textbf{20/4} &
      25/5 & 30/6\\
      \midrule
      \textbf{WS Improvement (\%)} & 14.0 & 13.9 & 13.5 & \textbf{12.4} & 11.9 & 10.3\\
      \bottomrule
    \end{tabular}
\caption{Performance improvement due to \ispb over \caprefpb with various
\tfaw and \trrd values.}
\label{table:faw_sweep}
\end{table}

\vspace{-0.08in}

\subsection{Effect of Subarrays-Per-Bank}
\label{sec:sense_sa}
Table~\ref{table:sa_sweep} shows that the average performance
gain of \ispb over \refpb increases as the number of subarrays
increases in 32Gb DRAM. This is because with more subarrays, the
probability of memory requests to a refreshing subarray reduces.

\begin{table}[h]
\small
\centering
    \setlength{\tabcolsep}{.33em}
    \begin{tabular}{lrrrrrrr}
      \toprule
      \textbf{Subarrays-per-bank} & 1 & 2 & 4 & \textbf{8} & 16 & 32 & 64\\
      \midrule
      \textbf{WS Improvement (\%)} & 0 & 3.8 & 8.5 & \textbf{12.4} & 14.9 & 16.2
      & 16.9\\
      \bottomrule
    \end{tabular}
\caption{Effect of number of subarrays per bank.}
\label{table:sa_sweep}
\end{table}

\vspace{-0.08in}

\subsection{Effect of Refresh Interval}
\label{sec:sense_ref}

For our studies so far, we use 32ms retention time (i.e., \trefi =
3.9$\mu$s) that represents a typical setting for a server environment
and LPDDR DRAM~\cite{jedec-lpddr3}. Table~\ref{table:64ms_summary}
shows the performance improvement of \combo over two baseline refresh
schemes using retention time of \emph{64ms} (i.e., \trefipb =
7.8$\mu$s). \combo consistently provides performance gains over both
refresh schemes. The maximum performance improvement over \refpb is
higher than that over \refab at 32Gb because \refpb actually degrades
performance compared to \refab for some workloads, as discussed in the
32ms results (Section~\ref{sec:multi-core-results}).

\begin{table}[h]
\small
\centering
\renewcommand{\arraystretch}{0.65}
    \begin{tabular}{lrrrr}
      \toprule
      \multirow{2}{*}{\textbf{Density}} &
      \multicolumn{2}{c}{\textbf{Max (\%)}} &
      \multicolumn{2}{c}{\textbf{Gmean (\%)}} \\
      & \refpb & \refab & \refpb & \refab \\
      \midrule

      {8Gb} &  2.5   & 5.8 & 1.0& 3.3 \\
      {16Gb}&  4.6   & 8.6 & 2.6& 5.3 \\
      {32Gb} & 18.2  & 13.6& 8.0& 9.1 \\

      \bottomrule
    \end{tabular}
\caption{Maximum and average WS improvement due to \combo.}
\label{table:64ms_summary}%
\end{table}

\subsection{DDR4 Fine Granularity Refresh}
\label{sec:ddr4}

DDR4 DRAM supports a new refresh mode called \emph{fine granularity
  refresh (FGR)} in an attempt to mitigate the increasing refresh
latency (\trfc)~\cite{jedec-ddr4}. \fgr trades off shorter \trfc with
faster refresh rate (\sfrac{1}{\trefi}) that increases by either 2x
or 4x. \figref{ddr4_ref} shows the effect of \fgr in comparison to
\refab, \emph{adaptive refresh policy (AR)}~\cite{mukundan-isca2013},
and \combo. 2x and 4x \fgr actually reduce average system performance
by 3.9\%/4.0\%/4.3\% and 8.1\%/13.7\%/15.1\% compared to \refab with
8/16/32Gb densities, respectively. As the refresh rate increases by
2x/4x (higher refresh penalty), \trfc does not scale down with the
same constant factors. Instead, \trfc reduces by 1.35x/1.63x with
2x/4x higher rate~\cite{jedec-ddr4}, thus increasing the worst-case
refresh latency by 1.48x/2.45x. This performance degradation due to
\fgr has also been observed in Mukundan et
al.~\cite{mukundan-isca2013}.  AR~\cite{mukundan-isca2013} dynamically
switches between 1x (i.e., \refab) and 4x refresh modes to mitigate
the downsides of \fgr.  AR performs slightly worse than \refab (within
1\%) for all densities. Because using 4x \fgr greatly degrades
performance, AR can only mitigate the large loss from the 4x mode and
cannot improve performance over \refab. On the other hand, \combo is a
more effective mechanism to tolerate the long refresh latency than
both \fgr and AR as it overlaps refresh latency with access latency
without increasing the refresh rate.


\figputGHS{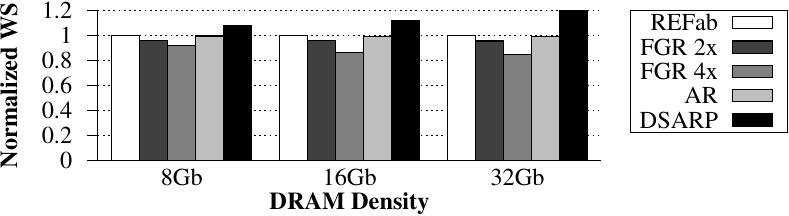}{1.3}{Performance comparisons to FGR and AR~\cite{mukundan-isca2013}.}

\section{Summary}

We introduced two new complementary techniques, \ib (\darplong) and
\is (\intersub), to mitigate the DRAM refresh penalty by enhancing
\emph{refresh-access parallelization} at the bank and subarray levels,
respectively.  \darp 1) issues per-bank refreshes to idle banks in an
out-of-order manner instead of issuing refreshes in a strict
round-robin order, 2) proactively schedules per-bank refreshes during
intervals when a batch of writes are draining to DRAM.  \is enables a
bank to serve requests from idle subarrays in parallel with other
subarrays that are being refreshed. Our extensive evaluations on a
wide variety of systems and workloads show that these mechanisms
significantly improve system performance and outperform
state-of-the-art refresh policies, approaching the performance of
ideally eliminating all refreshes. We conclude that \ib and \is are
effective in hiding the refresh latency penalty in modern and
near-future DRAM systems, and that their benefits increase as DRAM
density increases.


\chapter{FLY-DRAM: Understanding and Exploiting Latency Variation in DRAM}
\label{chap:latvar}

\vspace{-0.1in}
DRAM standards define a fixed value for each of the timing
parameters, which determine the latency of DRAM operations.  Unfortunately,
these latencies do \emph{not} reflect the \emph{actual} time the DRAM operations
take for each cell. This is because the true access latency varies for each
cell, as every cell is different in size and strength due to manufacturing
process variation effects. For simplicity, and to ensure that DRAM yield remains
high, DRAM manufacturers define a single set of latencies that guarantees
reliable operation, based on the \emph{slowest} cell in {\em any} DRAM chip
across \emph{all} DRAM vendors. As a result, there is a significant opportunity
to reduce DRAM latency if, instead of always using worst-case latencies, we
employ the true latency for each cell \newt{that enables the three operations
  reliably}.

\new{\textbf{Our goal} in this chapter is to {\em (i)}~understand the
impact of cell variation in the three fundamental DRAM operations for cell
access (activation, precharge, and restoration); {\em (ii)}~experimentally
characterize the latency variation in these operations; and {\em (iii)}~develop
new mechanisms that take advantage of this variation to reduce the latency of
these three operations.}


\section{Motivation}

The latencies of the three DRAM operations (\emph{activation}, \emph{precharge},
and \emph{restoration}), as defined by vendor specifications, have \emph{not}
improved significantly in the past 18 years, as depicted in
Figure~\ref{fig:latency_trend}. This is especially true when we compare
\new{latency improvements} to the capacity (128$\times$=$\frac{16Gb}{128Mb}$) and
bandwidth improvements
(20$\times$$\approx\frac{2666MT/s}{133MT/s}$)~\cite{jedec-ddr3, jedec-ddr4,
  lee-hpca2013, son-isca2013, lee-hpca2015} commodity DRAM chips experienced in
the past 18 years. In fact, the activation and precharge latencies
\emph{increased} from 2013 to 2015, when DDR \new{DRAM} transitioned from the
third generation (12.5ns for DDR3-1600J~\cite{jedec-ddr3}) to the fourth
generation (14.06ns for DDR4-2133P~\cite{jedec-ddr4}). As the latencies
specified by vendors have not reduced over time, the system performance
bottleneck caused by raw main memory latency remains largely unaddressed in
modern systems.

\figputHSL{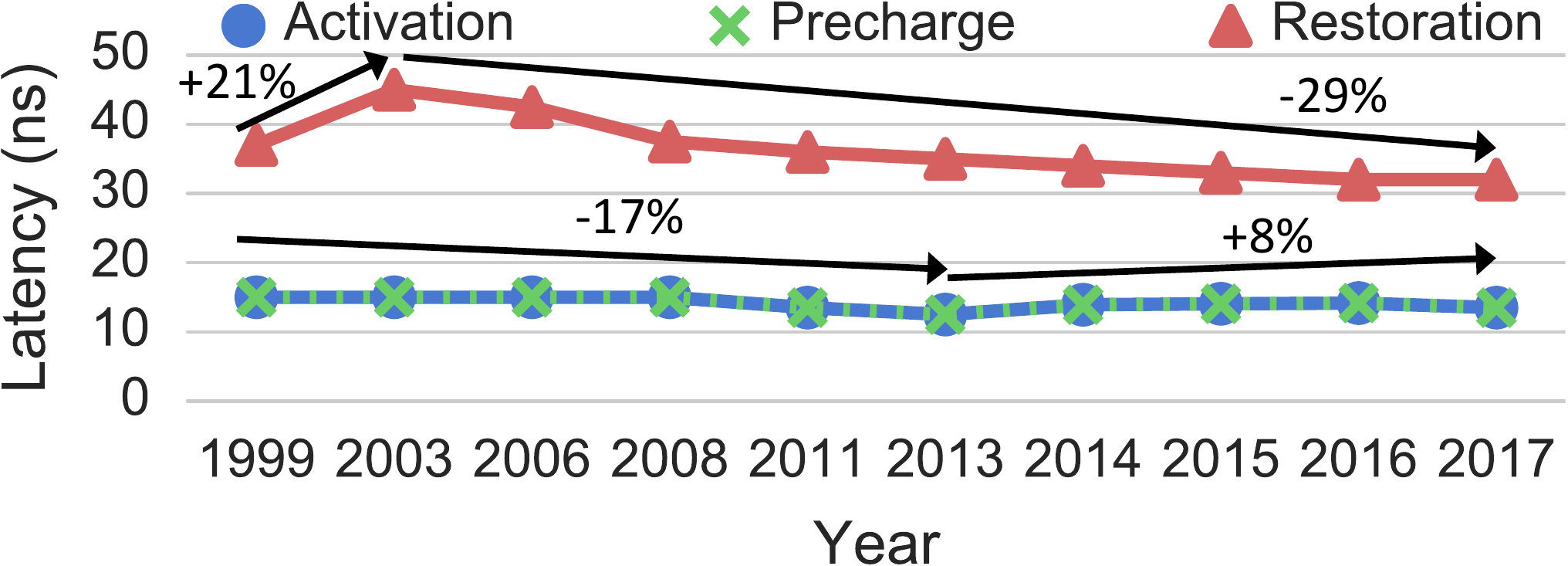}{0.6}{DRAM latency trends over
time~\cite{micronSDR_128Mb,jedec-ddr2,jedec-ddr3,jedec-ddr4}.}{latency_trend}

In this chapter, we observe that \new{the three fundamental DRAM} operations can
\emph{actually} complete with a much lower latency for many DRAM cells than the
specification, because \emph{there is inherent latency variation present across
  the DRAM cells within a DRAM chip}.  This is a result of manufacturing process
variation, which causes the \emph{sizes} and \emph{strengths} of cells to be
different, thus making some cells inherently faster and other cells inherently
slower to be accessed
reliably~\cite{li11}. The speed gap between the fastest and the slowest DRAM
cells is getting worse~\cite{kanad_dram_book, pvt_book}, as the technology node
continues to scale down to sub-20nm feature sizes. Unfortunately, instead of
optimizing the latency specifications for the common case, DRAM vendors use a
single set of standard access latencies, which provide reliable operation
guarantees for the \emph{worst case} (i.e., the slowest cells), to maximize
manufacturing yield.

We find that the (widening) speed gap among DRAM cells presents an opportunity
to reduce DRAM access latency.  If we can understand and characterize the
inherent variation in cell latencies, we can use the resulting understanding to
reduce the access latency for those rows that contain faster cells.
\textbf{The goal of this chapter} is to {\em (i)}~experimentally characterize and
understand the impact of latency variation in the three fundamental DRAM
operations for cell access (activation, precharge, and restoration), and
{\em(ii)}~develop new mechanisms that take advantage of this variation to
improve system performance.

To this end, we build an FPGA-based DRAM testing infrastructure and characterize
240 DRAM chips from three major vendors. We analyze the variations in the
latency of the three fundamental DRAM operations by operating DRAM at multiple
reduced latencies.

\section{Experimental Methodology}
\label{sec:exp_meth}

To study the effect of using different timing parameters on modern DDR3 DRAM
chips, we developed a DRAM testing platform that allows us to precisely control
the value of timing parameters and the tested DRAM location (i.e., banks, rows,
and columns) within a module. The testing platform, shown in
\figref{dram_fpga_white}, consists of Xilinx FPGA boards~\cite{xilinx-ml605}
and host PCs. We use the RIFFA~\cite{jacobsen-rts2015} framework to communicate
data over the PCIe bus from our customized \emph{testing software} running on
the host PC to our customized \emph{test engine} on the FPGA. Each DRAM module
is tested on an FPGA board, and is located inside a heat chamber that is
connected to a temperature controller. Unless otherwise specified, we test
modules at an ambient temperature of 20$\pm$1\celsius. \new{We examine various
    temperatures in Section~\ref{sec:rcd_temp}.}

\figputHS{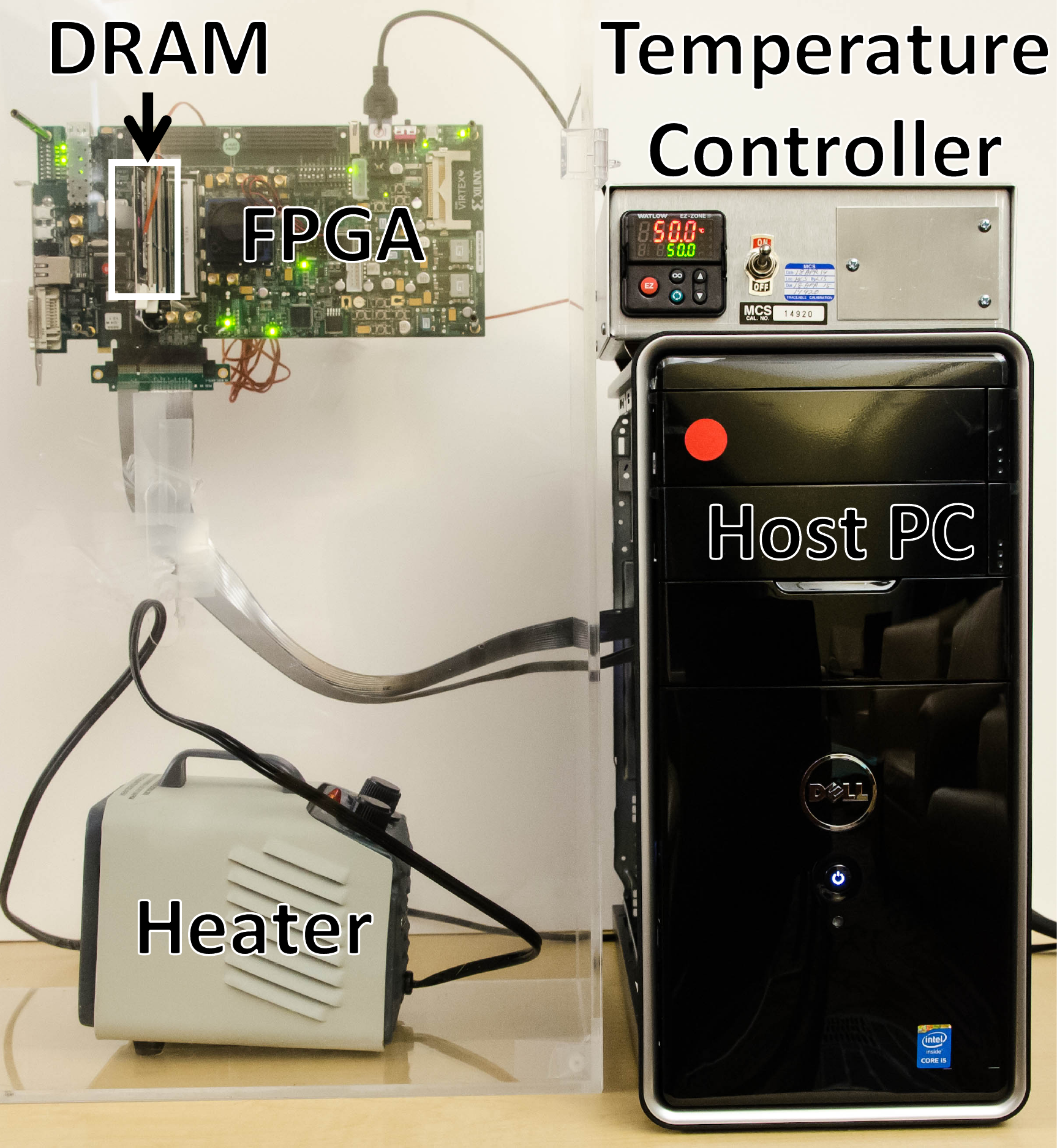}{0.4}{FPGA-based DRAM testing infrastructure.}

\subsection{DRAM Test}

To achieve the goal of controlling timing parameters, our FPGA test engine
supports a list of DRAM commands that get processed directly by the memory
controller on the FPGA. Then, on the host PC, we can write a \emph{test} that
specifies a sequence of DRAM commands along with the delay between the commands
(i.e., timing parameters).  The test sends the commands and delays from the
host PC to the FPGA test engine.

Test~\ref{readcl} shows the pseudocode of a test that reads a cache line from a
particular bank, row, and column with timing parameters that can be specified by
the user. The test first sends an \act to the target row (line 2). After a
\trcd delay that we specify (line 3), it sends a \crd (line 4) to the target
cache line. Our test engine enables us to specify the exact delay between two
DRAM commands, thus allowing us to tune certain timing parameters. The read
delay (\tcl) and data transfer latency (\newthree{\tbl}) are two DRAM internal
timings that \emph{cannot} be changed using our infrastructure. After our test
waits for the data to be fully transferred (line 5), we precharge the bank (line 6) with our
specified \trp (line 7). We describe the details of the tests that we created
to characterize latency variation of \trcd, \trp, and \tras in the next few
sections.


\floatname{algorithm}{Test}

\begin{algorithm}

\begin{algorithmic}[1]
\algrenewcommand\algorithmicfunction{}
\algrenewcommand\algorithmicdo{}
\algrenewcommand\alglinenumber[1]{\footnotesize\texttt{#1}}
\algrenewcommand\algorithmicindent{1.2em}
\small

\Function{ReadOneCacheLine}{$\mathit{my\_tRCD}, \mathit{my\_tRP}, \mathit{bank},
\mathit{row}, \mathit{col}$}
\State {\tt ACT(}$\mathit{bank}, \mathit{row}${\tt )}
\State {\tt cmdDelay(}$\mathit{my\_tRCD}${\tt )} \Comment{Set activation latency
(\trcd)}
\State {\tt READ(}$\mathit{bank}, \mathit{row}, \mathit{col}${\tt )}
\State {\tt cmdDelay(}\tcl+\tbl{\tt )} \Comment{Wait for
read to finish}
\State {\tt PRE(}$\mathit{bank}${\tt )}
\State {\tt cmdDelay(}$\mathit{my\_tRP}${\tt )} \Comment{Set precharge latency
(\trp)}

\State {\tt readData()} \Comment{Send the read data from FPGA to PC}

\EndFunction
\end{algorithmic}

\caption{Read a cache line with specified timing parameters.}
\label{readcl}
\end{algorithm}

\ignore{
\subsection{Tested Timing Parameters}

Although we have discussed four timing parameters, in this work, we present our
study on the following three timing parameters: 1) \trcd the activation latency,
2) \trp the precharge latency, and 3) \tras the restoration latency. The reason
we didn't examine \tcl is that it is a DRAM internal timing that is not
controlled by the memory controller. It does not define the latency between two
DRAM commands, but the time for the data to be read/written for a column
\crd/\cwr command. As a result, we cannot control \tcl for every memory request
at the memory controller.
}

\subsection{Characterized DRAM Modules}

We \new{characterize} latency variation on a total of 30 DDR3 DRAM modules,
comprising 240 DRAM chips, from the three major DRAM vendors that hold more than
90\% of the market share~\cite{bauer-mckinsey2016}. \tabref{dimm_list_latvar} lists the
\new{relevant} information about the tested DRAM modules. All of these modules are
\emph{dual in-line} (i.e., 64-bit data bus) with a single rank of DRAM chips.
Therefore, we use the terms \emph{DIMM} (dual in-line memory module) and module
interchangeably. In the rest of the chapter, we refer to a specific DIMM using the
label \dimm{n}{v}, where $n$ and $v$ stand for the DIMM number and vendor,
respectively. In the table, we group the DIMMs based on their model number,
which provides certain information on the process technology and array design
used in the chips.


\renewcommand{\arraystretch}{1.5}
\begin{table}[h]
\small
  \centering
    \setlength{\tabcolsep}{.5em}
    \begin{tabular}{ccccc}
        \toprule
        \multirow{2}{*}{Vendor} & DIMM &
        \multirow{2}{*}{Model} & Timing (ns) & Assembly\\
                               & Name & & (\trcd/\trp/\tras) & Year \\
        \midrule

        \multirow{4}{5em}{\centering A \\ \tiny \quad \\ \centering Total of \\
        \centering 8 DIMMs}
         & \dimm{0-1}{A} & M0 & 13.125/13.125/35 & 2013\\
         & \dimm{2-3}{A} & M1 & 13.125/13.125/36 & 2012 \\
         & \dimm{4-5}{A} & M2 & 13.125/13.125/35 & 2013 \\
         & \dimm{6-7}{A} & M3 & 13.125/13.125/35 & 2013 \\

        \midrule

        \multirow{2}{5em}{\centering B \\ \tiny \centering Total of \\
        \centering 9 DIMMs}
         & \dimm{0-5}{B} & M0 & 13.125/13.125/35 & 2011-12 \\
         & \dimm{6-8}{B} & M1 & 13.125/13.125/35 & 2012 \\

        \midrule

        \multirow{2}{5em}{\centering C \\ \tiny \centering Total of \\
        \centering 13 DIMMs}
         & \dimm{0-5}{C} & M0 & 13.125/13.125/34 & 2012 \\
         & \dimm{6-12}{C} & M1 & 13.125/13.125/36 & 2011 \\

        \bottomrule
    \end{tabular}
  \caption{\newt{Properties of tested DIMMs}.}
  \label{tab:dimm_list_latvar}%
\end{table}

%
%

%


\section{Activation Latency Analysis}
\label{sec:act_lat_analysis}

In this section, we present our methodology and results on varying the
activation latency, which is expressed by the \trcd timing
parameter. We first describe the nature of errors caused by \trcd
reduction in Section~\ref{sec:trcd_errors}. Then, we describe the FPGA
test we conducted on the DRAM modules to characterize \trcd variation in
Section~\ref{sec:trcd_tests}.  The remaining sections describe
different major observations we make based on our results.

\subsection{Behavior of Activation Errors}
\label{sec:trcd_errors}

As we discuss in Section~\ref{sec:dram_access}, \trcd is defined as the minimum
amount of time between the \act and the first column command (\crd/\cwr).
Essentially, \trcd represents the time it takes for a row of sense amplifiers
(i.e., the row buffer) to sense and latch a row of data. By employing a lower
\trcd value, a column \crd command may potentially read data from sense
amplifiers that are still in the \emph{sensing and amplification} phase, during
which the data has not been fully latched into the sense amplifiers. As a
result, reading data with a lowered \trcd can induce timing errors (i.e.,
flipped bits) in the data.

To further understand the nature of activation errors, we perform experiments to
answer two fundamental questions: \romnum{i}~Does lowering \trcd incur errors on
\new{\emph{all}} cache lines read from a sequence of \crd commands on an opened row?
\romnum{ii}~Do the errors propagate back to the DRAM cells, causing
\new{\emph{permanent}} errors for all future accesses?

\subsubsection{Errors Localized to First Column Command}
\label{sec:trcd_errors:first_col}

To answer the first question, we conduct Test~\ref{trcdlocal} that first
activates a row with a specific \trcd value, and then \new{reads} every
cache line in the entire row.
By conducting the test on every row in a number of DIMMs from all three vendors,
we make the following observation.

\floatname{algorithm}{Test}

\begin{algorithm}[h]

\algnewcommand\algorithmicto{\textbf{to}}
\algrenewtext{For}[3]{\algorithmicfor\ #1 $\gets$ #2 \algorithmicto\ #3 \algorithmicdo}

\algrenewcommand\algorithmicfunction{}
\algrenewcommand\algorithmicdo{}
\algrenewcommand\algorithmicindent{1.2em}
\algrenewcommand\alglinenumber[1]{\footnotesize\texttt{#1}}
\small

\begin{algorithmic}[1]
\Function{ReadOneRow}{$\mathit{my\_tRCD}$, $\mathit{bank}, \mathit{row}$}

%

\State {\tt ACT(}$\mathit{bank}, \mathit{row}${\tt )}
\State {\tt cmdDelay(}$\mathit{\textbf{my\_tRCD}}${\tt )} \Comment{Set activation latency}

\For{c}{1}{$\mathit{Col}_{\mathit{MAX}}$}
\State {\tt READ(}$\mathit{bank}, \mathit{row}, \mathit{c}${\tt )} \Comment{Read
one cache line}
\State {\tt findErrors()} \Comment{Count errors in a cache line}
\EndFor
\State {\tt cmdDelay(}\tcl+ \tbl{\tt )}
\State {\tt PRE(}$\mathit{bank}${\tt )}
\State {\tt cmdDelay(}\trp{\tt )}


\EndFunction
\end{algorithmic}

\caption{Read one row with a specified \trcd value.}
\label{trcdlocal}
\end{algorithm}

\obs{Activation errors are isolated to the cache line from
the first \crd command, and do not appear in subsequently-read cache lines from
the same row.\label{obs:rcd_local}}

There are two reasons why errors do \new{\emph{not}} occur in the subsequent cache line
reads. First, a \crd accesses \new{only} its corresponding sense amplifiers, without
accessing the other columns. Hence, a \newt{\crd's effect is isolated to} its
target cache line. Second, by the time the second \crd is issued, a sufficient
amount of time has passed for the sense amplifiers to properly latch the data.
Note that this observation is independent of DIMMs and vendors as the
fundamental DRAM structure is similar across different DIMMs. We discuss
the number of activation errors due to different \trcd values for each DIMM in
\new{\secref{trcd_ber}.}

\subsubsection{Activation Errors Propagate into DRAM Cells}

To answer our second question, we run two iterations of Test~\ref{trcdlocal}
(i.e., reading a row that is activated with a specified \trcd value) on the same
row. The first iteration reads a row that is activated with a lower \trcd value,
then closes the row. The second iteration re-opens the row using the standard
\trcd value, and reads the data to confirm if the errors remain in the cells.
Our experiments show that if the first iteration observes activation errors
within a cache line, the second iteration observes the same errors. This
demonstrates that activation errors not only happen at the sense amplifiers but
also propagate back into the cells.

We hypothesize this is because reading a cache line early causes the sense
amplifiers to latch the data based on the \emph{current} bitline voltage. If the
bitline voltage has not yet fully developed into \vdd or \zerov, the sense
amplifier latches in unknown data and amplifies this data to the bitline, which
is then restored back into the cell during the restoration phase.

\obs{Activation errors occur at the sense amplifiers and propagate back into the
cells. The errors persist until the data is overwritten.}

\new{After observing that reducing activation latency results in timing errors,
    we now consider \newt{two new questions.}} First, after how much
activation latency reduction do DIMMs start observing timing errors? Second, how
many cells experience activation errors at each latency reduction step?

\subsection{FPGA Test for Activation \newt{Latency}}
\label{sec:trcd_tests}

To characterize activation errors across every cell in DIMMs, we need to perform
an \act and a \crd on one cache line at a time since activation errors only
occur in one cache line per activation. \new{To achieve this}, we use
Test~\ref{trcdtest}, \newt{whose pseudocode is below,} for every cache line within a row.

\floatname{algorithm}{Test}

\begin{algorithm}[h]

\algnewcommand\algorithmicto{\textbf{to}}
\algrenewtext{For}[3]{\algorithmicfor\ #1 $\gets$ #2 \algorithmicto\ #3 \algorithmicdo}

\algrenewcommand\algorithmicfunction{}
\algrenewcommand\algorithmicdo{}
\algrenewcommand\algorithmicindent{1.2em}
\algrenewcommand\alglinenumber[1]{\footnotesize\texttt{#1}}
\small

\begin{algorithmic}[1]
\Function{tRCDColOrderTest}{$\mathit{my\_tRCD}, \mathit{data}$}

\For{b}{1}{$\mathit{Bank}_{\mathit{MAX}}$}
\For{c}{1}{$\mathit{Col}_{\mathit{MAX}}$} \Comment{Column first}
\For{r}{1}{$\mathit{Row}_{\mathit{MAX}}$}

\State {\tt WriteOneCacheLine(}$\mathit{b}, \mathit{r}, \mathit{c}, \mathit{data}${\tt)}
\State {\tt ReadOneCacheLine(}\trcd, \trp, $\mathit{b}, \mathit{r}, \mathit{c}${\tt)}
\State {\tt assert findErrors() == 0} \Comment{Verify data}
\State {\tt ReadOneCacheLine(}$\mathit{\textbf{my\_tRCD}}$, \trp, $
\mathit{b}, \mathit{r}, \mathit{c}${\tt)}
\State {\tt findErrors()} \Comment{Count errors in a cache line}

\EndFor
\EndFor
\EndFor

\EndFunction
\end{algorithmic}

\caption{Read each cache line with a specified \trcd value.}
\label{trcdtest}
\end{algorithm}

The test iterates through each cache line (lines 2-4) and performs the following
steps to test the cache line's reliability under a \new{reduced} \trcd value. First, it opens the
row that contains the target cache line, writes a specified data pattern into
the cache line, and then precharges the bank (line 5). Second, the test \new{re-opens
the row to read the cache line with the standard \trcd (line 6), and verifies if
the value was written properly (line 7).}
Then it precharges the bank again to
prepare for the next \act. Third, it re-activates the row using the reduced
\trcd value ($my\_tRCD$ in Test~\ref{trcdtest}) to read the target cache line
\new{(line 8)}. It records the number of timing errors (i.e., bit flips) out of the
64-byte (512-bit) cache line \new{(line 9)}.


In total, we have conducted more than 7500 rounds of tests on the DIMMs shown in
\tabref{dimm_list_latvar}, accounting for at least 2500 testing hours. For each
round of tests, we conducted Test~\ref{trcdtest} with a different \trcd value
and data pattern. We tested five different \trcd values: 12.5ns, 10ns, 7.5ns,
5ns, and 2.5ns. Due to the slow clock frequency of the FPGA, we can adjust
timings only at a 2.5ns granularity. We used a set of four different data
patterns: \texttt{0x00}, \texttt{0xaa}, \texttt{0xcc}, and \texttt{0xff}. Each
data pattern represents the value that was written into each byte of the entire
cache line.

\cc{ In this dissertation, we do not examine the latency behavior of each cell
over a controlled period of time, except for the fact that we perform the tests
for multiple rounds per DIMM. The latency of a cell could potentially change
over time, within a short period of time (e.g., similar effect as Variable
Retention Time) or long period of time (e.g., aging and wearout). However, we
leave comprehensive characterization of latency behavior due to time variation
as part of future work. }


\subsection{Activation Error Distribution}

In this section, we first present the distribution of activation errors
collected from all of the tests conducted on every DIMM. Then, we categorize the
results by DIMM model to investigate variation across models from different
vendors.

\subsubsection{Total Bit Error Rates}
\label{sec:trcd_ber}

\figref{violin} shows the box plots of the \emph{bit error rate} (BER)
observed on every DIMM as \trcd varies. The BER is defined as the fraction of
activation error bits in the total population of tested bits. For each box, the
bottom, middle, and top lines indicate the 25th, 50th, and 75th percentile of
the population. The ends of the whiskers indicate the minimum and maximum BER of
all DIMMs for a given \trcd value. Note that the y-axis is in log scale to
show low BER values. As a result, the bottom whisker at \trcde7.5ns cannot be
seen due to a minimum value of 0. In addition, we show \emph{all}
observation points for each specific \trcd value by overlaying them on top of
their corresponding box. Each point shows a BER collected from one round of
Test~\ref{trcdtest} on one DIMM with a specific data pattern and a \trcd value.
Based on these results, we make several observations.


\figputHSL{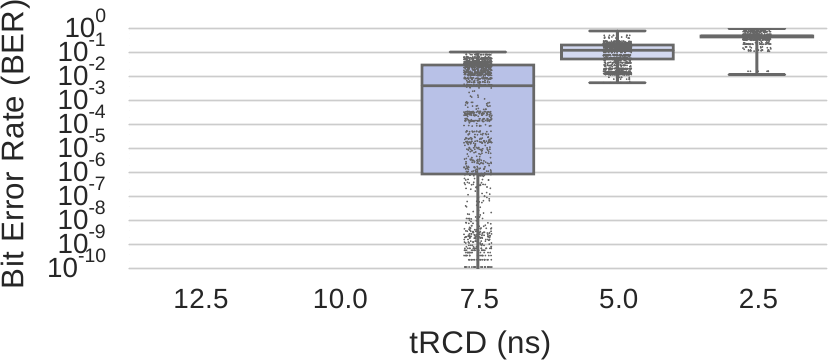}{1.5}{Bit error
    rate \new{of all DIMMs with reduced \trcd}.}{violin}

First, we observe that BER exponentially increases as \trcd decreases. With a
lower \trcd, fewer sense amplifiers are expected to have enough strength to
properly sense the bitline's voltage value and latch the correct data. Second,
at \trcd values of 12.5ns and 10ns, we observe no activation errors on any DIMM.
This shows that the \trcd latency of the slowest cells in our tested DIMMs
likely falls between 7.5 and 10ns, which are lower than the standard value
(13.125ns). The manufacturers use the extra latency as \new{a \emph{guardband}} to
provide additional protection against process variation.

Third, the BER variation among DIMMs becomes smaller as \trcd value decreases.
The reliability of DIMMs operating at \trcde7.5ns varies significantly depending
on the DRAM models and vendors, \new{as we} demonstrate in
\newt{\secref{rcd_model}}.
In fact, some DIMMs have no errors at \trcde7.5ns, which cannot be seen in the
plot due to the log scale. When \trcd reaches 2.5ns, most DIMMs become rife with
errors, with a median BER of 0.48, similar to the probability of a coin toss.

\subsubsection{Bit Error Rates by DIMM Model}
\label{sec:rcd_model}

Since the \new{performance of a DIMM can vary across different models}, vendors,
and fabrication processes, we provide a detailed analysis by breaking down the
\new{BER results by DIMM model} (listed in \tabref{dimm_list_latvar}).
\figref{rcd_ber_20c} presents the distribution of every DIMM's BER grouped by
each \new{vendor and model combination}. Each box shows the quartiles and
median, along with the whiskers indicating the minimum and maximum BERs. Since
all of the DIMMs work reliably at 10ns and above, we show the BERs for
\trcde7.5ns and \trcde5ns.

%

\figputHSL{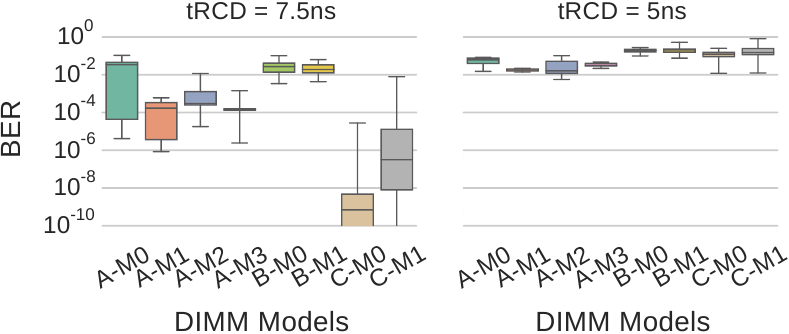}{1.5}{\newt{BERs} of DIMMs grouped
by model, when tested with different \trcd \newt{\xspace values}.}{rcd_ber_20c}


By comparing the BERs across models and vendors, we observe that BER variation
exists not only across DIMMs from different vendors, but also on DIMMs
manufactured from the same vendor. For example, for DIMMs manufactured by vendor
C, \emph{Model 0} DIMMs have fewer errors than \emph{Model 1} DIMMs. This
result suggests that different DRAM models have different circuit architectures
or process technologies, causing \new{latency variation} between them.

Similar to the observation we made across different DIMM models, we observe
variation across DIMMs that have the same model. The variation across DIMMs with
the same model can be attributed to process variation \new{due to} the imperfect
manufacturing process~\cite{nassif-isscc2000,li11,kanad_dram_book, pvt_book}.
Our tested results for every DIMM are available online~\cite{safari-github}.



\subsection{Impact of Data Pattern}
\label{sec:rcd_ber_patt}

In this section, we investigate the impact of reading different data patterns
under different \trcd values. \figref{rcd_ber_pattern_20c} shows the average BER
of test rounds for three representative DIMMs, one from each vendor, with four
data patterns. We do not show the BERs at \trcde2.5ns, as rows cannot be
reliably activated at that latency. We observe that pattern \texttt{0x00} is
susceptible to more errors than pattern \texttt{0xff}, while the BERs for
\new{patterns} \texttt{0xaa} and \texttt{0xcc} lie in between.\footnote{In a
    cache line, we write the 8-bit pattern to every byte.} This can be clearly
    seen on \dimm{0}{C}, where we observe that \texttt{0xff} incurs 4 orders of
    magnitude fewer errors than \patt{0x00} on average at \trcde7.5ns. We make a
    similar observation for the rest of the 12 DIMMs from vendor C.

With patterns \texttt{0xaa} and \texttt{0xcc}, we observe that bit \texttt{0} is
more likely to be misread than bit \texttt{1}. In particular, we examined the
flipped bits on three DIMMs that share the same model as \dimm{0}{C}, and
observed that \emph{all of the flipped bits} are due to bit \texttt{0} flipping
to \texttt{1}. From this observation, we can infer that there is a bias towards
bit \texttt{1}, which can be more reliably read under a shorter activation
latency than bit \texttt{0}.

\figputHSL{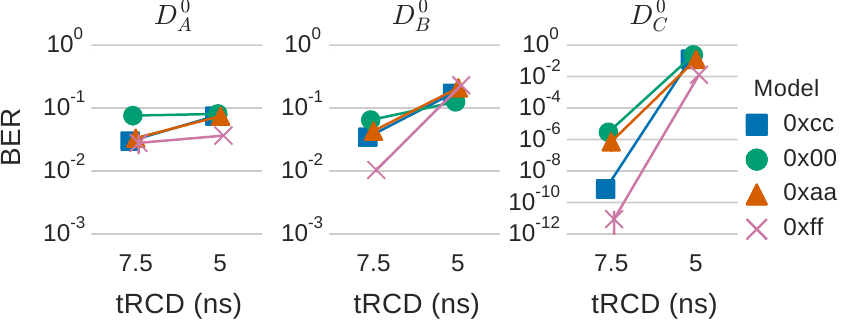}{1.5}{BERs due to four different data
patterns on three different DIMMs as \trcd varies.}{rcd_ber_pattern_20c}

We believe this bias is due to the sense amplifier design. One major DRAM vendor
presents a circuit design for a contemporary sense amplifier, and observes that
it senses the \vdd value on the bitline faster than
\zerov~\cite{kyunam-isscc2012}. Hence, the sense amplifier is able to sense and
latch bit \texttt{1} faster than \texttt{0}. Due to this pattern dependence, we
believe that it is promising to investigate asymmetric data encoding or error
correction mechanisms that favor \texttt{1}s over \texttt{0}s.



\obs{\newt{Errors caused by reduced activation latency are dependent on the
stored data pattern. Reading bit \texttt{1} is significantly more reliable than
bit \texttt{0} at reduced activation latencies.}}

\subsection{Effect of Temperature}
\label{sec:rcd_temp}

Temperature is an important external factor that may affect the reliability of
DIMMs~\cite{schroeder-sigmetrics2009,el-sayed-sigmetrics2012,
liu-isca2013,khan-sigmetrics2014}. In particular, Schroeder et
al.~\cite{schroeder-sigmetrics2009} and El-Sayed et
al.~\cite{el-sayed-sigmetrics2012} do not observe clear evidence for increasing
DRAM error rates with increased temperature in data centers. \new{Other works
    find that data retention time strongly depends on
    temperature~\cite{khan-sigmetrics2014,liu-isca2013,qureshi-dsn2015}}. However, none
    of these works have studied the effect of temperature on DIMMs when
    they are operating with a lower activation latency.

\begin{figure}[!h]
    \centering
    \captionsetup[subfigure]{justification=centering}
    \subcaptionbox{\trcde7.5ns\label{fig:temp_rcd75}}[\linewidth]
    {
        \includegraphics[scale=1.3]{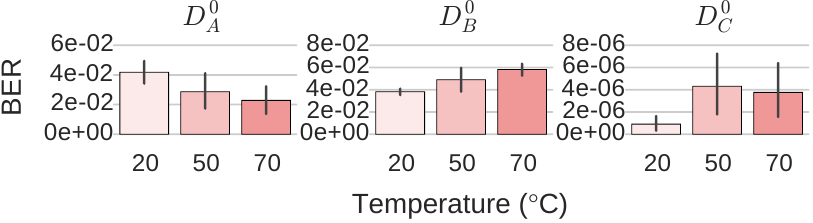}
    }

    \subcaptionbox{\trcde5ns\label{fig:temp_rcd50}}[\linewidth]
    {
        \includegraphics[scale=1.3]{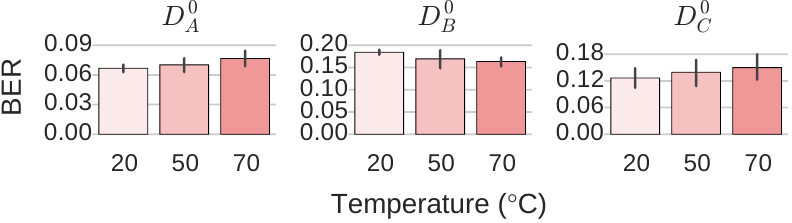}
    }

    \caption{BERs of \new{three \newt{example}} DIMMs operating under different temperatures.}
    \label{fig:rcd_ber_temp}
\end{figure}

To investigate the impact of temperature on DIMMs operating with an activation
latency lower than the standard value, we perform experiments that adjust the
\emph{ambient temperature} using a closed-loop temperature controller (shown in
\figref{dram_fpga_white}). \figref{rcd_ber_temp} shows the average BER of
\new{three example} DIMMs under three temperatures: 20\celsius, 50\celsius,
and 70\celsius~for \trcde7.5/5ns. We include error bars, which are computed
using 95\% confidence intervals.

We make two observations. First, at \trcde7.5ns (\figref{temp_rcd75}), every
DIMM shows a different BER trend as temperature increases. By calculating the
\emph{p-value} between the BERs of different temperatures, we find that the
change in BERs is not statistically significant from one temperature to another
for \new{two out of the three tested} DIMMs, meaning that we cannot conclude
that BER \new{increases at} higher temperatures. For instance, the p-values
between the BERs at 20\celsius~and 50\celsius~for \dimm{0}{A}, \dimm{0}{B}, and
\dimm{0}{C} are 0.084, 0.087, and 0.006, respectively. Two of the three DIMMs
have p-values greater than an $\alpha$ of 0.05, meaning that the BER change is
statistically insignificant. Second, at lower \trcd values~(5ns), the difference
between the BERs due to \new{temperature} becomes even smaller.

\obs{Our study does not show enough evidence
to conclude that activation errors increase with higher temperatures.}


\subsection{Spatial Locality of Activation Errors}
\label{sec:trcd_loc}

To understand the locations of activation errors within a DIMM, we show the
probability of experiencing at least one bit error in each cache line over a
large number of experimental runs. We present results of two representative
DIMMs from our experiments. Our results on the some other DIMMs are available
online~\cite{safari-github}.


\afterpage{
\begin{figure}[p]
    \centering
    \subcaptionbox{Bank 0.}[0.43\linewidth][l]
    {
        \includegraphics[width=0.43\linewidth]{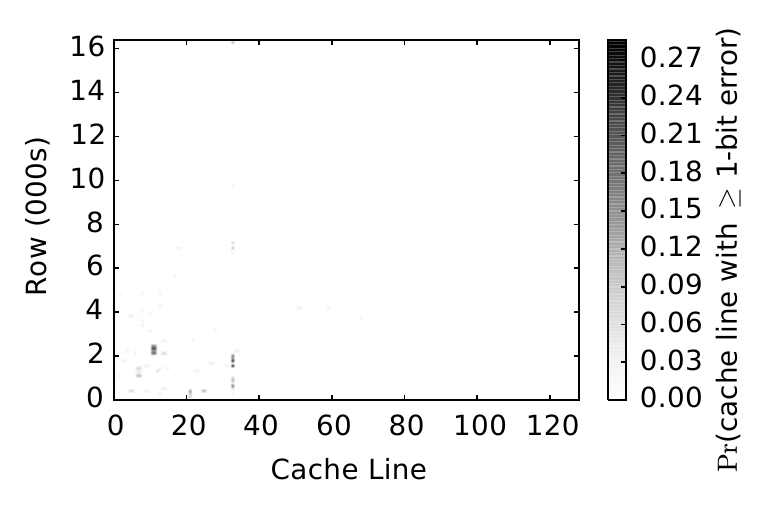}
    }
    \subcaptionbox{Bank 1.}[0.43\linewidth][r]
    {
        \includegraphics[width=0.43\linewidth]{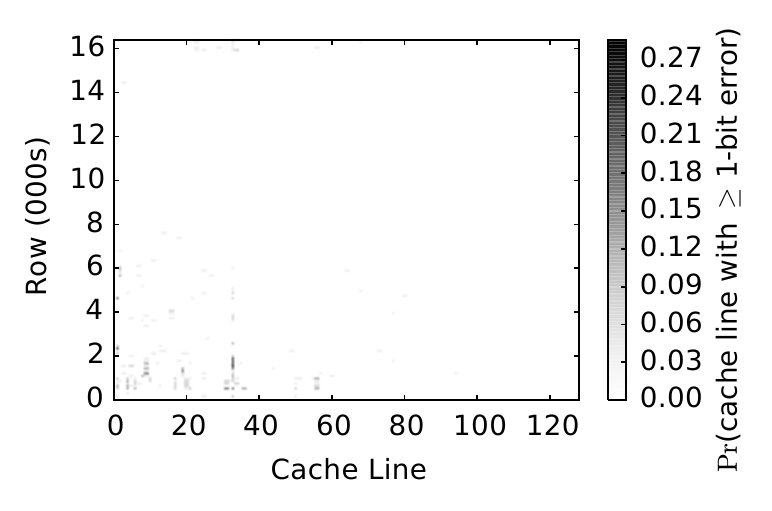}
    }

    \subcaptionbox{Bank 2.}[0.43\linewidth][l]
    {
        \includegraphics[width=0.43\linewidth]{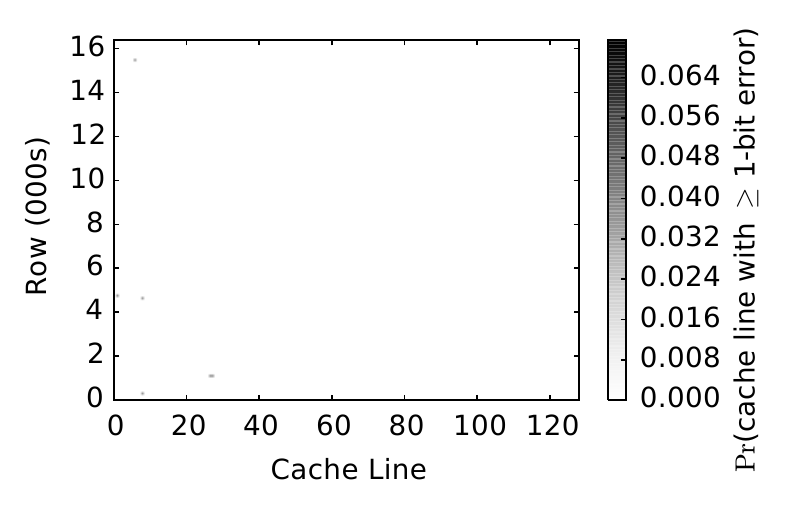}
    }
    \subcaptionbox{Bank 3.}[0.43\linewidth][r]
    {
        \includegraphics[width=0.43\linewidth]{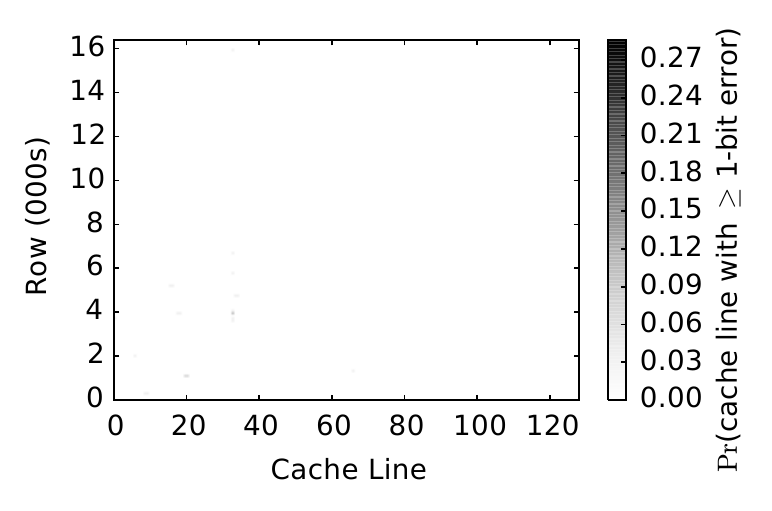}
    }

    \subcaptionbox{Bank 4.}[0.43\linewidth][l]
    {
        \includegraphics[width=0.43\linewidth]{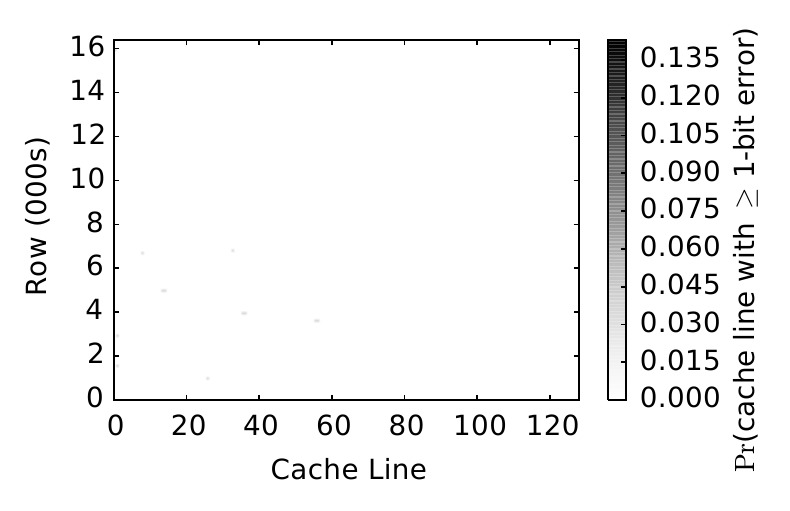}
    }
    \subcaptionbox{Bank 5.}[0.43\linewidth][r]
    {
        \includegraphics[width=0.43\linewidth]{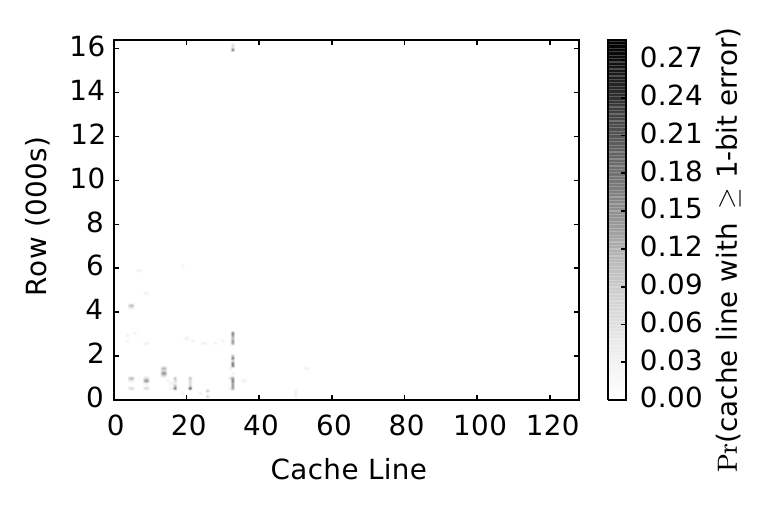}
    }

    \subcaptionbox{Bank 6.}[0.43\linewidth][l]
    {
        \includegraphics[width=0.43\linewidth]{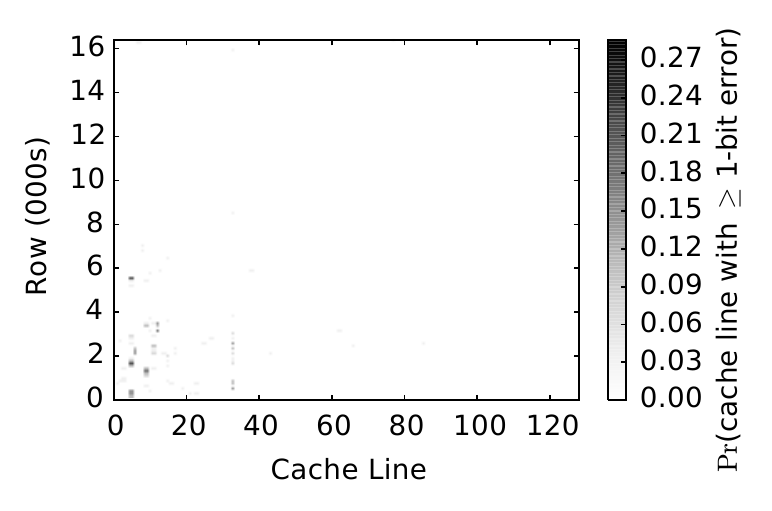}
    }
    \subcaptionbox{Bank 7.}[0.43\linewidth][r]
    {
        \includegraphics[width=0.43\linewidth]{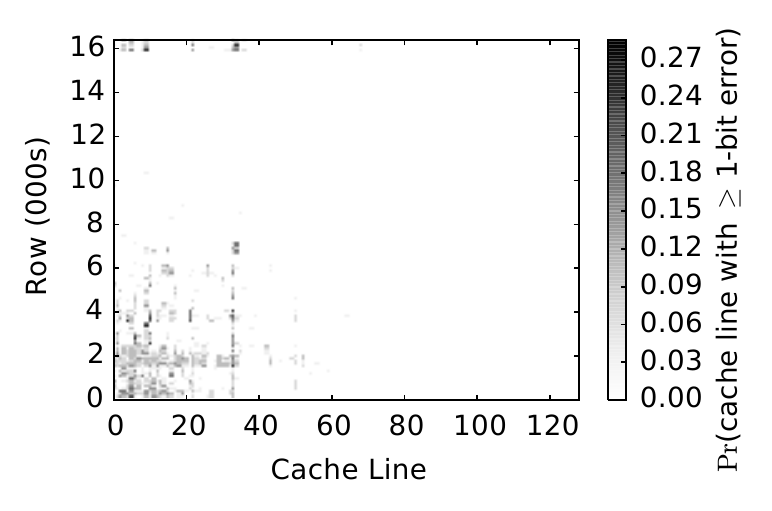}
    }

    \caption{Probability of \newt{observing activation} errors in all banks of \dimm{0}{C}.}
    \label{fig:rcd_col_err}
\end{figure}
}

\afterpage{
\begin{figure}[p]
    \centering
    \subcaptionbox{Bank 0.}[0.43\linewidth][l]
    {
        \includegraphics[width=0.43\linewidth]{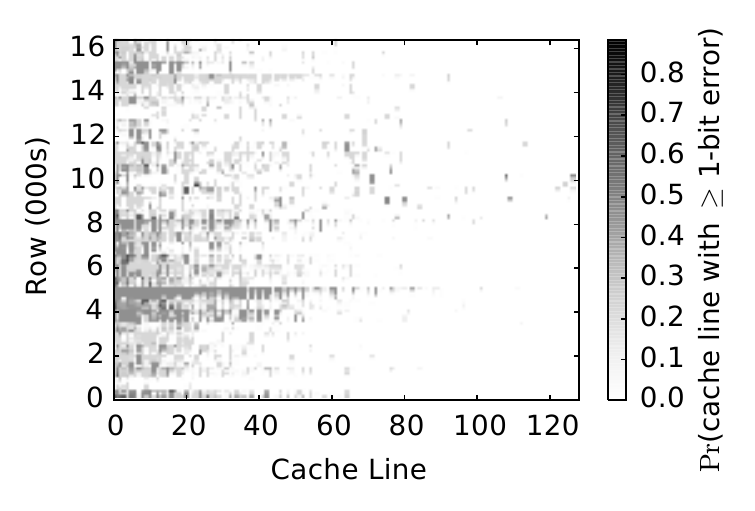}
    }
    \subcaptionbox{Bank 1.}[0.43\linewidth][r]
    {
        \includegraphics[width=0.43\linewidth]{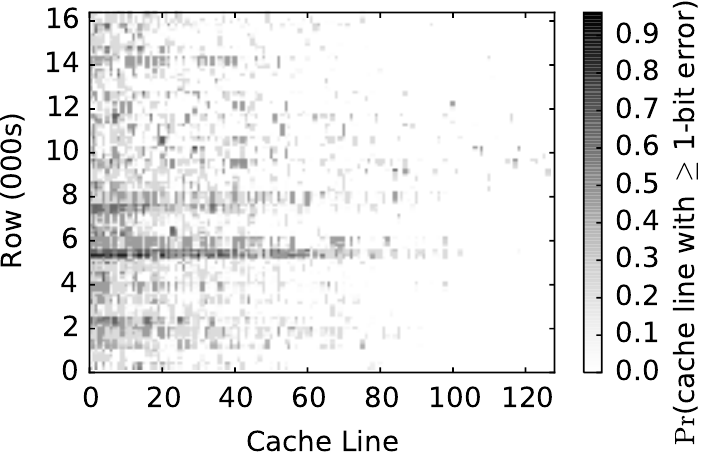}
    }

    \subcaptionbox{Bank 2.}[0.43\linewidth][l]
    {
        \includegraphics[width=0.43\linewidth]{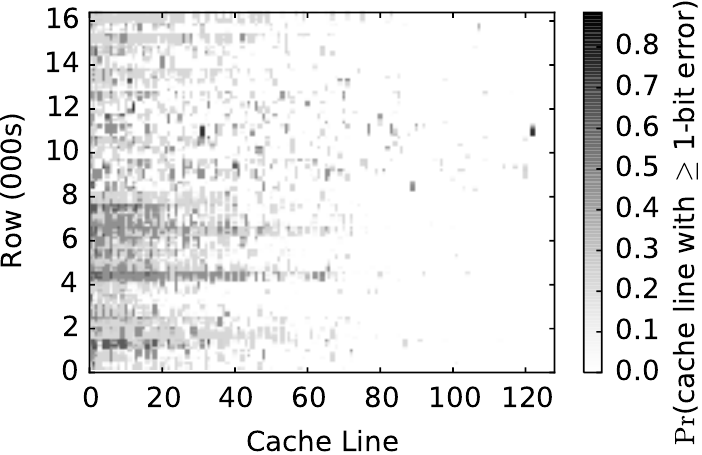}
    }
    \subcaptionbox{Bank 3.}[0.43\linewidth][r]
    {
        \includegraphics[width=0.43\linewidth]{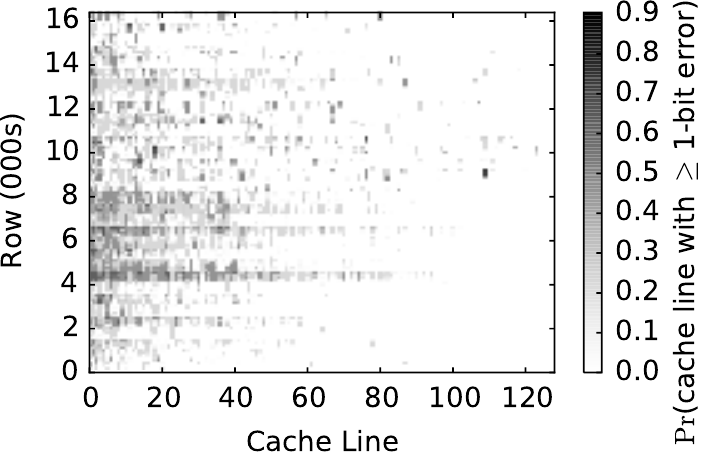}
    }

    \subcaptionbox{Bank 4.}[0.43\linewidth][l]
    {
        \includegraphics[width=0.43\linewidth]{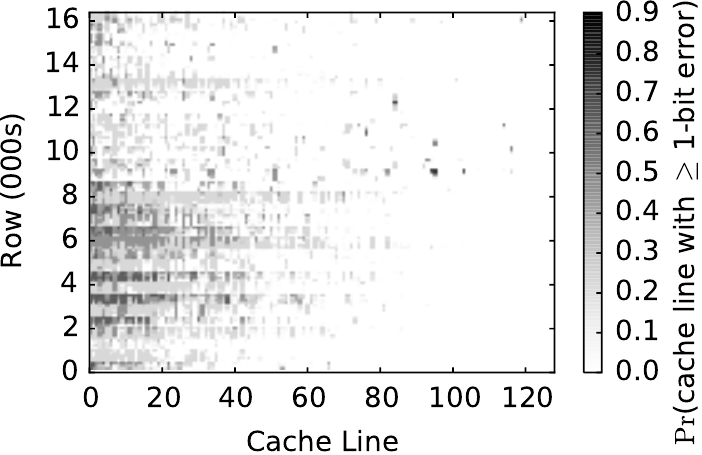}
    }
    \subcaptionbox{Bank 5.}[0.43\linewidth][r]
    {
        \includegraphics[width=0.43\linewidth]{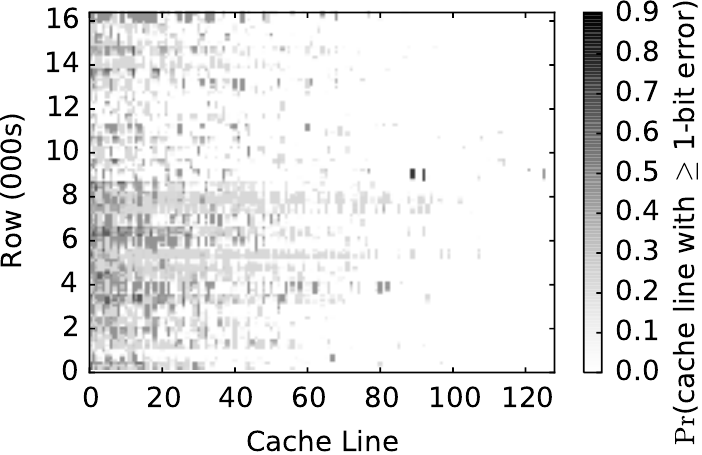}
    }

    \subcaptionbox{Bank 6.}[0.43\linewidth][l]
    {
        \includegraphics[width=0.43\linewidth]{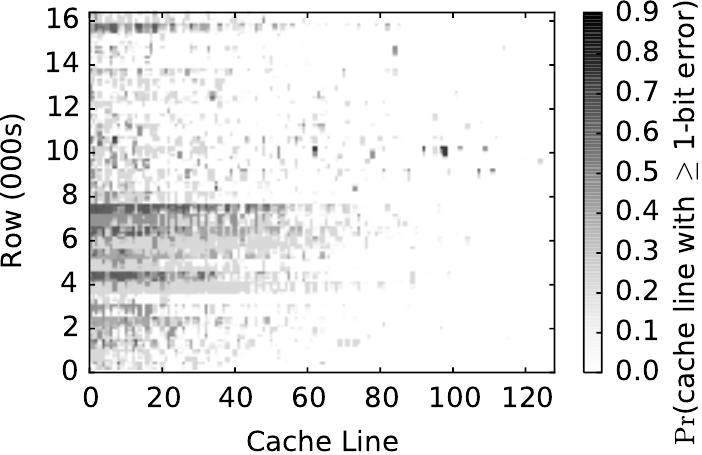}
    }
    \subcaptionbox{Bank 7.}[0.43\linewidth][r]
    {
        \includegraphics[width=0.43\linewidth]{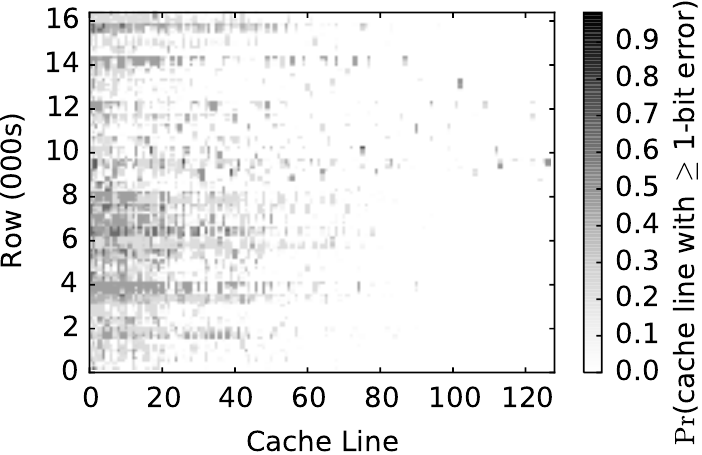}
    }

    \caption{Probability of \newt{observing activation} errors in all eight
      banks of \dimm{3}{A}.}
    \label{fig:rcd_err_region}
\end{figure}
}

Figures~\ref{fig:rcd_col_err} and \ref{fig:rcd_err_region} show the locations of
activation errors in all eight banks of two different DIMMs (\dimm{3}{A} and
\dimm{0}{C}, respectively) using \trcde7.5ns. The x-axis and y-axis indicate the
cache line number and row number (in thousands), respectively. In our tested
DIMMs, a row size is 8KB, \new{comprising} 128 cache lines (64 bytes).
\new{Results} are gathered from 40 and 52 iterations of tests for \dimm{0}{C}
and \dimm{3}{A}, respectively. We make two observations.

The first main observation is that errors tend to concentrate at certain
regions. Errors in \dimm{0}{C} (\figref{rcd_col_err}) cluster at certain
\emph{columns of cache lines}. For the majority of the remaining cache lines in
each bank, we observe no errors throughout the experiments. \dimm{3}{A}
(\figref{rcd_err_region}) shows that the activation errors cluster at certain
regions, repeatedly occurring within the first half of the majority of rows. We
hypothesize that the cause of \new{such spatial locality of errors is due to the
  locality of} variation in the fabrication process during manufacturing:
certain cache line locations can end up with less robust components, such as
weaker sense amplifiers, weaker cells, or higher resistance bitlines.

Second, although banks from the same DIMM demonstrate similar patterns of
spatial locality of errors, the fraction of errors varies across banks. Some
banks observe more errors than other banks. For example, in \dimm{0}{C}, Bank 7
experiences more errors than \mbox{Bank 2}. We hypothesize that the variation across
banks can be a result of manufacturing process variation, causing certain banks
to be more susceptible to reduced activation latency.



\obs{Activation errors do not occur uniformly within \new{DRAM. They instead}
exhibit strong spatial concentration at certain regions.}

\subsection{Density of Activation Errors}
\label{sec:density_err}

In this section, we investigate how errors are distributed within the erroneous
cache lines. We present the distribution of error bits at the granularity of
\emph{data beats}\new{, as conventional error-correcting codes (ECC)
work at the same granularity. We discuss the effectiveness of employing
ECC in Section~\ref{sec:rcd_ecc}}. Recall from \secref{dram_access} that a cache line
transfer consists of eight 64-bit data beats.

\begin{figure*}[!h]
    \centering
    \captionsetup[subfigure]{justification=centering}
    \subcaptionbox{A-M1}[\linewidth]
    {
        \includegraphics[scale=1.0]{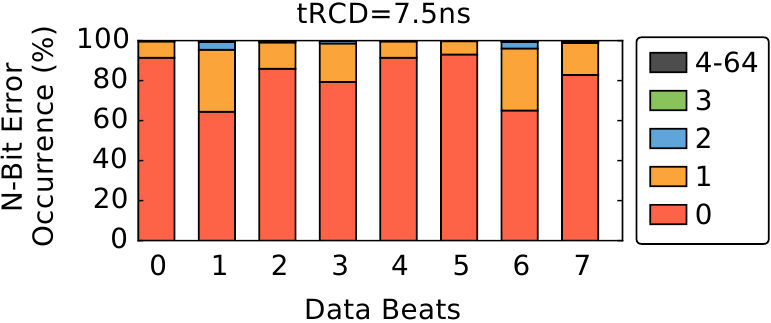}
    }

    \vspace{0.2in}

    \subcaptionbox{B-M1}[\linewidth]
    {
        \includegraphics[scale=1.0]{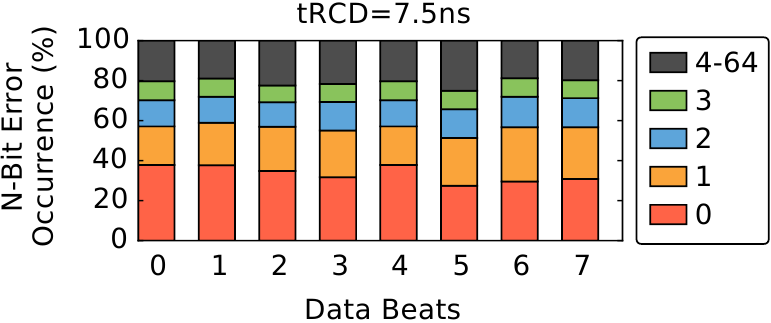}

    }

    \vspace{0.2in}

    \subcaptionbox{C-M0}[\linewidth]
    {
        \includegraphics[scale=1.0]{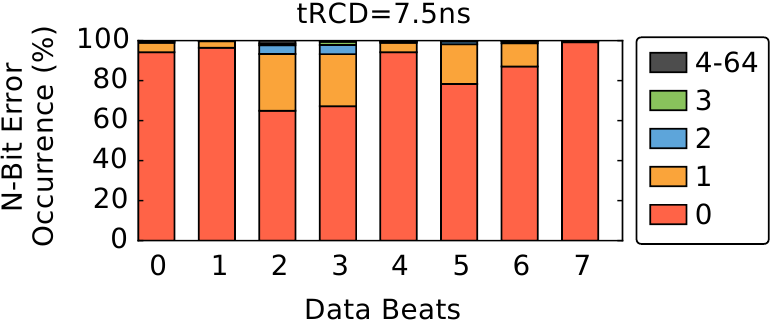}
    }

    \caption{Breakdown of the number of error bits observed in each data
    beat of erroneous cache lines \newt{at \trcde 7.5ns}.}
    \vspace{5pt}
    \label{fig:burst_rcd3_20c}
\end{figure*}

\figref{burst_rcd3_20c} shows the distribution of error bits observed in each
data beat of all erroneous cache lines when using \trcde7.5ns. We show
experiments from 9 DIMMs, categorized into three DIMM models (one per vendor).
We select the model that experiences the lowest average BER from each vendor,
and show the frequency of observing 1, 2, 3, and $\ge$4 error bits in each data
beat. The results are aggregated from all DIMMs of the selected models. We make
two observations.

First, most data beats experience fewer than 3 error bits at \trcde 7.5ns. We
observe that more than 84\%, 53\%, and 91\% of all the recorded activation
errors are just 1-bit errors for DIMMs in A-M1, B-M1, and C-M0, respectively.
\new{Across all of the cache lines that contain at least one error bit, 82\%,
41\%, and 85\% of the data beats that make up each cache line have no errors for
A-M1, B-M1, and C-M0, respectively.} Second, when \trcd is reduced to 5ns, the
number of errors increases. The distribution of activation errors in data beats
when using \trcde5ns is \newt{available online~\cite{safari-github}}, and it
shows that 68\% and 49\% of data beats in A-M1 and C-M0 still have no more than
one error bit.


\obs{\newt{For cache lines that experience activation errors, the majority of
their constituent data beats contain either no errors or just a 1-bit error.}}


\subsection{Effect of Error Correction Codes}
\label{sec:rcd_ecc}

As shown in the previous section, a majority of data beats in erroneous cache
lines contain only a few error bits. In contemporary DRAM, ECC is used to detect
and correct errors at the granularity of data beats. Therefore, this creates an
opportunity for applying error correction codes (ECC) to correct \new{activation
errors}. To study of the \new{effect of ECC}, we perform an analysis that
uses various strengths of ECC to correct activation errors.

\figref{err_free_rcd3_new} shows the percentage of cache lines that do
\new{\emph{not}} observe any activation errors when using \trcde 7.5ns at
various ECC strengths, ranging from single to triple error bit correction. These
results are gathered from the same 9 DIMMs used in
Section~\ref{sec:density_err}. The first bar of each group is the percentage of
cache lines that do not exhibit any activation errors in our experiments. The
following data bars show the fraction of error-free cache lines after applying
single, double, and triple error correction codes.

\figputGHS{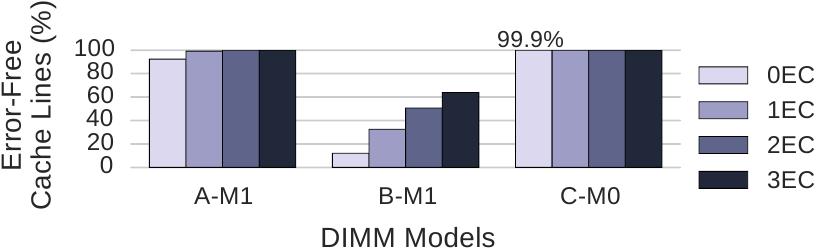}{1.5}{Percentage of error-free cache lines with various
strengths of error correction (EC), with \trcde7.5ns.}


We make two observations. First, without any ECC support, a large fraction of
cache lines can be read reliably without any errors in many of the DIMMs we
study. Overall, 92\% and 99\% of cache lines can be read without any activation
errors from A-M1 and C-M0 DIMMs, respectively. On the other hand, B-M1 DIMMs are
\new{more} susceptible to reduced activation latency\new{: only 12\% of their
cache lines can be read without any activation errors.}

\obs{A majority of cache lines can be read without any activation errors in most
    of our tested DIMMs. However, some \newt{DIMMs} are very susceptible to
activation errors, resulting in a small fraction of error-free cache lines.}

Second, ECC is effective in correcting the activation errors.  For example,
with a single error correction code (1EC), which is widely deployed in many
server systems, the fraction of reliable cache lines improves from 92\% to 99\%
for A-M1 DIMMs. Even for B-M1 DIMMs, which exhibit activation errors in a large
fraction of cache lines, \new{the} triple error correcting code is able to
improve the percentage of \new{error-free} cache lines from 12\% to 62\%.

\obs{ECC is an effective mechanism to correct activation errors, even in
modules with a large fraction of erroneous cache lines.}


\section{Precharge Latency Analysis}
\label{sec:pre_lat_analysis}

In this section, we present the methodology and results on varying the precharge
latency, represented by the \trp timing parameter. We first describe the nature
of timing errors caused by reducing the precharge latency in
Section~\ref{sec:trp_errors}. Then, we describe the FPGA test we conducted to
characterize \trp variation in Section~\ref{sec:trp_tests}. In the remaining
sections, we describe four major observations from our result analysis.

\subsection{Behavior of Precharge Errors}
\label{sec:trp_errors}

In order to access a new DRAM row, a memory controller issues a \cpre command,
which performs the following two functions in sequence: {\em (i)}~it closes the
currently-activated row in the array (i.e., it disables the \newt{activated} row's
wordline); and {\em (ii)}~it reinitializes the voltage value of every bitline
inside the array back to \halfvdd, to prepare for a new activation.

\newt{Reducing the precharge latency by a small amount} affects only the
reinitialization process of the bitlines \newt{without interrupting the process
  of closing the row}. The latency of this process is determined by the
\emph{precharge unit} that is placed \newt{by} each bitline, next to the sense
amplifier. By using a \trp value lower than the standard specification, the
precharge unit may not have \newt{sufficient time} to reset the bitline voltage
from either \vdd (\texttt{bit 1}) or \zerov (\texttt{bit 0}) to \halfvdd,
thereby causing the bitline to float at some other intermediate voltage value.
As a result, in the \emph{subsequent} activation, the sense amplifier can
incorrectly sense the wrong value \newt{from the DRAM cell} due to the extra
charge left (or lack thereof) on the bitline. \newt{We define precharge errors
  to be timing errors due to reduced precharge latency.}

To further understand the nature of precharge errors, we use a test similar to
the one for reduced activation latency in \secref{trcd_errors}. The
test reduces only the precharge latency, while keeping the activation latency at
the standard value, to isolate the effects that occur due to a reduced precharge
latency. We answer two fundamental questions: \romnum{i} Does
lowering the precharge latency incur errors on multiple cache lines in the row
activated \emph{after} the precharge? \romnum{ii} Do these errors propagate back
to the respective DRAM cells, causing permanent errors for all future accesses
to those cells?

\subsubsection{Precharge Errors \newt{Are} Spread Across a Row}

Throughout repeated test runs on DIMMs from all three vendors, we observe that
\newt{reducing} the precharge latency induces errors that are spread across
\emph{multiple} cache lines in the row activated after the precharge. This is
because reducing the \trp value affects the latency between two \emph{row-level}
DRAM commands, \cpre and \act. As a result, having an insufficient amount of
precharge time for the array's bitlines affects the \emph{entire} row.

\obs{\newt{Timing} errors occur in multiple cache lines in the row activated after a
    precharge with \newt{reduced} latency.}

Furthermore, these precharge errors are due to the sense amplifiers sensing the
wrong voltage on the bitlines, causing them to latch incorrect data. Therefore,
as the restoration operation reuses the data latched in the sense amplifiers,
the wrong data is written back into the cells, causing the data values in the
cells to be corrupted.

\subsection{FPGA Test for Precharge \newt{Latency}}
\label{sec:trp_tests}

In contrast to activation errors, precharge errors are spread across an
\emph{entire} row. As a result, we use a test that varies \trp at the row level.
The pseudocode of the test, Test~\ref{trptest}, is shown below.

\floatname{algorithm}{Test}

\begin{algorithm}[h]

\algnewcommand\algorithmicto{\textbf{to}}
\algrenewtext{For}[3]{\algorithmicfor\ #1 $\gets$ #2 \algorithmicto\ #3 \algorithmicdo}

\algrenewcommand\algorithmicfunction{}
\algrenewcommand\algorithmicdo{}
\algrenewcommand\algorithmicindent{1.2em}
\algrenewcommand\alglinenumber[1]{\footnotesize\texttt{#1}}
\small

\begin{algorithmic}[1]
\Function{tRPRowOrderTest}{$\mathit{my\_tRP}, \mathit{data}$}

\For{b}{1}{$\mathit{Bank}_{\mathit{MAX}}$}
\For{r}{1}{$\mathit{Row}_{\mathit{MAX}}$} \Comment{Row order}

\State {\tt WriteOneRow(}$\mathit{b}, \mathit{r}, \mathit{data}${\tt)}
\State {\tt ReadOneRow(}\trcd, \trp, $\mathit{b}, \mathit{r} ${\tt)}
\State {\tt WriteOneRow(}$\mathit{b}, \mathit{r+1}, \mathit{data\_bar}${\tt)} \Comment{Inverted data}
\State {\tt ReadOneRow(}\trcd, \trp, $\mathit{b}, \mathit{r+1} ${\tt)}
\State {\tt assert findErrors() == 0} \Comment{Verify data, data\_bar}
\State {\tt ReadOneRow(}\trcd, $\mathit{\textbf{my\_tRP}},
\mathit{b}, \mathit{r}${\tt)}
\State {\tt findErrors()} \Comment{Count errors in row r}

\EndFor
\EndFor

\EndFunction
\end{algorithmic}

\caption{Read each row with a specified \trp value.}
\label{trptest}
\end{algorithm}


In total, we have conducted more than 4000 rounds of tests on the DIMMs shown in
\tabref{dimm_list_latvar}, which accounts for at least 1300 testing hours. We use three
groups of different data patterns: (\texttt{0x00}, \texttt{0xff}),
(\texttt{0xaa}, \texttt{0x33}), and (\texttt{0xcc}, \texttt{0x55}). Each group
specifies two different data patterns, which are the inverse of each other,
placed in consecutive rows in the same array. This ensures that as we iterate
through the rows in order, the partially-precharged state of the bitlines will
not favor the data pattern in the \newt{adjacent row to be activated}.

\subsection{\newt{Precharge Error Distribution}}

\newt{In this section, we first show the distribution of precharge errors collected
from all of the tests conducted on every DIMM. Then, we categorize the results
by DIMM model to investigate variation across models from different vendors.}

\subsubsection{\newt{Total Bit Error Rates}}

\figref{trp_violin} shows the box plots of the BER observed for every DIMM as
\trp is varied from 12.5ns down to 2.5ns. Based on these results, we make
several observations.
\figputHSL{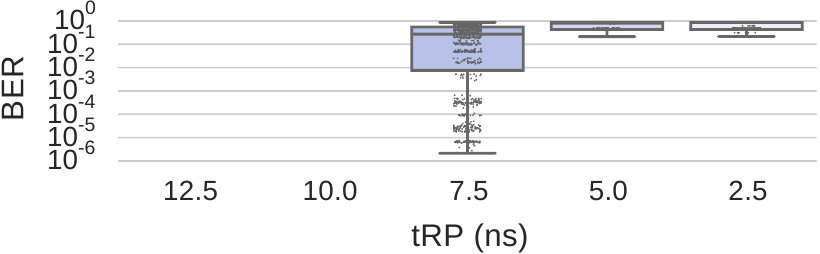}{1.5}{Bit error rate
    \newt{of all DIMMs with reduced \trp}.}{trp_violin}

First, similar to the observation made for activation latency, we do not observe
errors when the precharge latency is reduced to 12.5 and 10ns, as the \newt{reduced
latencies} are still within the guardband provided. Second, the precharge BER is
significantly higher than the activation BER when errors start appearing at
7.5ns -- the median of the precharge BER is 587x higher than that of the
activation BER \newt{(shown in \figref{violin})}. This is partially due to the
fact that reducing the precharge latency causes the errors to span across
multiple cache lines in an \newt{\emph{entire row}}, whereas reducing the
activation latency affects \newt{\emph{only the first cache line}} read from the
row. Third, once \trp is set to 5ns, the BER exceeds the tolerable range,
resulting in a median BER of 0.43. In contrast, the activation BER does not
reach this high an error rate until the activation latency is lowered down to
2.5ns.


\obs{With the same amount of latency reduction, the number of precharge errors
is significantly higher than the number of activation errors.}

\subsubsection{Bit Error Rates by DIMM Model}

To examine the precharge error trend for individual DIMM models, we show the BER
distribution of every DIMM categorized by its model in \figref{rp_ber_20c}.
Similar to the observation we made \newt{for} activation errors in \secref{trcd_errors},
variation exists across different DIMM models. These results provide further
support for the existence \newt{and prevalence} of latency variation in modern DRAM chips.

\figputHSL{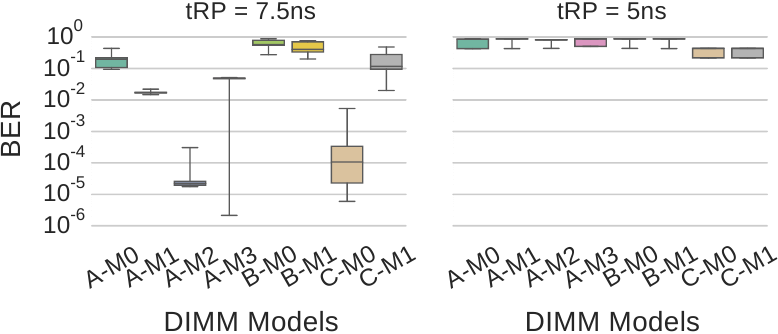}{1.5}{\newt{BERs of DIMMs grouped by
model, when tested with different \trp values.}}{rp_ber_20c}

\subsection{Spatial Locality of Precharge Errors}
\label{sec:trp_spatial}

In this section, we investigate the location and distribution of precharge
errors. Due to the large amount of available data, we show representative
results from two DIMMs, \dimm{0}{C} and \dimm{1}{C} from model C-M0. Our results
for some DIMMs are available publicly~\cite{safari-github}. \figref{trp_err_loc}
and \figref{trp_err_loc_2} show the probability of each cache line seeing at
least a one-bit precharge error in all banks of \dimm{0}{C} and \dimm{1}{C},
respectively, when we set \trp to 7.5ns. The x-axis indicates the cache line
number, and the y-axis indicates the row number (in thousands). The results are
gathered from 12 iterations of tests. We make several observations based on our
results.


\afterpage{
\begin{figure}[p]
    \centering
    \subcaptionbox{Bank 0.}[0.40\linewidth]
    {
        \includegraphics[width=0.40\linewidth]{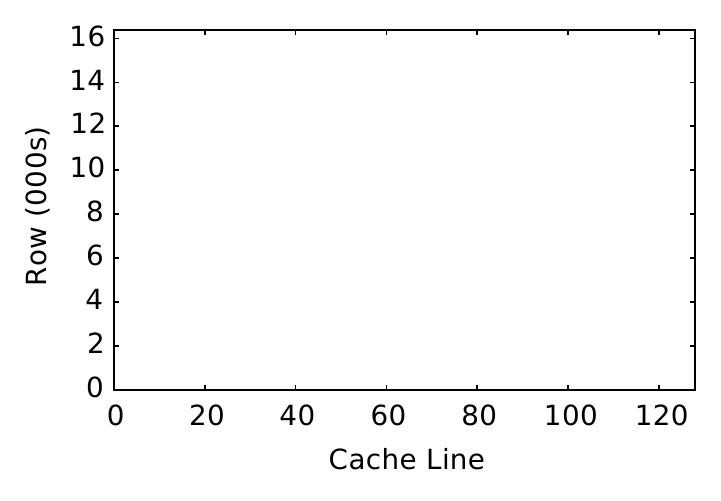}
    }
    \subcaptionbox{Bank 1.}[0.40\linewidth]
    {
        \includegraphics[width=0.40\linewidth]{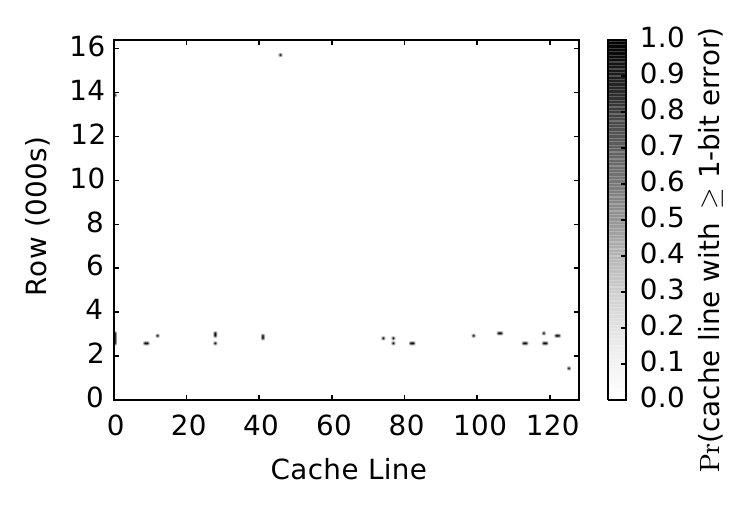}
    }

    \subcaptionbox{Bank 2.}[0.40\linewidth]
    {
        \includegraphics[width=0.40\linewidth]{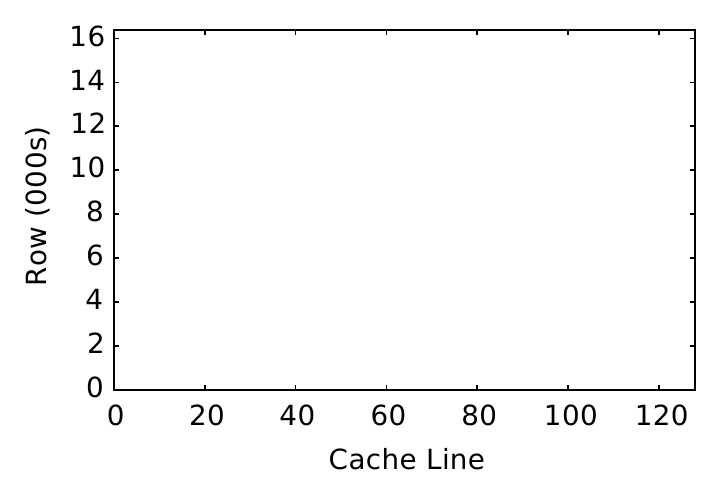}
    }
    \subcaptionbox{Bank 3.}[0.40\linewidth]
    {
        \includegraphics[width=0.40\linewidth]{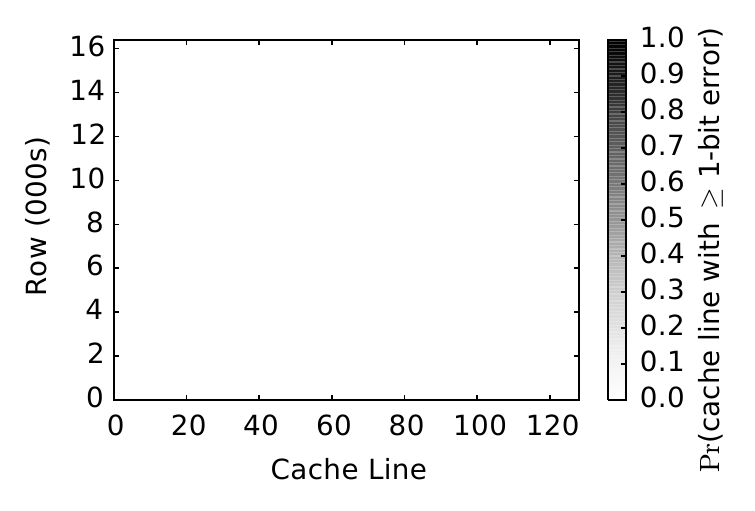}
    }

    \subcaptionbox{Bank 4.}[0.40\linewidth]
    {
        \includegraphics[width=0.40\linewidth]{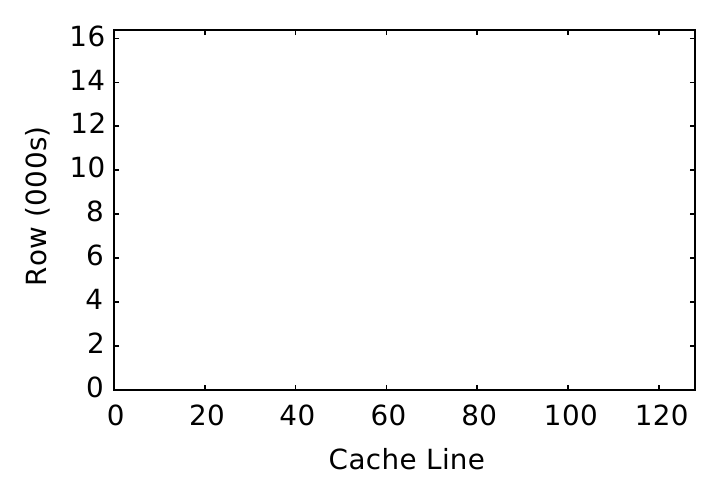}
    }
    \subcaptionbox{Bank 5.}[0.40\linewidth]
    {
        \includegraphics[width=0.40\linewidth]{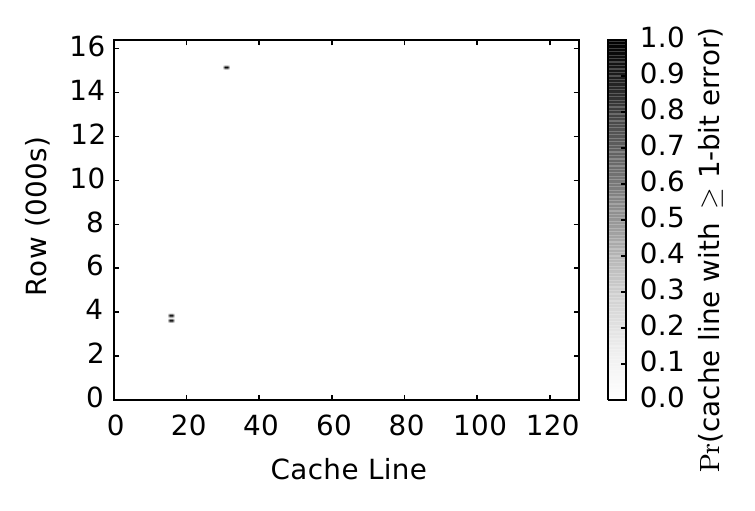}
    }

    \subcaptionbox{Bank 6.}[0.40\linewidth]
    {
        \includegraphics[width=0.40\linewidth]{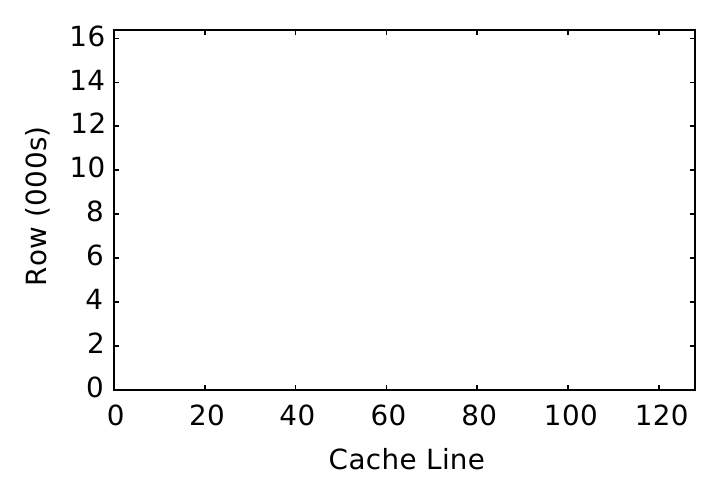}
    }
    \subcaptionbox{Bank 7.}[0.40\linewidth]
    {
        \includegraphics[width=0.40\linewidth]{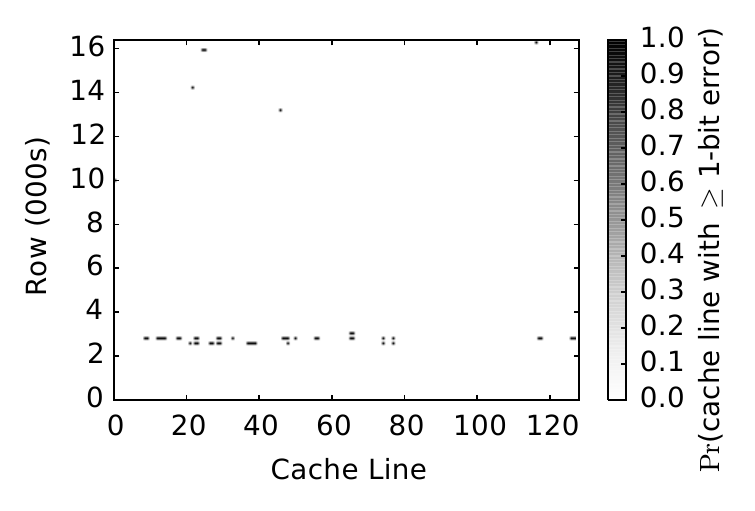}
    }

    \caption{Probability of \newt{observing precharge} errors in all eight banks of \dimm{0}{C}.}
    \label{fig:trp_err_loc}
\end{figure}
}

\afterpage{
\begin{figure}[p]
    \centering
    \subcaptionbox{Bank 0.}[0.40\linewidth]
    {
        \includegraphics[width=0.40\linewidth]{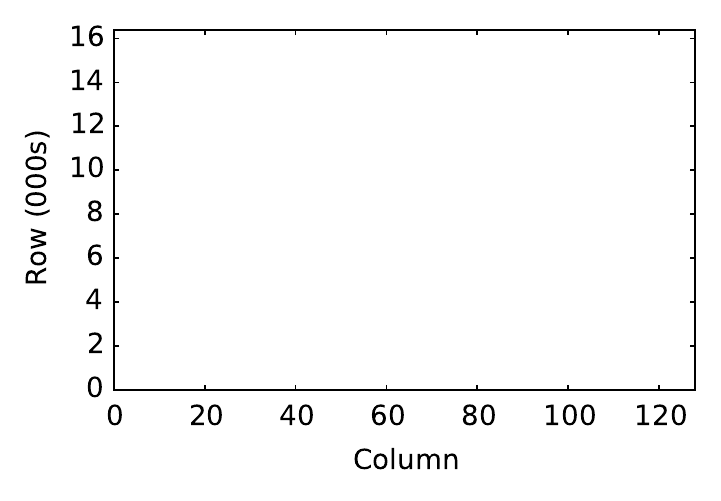}
    }
    \subcaptionbox{Bank 1.}[0.40\linewidth]
    {
        \includegraphics[width=0.40\linewidth]{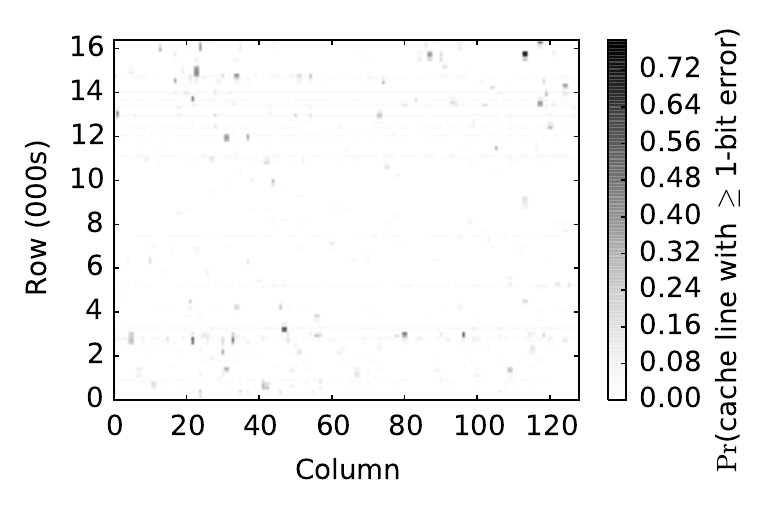}
    }

    \subcaptionbox{Bank 2.}[0.40\linewidth]
    {
        \includegraphics[width=0.40\linewidth]{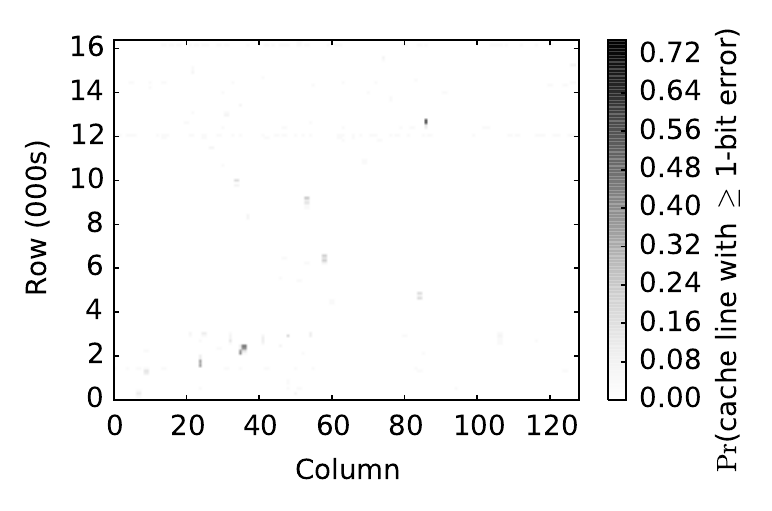}
    }
    \subcaptionbox{Bank 3.}[0.40\linewidth]
    {
        \includegraphics[width=0.40\linewidth]{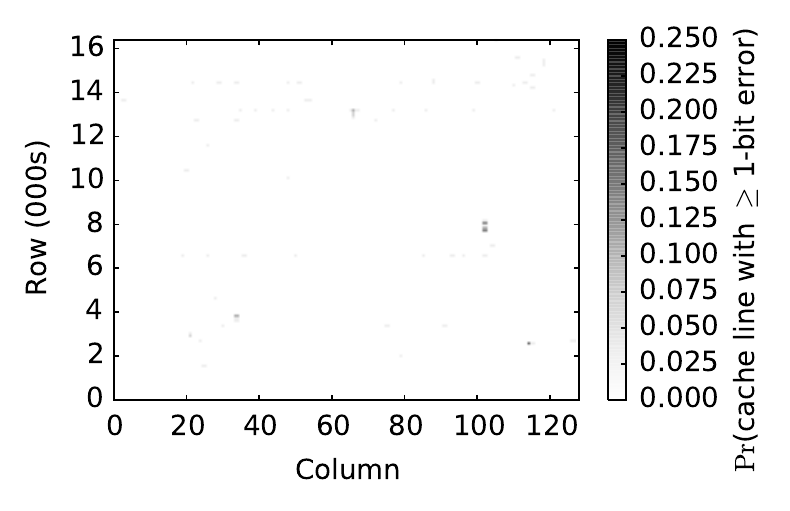}
    }

    \subcaptionbox{Bank 4.}[0.40\linewidth]
    {
        \includegraphics[width=0.40\linewidth]{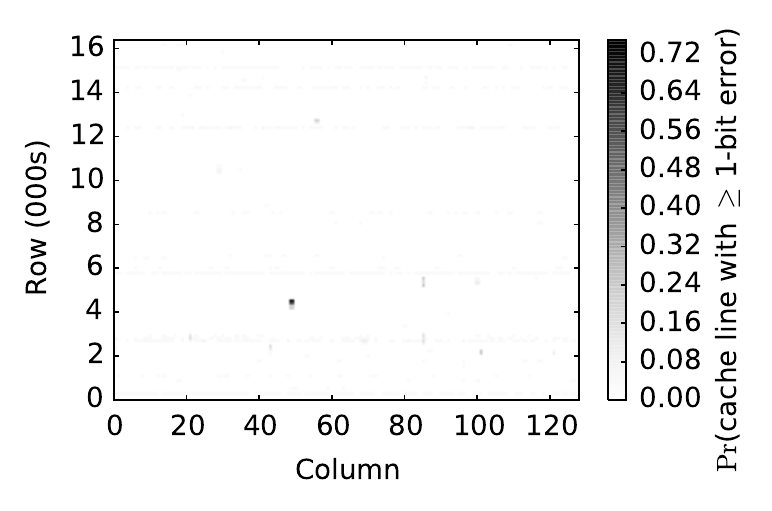}
    }
    \subcaptionbox{Bank 5.}[0.40\linewidth]
    {
        \includegraphics[width=0.40\linewidth]{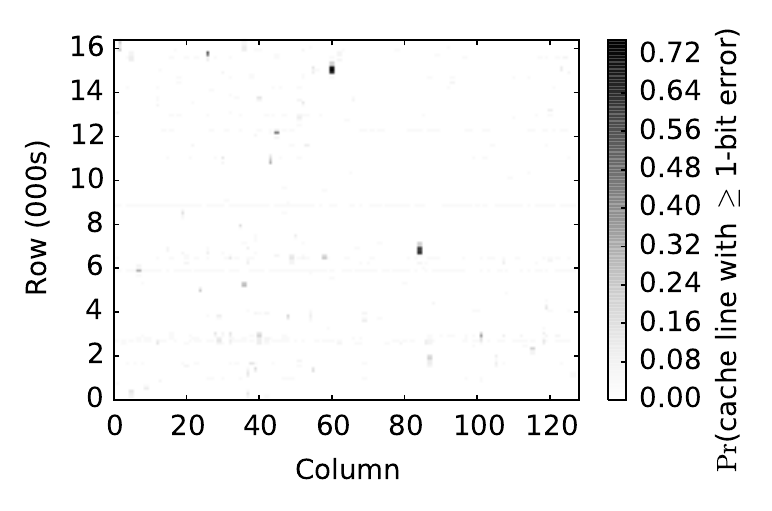}
    }

    \subcaptionbox{Bank 6.}[0.40\linewidth]
    {
        \includegraphics[width=0.40\linewidth]{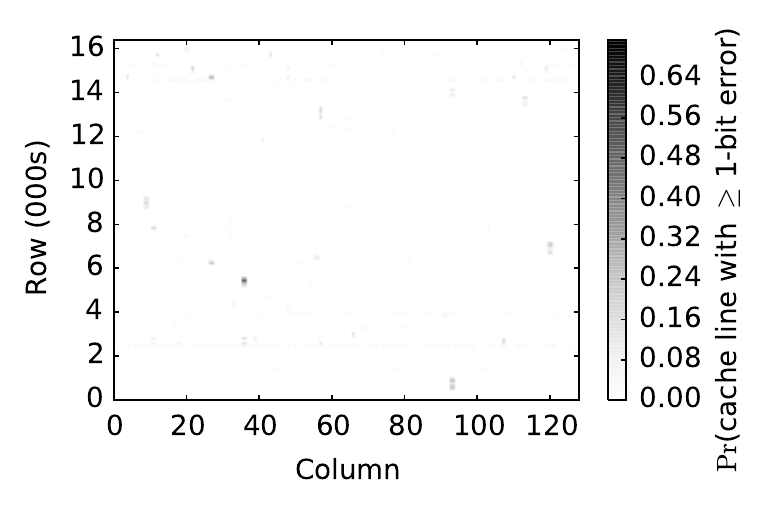}
    }
    \subcaptionbox{Bank 7.}[0.40\linewidth]
    {
        \includegraphics[width=0.40\linewidth]{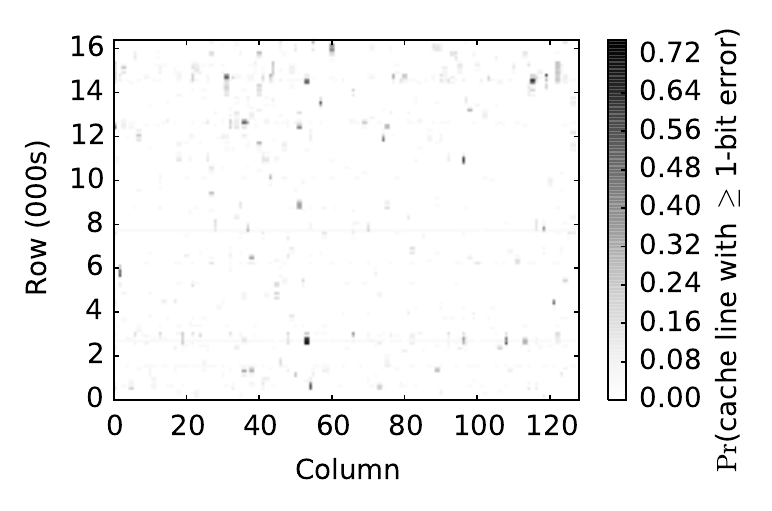}
    }

    \caption{Probability of \newt{observing precharge} errors in all eight banks
      of \dimm{1}{C}.}
    \label{fig:trp_err_loc_2}
\end{figure}
}


First, some banks do not have any precharge errors throughout the experiments,
such as Bank~0 (hence \newt{the plot is all white}). Similar to the activation
errors, precharge errors are \newt{\emph{not}} distributed uniformly across
locations within DIMMs. Second, some banks (e.g., bank~1 in \dimm{0}{C} and
bank~4 in \dimm{1}{C}) show that the errors concentrate on a certain region of
rows, \newt{while} the other regions experience much fewer or no errors. This
demonstrates that certain sense amplifiers, or cells at certain locations are
more robust than others, allowing them to work reliably under a reduced
precharge latency. Third, similar as our results on activation errors
(\secref{density_err}), the fraction of errors varies across banks. We
hypothesize the cause of such variation is the manufacturing process variation,
leading to different error behavior across banks within the same chip.

\obs{Precharge errors do not occur uniformly within DIMMs, but exhibit strong
spatial concentration at certain regions.}

Overall, we observe that 71.1\%, 13.6\%, and 84.7\% of cache lines contain no
precharge errors when they are read from \mbox{A-M1}, B-M1, and C-M0 model
DIMMs, respectively, \newt{with \trpe7.5ns}. Similar to the trend discussed in
\secref{rcd_ecc}, C-M0 DIMMs have the highest fraction of reliable cache lines
among the DIMMs tested, while B-M1 DIMMs experience the largest amount of
errors. Even though the number of error-free cache lines at \trpe7.5ns is lower
than that at \trcde7.5ns, the portion is still significant enough to show the
\newt{prevalence} of precharge latency variation in modern DIMMs.

\obs{When precharge latency is reduced, a majority of cache lines can be read without any timing errors in some
of our tested DIMMs. However, other DIMMs are largely susceptible to
precharge errors, resulting in a small fraction of error-free cache lines.}

\section{Restoration Latency Analysis} \label{sec:ras_lat_analysis}

In this section, we present \newt{a} methodology and findings on varying the
restoration latency, defined \newt{by} the \tras timing parameter. First, we
elaborate on the impact of reducing \tras on performance and reliability in
\secref{tras_effect}. Then, we explain our FPGA test conducted to characterize
\tras variation, and present our observations.

\subsection{Impact of Reduced tRAS} \label{sec:tras_effect}

As mentioned in Section~\ref{sec:dram_access}, \tras specifies the minimum
amount of time between issuing an \act and a \cpre command to a bank. By
reducing \tras, we can complete an access to one row faster, and quickly switch
to \newt{access} the next row. From the perspective of {\em reliability}, reducing
the restoration latency may potentially induce errors in the cells due to having
insufficient time to restore the lost charge back to the cells. When a row of
cells is activated, the cells temporarily lose their charge to the bitlines, so
that the sense amplifiers can sense the charge. During the restoration phase,
the sense amplifiers restore charge back into the cells, bringing them back
to the fully-charged state. By reducing the restoration latency, the amount of
restored charge reduces, and the cells may not reach the fully-charged state. As
a result, a subsequent access to the same row may not able to sense the correct
value, \newt{thereby leading to} errors.

\subsection{Test Methodology and Results} \label{sec:tras_tests_results}

To characterize the variation in restoration latency (\tras), we consider
another important factor that affects the amount of charge stored in DRAM cells,
which is \emph{leakage}. DRAM cells lose charge over time, thus requiring a
periodic {\em refresh} operation to restore the charge. \newt{Reducing} the
restored charge in the cells can cause them to lose too much charge before the
next refresh, generating an error.

To perform a conservative characterization, we integrate this leakage factor
into our test methodology. We access each row by issuing a pair of commands,
\act and \cpre, with a specific \tras value between these two commands. Then, we
wait for a full refresh period (defined as 64ms in the DRAM
standard~\cite{jedec-ddr3, jedec-ddr4}) before we access the row again to verify
\newt{the correctness of its data}. We test this sequence on a representative
set of DIMMs from all three DRAM vendors and we use four data patterns:
\texttt{0x00}, \texttt{0xff}, \texttt{0xaa}, and \texttt{0xcc}.

In our previously described tests on activation and precharge variation, we test
every time step from the default timing value to a minimum value of 2.5ns, with
a reduction of 2.5ns per step. Instead of reducing \tras all the way down to
2.5ns from its standard value of 35ns, we lower it until \newthree{$t\textsc{ras}_{min}=
\trcd + \tcl + \tbl$}, which is the latency of activating a row and
reading a cache line from it. In a typical situation where the memory controller
reads or writes a piece of data after opening a row, lowering \tras below
$t\textsc{ras}_{min}$ means that the memory controller can issue a \cpre while
the data is still being read or written. Doing so risks terminating \crd or \cwr
operations prematurely, causing unknown behavior.

In order to test \tras with a reasonable range of values, we iterate \tras from
35ns to $t\textsc{ras}_{min}$. Our $t\textsc{ras}_{min}$ is calculated by using
the standard \tcle13.125ns and \newthree{\tble5ns} along with a fast \trcde5ns.
$t\textsc{ras}_{min}$ is rounded up to the nearest multiple of 2.5ns, which
is \emph{22.5ns}.


We do not observe errors across the range of \tras values we tested \newt{in any
of our experiments}. This implies that charge restoration in modern DRAMs
completes within the duration of an activation and a read. Therefore, \tras can
be reduced aggressively without affecting data integrity.

\obs{Modern DIMMs have sufficient timing margin to complete charge
restoration within the period of an \act and a \crd. Hence, \tras can be
reduced without introducing any errors.}



\section{Exploiting Latency Variation}

Based on our extensive experimental characterization, we propose two new
mechanisms to reduce DRAM latency for better system performance. Our mechanisms
exploit the key observation that different DIMMs have different \newt{amounts of
tolerance} for lower DRAM latency, and there is a strong correlation between the
location of the cells and the lowest latency that the cells can tolerate. The
first mechanism (Section~\ref{sec:var_latency_mc}) is a pure hardware approach
to reducing DRAM latency. The second mechanism
(Section~\ref{sec:dram_aware_page_alloc}) leverages OS support to maximize the
benefits of the first mechanism.

\subsection{Flexible-Latency DRAM}
\label{sec:var_latency_mc}

As we discussed in Sections~\ref{sec:trcd_loc} and~\ref{sec:trp_spatial}, the
timing errors caused by reducing the latency of the activation/precharge
operations are concentrated on certain DRAM regions, which implies that the
latency heterogeneity among DRAM cells exhibits strong locality. Based on this
observation, we propose \emph{\mechlong (\mech)}, a software-transparent design
that exploits this heterogeneity in cells to reduce the overall DRAM latency.
The key idea of \mech is to determine the shortest reliable access latency of
each DRAM region, and to use the memory controller to apply that latency to the
corresponding DRAM region at runtime. There are two key design challenges of
\mech, as we discuss below.

The first challenge is determining the shortest access latency.  This can be
done using a \emph{latency profiling} procedure, which \romnum{i}~runs
Test~\ref{trcdtest} (Section~\ref{sec:trcd_tests}) with different timing values
and data patterns, and \romnum{ii}~records the smallest \newt{latency that
enables reliable access to} each region.  This procedure can be performed at one
of two times. First, the system can run the procedure the very first time the
DRAM is initialized, and store the profiling results to non-volatile memory
(e.g., \newt{disk} or flash memory) for future reference. Second, DRAM vendors
can run the procedure at manufacturing time, and embed the results in the Serial
Presence Detect (SPD) circuitry (a ROM present in each DIMM)~\cite{jedec-spd}.
The memory controller can read the profiling results from the SPD circuitry
during DRAM initialization, and apply the correct latency for each DRAM region.
While the second approach involves a slight modification \newt{to the DIMM}, it
can provide better latency information, as DRAM vendors have detailed knowledge
on DRAM cell variation, and can use this information to run more thorough tests
to determine a lower bound on the latency of each DRAM region.

The second design challenge is limiting the storage overhead of the latency
profiling results. Recording the shortest latency for each cache line can incur
a large storage overhead. For example, supporting four different \trcd and \trp
timings requires 4~bits per 512-bit cache line, which is almost 0.8\% of the
entire DRAM storage. Fortunately, the storage overhead can be reduced based on
\newt{a new} observation of ours. As shown in Figures~\ref{fig:rcd_col_err} and
\ref{fig:rcd_err_region}, timing errors typically concentrate on certain DRAM
\emph{columns}. Therefore, \mech records the shortest latency \emph{at the
granularity of DRAM columns}. Assuming we still need 4~bits per DRAM cache line,
we need only 512~bits per DRAM bank, or an insignificant \newt{0.00019\%}
storage overhead for the DIMMs we evaluated. One can imagine using more
sophisticated structures, such as Bloom Filters~\cite{bloom-cacm70}, to provide
finer-grained latency information within a reasonable storage overhead, as shown
in prior work on variable DRAM refresh time~\cite{liu-isca2012,qureshi-dsn2015}.
We leave this for future work.

\newt{The} \mech memory controller \romnum{i} loads the latency profiling results
into on-chip SRAMs at system boot time, \romnum{ii} looks up the profiled latency for
each memory \newt{request} based on its memory address, and \romnum{iii} applies the
\newt{corresponding latency to the request}. By reducing the latency values of
\trcd, \tras, and \trp for some memory requests, \mech improves overall system
performance, which we quantitatively demonstrate in the next two sections.


\subsection{Evaluation Methodology}

We evaluate the performance of \mech on an eight-core system using
Ramulator~\cite{kim-cal2015,ramulator}, an open-source cycle-level DRAM
simulator, driven by CPU traces generated from the Pin dynamic binary
instrumentation tool~\cite{luk-pldi2005}. Our source code is publicly
available~\cite{safari-github, ramulator}. Table~\ref{tab:sys-config} summarizes
the configuration of our evaluated system. We use the standard DDR3-1333H timing
parameters~\cite{jedec-ddr3} as our baseline.

\begin{table}[h]
    \renewcommand{\arraystretch}{0.9}
    \small
  \centering
    \setlength{\tabcolsep}{.45em}
    \begin{tabular}{ll}
        \toprule
        \textbf{Processor}   & 8 cores, 3.3 GHz, OoO 128-entry window \\
        \midrule

        \textbf{LLC} & 8 MB shared, 8-way set associative \\

        \midrule


        \multirow{3}{*}{\textbf{DRAM}} & DDR3-1333H~\cite{jedec-ddr3},
        open-row policy~\cite{rixner-isca2000}, \\
        & 2 channels,  1 rank per channel, 8 banks per rank, \\
        & Baseline: \trcd/\tcl/\trp= 13.125ns, \tras= 36ns \\



        \bottomrule
    \end{tabular}
  \caption{Evaluated system configuration.}
  \label{tab:sys-config}%
\end{table}


\noindent\textbf{\mech Configuration.} To conservatively evaluate \mech, we use
a randomizing page allocator that maps each virtual page to a randomly-located
physical page in memory. This allocator essentially distributes memory accesses
from an application to different latency regions at random, and is thus unaware
of \mech regions.

Because each DIMM has a different \newt{fraction} of fast cache lines, we
evaluate \mech on three different yet representative real DIMMs that we
characterized. We select one DIMM from each vendor. \tabref{vldram_config} lists
the distribution of cache lines that can be read \newt{reliably} under different
\trcd and \trp values, based on our characterization. For each DIMM, we use its
distribution as listed in the table to model the percentage of cache lines with
different \trcd and \trp values. For example, for \dimm{2}{A}, we set 93\% of
its cache lines to use a \trcd of 7.5ns, and the remaining 7\% of cache lines to
use a \trcd of 10ns. Although these DIMMs have a small fraction of cache lines
($<$10\%) that can be read using \trcde5ns, we conservatively set \trcde7.5ns for
them to ensure high reliability. \mech dynamically sets \trcd and \trp to either
7.5ns or 10ns for each memory \newt{request}, based on which cache line
\newt{the request is to}. For the \tras timing parameter, \mech uses 27ns
($\lceil{}${\small t}\textsc{rcd}+{\small t}\textsc{cl}$\rceil{}$) for all cache
lines in these three tested DIMMs, as we observe no errors in any of the tested
DIMMs due to lowering \tras (see \secref{tras_tests_results}).

\begin{table}[h]
    \renewcommand{\arraystretch}{1.2}
    \small
    \centering
    \begin{tabular}{cccrrrr}
        \toprule
        \multirow{2}{*}{DIMM} &
        \multirow{2}{*}{Vendor} & \multirow{2}{*}{Model} &
        \multicolumn{2}{c}{\trcd Dist. (\%)}  & \multicolumn{2}{c}{\trp Dist. (\%)} \\
        \cmidrule(l{2pt}r{2pt}){4-5}
        \cmidrule(l{2pt}r{2pt}){6-7}
        Name & & & 7.5ns & 10ns & 7.5ns & 10ns\\
        \midrule

        \dimm{2}{A} & A & M1 & 93 & 7 & 74 & 26 \\
        \dimm{7}{B} & B & M1 & 12 & 88 & 13 & 87 \\
        \dimm{2}{C} & C & M0 & 99 & 1 & 99 & 1 \\

        \bottomrule
    \end{tabular}
    \caption{Distribution of cache lines under various \trcd and \trp values for three
    characterized DIMMs.}
    \label{tab:vldram_config}%
\end{table}



\noindent\textbf{\mech Upper-Bound Evaluation.} We also evaluate the upper-bound
performance of \mech by assuming that \emph{every} DRAM cell is fast (i.e.,
100\% of cache lines can be accessed using \trcd/\trpe7.5ns).

\noindent\textbf{Applications and Workloads.} To demonstrate the benefits of
\mech in an 8-core system, we generate 40 8-core multi-programmed workloads by
assigning one application to each core. For each 8-core workload, we randomly
select 8 applications from the following benchmark suites: SPEC
CPU2006~\cite{spec2006}, TPC-C/H~\cite{tpc}, and STREAM~\cite{stream}. We use
PinPoints~\cite{patil-micro2004} to obtain the representative phases of each
application. Our simulation executes at least 200 million instructions on each
core, as done in prior works on multi-core performance
evaluation~\newt{\cite{subramanian-hpca2013,muralidhara-micro2011,kim-isca2012,lee-hpca2013,hassan-hpca2016,chang-hpca2016}.}

\noindent\textbf{Performance Metric.} We measure system performance with the
\emph{weighted speedup (WS)} metric~\cite{snavely-asplos2000}, which is a
measure of job throughput on a multi-core system~\cite{eyerman-ieeemicro2008}.
Specifically, $WS = \sum_{i=1}^N{\frac{IPC_{i}^{shared}}{IPC_{i}^{alone}}}$. $N$
is the number of cores in the system. ${IPC_{i}^{shared}}$ is the IPC of an
application that runs on $core_i$ while other applications are running on the
other cores. ${IPC_{i}^{alone}}$ is the IPC of an application when it runs alone
in the system without any other applications. Essentially, WS is the sum of
every application's slowdown compared to when it runs alone on the same system.

\subsection{Multi-Core System Results}
\label{sec:var_latency_mc:results}

\begin{figure}[h]
\centering
\includegraphics[scale=1.3]{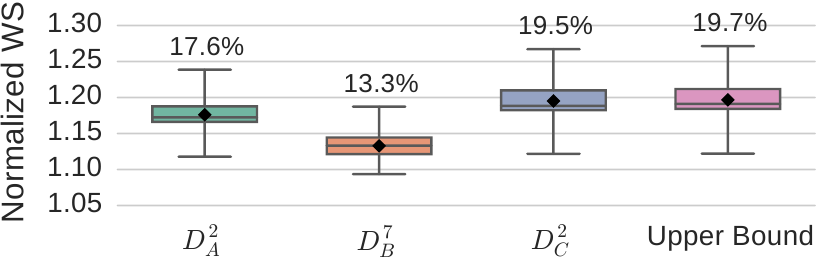}%
\caption{System performance \newt{improvement} of \mech for various DIMMs (listed in
\tabref{vldram_config}).\label{fig:vldram}}
\end{figure}

\figref{vldram} illustrates the system performance improvement of \mech over the
baseline for 40 workloads. The x-axis indicates the evaluated DRAM
configurations, as shown in \tabref{vldram_config}. The percentage value on top
of each box is the average performance improvement over the baseline. We use box
plots to show the performance distribution among all workloads. For each box,
the bottom, middle, and top lines indicate the 25th, 50th, and 75th percentile
of the population. The ends of the whiskers indicate the minimum and maximum
performance improvements. The black dot indicates the means.

We make the following observations. First, \mech improves system performance
significantly, by 17.6\%, 13.3\%, and 19.5\% on average across all 40 workloads
for the three real DIMMs that we characterize. This is because \mech reduces the
latency of \trcd, \trp, and \tras by 42.8\%, 42.8\%, and 25\%, respectively, for
many cache lines.
In particular, DIMM \dimm{2}{C}, whose great majority of cells are reliable at
low \trcd and \trp, performs within 1\% of the upper-bound performance (19.7\%
on average). Second, although DIMM \dimm{7}{B} has only a small fraction of
cells that can operate at 7.5ns, \mech still attains significant system
performance benefits by using low \trcd and \trp latencies (10ns), which are
23.8\% lower than the baseline, for the majority of cache lines. We conclude
that \mech is an effective mechanism to improve system performance by exploiting
the widespread latency variation present across DRAM cells.

\subsection{Discussion: DRAM-Aware Page Allocator}
\label{sec:dram_aware_page_alloc}

While \mech significantly improves system performance in a software-transparent
manner, we can take better advantage of it if we expose the \newt{different
latency regions of \mech to} the software stack.
\newt{We} propose the idea of a DRAM-aware page allocator in the OS, whose goal is to
better take advantage of \mech by intelligently mapping application pages to
different-latency DRAM regions in order to improve performance.

Within an application, there is heterogeneity in the access frequency of
different pages, where some pages are accessed much more frequently than other
pages, as shown in prior works~\cite{yu-micro2017,verghese-asplos1996,
  bhattacharjee-isca2009,ramos-ics2011,
  son-isca2013,sudan-asplos2010,yoon-iccd2012}. Our DRAM-aware page allocator
places more frequently-accessed pages into lower-latency regions in DRAM. This
\emph{access frequency aware page placement} allows a greater number of DRAM
accesses to \newt{experience} a reduced latency than a page allocator that is
oblivious to DRAM latency variation, thereby likely increasing system
performance.


For our page allocator to work effectively, it must know which pages are
expected to be accessed frequently.  In order to convey this information to the
page allocator, we extend the OS system calls for memory allocation to take in a
Boolean value, which states whether the memory being allocated is expected to be
accessed frequently.  This information either can be annotated by the
programmer, or can be estimated by various dynamic profiling
techniques~\cite{jiang-hpca2010, sudan-asplos2010, verghese-asplos1996,
  marathe-ppopp2006, ramos-ics2011, agarwal-asplos2015, yu-micro2017,
  chandra-asplos1994,yoon-iccd2012}. The page allocator uses this information to
find a free physical page in DRAM that suits \newt{the expected access frequency
  of the} application page that is being allocated.

We expect that by using our proposed page allocator, \mech can perform close to
the upper-bound performance reported in
Section~\ref{sec:var_latency_mc:results}, even for DIMMs that have a smaller
fraction of fast regions. We leave a thorough study of such a page allocator to
future work.

\section{Impact of Technology Scaling}
\label{sec:scaling}

In the previous sections, we demonstrate the existence and various behavior of
latency variation in many DIMMs manufactured from year 2011 to 2013. To lower
the cost-per-bit and power of DRAM chips, DRAM manufacturers scale down the size
of DRAM cells every two years to increase the chip density~\cite{itrs2-2015}.
Because the performance of DRAM changes with scaling, we set to perform a
first-order study on understanding how technology scaling affects the
characteristics of latency variation and reliability in DRAM chips.


\subsection{Methodology}

\noindent\textbf{DRAM DIMMs.} To study the trend for the three major vendors, we
purchase 30 new DDR3L (low-voltage) modules that were manufactured 1) in recent
years, between 2014 and 2016, and 2) with double amount of chip density, i.e.,
4Gb per chip over the 2Gb chips used in the previous sections.
\tabref{scaling_dimm_list} lists the other related parameters of these DIMMs.
For DRAMs manufactured by vendor B, we are able to obtain their technology node
size. However, technology node information is not publicly disclosed from the
other two vendors.

\renewcommand{\arraystretch}{0.9}
\begin{table}[h]
\small
  \centering
    \setlength{\tabcolsep}{1em}
    \begin{tabular}{ccccc}
        \toprule
        \multirow{2}{*}{Vendor} & Tech.& Chip & Assembly & DIMM \\
               & Node & Density & Year & Counts \\
        \midrule

        A & -    & 4Gb & 2016    & 8 \\
        B & 29nm & 4Gb & 2015    & 5 \\
        B & 28nm & 4Gb & 2015    & 4 \\
        C & -    & 4Gb & 2014-15 & 13 \\

        \bottomrule

    \end{tabular}
  \caption{Tested DIMMs manufactured from recent years.}
  \label{tab:scaling_dimm_list}%
\end{table}

\noindent\textbf{DRAM Tests.} We test the DIMMs with Test~\ref{trcdtest} for
\trcd and Test~\ref{trptest} for \trp on our FPGA platform to study latency
variation in these DRAM chips. For each DIMM, we performed at least 36 rounds of
Test~\ref{trcdtest} and Test~\ref{trptest}. The data patterns used in
Tests~\ref{trcdtest} and \ref{trptest} are described in
Sections~\ref{sec:trcd_tests} and~\ref{sec:trp_tests}, respectively. Unless
otherwise specified, the testing environment is under an ambient temperature of
20$\pm$1\celsius.

\subsection{Impact on Activation}

\begin{figure}[!h]
    \centering
    \captionsetup[subfigure]{justification=centering}
    \subcaptionbox{Vendor A\label{fig:rcd_scaling_A}}[\linewidth]
    {
        \includegraphics[scale=1.1]{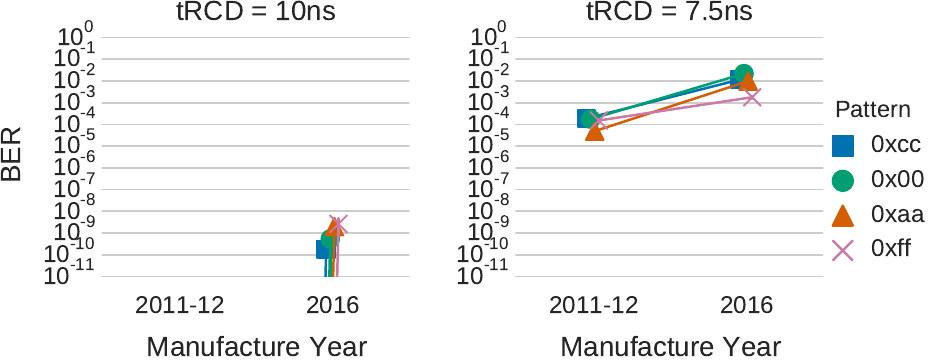}
    }

    \subcaptionbox{Vendor B\label{fig:rcd_scaling_B}}[\linewidth]
    {
        \includegraphics[scale=1.1]{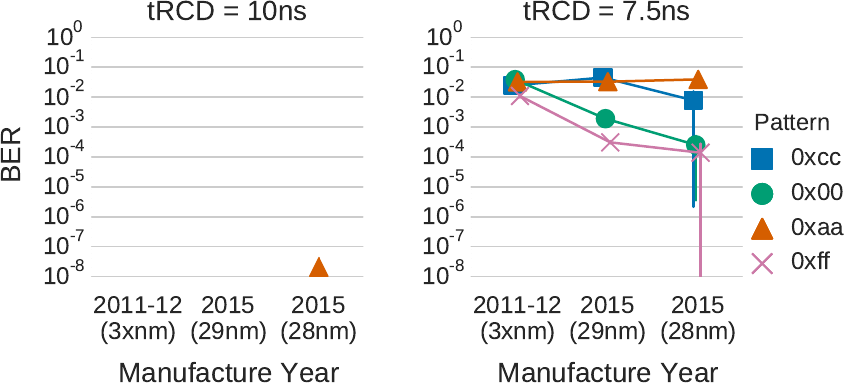}
    }

    \subcaptionbox{Vendor C\label{fig:rcd_scaling_C}}[\linewidth]
    {
        \includegraphics[scale=1.1]{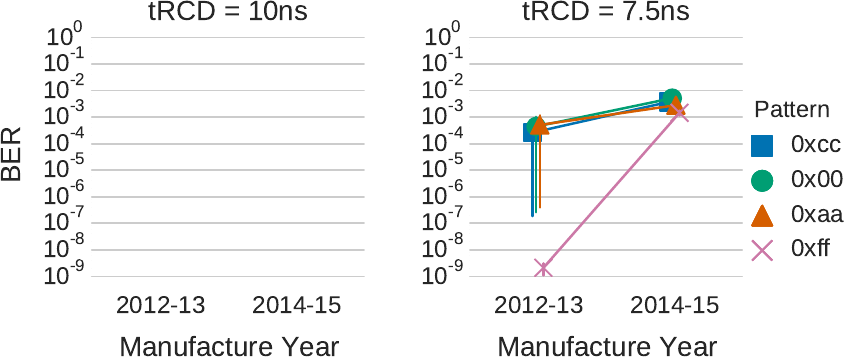}
    }

    \caption{BERs of DIMMs grouped by manufacture year, when tested under
    redueced activation latency.}
    \label{fig:rcd_scaling}
\end{figure}

\figref{rcd_scaling} shows the average BER of DIMMs at each corresponding
manufacture year due to various data patterns under a specific \trcd from three
major vendor. The error bars denote the 95\% confidence intervals. Note that we
do not show the results under \trcd=12.5ns because there are no observed errors.
We make several major observations.

First, Figures~\ref{fig:rcd_scaling_A}/\ref{fig:rcd_scaling_B} show that
recently manufactured DIMMs of vendors A and B observe a small amount of errors
when activation latency is set to 10ns. In contrast, older generation DIMMs do
not observe any errors at such activation latency. We believe that this is
because DRAM cells fabricated with the latest generations of process technology
are becoming more susceptible to process variation and electrical noise as their
dimensions are smaller~\cite{kim-thesis}. This means that post-2015 DIMMs
consist of more slow cells than DIMMs manufactured in years 2011 and 2012. In
contrast to capacity and bandwidth, which improve in newer generation of DIMMs,
DRAM access latency needs to increase to ensure data correctness. Our
observation shows that DRAM latency is not improving with DRAM scaling, thus
becoming a more critical performance bottleneck.

\begin{observation}
    DIMMs manufactured with the latest generation of technology node consist of
    slower cells that require longer activation latency than the pre-2013 DIMMs.
\end{observation}

Second, when activation latency is reduced to 7.5ns, new DIMMs by vendor A and C
observe higher BER across all patterns than the old DIMMs, respectively. This
implies that newer generation of DIMMs are likely facing worsened activation
reliability, thus increasing the number of slow cells.

Third, the BER gaps between the four data patterns become wider as technology
scales for vendor B, as shown under \trcd=7.5ns in \figref{rcd_scaling_B}. The
average BER due to \patt{0xaa} is at least an order of magnitude higher than
that of \patt{0xff} for 29nm DIMMs operating at 7.5ns. This pattern dependence
is similarly observed for another model of 2015 DIMMs that are manufactured with
28nm -- \patt{0xaa} induces an order of magnitude higher BER than \patt{0xff}.
Furthermore, \patt{0xaa} is the only pattern that induces errors when rows are
activated with \trcde10ns.

We believe the strong pattern dependence on \patt{0xaa} is because storing
alternating values (i.e., \patt{0xaa}) in consecutive bitlines generates higher
\emph{bitline-to-bitline coupling noise} than the other patterns, as suggested
and studied by many prior
works~\cite{koob-transvlsi2011,al-ars-vts2004,al-ars-ddecs2007,
itoh-ultra-low-memories,itoh-vlsi,liu-isca2013,kim-isca2014}. Scaling
exacerbates bitline coupling noise during row activation because scaling
physically shortens the distance between bitlines, thus amplifying the magnitude
of coupling noise.


\begin{observation}
    Data pattern dependence on row activation becomes stronger as technology scales.
\end{observation}

\subsection{Effect of High Temperature}

\begin{figure}[!h]
    \centering
    \captionsetup[subfigure]{justification=centering}
    \subcaptionbox{\trcde10ns\label{fig:rcd_temp_scaling_10}}[\linewidth]
    {
        \includegraphics[scale=1.1]{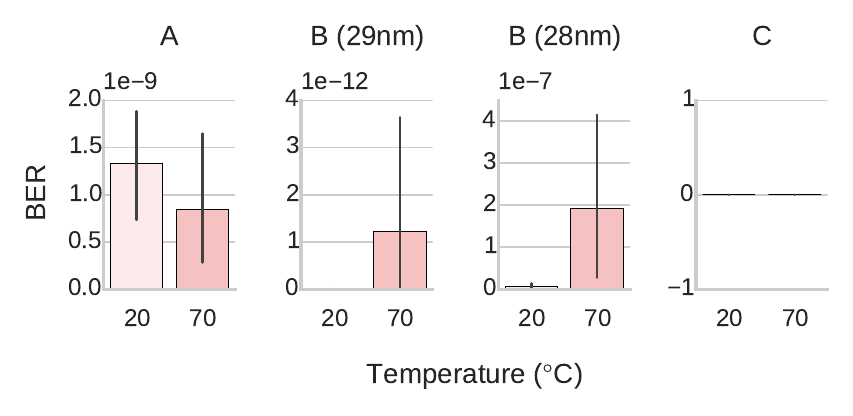}
        \vspace{-0.2in}
    }

    \subcaptionbox{\trcde7.5ns\label{fig:rcd_temp_scaling_7}}[\linewidth]
    {
        \includegraphics[scale=1.1]{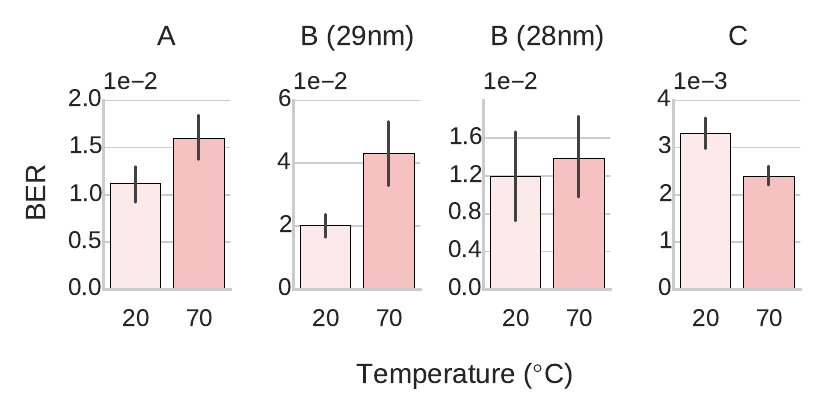}
        \vspace{-0.2in}
    }

    \caption{Effect of high temperature on the BERs of new generation DIMMs
    across vendors at different activation latency (\trcd) values.}
    \label{fig:rcd_temp_scaling}
\end{figure}

To study the effect of high temperature on new generation DIMMs operating with
reduced activation latency, we repeat the same set of DRAM tests on DIMMs placed
in high \emph{ambient temperature} of 70\celsius. \figref{rcd_temp_scaling}
shows the average BER across all DIMMs, categorized by vendors. The error bars
indicate the 95\% confidence interval. Since vendor B has two models of DIMMs
manufactured with different process technology (i.e., 29 and 28nm), we split
vendor B's DIMMs based on the corresponding technology node information.
\figref{rcd_temp_scaling_10} and \figref{rcd_temp_scaling_7} show the BER
results when DIMMs are tested with activation latency (\trcd) of 10ns and 7.5ns,
respectively. Several observations can be made from these results.

First, when \trcde10ns (\figref{rcd_temp_scaling_10}), DIMMs from both vendor
B's models observe an increase of errors when temperature raises from
20\celsius\xspace to 70\celsius. The fraction of slow cells that cannot be
activated with 10ns of latency increases due to higher temperature. In
particular, raising the temperature to 70\celsius starts inducing a very small
fraction of errors in vendor B's DIMMs that are made with the 29nm node -- only
$4.8e^{-3}$ ($=4Gb \times 1.2e^{-12}$) of cells experience bit flips when
activated with 10ns on average across all the experiments. For DIMMs made with
28nm, the BER increases by 17x on average under 70\celsius. In contrast, vendor
A observes no statistical significant difference in errors induced by low and
high temperature -- \emph{p-value}$=0.296 > 0.05$. We do not identify a single
bit of error in vendor C's DIMMs.

Second, under \trcde7.5ns (\figref{rcd_temp_scaling_7}), DIMMs from vendor A, B
(29nm) and C observe statistical significant difference in BER when temperature
is raised to 70\celsius. The results indicate that the BER increases for vendor
A and B. However, vendor C actually sees a small decrease in BER. This anomaly
requires further understanding of the DRAM design, which is undisclosed, by
vendor C in order to identify the root cause. On the other hand, Vendor B's 28nm
DIMMs observe no statistical difference in BER.

\begin{observation}
    Certain commodity DIMMs require longer activation latency in order for a
    small fraction of cells to operate reliably under high temperature.
\end{observation}

\subsection{Impact on Precharge}

\begin{figure}[!h]
    \centering
    \captionsetup[subfigure]{justification=centering}
    \subcaptionbox{Vendor A\label{fig:rp_scaling_A}}[0.32\linewidth]
    {
        \includegraphics[scale=1.1]{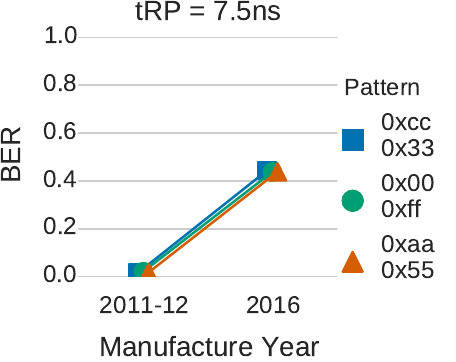}
    }
    \subcaptionbox{Vendor B\label{fig:rp_scaling_B}}[0.32\linewidth]
    {
        \includegraphics[scale=1.1]{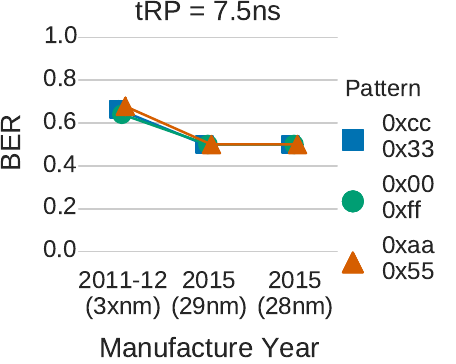}
    }
    \subcaptionbox{Vendor C\label{fig:rp_scaling_C}}[0.32\linewidth]
    {
        \includegraphics[scale=1.1]{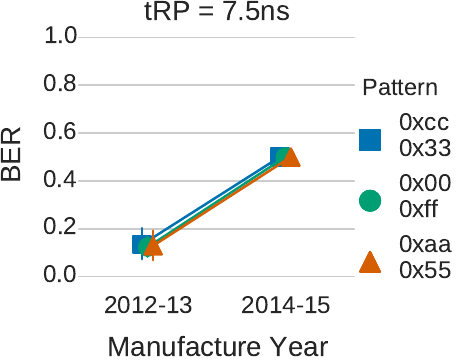}
    }

    \caption{BERs of DIMMs grouped by manufacture year, when tested under redueced precharge latency.}
    \label{fig:rp_scaling}
\end{figure}

\figref{rp_scaling} shows the average BER of DIMMs at each corresponding
manufacture year due to various data patterns under a \trp value of 7.5ns from
three major vendor. The error bars denote the 95\% confidence intervals. Note
that we do not show the results under \trcd=12.5ns because there are no observed
errors.

Compared to the older generation, the recently-manufactured DIMMs observe
significantly higher average BER at 48\%, 50\%, and 49\% for vendor A, B, and C,
respectively. These BER values indicate that the DIMMs become rife with errors,
similar to the probability of a coin toss, which exceeds the tolerable range.
Furthermore, applying different patterns or high temperature (not shown in the
figure) experiences similar BER without significant differences. One reason that
reducing \trp has much higher error rate than \trcd is that reducing the \trp
value affects the latency between two row-level DRAM commands, precharge and
activate. As a result, having an insufficient amount of precharge time for the
array’s bitlines affects the entire row, whereas lowering \trcd induces errors
only at a cache line level. However, it is unclear why post-2014 generation
DIMMs incur much higher error rates than the older generation DIMMs.

\begin{observation}
    The number of precharge errors is significantly higher in the recently
    manufactured DIMMs than the pre-2014 DIMMs.
\end{observation}

\section{Summary}

This chapter provides the first experimental study that comprehensively
characterizes and analyzes the latency variation within modern DRAM chips for
three fundamental DRAM operations (activation, precharge, and restoration).  We
find that significant latency variation is present across DRAM cells in all 240
of our tested DRAM chips, and that a large fraction of cache lines can be read
reliably even if the activation/restoration/precharge latencies are reduced
significantly. Consequently, exploiting the latency variation in DRAM cells can
greatly reduce the DRAM access latency.
Based on the findings from our experimental characterization, we propose and
evaluate a new mechanism, \mech (\mechlong), which reduces DRAM latency by
exploiting the inherent latency variation in DRAM cells.  \mech reduces DRAM
latency by categorizing the DRAM cells into fast and slow regions, and accessing
the fast regions with a reduced latency.  We demonstrate that \mech can greatly
reduce DRAM latency, leading to significant system performance
\newt{improvements} on a variety of workloads.

We conclude that it is promising to understand and exploit the inherent latency
variation within modern DRAM chips. We hope that the experimental
characterization, analysis, and optimization techniques presented in this
chapter will enable the development of other new mechanisms that exploit the
latency variation within DRAM to improve system performance and perhaps
reliability.

%
%
%


\chapter{Voltron: Understanding and Exploiting the Trade-off Between Latency and
Voltage in DRAM}
\label{chap:voltron}

In the previous chapter, we present our experimental study on characterizing
memory latency variation and its reliability implication in real DRAM chips. One
important factor that we have not discussed is \emph{supply voltage}, which
significantly impacts DRAM performance and DRAM energy consumption.
\changes{\textbf{Our goal} in this chapter is to characterize and understand the
relationship between supply voltage and DRAM latency. Furthermore, we study
the trade-off with various other characteristics of DRAM, including
reliability and data retention.


%


\section{Background and Motivation}
\label{sec:background}

In this section, we first provide necessary DRAM background and terminology. We
then discuss related work on reducing the voltage and/or frequency of DRAM, to
motivate the need for our study. \figref{dram-organization} shows a high-level
overview of \fixII{a modern} memory system organization. A processor (CPU) is
connected to a DRAM module via a \emph{memory channel}, which is a bus used to
transfer data and commands between the processor and DRAM. A DRAM module is also
called a \emph{dual in-line memory module} (\fixII{DIMM}) and it consists of
multiple \emph{DRAM chips}, which are controlled
together\fixIV{.\footnote{\fixIV{In this chapter, we study DIMMs that contain a
      single \emph{rank} (i.e., a group of chips in a single DIMM that operate
      in lockstep).}}} \changes{Within each DRAM chip, illustrated in
  \figref{chip-organization}, we categorize the internal components into two
  broad categories: \myitem{i} the \emph{DRAM array}, which consists of multiple
  banks of DRAM cells organized into rows and columns, and \myitem{ii}
  \emph{peripheral circuitry}, which consists of the circuits that sit outside
  of the DRAM array.


\begin{figure}[h]
  \captionsetup[subfigure]{justification=centering}
  
  \subcaptionbox{DRAM System\label{fig:dram-organization}}[0.4\linewidth] {
    \vspace{0.6cm}
    \includegraphics[scale=0.95]{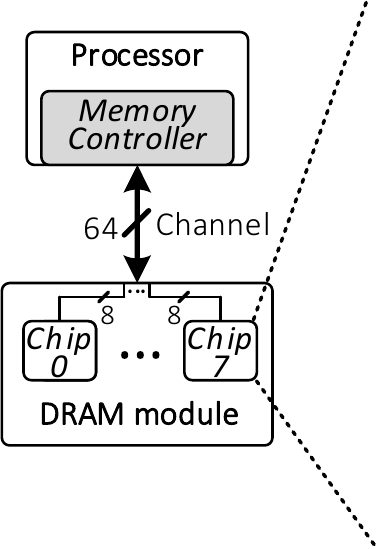}
}\hspace{-1.4cm}
  \subcaptionbox{DRAM Chip\label{fig:chip-organization}}[0.48\linewidth]
  {
    \includegraphics[scale=0.95]{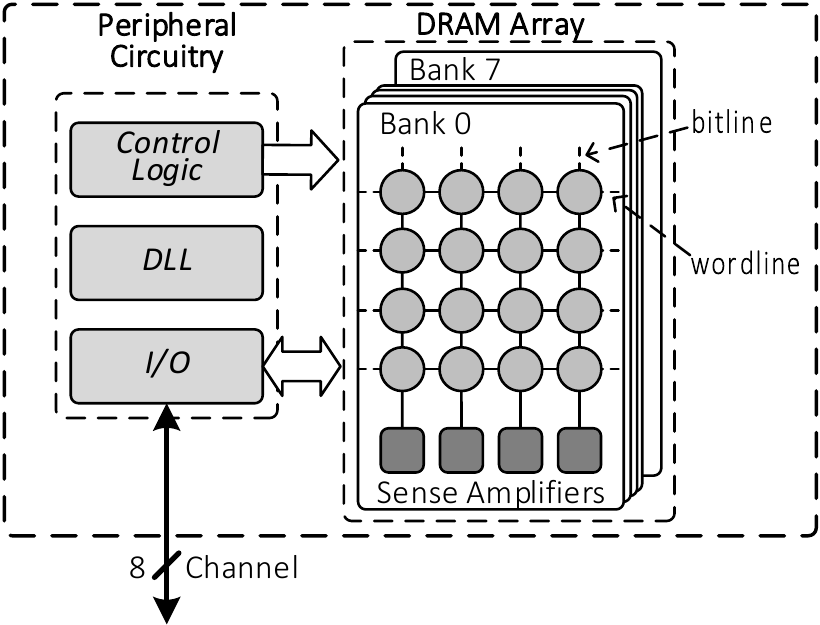}
  }%
  \caption{DRAM system and chip organization.}
  \label{fig:background}
\end{figure}


A DRAM array is divided into multiple banks (typically eight \fixIV{in DDR3
DRAM~\cite{jedec-ddr3l, jedec-ddr3}}) that can process DRAM commands
independently from each other to increase parallelism. A bank \fixIV{contains} a
2-dimensional array of DRAM cells. Each cell uses a capacitor to store \fixIV{a
  single bit} of data. Each array of cells is connected to a row of sense
amplifiers via vertical wires, called \emph{bitlines}. \fixII{This} row of sense
amplifiers \fixII{is called the} \emph{row buffer}. The row buffer senses the
data stored in one row of DRAM cells and serves as a temporary buffer for the
data. A typical row \fixIV{in a DRAM module (i.e., across all of the DRAM chips
  in the module)} is 8KB wide, comprising 128 \fixIV{64-byte} cache lines.


The peripheral circuitry has three major components. First, the I/O component is
used to receive commands or transfer data between the DRAM \fixIV{chip} and the
processor via the memory channel. Second, a typical DRAM chip uses a delay-lock
loop (DLL) to synchronize its data signal with the external clock to coordinate
data transfers on the memory channel.
Third, the control logic decodes DRAM commands sent across the memory channel
and selects the row \fixIV{and column} of cells to read data from or write data
into. For a more detailed view of the components in a DRAM chip and how to
access data stored in DRAM, we refer the reader to \chapref{background}.

\subsection{Effect of DRAM Voltage and Frequency on Power \fixII{Consumption}}
\label{sec:dram_power}

DRAM power \fixII{is divided} into dynamic and static power. \fixII{Dynamic}
power \fixII{is} the power consumed by executing the access commands: \act,
\pre, and \crd/\cwr. \fixIV{Each} \act and \pre consumes \fixIV{power in the
  DRAM array and the peripheral circuitry due to the activity in the DRAM
  array and control logic.} \fixII{Each} \crd/\cwr consumes power in the DRAM
array by accessing data in the row buffer, and in the peripheral circuitry by
driving data on the channel. On the other hand, static power \fixII{is} the
power that is consumed \emph{regardless} of the DRAM accesses, and it is mainly
due to transistor leakage. DRAM power is governed by both the supply voltage and
\fixII{operating clock} frequency: $Power \propto Voltage^2 \times
Frequency$~\cite{david-icac2011}. As shown in \fixIV{this} equation, power consumption
scales quadratically with supply voltage, and linearly with frequency.

DRAM supply voltage is distributed to both the DRAM array and the
peripheral circuitry through \fixII{respective} power pins on the
DRAM chip, \fixII{dedicated separately to the DRAM array and the peripheral
circuitry}. We call the voltage supplied to the DRAM array, \varr, and the
voltage supplied to the peripheral circuitry, \vperi. Each DRAM standard
requires a specific nominal supply voltage value, which depends on many
factors, such as the architectural design and process technology. In this
chapter, we focus on the widely used DDR3L DRAM design that requires a nominal
supply voltage of 1.35V~\cite{jedec-ddr3l}. To remain operational when the
\fixII{supply} voltage is unstable, DRAM can tolerate a small amount of
\fixIV{deviation from the nominal supply}
voltage. In particular, DDR3L DRAM \fixII{is specified to} operate
with a supply voltage ranging from 1.283V to 1.45V~\cite{micronDDR3L_2Gb}.

The DRAM \fixII{channel} frequency value of a DDR DRAM chip is typically
specified \fixIV{using the} \emph{channel data rate}, measured in mega-transfers per second
(MT/s). The size of each data transfer is dependent on the width of the data
bus, which ranges from 4 to 16~bits for a DDR3L chip~\cite{micronDDR3L_2Gb}.
Since a modern DDR channel transfers data on both the positive and \fixII{the}
negative clock edge\fixII{s (hence the term \emph{double data rate}, or DDR)},
the channel frequency \fixII{is} \emph{half of the data rate}. For example, a
DDR data rate of 1600 MT/s means that the frequency \fixII{is} 800 MHz. To run
the channel at a specified data rate, the peripheral circuitry requires a
certain minimum voltage (\vperi) for stable operation. As a result, the supply
voltage scales directly \fixIV{(i.e., linearly)} with DRAM frequency, and it
determines the maximum operating
frequency~\cite{deng-asplos2011,david-icac2011}.

\ignore{
Modern DDR DRAM chips are \emph{synchronous} state machines, driven by an
external clock signal from the memory controller. This allows the memory
controller to have timing control on when to send commands to or receive data
from DRAM. The DRAM frequency value of a DDR DRAM chip is typically specified in
channel data rate, measured in mega-transfers per second (MT/s). The size of
each data transfer is dependent on the width of the data bus, which ranges from
4 to 16 bits for a DDR3L chip~\cite{micronDDR3L_2Gb}. Since a modern DDR channel
transfers data on both the positive and negative clock edge, the channel
frequency operates at half of the data rate. For example, a DDR data rate of
1600 MT/s means that the channel operates at 800 MHz.\footnote{\reftiny Note that
  several prior works (imprecisely) use MT/s and MHz interchangeably to discuss
  channel frequency.} A DRAM chip typically supports several frequency values
(e.g., 1066, 1333, 1600 MT/s) that can be chosen by the processor only during the
system boot time. The DRAM frequency not only affects the data throughput, but
also the DRAM power. Except for the dynamic power of \act and \pre, which are
\emph{asynchronous operations}, the remaining DRAM power scales \emph{linearly} with
frequency: $Power \propto Frequency$.
}
}

\subsection{Memory Voltage and Frequency Scaling}
\label{ssec:dvfs}

\ignore{The memory system has become a major energy consumer in modern computing
systems, consuming 40\% of the total system energy in
servers~\cite{hoelzle-book2009,ibmpower7-hpca}, and 40\% of the total system
power in GPUs~\cite{paul-isca2015}).}

One \fixII{proposed} approach to reducing memory energy consumption is to scale
the voltage and/or the frequency of DRAM based on the \fixIV{observed memory
  channel} utilization. We briefly describe two different ways of scaling
frequency and/or voltage below.



\paratitle{Frequency Scaling} To \fix{enable the power reduction} \fixII{that
comes with reduced} \fixIV{DRAM} frequency, prior works propose to apply
\emph{dynamic frequency scaling} (DFS) by adjusting the DRAM \fixIV{channel}
frequency based on \fixIV{the} memory bandwidth demand \fixIV{from the DRAM
  channel}~\cite{deng-asplos2011,deng-islped2012,deng-micro2012, paul-isca2015,
  begum-iiswc2015,sundriyal2016}. A major \fixIV{consequence of} lowering the
frequency is the \fixII{likely} performance loss that occurs, as it takes a
longer time to transfer data across the DRAM channel \fixIV{while operating at a
  lower frequency}. The clocking logic within the peripheral circuitry requires
a \emph{fixed number of DRAM cycles} to transfer the data, \fixII{since} DRAM
sends data on each edge of the clock cycle. For a 64-bit memory channel with a
64B~cache line size, the transfer typically takes four DRAM
cycles~\cite{jedec-ddr3}. Since \fixII{lowering} the frequency increases the
time required for each cycle, the total amount of time spent on data transfer,
in nanoseconds, increases accordingly. As a result, \fix{not only does memory
  latency increase, but} \fixIII{also} memory \fix{data} throughput decreases,
\fixII{making} DFS \fixII{undesirable to use} when the running workload's memory
bandwidth \fixII{demand} \fixIV{or memory latency sensitivity} is high. The
extra transfer \fix{latency from DRAM} can also cause longer queuing times for
requests waiting at the memory controller\fixIII{~\cite{lee-tech2010,
    ipek-isca2008,
    kim-rtas2014,subramanian-iccd2014,subramanian-tpds2016,kim-rts2016}},
further exacerbating the performance loss \fixII{and potentially delaying
  latency-critical applications}~\cite{deng-asplos2011,david-icac2011}.


\paratitle{Voltage and Frequency Scaling} While decreasing the channel
frequency \fixII{reduces} the peripheral circuitry power and static power, it
does \fix{\emph{not}} affect the dynamic power consumed by the operations performed on
the DRAM array (i.e., activation, restoration, precharge). This is because
\fixII{DRAM array operations} are asynchronous\fix{, i.e., independent of}
the channel frequency~\cite{micron-tr}. As a result, these operations
require a fixed \fixIII{time (in nanoseconds)} to complete.  For example, the
activation latency in a DDR3L DRAM module \fixII{is} 13ns, regardless of the DRAM
frequency~\cite{micronDDR3L_2Gb}.  If the channel frequency is \fixII{doubled
from 1066~MT/s to 2133~MT/s}, the memory controller \fixII{doubles} the
number of cycles for the \act timing parameter (i.e., \trcd) (from 7~cycles
to 14~cycles), to maintain the 13ns latency.


\changes{ In order to reduce the dynamic power consumption of the DRAM
array as well, prior work \fixII{proposes} \emph{dynamic voltage and
frequency scaling} (DVFS) for DRAM, which reduces the supply voltage along
with the \fixII{channel} frequency~\cite{david-icac2011}. This mechanism
selects a DRAM frequency based on the current memory bandwidth utilization
and finds the \emph{minimum operating voltage} (\vmin) for that frequency.
\vmin is defined to be the lowest voltage that still provides ``stable
operation'' for DRAM (i.e., no errors occur within the data). There are two
significant limitations for this proposed DRAM DVFS mechanism. The first
limitation is due to a lack of understanding of how voltage scaling affects
the DRAM behavior. No prior work \fixIV{provides} experimental characterization
or analysis of \fixII{the effect of} reducing \fixIV{the DRAM supply} voltage \fixII{on latency,
reliability, and data retention} in real DRAM chips. As the DRAM behavior \fix{under reduced-voltage operation}
is unknown \fixII{to \fix{satisfactorily} maintain the latency and reliability of DRAM}, the
\fixII{proposed DVFS} mechanism~\cite{david-icac2011} can reduce \fixII{supply}
voltage only \emph{very conservatively}. The second limitation is that this
prior work
\fixII{reduces the
supply} voltage only when it \fixII{reduces the channel frequency, since
a lower channel frequency requires a lower supply voltage for stable operation.}
As a result, \fixII{DRAM} DVFS results in the same
performance issues experienced by \fixII{the DRAM} DFS mechanisms. In
\ssecref{pavc_eval}, we evaluate \fixII{the main} prior
work~\cite{david-icac2011} on memory DVFS to quantitatively demonstrate its
\fixII{benefits and limitations}.}


\subsection{Our Goal} The goal of this chapter is to \myitem{i}
\fixII{experimentally} characterize and analyze \emph{real modern DRAM chips}
operating at different supply voltage levels, \fixII{in order to develop a solid
  and thorough understanding of how \fix{reduced-voltage} operation affects
  latency, reliability, and data retention in DRAM;} and \myitem{ii} develop a
mechanism that can reduce DRAM energy consumption \fix{by reducing DRAM
  voltage,} without having to sacrifice memory \fixII{data} throughput,
\fixII{based on the insights obtained from \fixIV{comprehensive} experimental}
characterization. Understanding how DRAM characteristics change at different
voltage levels is imperative not only for enabling memory DVFS in real systems,
but also for developing other low-power \fixII{and low-energy} DRAM designs that
can effectively reduce the DRAM voltage. We \fixII{experimentally analyze} the
\fixII{effect} of reducing \fixII{supply} voltage \fixII{of modern} DRAM chips
in \secref{dram_exp}, and \fixII{introduce} our \fixII{proposed new} mechanism
\fixII{for reducing DRAM energy} in \secref{varray}.


\section{Experimental Methodology}
\label{sec:fpga}

To study the behavior of real DRAM chips under \fixII{reduced} voltage, we build
an FPGA-based infrastructure based on SoftMC~\cite{hassan-hpca2017}, which
allows us to have precise control over the DRAM modules. This method was used in
many previous works\fixIII{~\cite{jung-memsys2016, jung-patmos2016,
    kim-isca2014, chang-sigmetrics2016, khan-sigmetrics2014,
    khan-dsn2016,hassan-hpca2017,lee-sigmetrics2017,mathew-rapido2017,
    lee-hpca2015, qureshi-dsn2015, khan-cal2016, lee-thesis2016, kim-thesis,
    liu-isca2013}} as an effective way to explore different DRAM characteristics
(e.g., latency, reliability, \fixVI{and} \fixIV{data retention time}) that have
not been known or exposed to the public by DRAM manufacturers. Our testing
platform consists of a Xilinx ML605 FPGA board and a host PC that communicates
with the FPGA via a PCIe bus (\figref{fpga}). We adjust the supply voltage to
the DRAM by using a USB interface adapter~\cite{ti-usb} that enables us to tune
the power rail connected to the DRAM module directly. The power rail is
connected to all the power pins of every chip on the module (as shown in
Appendix~\ref{sec:pin_layout}).

\figputHS{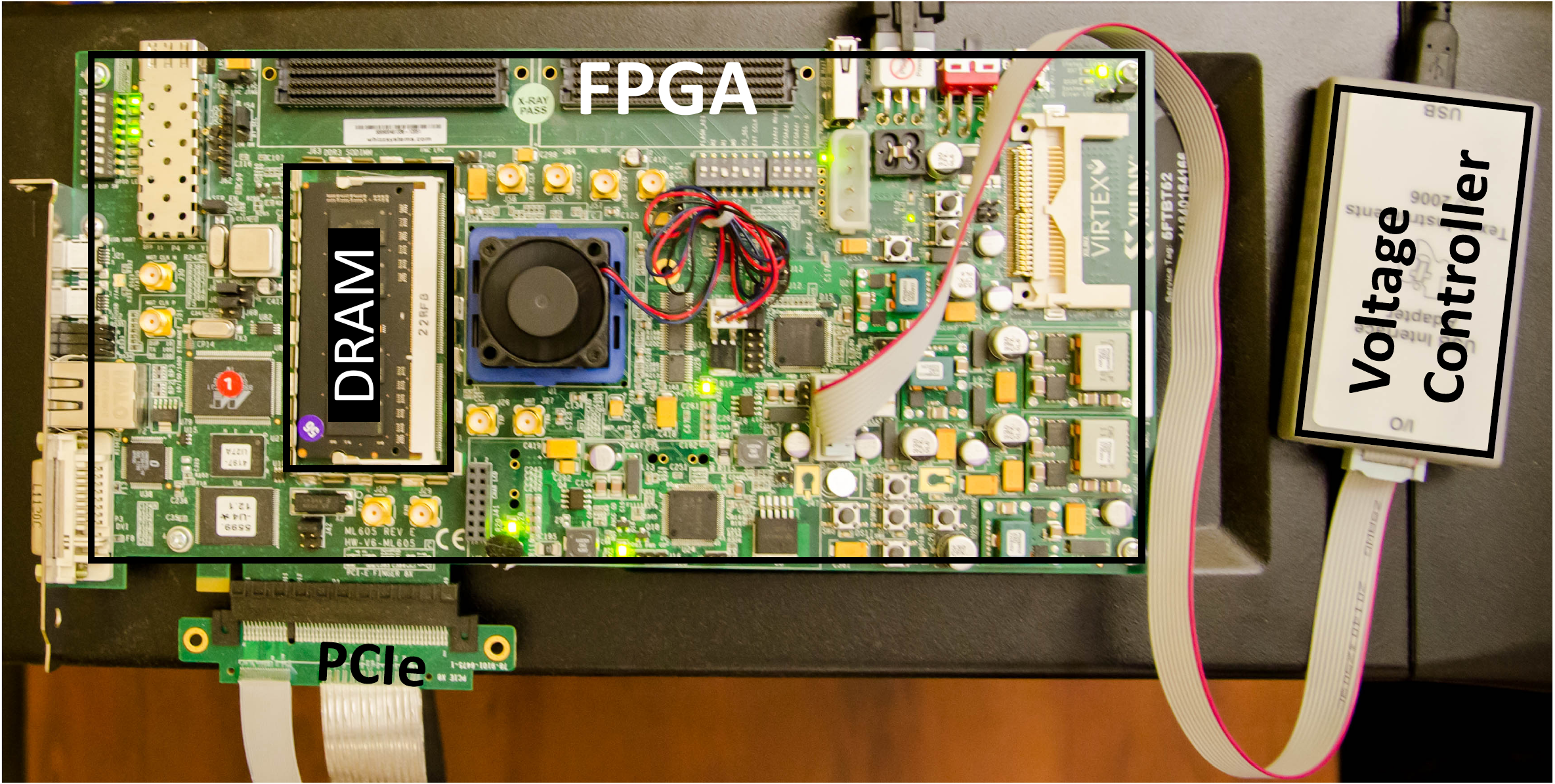}{0.5}{FPGA-based DRAM testing platform.}

\paratitle{Characterized DRAM Modules} In total, we tested 31 DRAM DIMMs,
comprising \fix{of} 124 DDR3L (low-voltage) chips, from the three \fixII{major}
DRAM chip vendors that hold more than 90\% of the DRAM market
share~\cite{bauer-mckinsey2016}.  Each chip has a 4Gb density. Thus,
\fixII{each} of our DIMMs \fixII{has} a 2GB capacity. The DIMMs support up to a
1600 MT/s channel frequency. Due to \fixII{our} FPGA's maximum operating
frequency limitations, all of our tests are conducted at 800 MT/s.  Note that
the experiments we perform \fixII{do \emph{not} require us to adjust} the
channel frequency. \tabref{dimm_list} describes the relevant information about
\fixIV{the} tested DIMMs. \fixIV{Appendix~\ref{sec:dimm_info} provides
  detailed information on each DIMM}. Unless otherwise specified, we test our
DIMMs at an ambient temperature of 20$\pm$1\celsius. We examine the effects of
high \fixIV{ambient} temperature \fixIV{(i.e., 70$\pm1\celsius$)} in
Section~\ref{ssec:temperature}.


\begin{table}[h]
  \centering
    \setlength{\tabcolsep}{.35em}
    \begin{tabular}{cccc}
        \toprule
        \multirow{2}{*}{Vendor} & Total Number & Timing (ns) &
        Assembly  \\
        & of Chips & (\trcd/\trp/\tras) & Year  \\
        \midrule
        A (10 DIMMs) & 40 & 13.75/13.75/35 & 2015-16\\
        B (12 DIMMs) & 48 & 13.75/13.75/35 & 2014-15 \\
        C (9 DIMMs) & 36 & 13.75/13.75/35 & 2015 \\
        \bottomrule
    \end{tabular}
  \caption{\fixIII{Main} properties of the tested DIMMs.}
  \label{tab:dimm_list}
\end{table}


\paratitle{DRAM Tests} \label{ssec:dramtest} At a high level, we develop a test
(Test~\ref{test}) that writes/reads data to\fixII{/from} \emph{every} row in the
\emph{entire} DIMM, \fixIV{for a given supply voltage}. The test takes in
several different input parameters: activation latency (\trcd), precharge
latency (\trp), and data pattern. The goal of the test is to examine if any
errors occur under \fixIV{the given supply voltage with} the different input
parameters.

\floatname{algorithm}{Test}

\begin{algorithm}[h]

\algnewcommand\algorithmicto{\textbf{to}}
\algrenewtext{For}[3]{\algorithmicfor\ #1 $\gets$ #2 \algorithmicto\ #3 \algorithmicdo}

\algrenewcommand\algorithmicfunction{}
\algrenewcommand\algorithmicdo{}
\algrenewcommand\algorithmicindent{1.2em}
\algrenewcommand\alglinenumber[1]{\footnotesize\texttt{#1}}
\small

\begin{algorithmic}[1]
\Function{VoltageTest}{$\mathit{DIMM}, \mathit{tRCD}, \mathit{tRP},
\mathit{data}, \overline{\mathit{data}}$}

\For{bank}{1}{$\mathit{DIMM.Bank}_{\mathit{MAX}}$}
\For{row}{1}{$\mathit{bank.Row}_{\mathit{MAX}}$} \Comment{Walk through every
  row within the current bank}
\State {\tt WriteOneRow(}$\mathit{bank}, \mathit{row}, \mathit{data}${\tt)}
\Comment{Write the data pattern into the current row}
\State {\tt WriteOneRow(}$\mathit{bank}, \mathit{row+1},
\overline{\mathit{data}}${\tt)} \Comment{Write the inverted data pattern into
  the next row}
\State {\tt ReadOneRow(}\trcd, \trp, $\mathit{bank}, \mathit{row} ${\tt)}
\Comment{Read the current row}
\State {\tt ReadOneRow(}\trcd, \trp, $\mathit{bank}, \mathit{row+1} ${\tt)}
\Comment{Read the next row}
\State {\tt RecordErrors()} \Comment{Count errors in both rows}

\EndFor
\EndFor

\EndFunction
\end{algorithmic}

    \caption{Test DIMM with specified tRCD/tRP and data \fix{pattern}.}
\label{test}
\end{algorithm}


In the test, we iteratively test two consecutive rows at a time. The two rows
hold data that are the inverse of each other (i.e., $\mathit{data}$ and
$\overline{\mathit{data}}$). Reducing \trp lowers the amount of time the
precharge unit has to reset the bitline voltage from either \vddbl (bit value 1)
or \vddzero (bit value 0) to \vddblhalf. If \trp were reduced too much, the
bitlines would float at some other intermediate voltage value between \vddblhalf
and \emph{full/zero voltage}. As a result, the next activation can potentially
start before the bitlines are fully precharged. If we were to use the same data
pattern in both rows, the sense amplifier would require \emph{less} time to
sense the value during the next activation, as the bitline is already
\emph{biased} \fixII{toward} those values. By using the \emph{inverse} of
\fixII{the data pattern in} the row \fixII{that is} precharged for the next row
\fixII{that is} activated, we ensure that the partially-precharged state of the
bitlines does \emph{not} unfairly favor the \fix{access to the} next
row~\cite{chang-sigmetrics2016}.
In total, we use three different groups of data patterns for our test: (0x00,
0xff), (0xaa, 0x33), and (0xcc, 0x55). Each specifies the $\mathit{data}$ and
$\overline{\mathit{data}}$, placed in consecutive rows in the same bank.

\section{Characterization of DRAM Under Reduced Voltage}
\label{sec:dram_exp}

In this section, we present our major observations from our detailed
experimental characterization of 31 commodity DIMMs \changes{(124 chips)} from
three vendors, when \changes{the DIMMs} operate under \fix{reduced} supply
voltage (i.e., below the nominal voltage level specified by \fixIV{the DRAM
  standard}).
First, we analyze the reliability of DRAM chips as we reduce the \changes{supply}
voltage without changing the DRAM access latency (\ssecref{volt_sensitivity}).
Our experiments \changes{are designed} to identify if lowering \changes{the
  supply} voltage induces bit errors (i.e., \emph{bit flips}) in data.
\fix{Second}, we present our findings on the effect of increasing the activation
and precharge latencies for DRAM operating under \fixIV{reduced} supply voltage
(\ssecref{low_volt_latency}). The purpose of this experiment is to understand
the trade-off between access latencies (which \fix{impact} performance) and the
supply voltage of DRAM \fixIII{(which impacts energy consumption)}. We use detailed circuit-level DRAM simulations to
validate \changes{and explain} our observations on the relationship between
access latency and supply voltage. \fix{Third}, we examine the spatial locality
of errors \changes{induced due to reduced-voltage operation} (\ssecref{spatial})
and the distribution of errors in the data sent across the memory channel
(\ssecref{ecc}).  \fix{Fourth}, we study the effect of temperature on
\changes{reduced}-voltage operation (\ssecref{temperature}). \fix{Fifth}, we
study \fix{the effect of reduced voltage on the} data retention times within
DRAM (\ssecref{refresh}). We present a summary of our findings in
\ssecref{char_summary}.

\subsection{DRAM Reliability as Supply Voltage Decreases}
\label{ssec:volt_sensitivity}

We first study the reliability of DRAM chips under low voltage, which was not
studied by prior works on DRAM voltage scaling
\fix{(e.g.,~\cite{david-icac2011}).} For these experiments, we use the minimum
activation and precharge latencies that we experimentally determine to be
reliable (i.e., they do not induce any errors) under the nominal voltage of
1.35V \changes{at 20$\pm$1$\celsius$ temperature}. As shown in \fix{prior
  works~\cite{chang-sigmetrics2016, lee-hpca2015, chandrasekar-date2014,
    lee-sigmetrics2017, hassan-hpca2017, lee-thesis2016,
liu-isca2012, agrawal-hpca2014, qureshi-dsn2015,
venkatesan-hpca2006,bhati-isca2015,lin-iccd2012,ohsawa-islped1998,
patel-isca2017, khan-sigmetrics2014, khan-dsn2016, khan-cal2016}},
\changes{DRAM} manufacturers adopt a pessimistic standard latency that
\fix{incorporates} a large margin as a \fix{safeguard to ensure that each chip
  \fixIII{deployed in the field} operates correctly under a wide range of
  conditions.  Examples of these conditions include process variation, which
  causes some chips or some cells within a chip to be slower than others, or
  high operating temperatures, which can affect the time required to perform
  various operations within DRAM.}
Since our goal is to understand how the inherent DRAM latencies vary with
voltage, we \fixIII{conduct} our experiments \emph{without} such an excessive
margin. We identify that the reliable \trcd and \trp latencies are
\changes{both} 10ns (instead of \fix{the 13.75ns latency} specified by the
\fix{DRAM} standard) at \fix{20$\celsius$}, which agree with the values reported
by prior work on DRAM latency characterization\fix{~\cite{chang-sigmetrics2016,
    lee-hpca2015, lee-thesis2016}}.

Using the \fix{\emph{reliable minimum latency values} (i.e., 10ns for all of the
DIMMs),} we run Test~\ref{test}, which accesses every bit within a DIMM at the
granularity of a 64B cache line. In total, there are 32~million cache lines in a \fix{2GB}
DIMM. We vary the supply voltage from the nominal voltage of 1.35V down to
1.20V, using a step size of 0.05V (50mV). Then, we change to a smaller step \fix{size} of
0.025V (25mV), until we reach the lowest voltage \fix{at which} the DIMM can
\fix{operate reliably (i.e., without any errors) while employing the reliable minimum latency values.}
\fix{(We examine methods to further reduce the \fixIII{supply} voltage in
  \ssecref{low_volt_latency}.)} For each voltage step, we run 30~rounds of
Test~\ref{test} for each DIMM.  \figref{dimm_errors_all} shows the fraction of
cache lines that \changes{experience} at least 1~bit of error (i.e., 1~bit flip)
in each DIMM \changes{(represented by each curve)}, \fixIV{categorized based
  on vendor.}

\figputHS{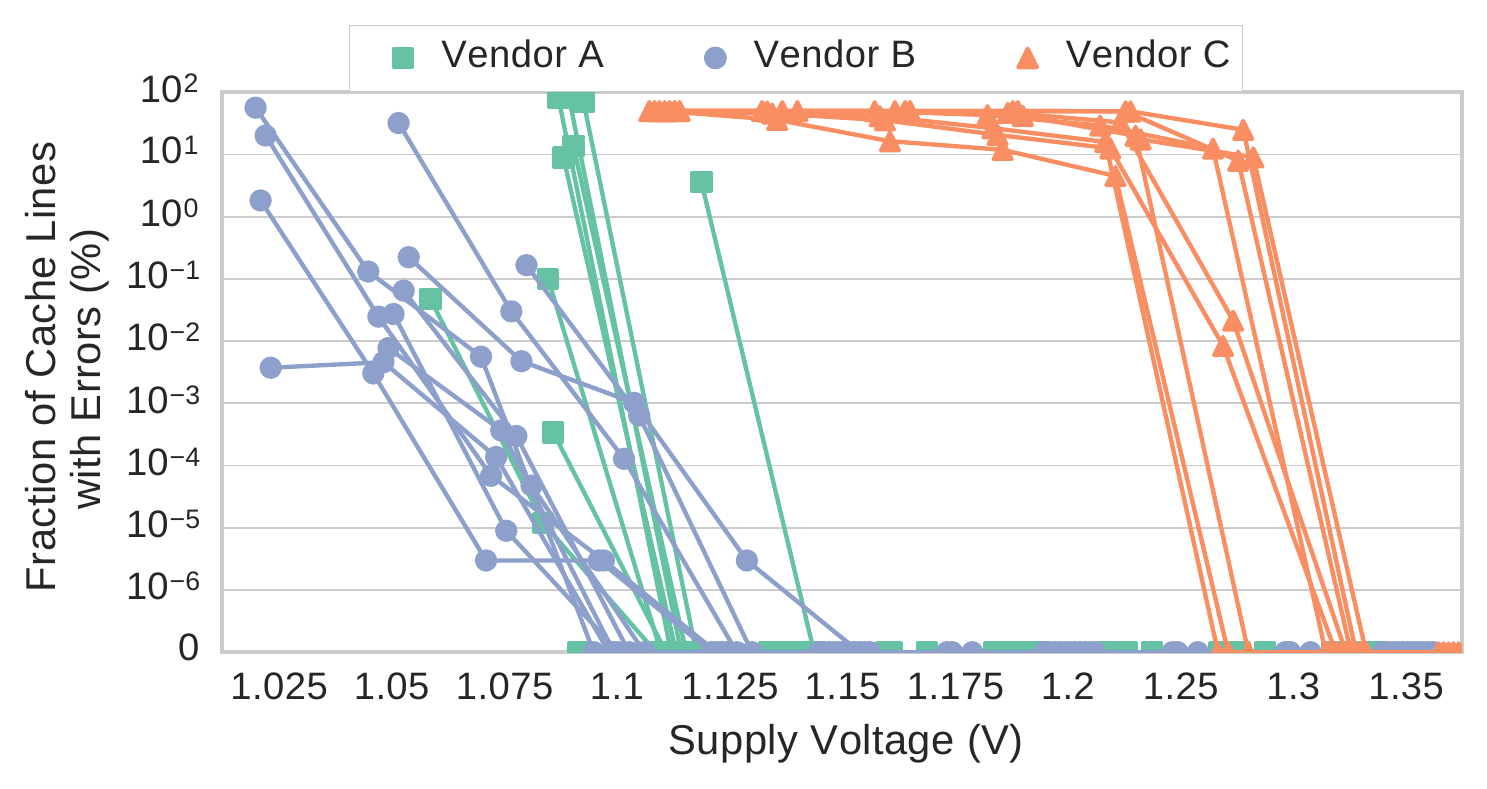}{0.8}{The fraction of erroneous cache lines in each DIMM as
we reduce the supply voltage, with a fixed access latency.}

We make three observations. First, when each DIMM runs below a certain voltage
level, errors start occurring. We refer to the \emph{minimum voltage level} of
each DIMM that allows error-free \changes{operation} as \vmin. For example, most
DIMMs from Vendor~C have $V_{min}=1.30V$.
\changes{Below \vmin, we observe errors because the fundamental DRAM array
operations (i.e., activation, restoration, precharge) \emph{cannot} fully
complete within the time interval specified by the latency parameters (e.g.,
\trcd, \tras) at low voltage.}
%
Second, not all cache lines exhibit errors for all supply voltage values below
\vmin. Instead, the number of erroneous cache lines \changes{for each DIMM}
increases as we reduce the voltage further below \vmin. Specifically, Vendor A's
DIMMs \fix{experience a} near-exponential increase \fix{in} errors as \fix{the} supply voltage reduces
below \vmin. This is mainly due to the \emph{manufacturing process and
  architectural variation}, which introduces strength and size variation across
the different DRAM cells within a chip\fixIII{~\cite{kim-edl2009, li11,
    chang-sigmetrics2016, lee-sigmetrics2017, lee-hpca2015,
    chandrasekar-date2014, lee-thesis2016, kim-thesis}}. \changes{Third, variation in \vmin exists not only
  across DIMMs from different vendors, but also across DIMMs from the same
  vendor.} However, the variation across DIMMs \fix{from} the same vendor is
much smaller compared to cross-vendor variation, since the fabrication process
and circuit designs can differ drastically across vendors. These results
demonstrate that reducing voltage beyond \vmin, without altering the access
latency, has a negative impact on \fix{DRAM} reliability.


We also conduct an analysis of \fixIII{storing different \emph{data patterns}} on the error rate
\fixIII{during reduced-voltage operation} (see Appendix~\ref{sec:datapatt}).  In summary, our results show that the data
pattern does \emph{not} \changes{have a} consistent effect on the rate of errors
induced by reduced-voltage operation. For most supply voltage values, the data
pattern does \emph{not} have a statistically significant effect on the error
rate.

\paratitle{Source of Errors} To understand why errors occur in data as the
supply voltage \fixIV{reduces} below \vmin, we \changes{perform} circuit-level SPICE
simulations~\cite{nagel-spice,massobrio1993semiconductor}, which reveal more
detail on how the cell arrays operate. \fix{We develop a} SPICE model of the DRAM array \fix{that} uses a
sense amplifier design from prior work~\cite{baker-dram} with the \SI{45}{\nano\meter}
transistor model from the Predictive Technology Model (PTM)~\cite{ptm,
  zhao-isqed2006}. \fix{Appendix~\ref{spice_model} provides a detailed description of our SPICE
simulation model, which we have open-sourced~\cite{volt-github}.}

We vary the supply voltage of the DRAM array (\vdd) \fix{in our SPICE simulations} from
1.35V to 0.90V.
\figref{spice_act+pre_v3} shows the bitline voltage during activation and
precharge for different \vdd values. Times 0ns and 50ns correspond to when the
DRAM receives the \act and the \pre commands, respectively. An \act causes the
bitline voltage to increase from \vddhalf to \vdd in order to sense the stored
data value of ``1''. A \pre resets the bitline voltage back to \vddhalf in order
\fix{to enable the issuing of a later}
\act to another row within the same bank. In the figure, we mark the
points where the bitline reaches \fix{the} \circled{1} \emph{ready-to-access} voltage,
which we assume to be 75\% of \vdd; \circled{2} \emph{ready-to-precharge}
voltage, which we assume to be 98\% of \vdd; and \circled{3}
\emph{ready-to-activate} voltage, which we assume to be within 2\% of \vddhalf.
These points \fix{represent the minimum \trcd, \tras, and \trp values, respectively,}
required for reliable DRAM operation.
For readers who wish to understand \fix{the bitline voltage behavior in more detail,}
\fix{we refer them to recent
works~\cite{lee-hpca2015,lee-hpca2013,hassan-hpca2016, lee-sigmetrics2017, lee-thesis2016} that provide
extensive background on how the bitline voltage changes during the three DRAM
operations.}

\figputHS{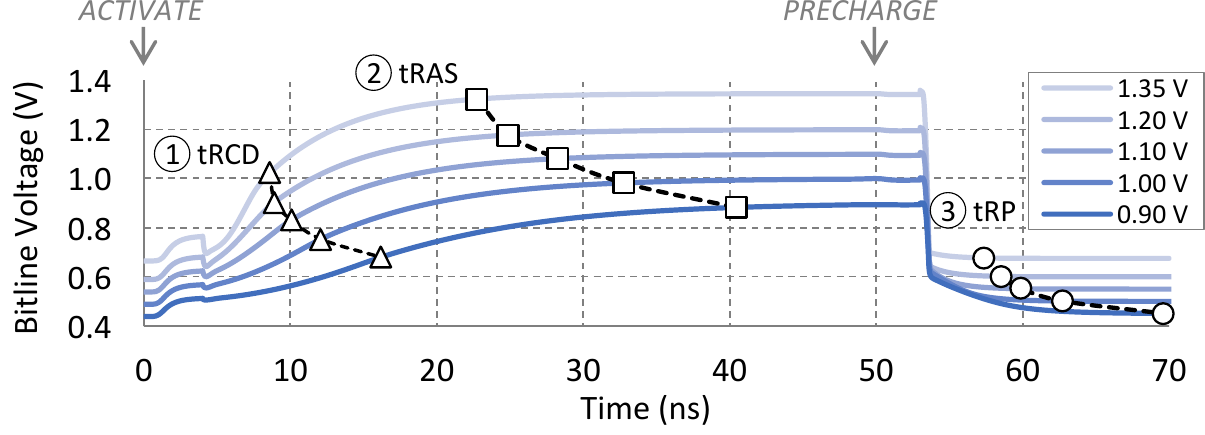}{1}{Effect of reduced array supply
voltage on activation, restoration, and precharge, from SPICE simulations.}

We make two observations from our SPICE simulations. First, we observe that the
bitline voltage during activation increases at a different rate depending on the
\changes{supply} voltage \fix{of} the DRAM array (\vdd). \changes{Thus},
\fix{the} \changes{supply voltage affects} the latency of the three DRAM
operations (activation, restoration, and precharge). When the nominal voltage
level (1.35V) is used \fix{for \vdd}, the time (\trcd) it takes for the sense
amplifier to drive the bitline to the \emph{ready-to-access voltage
  \changes{level}} (75\% of \vdd) is much shorter than the time \fix{to do so at
  a} lower \vdd. As \vdd decreases, the sense amplifier needs more time to latch
in the data, increasing the activation latency. \ignore{Similarly, the
  \changes{restoration latency} (\tras)\changes{, i.e., the} time needed to
  restore the cell to 98\% \changes{of} \vdd and the \changes{precharge latency}
  (\trp), \changes{i.e, the time} for the bitline to reset back to \vddhalf
  increase as \vdd decreases.} Similarly, the \changes{restoration latency}
(\tras) and the \changes{precharge latency} (\trp) increase as \vdd decreases.

Second, the latencies of the three fundamental DRAM array operations (i.e.,
activation, restoration, precharge) \changes{do} \emph{not} correlate with the
channel (or clock) frequency \fixIV{(not shown in \figref{spice_act+pre_v3})}. This is
because these operations are clock-independent \changes{asynchronous} operations
that are a function of the cell capacitance, bitline capacitance, and
\vdd~\cite{keeth-dram-tutorial}.\fix{\footnote{\fix{In
      Appendix~\ref{spice_model}, we show a detailed circuit schematic of a DRAM
      array that operates asynchronously\fixIII{, which forms the basis of our
        SPICE circuit simulation model~\cite{volt-github}}.}}} As a result, the
channel frequency is \emph{independent} of the three fundamental \changes{DRAM}
operations.

Therefore, we hypothesize that DRAM errors occur at lower supply voltages
because the \fix{three} DRAM \fix{array} operations have insufficient latency to
fully complete \fix{at lower voltage levels}. In the next section, we
\changes{experimentally} investigate the effect of increasing latency values as
we vary the supply voltage \changes{on real DRAM chips}.

\subsection{Longer Access Latency Mitigates Voltage-Induced Errors}
\label{ssec:low_volt_latency}

%

To confirm our hypothesis from \ssecref{volt_sensitivity} that a lower supply
voltage requires a longer access latency, we test our DIMMs at supply voltages
below the nominal voltage (1.35V) while incrementally increasing the activation
and precharge latencies \changes{to be} as high as 20ns (2x higher than the
tested latency in \ssecref{volt_sensitivity}). At each supply voltage value, we
call the minimum required activation and precharge latencies that do \emph{not}
exhibit any errors \trcdmin and \trpmin, respectively.

\figref{fpga_latency_range} shows the distribution of \trcdmin (top row) and
\trpmin (bottom row) measured for all DIMMs across three vendors as we vary the
supply voltage. Each circle represents a \trcdmin or \trpmin value. A circle's
size indicates the DIMM population size, with bigger circles representing more
DIMMs. The number above each circle indicates the fraction of DIMMs that work
reliably at the specified voltage and latency. Also, we
shade the range of potential \trcdmin and \trpmin values. Since our
infrastructure can adjust the latencies at a granularity of
2.5ns, a \trcdmin or \trpmin value of 10ns is only an approximation of the
minimum value, as the precise \trcdmin or \trpmin falls between 7.5ns and 10ns.
We make three major observations.

\begin{figure*}[ht!]
    \centering
    \captionsetup[subfigure]{justification=centering}
    {
        \includegraphics[width=\linewidth]{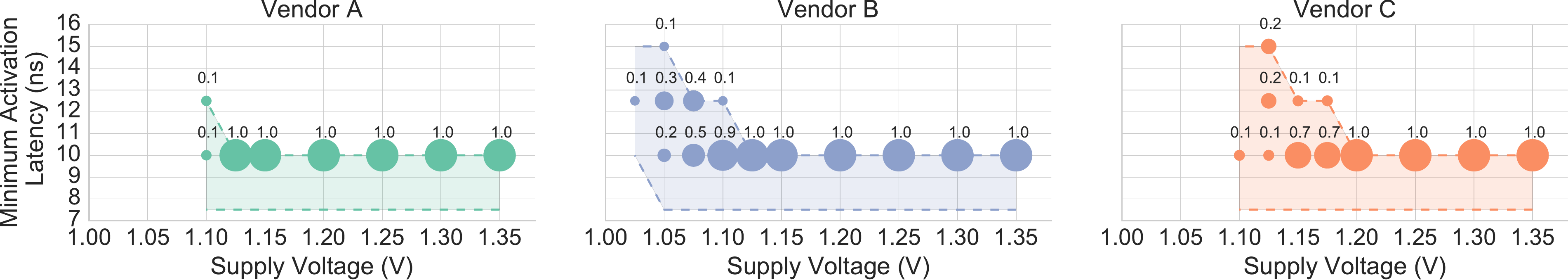}
    }

    {
        \includegraphics[width=\linewidth]{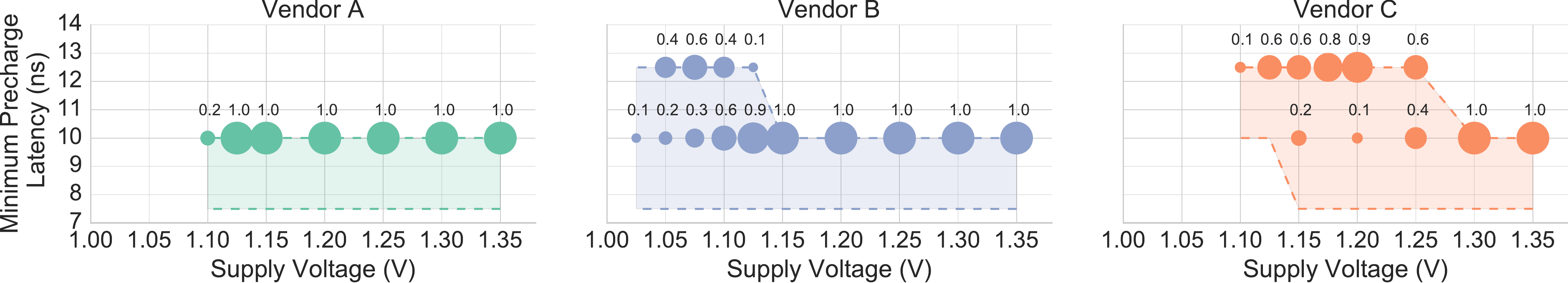}
        \vspace{-0.1in}
    }
    \vspace{-0.15in}
    \caption{Distribution of minimum reliable latency values as the supply
      voltage is decreased for 31 DIMMs. The number above each point indicates
      the fraction of DIMMs that \fixIV{work reliably at the specified voltage
      and latency}. Top row: \trcdmin; Bottom row: \trpmin.}
    \label{fig:fpga_latency_range}
\end{figure*}


First, when the supply voltage falls below \vmin\footnote{\fix{In
    \ssecref{volt_sensitivity}, we define \vmin as the minimum voltage level of
    each DIMM that allows error-free operation. \tabref{modules} in Appendix~\ref{sec:dimm_info}
shows the \vmin value we found for each DIMM.}},
the tested DIMMs show
that an increase of at least 2.5ns is needed for \trcdmin and \trpmin to read
data without errors. For example, some DIMMs require at least a 2.5ns increase
\fix{of} \trcdmin or \trpmin to read data without errors at 1.100V, 1.125V, and 1.25V
from \fix{Vendors}~A, B, and C, respectively. Since our testing platform can only
identify the minimum latency at a granularity of 2.5ns\fixIII{~\cite{hassan-hpca2017}}, we use circuit-level
simulations \fix{to obtain} a more precise latency measurement of \trcdmin and \trpmin
(\fix{which we describe} in the latter part of this section).

Second, DIMMs from different vendors \changes{exhibit} very different behavior
on how much \trcdmin and \trpmin need to increase for reliable operation as
supply voltage falls below \vmin. Compared to other vendors, many more of
Vendor~C's DIMMs require higher \trcdmin and \trpmin to operate at a lower \vdd.
This is particularly the case for the precharge latency, \fixIV{\trpmin}. For
instance, 60\% of Vendor~C's DIMMs require a \trpmin of 12.5ns to read data
without errors at 1.25V, whereas this increase is not necessary at all for
\fix{DIMMs from Vendor~A, which \emph{all} operate reliably at \fixIII{1.15V}.}
This reveals that different vendors may have different circuit architectures or
manufacturing process technologies, \fixIII{which lead to variations in} the
\fix{additional} latency required to compensate for a reduced \vdd in DIMMs.

%

Third, \changes{at very low supply voltages}, not all of the DIMMs have valid
\trcdmin and \trpmin values less than or equal to 20ns that \changes{enable
  error-free operation of the DIMM}. We see that the circle size gets smaller as
the \fixIV{supply} voltage reduces, indicating that the number of DIMMs
\fix{that can operate reliably (even at higher latency) reduces.} For example,
Vendor~A's DIMMs can no longer operate \changes{reliably (\fix{i.e.,}
  error-free)} when the voltage is below 1.1V.
We tested a small subset of DIMMs with latencies of more than 50ns and found
that these very high latencies still do \emph{not} prevent errors from
occurring. We hypothesize that this is because of signal integrity issues on the
channel, causing bits to flip during data transfer at very low supply voltages.

\fix{We correlate our characterization results with} our SPICE simulation results
from \ssecref{volt_sensitivity}, demonstrating that there is a direct
relationship between supply voltage and access latency. This \fix{new
  observation on the trade-off between supply voltage and access latency} is not
discussed or \fix{demonstrated} in prior work \changes{on DRAM voltage
  scaling}~\cite{david-icac2011}, where the access latency (in nanoseconds)
remains \emph{fixed} when performing memory DVFS. \changes{In conclusion, we
  demonstrate both experimentally and in circuit simulations that increasing the
  access latency (i.e., \trcd and \trp) allows us to lower the supply voltage
  while still reliably accessing data without errors.}


%
\paratitle{Deriving More Precise Access Latency Values}
One limitation of our experiments is that we cannot \emph{precisely} measure the
\emph{exact} \trcdmin and \trpmin values, due to the 2.5ns \fix{minimum latency} granularity of our
experimental framework~\cite{hassan-hpca2017}. Furthermore, supply voltage
is a continuous value, and it would take a prohibitively long time to study
the supply voltage \changes{experimentally} at a finer granularity.
We address these limitations by enriching our experimental results
with circuit-level DRAM SPICE simulations that model a DRAM array
(\fix{see} Appendix~\ref{spice_model} \fixIII{for details of our circuit simulation model}).

%
%

The SPICE simulation results highly depend on the specified transistor
parameters (e.g., transistor width). To fit our SPICE results with our
experimental results (for the supply voltage values that we studied
experimentally), we manually adjust the transistor parameters until the
simulated results fit within our \emph{measured} range of latencies.
\figref{spice+fpga_annotate} shows the latencies reported for activation and
precharge \fixIV{operations} using our final SPICE model\fix{,} based on the measured experimental data
for Vendor~B.

\figputHS{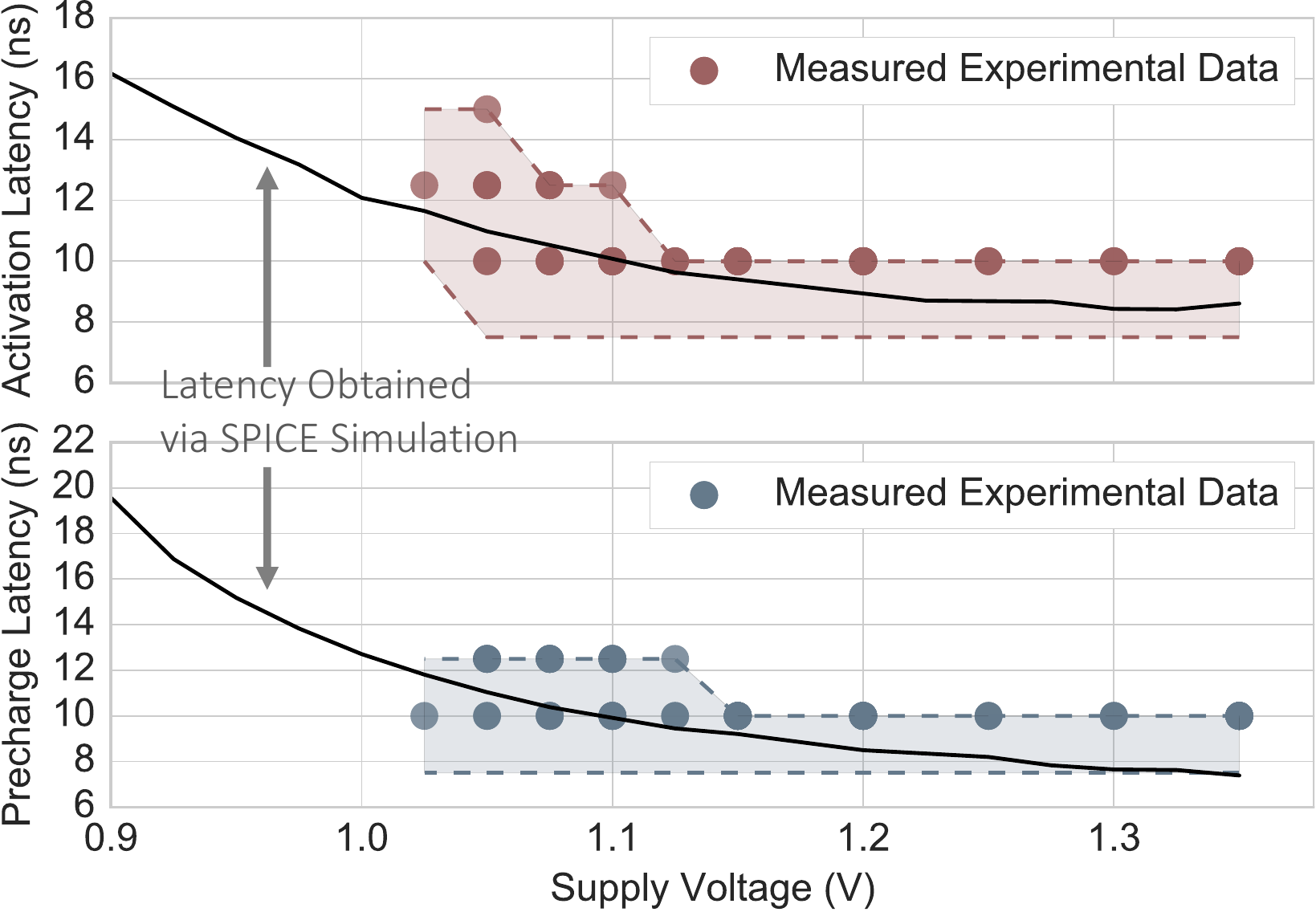}{0.6}{SPICE simulation results compared with
  \fixIV{experimental} measurements
from 12 DRAM DIMMs for Vendor~B.}

We make \fix{two} major observations. First, \fix{we see} that the SPICE
\fix{simulation} results fit within the \fixIII{range of latencies measured
  during our experimental characterization}, confirming that our \fix{simulated}
circuit behaves close to the real DIMMs. \fix{As a result, our circuit model
  allows us to derive a more precise minimum latency for reliable operation than
  our experimental data.\fixIII{\footnote{\fixIII{The circuit model can further
        serve as a common framework for studying other characteristics of
        DRAM.}}}} Second, DRAM arrays can operate at a wide range of voltage
\fix{values} without \fix{experiencing} errors. This aligns with our hypothesis
that errors at very low supply voltages (e.g., 1V) occur during data transfers
\fix{across the channel rather than during DRAM array \fixIII{operations.}}
Therefore, our SPICE simulations not only validate \fixIV{our observation} that
a lower supply voltage requires longer access latency, but also provide us with
\fixIII{a} more precise \fix{reliable minimum operating latency
  \fixIII{estimate} for a given supply voltage.}




\subsection{Spatial Locality of Errors}
\label{ssec:spatial}

While reducing the supply voltage induces errors when the DRAM latency is not
long enough, we also show that not all DRAM locations experience errors at all
supply voltage levels. To understand the locality of the errors induced by a low
supply voltage, we show the probability of each DRAM row in a DIMM
\fix{experiencing} at least one bit of error across all experiments.
\response{We present results for two \fixIV{representative} DIMMs from two
  different vendors\fix{, as} the observations from these two DIMMs are similar
  to those we make \fixIV{for} the other tested DIMMs. Our results collected
  from \fix{each of the 31} DIMMs are publicly available~\cite{volt-github}.}

\figref{locality_C} shows the probability of each row \fix{experiencing} at
least a one-bit error due to reduced voltage in the two representative DIMMs.
For each DIMM, we choose the supply voltage when errors start appearing (i.e.,
the voltage \fix{level} one step below \vmin), and we do \emph{not} increase the
DRAM access latency \fixIV{(i.e., 10ns for both \trcd and \trp)}. The x-axis
and y-axis indicate the bank number and row number (in thousands), respectively.
Our tested DIMMs are divided into eight banks, and each bank consists of
32K~rows of cells.\footnote{Additional results showing the error locations at
  different voltage \fix{levels} are in Appendix~\ref{spatial}.}

\begin{figure}[!h]
    \centering


    \subcaptionbox{DIMM B$_6$ of vendor~B at 1.05V.}[0.49\linewidth][l]
    {
        \includegraphics[width=0.48\linewidth]{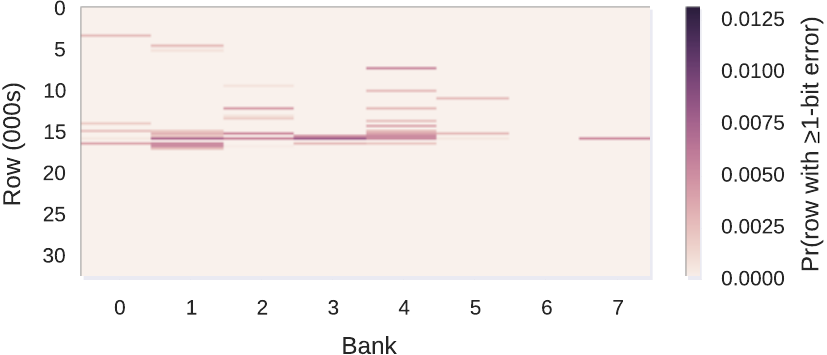}
    }
    \subcaptionbox{DIMM C$_2$ of vendor~C at 1.20V.}[0.49\linewidth][r]
    {
        \includegraphics[width=0.48\linewidth]{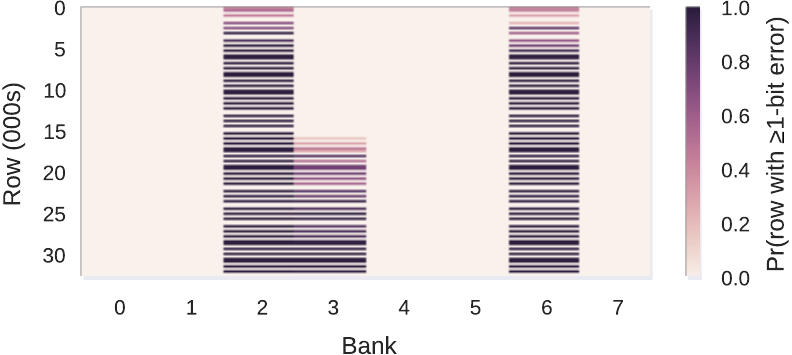}
    }


        \vspace{-0.1in}
		\caption{The probability of error occurrence for two representative
          DIMMs, \fixIV{categorized into different rows and banks}, due to reduced voltage.}

    \label{fig:locality_C}
\end{figure}


Our main observation is that errors tend to cluster at certain locations. For
our representative DIMMs, we see that errors tend to cluster at certain rows
across multiple banks for Vendor~B. On the contrary, Vendor~C's DIMMs exhibit
errors in certain banks but not in other banks. We hypothesize that the error
concentration can be a result of \myitem{i} manufacturing process variation,
resulting in less robust components at certain locations, as observed in
Vendor~B\fix{'s DIMMs}; or \myitem{ii} architectural design variations in the power delivery
network. However, it is hard to verify our hypotheses without knowing the
specifics of the DRAM circuit design, which is proprietary information that
varies across different DRAM models within and across vendors.

Another implication of the spatial concentration of errors under low voltage is
that \emph{only those regions with errors require a higher access latency to
  read or write data correctly}, whereas error-free regions can be accessed
reliably with the standard latency. \fix{In \ssecref{eval_el}, \fixIII{we
    discuss} and evaluate a technique that exploits this \fixIV{spatial}
  locality \fixV{of errors} to improve system performance.}


\subsection{Density of Errors}
\label{ssec:ecc}

In this section, we investigate the density (i.e., the number) of error bits
that occur within each \emph{data beat} (i.e., the unit of data transfer,
\fixIV{which is 64 bits,} through the data bus) read back from DRAM.
Conventional error-correcting codes (ECC) used in DRAM detect and correct errors
at the granularity of a data beat. For example, SECDED
ECC\fixIII{~\cite{sridharan-asplos2015, luo-dsn2014}} can \fix{correct a
  single-bit error and detect two-bit errors} within a data beat.
\figref{err_density} shows the distribution of data beats that contain no
errors, \fix{a single-bit} error, two-bit errors, \fixVI{or} \fix{more than two
  bits of} errors, under different supply voltages for all DIMMs. These
distributions are collected from \fix{30 rounds of experiments that were tested
  on each of the 31 DIMMs per voltage} level, using 10ns of activation and
precharge latency. \fixIV{A round of experiment refers to a single run of
  Test~\ref{test}, as described in \secref{fpga}, on a specified DIMM.}

\figputHS{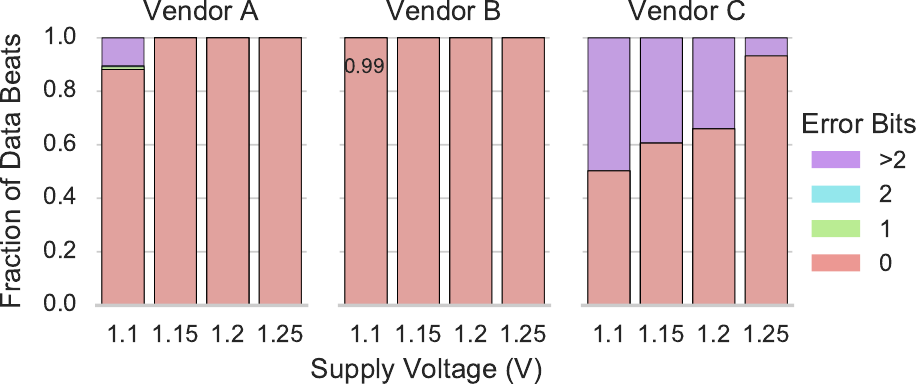}{1}{Distribution of bit \dhlfix{errors} in data beats.}

The results show that lowering the supply voltage increases the fraction of
beats that contain \fix{more than two bits of} errors. There are very few beats
that contain only one or two error bits. This implies that the most
commonly-used \fix{ECC scheme, SECDED, is} unlikely to alleviate errors induced
by a low supply voltage. \dhlfix{\fix{Another ECC mechanism,
    Chipkill\fixIII{~\cite{sridharan-asplos2015, luo-dsn2014}},} protects
  multiple bit failures within a DRAM chip. However, it \fixIII{cannot} correct
  errors in \emph{multiple} DRAM chips.} Instead, \fixIV{we believe that}
increasing the access latency, as shown in \ssecref{low_volt_latency}, is a more
effective way of eliminating errors under low supply voltages.


\subsection{Effect of Temperature}
\label{ssec:temperature}

Temperature is an important external factor that can affect the behavior of
DRAM\fixIII{~\cite{schroeder-sigmetrics2009,el-sayed-sigmetrics2012,
    liu-isca2013, khan-sigmetrics2014, lee-hpca2015, liu-isca2012, kim-isca2014,
    lee-thesis2016}}. Prior works have studied the impact of temperature on
reliability\fixIII{~\cite{schroeder-sigmetrics2009, el-sayed-sigmetrics2012,
    kim-isca2014,kim-thesis}},
latency\fixIII{~\cite{lee-hpca2015,chang-sigmetrics2016, lee-thesis2016}}, and
retention
time~\cite{khan-sigmetrics2014,liu-isca2013,liu-isca2012,qureshi-dsn2015} at the
nominal supply voltage. However, no prior work has studied \dhlfix{the effect of
  temperature on the latency \fix{at which DRAM operates reliably},} as the
supply voltage changes.

To reduce the test time, we test 13~representative DIMMs under a high ambient
temperature of 70$\celsius$ using a closed-loop temperature
controller~\cite{hassan-hpca2017}. \figref{temperature} shows \fix{the} \trcdmin
and \trpmin values of tested DIMMs, categorized by \fix{vendor, at} 20$\celsius$
and 70$\celsius$. \fix{The error bars indicate the minimum and maximum latency
  values across all \fixIII{DIMMs we} tested that are from the same vendor. We
  increase the horizontal spacing between the low and high temperature data
  points at each voltage level to improve readability.}

\figputHS{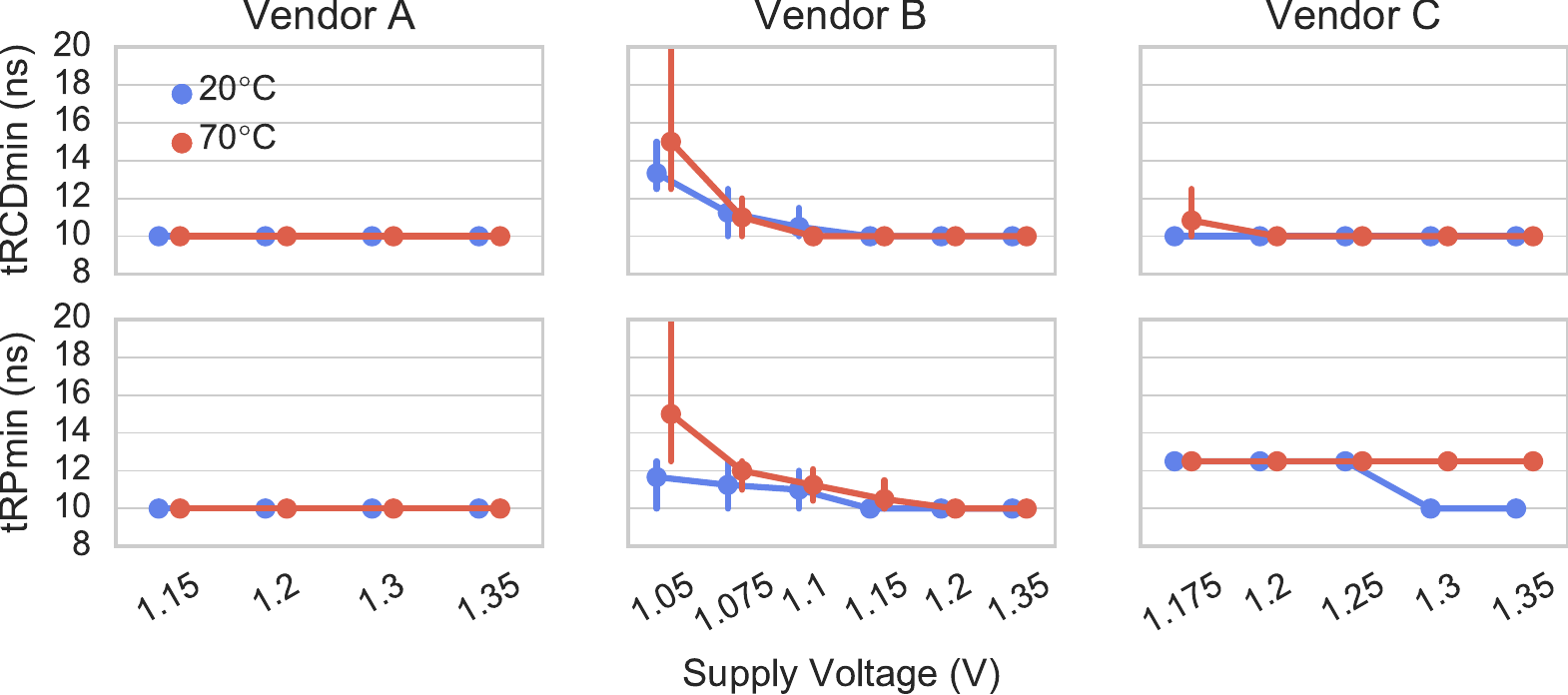}{0.8}{Effect of high \fix{ambient} temperature (70$\celsius$) on minimum
  reliable operation latency at reduced voltage.}

We make two observations. First, temperature impacts vendors differently. On
Vendor~A's DIMMs, temperature does not have an observable impact on the
\dhlfix{reliable operation} latencies. \fix{Since our platform can test
  latencies with  a step size of only 2.5ns, it is possible that the effect of high
  temperature on the reliable minimum operating latency for Vendor~A's DIMMs
  may be within 2.5ns.}
On the other hand, the temperature effect on latency is measurable on DIMMs from
Vendors~B and C. DIMMs from Vendor~B are not strongly affected by temperature
when the supply voltage is above 1.15V. \fix{The precharge latency for
Vendor~C's DIMMs is affected by high temperature} at supply voltages of 1.35V and
1.30V, \fix{leading to an increase in the minimum} latency from 10ns to 12.5ns.
When the voltage is below 1.25V, the impact \fix{of high temperature on
precharge latency is} not observable, as the precharge
latency already needs to be raised by 2.5ns, to 12.5ns, at 20$\celsius$. Second,
the precharge latency is more sensitive to temperature than the activation
latency. Across all of our tested DIMMs, \trp increases \fix{with high
  temperature} under a greater number of supply voltage \fix{levels}, whereas
\trcd is less likely to be perturbed by temperature.

Since temperature can affect latency behavior under different voltage levels,
techniques that compensate for temperature changes can be used to dynamically
adjust the activation and precharge latencies, as proposed by prior
work~\cite{lee-hpca2015, lee-thesis2016}.


\subsection{Impact on Refresh Rate}
\label{ssec:refresh}

Recall from \chapref{background} that a DRAM cell uses a capacitor to store data.
The charge in the capacitor leaks over time. To prevent data loss, DRAM
periodically performs an operation called \emph{refresh} to restore the charge
stored in the cells. The frequency of refresh is determined by the amount of
time a cell can retain enough charge without losing information, commonly
referred to as a cell's \emph{retention time}. \fix{For DDR3 DIMMs, the}
\fixIV{worst-case} retention time \fixIV{assumed}
for a DRAM cell is 64ms (or 32ms \fixIII{at temperatures above
  85$\celsius$~\cite{liu-isca2013, liu-isca2012}}). \fixIV{Hence, each cell is
  refreshed every 64ms, which is the \fixV{DRAM-standard} refresh interval.}


\fix{When we reduce the supply voltage of the DRAM array, we expect the
  retention time \fixIV{of a cell}}
to \emph{decrease}, as less charge \fix{is} stored in each cell. This could
\fixIV{potentially}
require a shorter refresh interval (i.e., more frequent refreshes). To
investigate the impact of low supply voltage on retention time, our experiment
writes all 1s to every cell, and reads out the data after a given amount of
retention time, with refresh disabled. We test a total of seven different
retention times (in ms): 64 (the standard time), 128, 256, 512, 1024, 1536, and
2048. We conduct the experiment for \fix{ten}~rounds on every DIMM from all three
vendors. \figref{retention_line_new} shows the average number of \emph{weak}
cells (i.e., cells that experience bit flips due to too much leakage \fixIV{at a
given retention time}) across all
tested DIMMs, for each retention time, under both 20$\celsius$ and 70$\celsius$.
We evaluate three voltage levels, 1.35V, 1.2V, and 1.15V, that allow
\fix{data to be read} reliably with \fix{a} sufficiently long latency. The error bars indicate the 95\%
confidence interval. \fix{We increase the horizontal spacing between the curves
  at each voltage level to improve readability.}


\ignore{
\figref{retention_line} shows two heatmaps indicating the average number of
\emph{weak} cells (i.e., cells that experience bit flips due to too much
leakage) across all 17 DIMMs, for each tested retention time, under both
20$\celsius$ and 70$\celsius$. The x-axis shows four different tested voltage
levels, and the y-axis shows the tested retention time.
A heatmap entry with zero weak cells means that every cell across all the tested
DIMMs fully retains data throughout the given retention time.
}

\figputHS{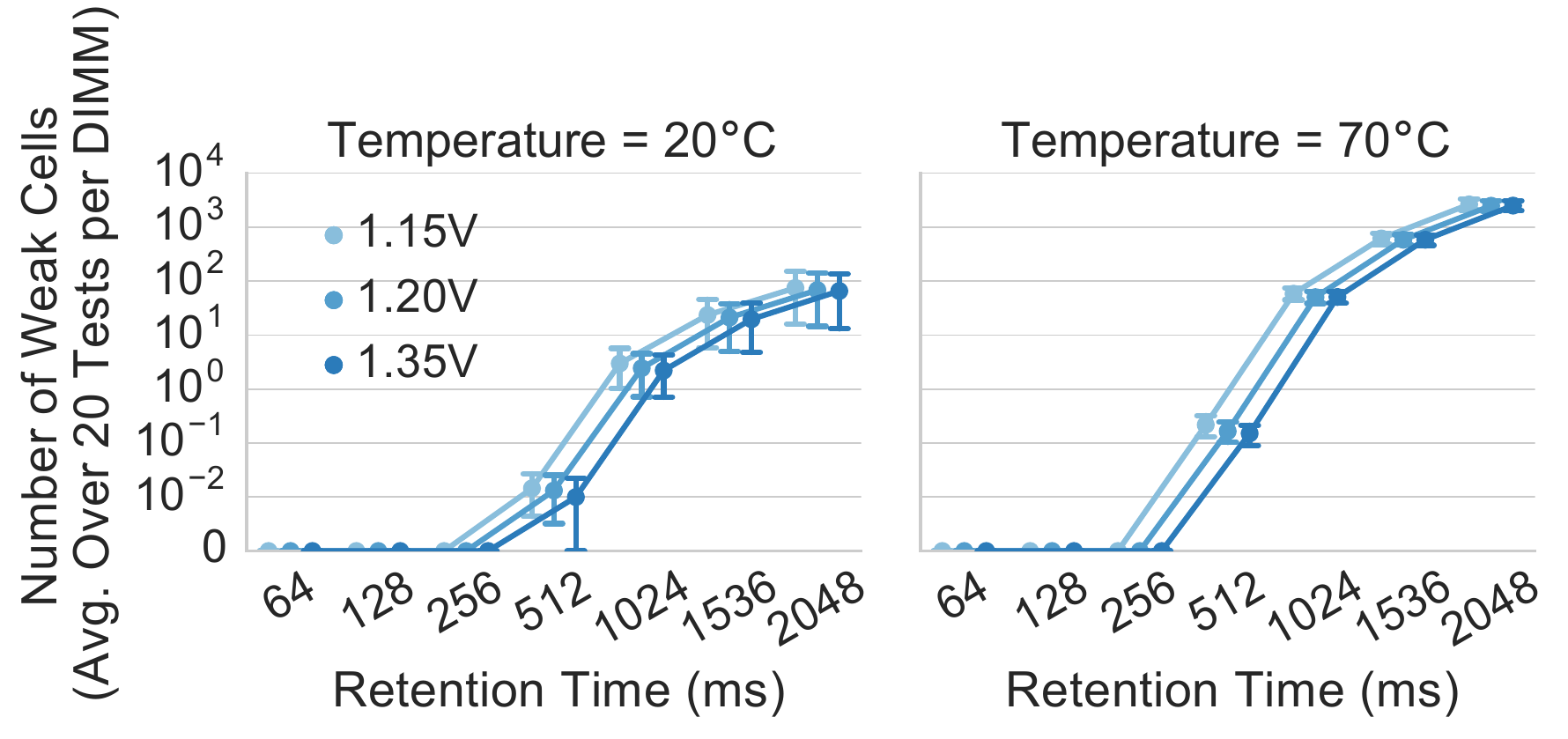}{0.6}{The number of weak cells that experience errors
  under different retention times as supply voltage varies.}

\changes{Our results show that every DIMM can retain data for at least 256ms
\fix{before requiring a refresh operation}, which is 4x higher than the standard
\fixIV{worst-case} specification. These results align with prior works, which
also experimentally demonstrate that commodity DRAM cells have much higher
retention times than the standard specification of
64ms\fixIII{~\cite{kim-edl2009,liu-isca2013,khan-sigmetrics2014,lee-hpca2015,hassan-hpca2017,
    lee-thesis2016, patel-isca2017, qureshi-dsn2015}}. Even though higher
retention times (i.e., longer times without refresh) reveal more weak cells, the
number of weak cells is still very small, e.g., tens of weak cells out of
billions of cells, on average across all DIMMs \fixIV{at} under 20$\celsius$. Again, this
corresponds closely \fixIII{to} observations from prior \fix{works} showing that
there are relatively few weak cells with low retention time in DRAM
chips, \fixIV{especially at lower temperatures}\fixIII{~\cite{kim-edl2009,liu-isca2013,khan-sigmetrics2014,lee-hpca2015,hassan-hpca2017,qureshi-dsn2015,
    lee-thesis2016, patel-isca2017}}. }

We observe that the effect of \fixIV{the supply voltage} on retention times is
\fixV{\emph{not}} statistically significant. For example, at a 2048ms retention time,
the average \emph{number} of weak cells \fixV{in a DRAM module} increases by only 9~cells (out of a
population of billions of cells) when the supply voltage drops from \fix{1.35V
  (66~weak cells) to 1.15V (75~weak cells)} at $20\celsius$. \fixIV{For the same
  2048ms retention time at $70\celsius$, the average number of weak cells
  increases by only 131~cells when the supply voltage reduces from 1.35V
  (2510~weak cells) to 1.15V (2641~weak cells).}

  \ignore{ the average number of weak cells does not increase, but decreases by
  only 55 cells when the supply voltage drops from 1.35V (2519~weak cells) to
  1.15V (2464~weak cells). Although the average number of weak cells is lower at
  1.15V than 1.35V, the difference is only 2\% of the average number of weak
  cells at 1.35V. This small decrease of weak cells is likely due to a
  well-known phenomenon, called Variable Retention Time
  (VRT)~\cite{khan-sigmetrics2014, khan-dsn2016, khan2016case,
    patel-isca2017,liu-isca2013,qureshi-dsn2015,restle-iedm1992,yaney-iedm1987},
  which causes the retention times of some cells to shift randomly over time.}

\ignore{ For instance, out of Y rounds of tests performed on DIMM X, the number
of weak cells varies from round to round at 2048ms. The average number of weak
cells across all rounds is X with a maximum and minimum of x and z,
respectively. This shows that the retention times of cells change over time, as
experimentally demonstrated in prior works~\cite{liu-isca2013, kim-edl2009}. }


\fix{When we lower the supply voltage, we do not observe \emph{any} weak cells
until a retention time of 512ms, which is 8x the standard refresh interval of
64ms. Therefore, we conclude that using a reduced supply voltage does not
require any changes to the standard refresh interval \fixIII{at 20$\celsius$ and
  70$\celsius$ ambient temperature}.}


\subsection{Summary}
\label{ssec:char_summary}

\changes{We have presented extensive characterization results and analyses on
DRAM chip \fix{latency, reliability, and data retention time} behavior under
various supply voltage levels. We summarize our findings in six key points.
First, DRAM reliability worsens \fixIV{(i.e., more errors start appearing)} as
we reduce the supply voltage below \vmin. Second, we discover that
voltage-induced errors occur mainly because, at low supply voltages, the DRAM
access latency is no longer sufficient to allow the fundamental DRAM operations
to complete. Third, via both experiments on real DRAM chips and SPICE
simulations, we show that increasing the latency of activation, restoration, and
precharge \fix{operations in DRAM} can mitigate errors under low supply voltage
levels until a certain voltage level. Fourth, we show that voltage-induced
errors \fixIV{exhibit} strong spatial locality \fixIV{in a DRAM chip},
clustering at certain locations \fixIV{(i.e., \fixV{certain} banks and rows)}. Fifth,
temperature affects the reliable access latency at low supply voltage levels
\fix{and the effect is very vendor-dependent}. Sixth, we find that reducing the
supply voltage does \emph{not} require increasing the standard DRAM refresh
\fix{rate} for reliable \fixIII{operation below 70$\celsius$}.}


\section{Voltron: Reducing DRAM Energy Without Sacrificing Memory Throughout}
\label{sec:varray}

\changes{Based on the extensive understanding we developed on reduced-voltage
operation of real DRAM chips in \secref{dram_exp}, we propose a new mechanism
called \emph{\voltron}, which reduces DRAM energy without sacrificing memory
throughput. \voltron exploits the fundamental observation that reducing the
supply voltage to DRAM requires increasing the latency of the three DRAM
operations in order to prevent errors. Using this observation, the key idea of
Voltron is to use a performance model to determine \fix{by how much to reduce
  the DRAM supply voltage}, without introducing errors and without exceeding a
user-specified threshold for performance loss. \voltron consists of two main
components: \myitem{i} \emph{array voltage scaling}, a hardware mechanism that
leverages our \fix{experimental observations} to scale \emph{only} the voltage supplied
to the DRAM array; and \myitem{ii} \emph{performance-aware voltage control},
\fixIII{a software} mechanism\footnote{\fixIV{Note that this mechanism can also
    be implemented in hardware, or as a cooperative hardware/software
    mechanism.}} that automatically chooses the minimum DRAM array voltage that
meets a \fixIV{user-specified} performance target.}


\subsection{Array Voltage Scaling}
\label{ssec:avs}

\changes{ As we discussed in \secref{dram_power}, the DRAM supply voltage to the
peripheral circuitry determines the maximum operating frequency. If we reduce
the supply voltage directly, the frequency needs to be lowered as well.
As more applications become more sensitive to memory bandwidth, reducing DRAM
frequency can result in a substantial performance loss due to lower data
throughput. In particular, we find that reducing the DRAM frequency from 1600
MT/s to 1066 MT/s significantly degrades performance of our evaluated
\fix{memory-intensive} applications by 16.1\%. Therefore, the design challenge
of \voltron is to reduce \fixIII{the} DRAM supply voltage \emph{without}
\fixIII{changing the} DRAM frequency.

To address this challenge, the key idea of Voltron's first component,
\emph{array voltage scaling}, is to reduce the voltage supplied to the
\emph{DRAM array} (\varr) \emph{without changing the voltage supplied to the
  peripheral circuitry}, thereby allowing the DRAM channel to maintain a high
frequency while reducing the power consumption of the DRAM array. To prevent
errors from occurring during reduced-voltage operation, \voltron increases the
latency of the three DRAM operations (activation, restoration, and precharge) in
every DRAM bank based on our observations in \secref{dram_exp}.


By reducing \varr, we effectively reduce \myitem{i} the dynamic DRAM power on
activate, precharge, and refresh \fixIV{operations}; and \myitem{ii} the portion
of the static power that comes from the DRAM array. These power components
decrease \emph{quadratically} with the square of \fix{the} array voltage
reduction in a modern DRAM chip~\cite{keeth-dram-tutorial, baker-dram}. The
trade-off is that reducing \varr requires \fixIV{increasing} the latency of the
three DRAM operations, \fixIV{for reliable operation}, \fix{thereby} leading to
some system performance degradation, which we quantify in our evaluation
(\secref{eval}). }



\subsection{Performance-Aware Voltage Control}
\label{ssec:pavc}


Array voltage scaling provides system users with the ability to decrease \varr
to reduce DRAM power. Employing a lower \varr provides greater power savings,
but at the cost of longer DRAM access latency, which leads to larger performance
degradation. This trade-off varies widely across different applications, as each
application has a different tolerance to the increased memory latency. This
raises the question of how to pick a ``suitable'' array voltage level for
different applications as a system user or designer. For this dissertation, we say that
an array voltage level is suitable if it does not degrade system performance by
more than a user-specified threshold. Our goal is to provide a simple \fixIII{technique}
that can automatically select \fix{a} suitable \varr \fix{value} for
different applications. \changes{To this end, we propose
  \emph{performance-aware voltage control}, a \fix{power-performance} management
  policy that selects a minimum \varr that satisfies a desired performance
  constraint. The key observation is that an application's performance loss (due
  to increased memory latency) scales linearly with the application's memory
  \fix{demand \fixIII{(e.g., memory intensity)}.}
  Based on this \fixIV{empirical} observation \fixIV{we make}, we build a
  \emph{performance loss predictor} that leverages a linear model to predict an
  application's performance loss based on its characteristics at runtime. Using
  the performance loss predictor, Voltron finds a \varr that \fix{can keep the}
  predicted performance within a user-specified target at runtime. }

%

\paratitle{Key Observation} We find that an application's performance loss due
to higher latency has a strong linear relationship with its memory \fix{demand
  \fixIII{(e.g., memory intensity)}}. \figref{voltron_mot} shows the
relationship between the performance loss of each application \fix{(due to
  reduced voltage)} and its memory \fix{demand} under two different
\fix{reduced-voltage} values (see \ssecref{meth} for our methodology). Each data
point represents a single application. \fixIV{\figref{mpki}} shows each
application's performance loss versus its \emph{memory intensity}, expressed
using the commonly-used metric MPKI (last-level cache misses per
kilo-instruction). \fixIV{\figref{memstall}} shows \fix{each application's
  performance loss versus its \emph{memory stall time}}, the fraction of
execution time for which memory requests stall the CPU's \fixIV{instruction
  window}  (i.e., reorder buffer). In \fixIV{\figref{mpki}}, we
see that the performance loss is a \fix{\emph{piecewise linear function}} based
on the MPKI. The observation that an application's \emph{sensitivity to memory
  latency} is correlated with MPKI has also been \fixV{made} and utilized by prior
\fixIV{works~\cite{kim-hpca2010, kim-micro2010, mutlu-isca2008, zheng-icpp2008,
    mutlu-micro2007, muralidhara-micro2011, das-hpca2013, das-micro2009,
    usui-taco2016, zhao-micro2014, das-isca2010}.}

\begin{figure}[!h]
    \centering
    \captionsetup[subfigure]{justification=centering}
    \subcaptionbox{\fixV{Performance loss vs.\ last-level cache MPKI.}\label{fig:mpki}}
    {
        \includegraphics[scale=0.8]{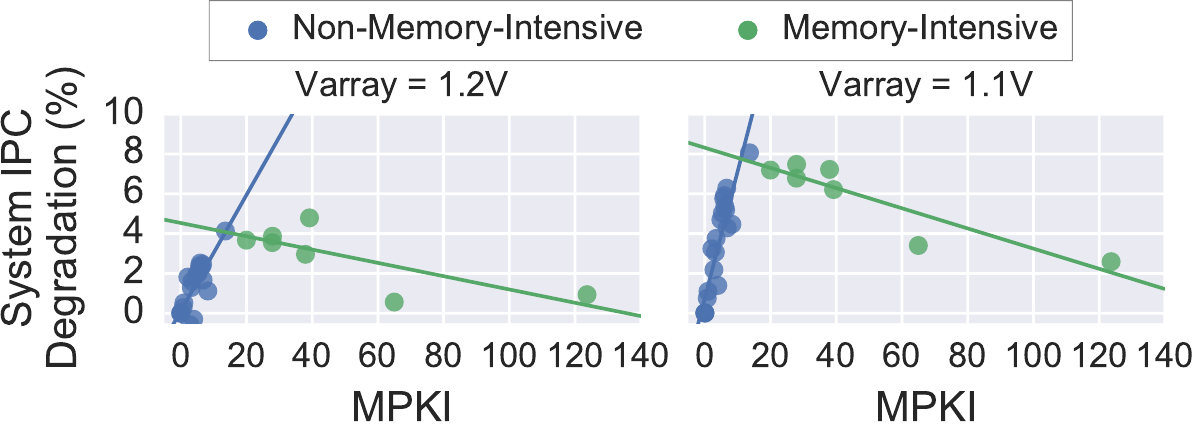}
    }

    \vspace{0.15in}
    \subcaptionbox{\fixV{Performance loss vs.\ memory stall time fraction.}\label{fig:memstall}}
    {
        \includegraphics[scale=0.8]{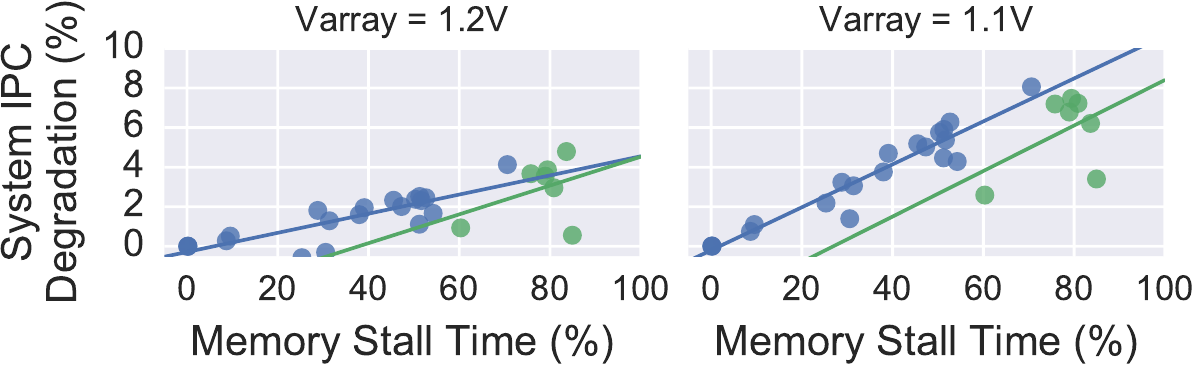}
    }
    \vspace{-0.15in}

    \caption{Relationship between performance loss (due to increased
    memory latency) and applications' characteristics: \fixIV{MPKI (a) and memory
    stall time fraction (b)}. Each data point
    represents a single application.}
    \label{fig:voltron_mot}
\end{figure}


When an application is \fix{\emph{not} memory-intensive} (i.e., \fixIV{has an} $MPKI < 15$), its performance
loss grows linearly with MPKI, becoming \emph{more sensitive} to memory latency.
\fix{ Latency-sensitive applications spend most of their time \fixIII{performing computation at the CPU cores} and
  issue memory requests infrequently. As a result, increasing the number of
  memory requests causes more stall cycles in the CPU. }

On the other hand, the performance of \fix{memory-intensive} applications (i.e.,
\fixIV{those with} $MPKI \ge 15$) is \emph{less sensitive} to memory latency as
the MPKI grows. This is because \fix{memory-intensive} applications experience
frequent cache misses and spend a large portion of their time waiting on pending
memory requests. As a result, their rate of progress is significantly affected
by the memory bandwidth, \fix{and therefore they are more} \emph{sensitive to
  memory throughput} instead of latency. \fix{With more \fixIII{outstanding}
  memory requests (i.e., higher MPKI), the memory \fixIII{system} is more likely
  to service them in parallel, leading to more \emph{memory-level
    parallelism}\fixIII{~\cite{kim-micro2010,mutlu-isca2008,glew-asploswac98,
      mutlu-hpca2003, mutlu-ieeemicro2006, lee-micro2009}}. Therefore, improved
  memory-level parallelism enables applications to tolerate \fixIII{higher
    latencies} more easily.}


\fixIV{\figref{memstall}} shows that an application's performance loss increases
with its \fixIV{instruction window (reorder buffer)} \emph{stall time
  \fixIV{fraction}} due to memory requests for both \fix{memory-intensive and
  non-memory-intensive} applications. \fix{A stalled \fixIV{instruction window}
  prevents the CPU from fetching or dispatching new
  instructions~\cite{mutlu-hpca2003}, thereby degrading the running
  application's performance.} This observation has also been \fixIII{made} and
utilized by prior
\fixV{works~\cite{ghose-isca2013,mutlu-ieeemicro2006,mutlu-isca2005,mutlu-hpca2003}}.

\paratitle{Performance Loss Predictor} Based on the observed linear
relationships \fixIV{between performance loss vs. MPKI and memory stall time
  fraction}, we use \emph{ordinary least squares (OLS)} regression to develop
a piecewise linear model for each application \fix{that can serve} as the performance loss
predictor for Voltron. Equation~\ref{eq:plm} shows the model, which takes the
following inputs: memory latency ($Latency=tRAS+tRP$), the application's MPKI, and
its memory stall time \fixIV{fraction}.

\vspace{-0.15in}
\begin{align}
    \mathrm{PredictedLoss}_{i} =
    \begin{cases}
        \label{eq:plm}
        \!\begin{aligned}
        &\alpha_{1} + \beta_{1}  \mathrm{Latency}_{i} + \beta_{2}
        \mathrm{App.MPKI}_{i} \\&+ \beta_{3}  \mathrm{App.StallTimeFraction}_{i} \\[1.5ex]
        \end{aligned} &\text{if } MPKI < 15 \\
        \!\begin{aligned}
        &\alpha_{2} + \beta_{4}  \mathrm{Latency}_{i} + \beta_{5}
        \mathrm{App.MPKI}_{i} \\&+ \beta_{6}  \mathrm{App.StallTimeFraction}_{i}
        \end{aligned} &\text{otherwise}
    \end{cases}
\end{align}
\vspace{-0.1in}
\begin{table}[h]
    \renewcommand{\arraystretch}{0.9}
    \centering
    \setlength{\tabcolsep}{.45em}
    \begin{tabular}{llll|llll}
        \toprule
        $\alpha_{1}$ & $\beta_{1}$ & $\beta_{2}$ & $\beta_{3}$  & $\alpha_{2}$ &
        $\beta_{4}$ & $\beta_{5}$ & $\beta_{6}$ \\
        \midrule
        -30.09 & 0.59 & 0.01 & 19.24 & -50.04 & 1.05 & -0.01 & 15.27 \\
        \bottomrule
    \end{tabular}
\end{table}

$\mathrm{PredictedLoss}_{i}$ is the predicted performance loss for the
application. The subscript $i$ refers to each data sample,
\fixIV{which describes a particular application's characteristics (MPKI and
  memory stall time fraction) and the memory latency associated with the
  selected voltage level.} \fixIV{To generate the data samples, we run a total
  of 27 workloads across 8 different voltage levels that range from 1.35V to
  \fixVI{0.90V}, at a 50mV step (see \ssecref{meth} for our methodology). In total, we
  generate 216 data samples for finding the coefficients (i.e., $\alpha$ and
  $\beta$ values) in our model. To avoid overfitting the model, we use
  \fixIII{the \emph{scikit-learn} machine learning toolkit}~\cite{scikit} to
  perform cross-validation, which randomly splits the data samples into a
  training set (151 samples) and a test set (65 samples)}. To assess the fit of
the model, we use a common metric, root-mean-square error (RMSE), which is 2.8
and 2.5 for the low-MPKI and high-MPKI pieces of the model, respectively.
Furthermore, we calculate the R\textsuperscript{2} value to be 0.75 and 0.90 for
the low-MPKI and high-MPKI models, respectively. \fixV{Therefore, the RMSE and
R\textsuperscript{2} metrics
  indicate that our model provides high accuracy for predicting the performance
  loss of applications under different \varr values.}

\paratitle{Array Voltage Selection} Using the performance loss predictor,
\voltron selects the minimum value of \varr that satisfies the given user target
for performance loss. \fix{Algorithm~\ref{voltron} depicts the array voltage
  selection component of Voltron.} The voltage selection \fixIII{algorithm is
  executed} at periodic intervals throughout the runtime of an application.
During each interval, the application's memory \fix{demand} is profiled. At the
end of an interval, \voltron uses the profile to iteratively compare the
\fix{performance loss} target to the predicted performance loss incurred by each
voltage level, starting from a minimum value of 0.90V. Then, \voltron selects
the minimum \varr that \fixIV{does not exceed} the performance loss target and
\fixIII{uses} this selected \varr\fixIII{\ as the DRAM supply voltage} in the
subsequent interval. In our evaluation, we provide \voltron with a total of 10
voltage levels (every 0.05V step from 0.90V to 1.35V) for selection.

\fix{

\begin{afloat}[h]

\algnewcommand\algorithmicto{\textbf{to}}
\algrenewtext{For}[3]{\algorithmicfor\ #1 $\gets$ #2 \algorithmicto\ #3 \algorithmicdo}

\algrenewcommand\algorithmicfunction{}
\algrenewcommand\algorithmicdo{}
\algrenewcommand\algorithmicindent{1.2em}
\algrenewcommand\alglinenumber[1]{\footnotesize\texttt{#1}}
  \footnotesize

\begin{algorithmic}[1]
\Function{SelectArrayVoltage}{$target\_loss$}

\ForEach {$interval$} \Comment{Enter at the end of an interval}
  \State $profile$ = GetMemoryProfile()
  \State $Next$\varr = 1.35
  \For{\varr}{0.9}{1.3} \Comment{\fixV{Search for the smallest \varr that satisfies the
    performance loss target}}
    \State $predicted\_loss$ = Predict(Latency(\varr), $profile$.MPKI,
    $profile$.StallTime) \Comment{\fixV{Predict performance loss}}
    \If{$predicted\_loss \leq target\_loss$} \Comment{\fixV{Compare the
        predicted loss to the target}}
    \State $Next$\varr = \varr \Comment{\fixV{Use the current \varr for the next interval}}
        \State \fixIII{\textbf{break}}
    \EndIf
  \EndFor
  \State ApplyVoltage($Next$\varr) \Comment{\fixV{Apply the new \varr for the
      next interval}}
\EndFor

\EndFunction
\end{algorithmic}
    \caption{Array Voltage Selection}
\label{voltron}
\end{afloat}

}


\subsection{Implementation}

\voltron's two components require modest modifications \fix{to} different parts of the
system. In order to support array voltage scaling, \voltron requires minor
changes to the power delivery network of DIMMs, as commercially-available DIMMs
currently \fixIII{use} a single supply voltage for both the DRAM array and \fixIII{the} peripheral
circuitry. Note that this supply voltage goes through \emph{separate} power
pins: $V_{DD}$ and $V_{DDQ}$ for the DRAM array and peripheral circuitry,
respectively, on a modern DRAM chip~\cite{micronDDR3L_2Gb}. Therefore, to enable
independent voltage adjustment, we propose to partition the power delivery
network on the DIMM into two domains: one domain to supply only the DRAM array
($V_{DD}$) and the other domain to supply only the peripheral circuitry
($V_{DDQ}$). 

Performance-aware voltage control requires \myitem{i} performance monitoring
hardware that records the MPKI and memory stall time of each application; and
\myitem{ii} a control algorithm block, which predicts the performance loss at
different \varr values and accordingly selects the smallest acceptable \varr.
\voltron utilizes the performance counters that exist in most modern CPUs to
perform performance monitoring, thus requiring no additional hardware overhead.
\voltron reads these counter values and feeds them into the \fix{array voltage
  selection algorithm, which is} implemented in the system software layer.
\fix{Although reading the performance monitors has a small amount of software
  overhead, we believe the overhead is negligible because we do so only at the
  end of each interval \fixV{(i.e., every four million cycles in most of our
    evaluations; see sensitivity studies in \ssecref{interval})}.}

Voltron periodically executes this performance-aware voltage control
\fix{mechanism} during the runtime of the target application. During each time
interval, Voltron monitors the application's behavior through hardware counters.
At the end of an interval, the \fixIII{system software} \fix{executes the array
  voltage selection algorithm} to select the predicted \varr and
\fixIV{accordingly adjust} the timing parameters stored in the memory controller
for activation, restoration, and precharge. \fix{Note that there could be other
  (e.g., completely hardware-based) implementations of Voltron. We leave a
  detailed explanation of different implementations to future work.}

%
%
%


\section{System Evaluation}
\label{sec:eval}

In this section, we evaluate the system-level \fixIV{performance and energy}
impact of \voltron.  We present our evaluation methodology in \ssecref{meth}.
Next, we study the energy savings and performance loss when we use array voltage
scaling without any control (\ssecref{scaling_eval}).  We study how
performance-aware voltage control delivers overall system energy reduction with
only a modest \fix{amount of} performance loss
(\fixIII{Sections~\ref{ssec:pavc_eval} and \ref{ssec:energy_breakdown}}). We
then evaluate \fix{an enhanced version of} \voltron\fix{, which exploits}
spatial error locality (\ssecref{eval_el}). Finally,
Sections~\ref{ssec:eval_hetero} to \ref{ssec:interval} present sensitivity
studies of \voltron\fix{\ to various system and algorithm parameters}.

\subsection{Methodology}
\label{ssec:meth}

We evaluate \voltron using Ramulator~\cite{kim-cal2015}, a detailed and
\fixV{and cycle-accurate}
  \fix{open-source DRAM simulator~\cite{ram-github}}, integrated with a
  multi-core performance simulator. We model a low-power mobile system that
  consists of 4 ARM cores and DDR3L DRAM. \tabref{sys-config} shows our system
  parameters. Such a system resembles existing commodity devices, such as the
  Google Chromebook~\cite{chromebook} \fix{or the NVIDIA} SHIELD
  tablet~\cite{shield}. To model power and energy consumption, we use
  McPAT~\cite{mcpat:micro} for the processor and DRAMPower~\cite{drampower} for
  the DRAM-based memory system. \fix{We open-source the code of Voltron
    \cite{volt-github}.}

\begin{table}[h]
  \renewcommand{\arraystretch}{0.9}
  \centering
    \setlength{\tabcolsep}{.6em}
    \begin{tabular}{ll}
        \toprule
        \multirow{2}{*}{Processor} & 4 ARM Cortex-A9 cores~\cite{arma9}, 2GHz, \\
        & 192-entry instruction window \\
        \midrule
        Cache & L1: 64KB/core, L2: 512KB/core, L3: 2MB shared \\
        \midrule
        Memory & \multirow{2}{*}{64/64-entry read/write request queue,
            FR-FCFS~\cite{rixner-isca2000,zuravleff-patent}} \\
        Controller & \\
        \midrule
        \multirow{2}{*}{DRAM} & DDR3L-1600~\cite{jedec-ddr3l}\\
        & 2 channels (1 rank and 8 banks per channel) \\
        \bottomrule
    \end{tabular}
  \caption{Evaluated system configuration.}
  \label{tab:sys-config}
\end{table}


\tabref{varr_latency} lists the latency values we evaluate for each DRAM array
voltage (\varr). The latency values are obtained from our SPICE model using data
from real devices (\ssecref{low_volt_latency}), \fix{which is available
  online~\cite{volt-github}}.\footnote{\fixIV{In this chapter, we do not have
  experimental data on the restoration latency (\tras) under reduced-voltage
  operation. This is because our reduced-voltage tests access cache lines
  sequentially from each DRAM row, and \tras overlaps with the latency of
  reading all of the cache lines from the row.  Instead of designing a separate
  test to measure \tras, we use our circuit simulation model
  (\ssecref{low_volt_latency}) to derive \tras values for reliable operation
  under different voltage levels. We leave the thorough experimental evaluation
  of \tras under reduced-voltage operation to future work.}} To account for
  manufacturing process variation,
we conservatively add in the same latency guardband (i.e., 38\%) used by
manufacturers at the nominal voltage level of 1.35V to each of our \fix{latency
  values}. We then round up each latency \fix{value} to the nearest clock cycle
time (i.e., 1.25ns).

\begin{table}[h]
  \centering
    \setlength{\tabcolsep}{.5em}
    \begin{tabular}{ll|ll}
        \toprule
        \varr & tRCD - tRP - tRAS (ns) & \varr & tRCD - tRP - tRAS (ns) \\
        \midrule

        1.35 & 13.75 - 13.75 - 36.25 &  1.10 & 15.00 - 16.25 - 40.00 \\
        1.30 & 13.75 - 13.75 - 36.25 &  1.05 & 16.25 - 17.50 - 41.25 \\
        1.25 & 13.75 - 15.00 - 36.25 &  1.00 & 17.50 - 18.75 - 45.00 \\
        1.20 & 13.75 - 15.00 - 37.50 &  0.95 & 18.75 - 21.25 - 48.75 \\
        1.15 & 15.00 - 15.00 - 37.50 &  0.90 & 21.25 - 26.25 - 52.50 \\
        \bottomrule
    \end{tabular}
  \caption{DRAM latency \fixIV{required for correct operation} for each evaluated \varr.}
  \label{tab:varr_latency}
\end{table}


\paratitle{Workloads} We evaluate 27~benchmarks from SPEC
CPU2006~\cite{spec2006} and YCSB~\cite{cooper-socc2010}, as shown in
\tabref{workload_list} \fix{along with each benchmark's \fixIII{L3 cache} MPKI,
  i.e., memory intensity}. \fix{We use the 27~benchmarks to form
  \emph{homogeneous} and \emph{heterogeneous} \fixIV{multiprogrammed}
  workloads.} \fix{For each \emph{homogeneous workload}, we replicate one of our
  benchmarks} by running one copy on each core to form a four-core
\fixIV{multiprogrammed} workload, as done in many past works \fixIII{that
  evaluate multi-core system performance}~\cite{lee-hpca2015,
  chang-sigmetrics2016, lee-hpca2013, nair-isca2013, nair-micro2014,
  shafiee-hpca2014, shevgoor-micro2013, chatterjee-micro2012}. \fixIV{Evaluating
  homogeneous workloads enables easier analysis and understanding of the
  system.} \fix{For each \emph{heterogeneous workload}, we combine four
  \emph{different} benchmarks \fix{to create a four-core workload}. We
  categorize the heterogeneous workloads by varying the fraction of
  \fix{memory-intensive} benchmarks in each workload (0\%, 25\%, 50\%, 75\%, and
  100\%). Each category consists of 10~workloads, resulting in a total of
  50~workloads across all categories.} Our simulation executes at least
500~million instructions on each core. \fix{We calculate system energy as} the
product of the average dissipated power (from both CPU and DRAM) and the
workload runtime. We measure system performance with the commonly-used
\emph{weighted speedup} (WS) metric~\cite{snavely-asplos2000}, which is a
measure of job throughput on a multi-core system~\cite{eyerman-ieeemicro2008}.


\begin{table}[h]
  \renewcommand{\arraystretch}{0.92}
  \centering
    \setlength{\tabcolsep}{.3em}
    \footnotesize
    \begin{tabular}{clrc|clrc|clr}
        \toprule
        Number & Name & \fixV{L3} MPKI & & Number & Name & L3 MPKI & & Number & Name & L3 MPKI\\
        \midrule
        0 & YCSB-a    & 6.66 & & 9  & calculix    & 0.01  & & 18 & milc     & 27.91 \\
        1 & YCSB-b    & 5.95 & & 10 & gamess      & 0.01  & & 19 & namd     & 2.76 \\
        2 & YCSB-c    & 5.74 & & 11 & gcc         & 3.20  & & 20 & omnetpp  & 27.87\\
        3 & YCSB-d    & 5.30 & & 12 & GemsFDTD    & 39.17 & & 21 & perlbench& 0.95 \\
        4 & YCSB-e    & 6.07 & & 13 & gobmk       & 3.94  & & 22 & povray   & 0.01 \\
        5 & astar     & 3.43 & & 14 & h264ref     & 2.14  & & 23 & sjeng    & 0.73 \\
        6 & bwaves    & 19.97& & 15 & hmmer       & 6.33  & & 24 & soplex   & 64.98\\
        7 & bzip2     & 8.23 & & 16 & libquantum  & 37.95 & & 25 & sphinx3  & 13.59\\
        8 & cactusADM & 6.79 & & 17 & mcf         & 123.65& & 26 & zeusmp   & 4.88 \\
        \bottomrule
    \end{tabular}

  \caption{Evaluated benchmarks with their \fixIII{respective} L3 MPKI values.}
  \label{tab:workload_list}
\end{table}

\vspace{-0.2in}




\ignore{
\subsection{Dynamic DRAM Energy Results}

Since higher DRAM access latency can potentially increase the energy
consumption, we perform a first order study to evaluate the \emph{dynamic}
energy consumption of the three fundamental access operations that we
demonstrate to be affected by \varr: \myitem{i} activation, \myitem{ii}
precharge, and \myitem{iii} restoration.

\figref{spice_energy_savings} shows the normalized dynamic energy consumption of
the three operations as voltage varies. As \varr lowers, the power consumption
of these operations reduces with high latency. We calculate their energy by
multiplying their power consumption with its latency at each \varr point. The
results show that even though latency increases, the power reduction still saves
\emph{dynamic} energy per operation. However, the dynamic energy savings have
diminished returns when \varr reduces below 1.1V since the latency is increasing
at an exponential rate (shown previously in Eq.~\ref{exp_decay},
\ssecref{spice}). For precharge, the dynamic energy of precharge starts to give
negative returns on energy savings. These results show that decreasing \varr
directly reduces the energy consumption of DRAM operations, but lower \varr does
not necessarily provide more energy savings. Now, we evaluate the system
performance and energy savings by taking into account other dynamic and static
energy consumption in both the DRAM and CPU.

\begin{figure}[!h]
    \centering
    \captionsetup[subfigure]{justification=centering}

    {
        \includegraphics[width=\linewidth]{plots/spice_dram_op_energy_over_vdd}
        \vspace{-0.1in}
    }
    \caption{Dynamic energy of DRAM operations as \varr varies.}
    \label{fig:spice_energy_savings}
\end{figure}

}

\subsection{Impact of Array Voltage Scaling}
\label{ssec:scaling_eval}

In this section, we evaluate how array voltage scaling (\ssecref{avs}) affects
the system energy consumption and application performance \fix{of our homogeneous workloads} at different \varr
values. We split our discussion into two parts: the results for \fix{memory-intensive}
workloads (i.e., applications where MPKI \fix{$\geq$} 15 for each core), and the results
for \fix{non-memory-intensive} workloads.


\paratitle{Memory-Intensive Workloads} \figref{vsweep_high_mpki} shows the
system performance (WS) loss, DRAM power reduction, and system energy reduction,
compared to a baseline DRAM with 1.35V, when we vary \varr from 1.30V to 0.90V.
We make three observations from these results.

\begin{figure}[ht]
    \centering
    \captionsetup[subfigure]{justification=centering}
    {
        \includegraphics[scale=0.75]{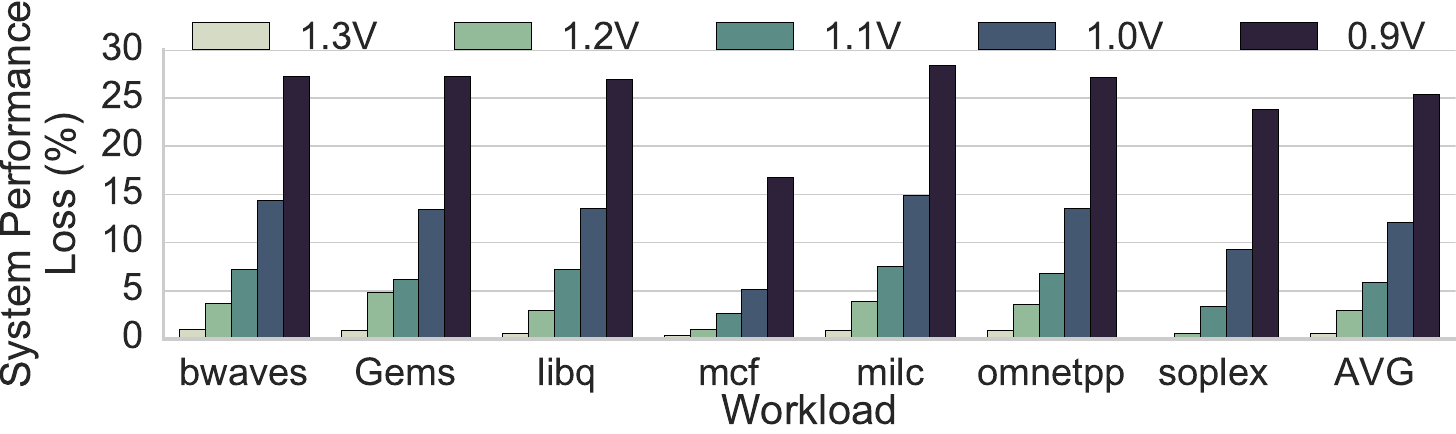}
        \vspace{0.1in}
    }

    {
        \includegraphics[scale=0.75]{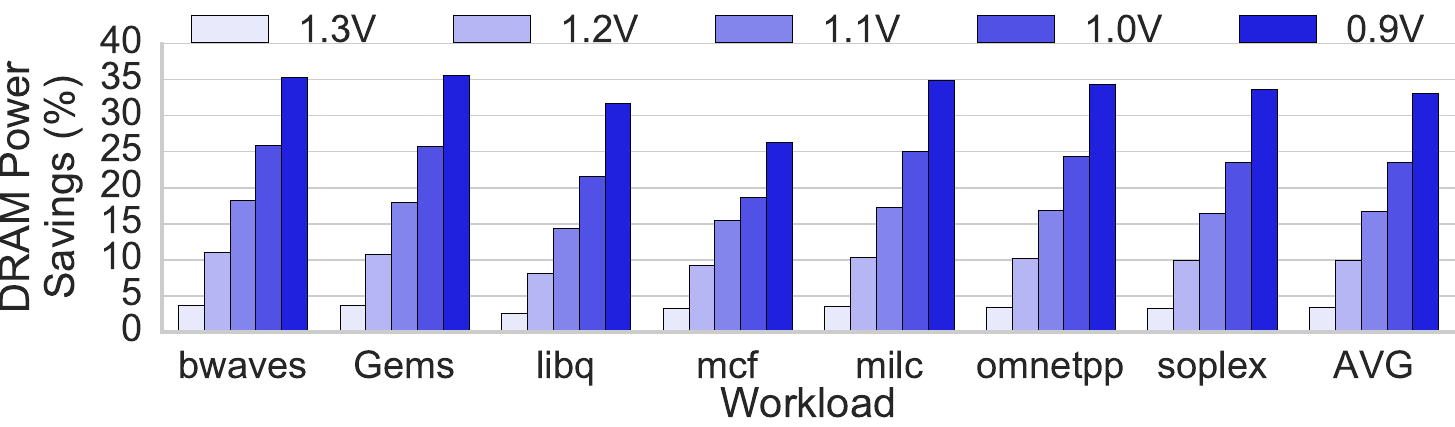}
        \vspace{0.1in}
    }

    {
        \includegraphics[scale=0.75]{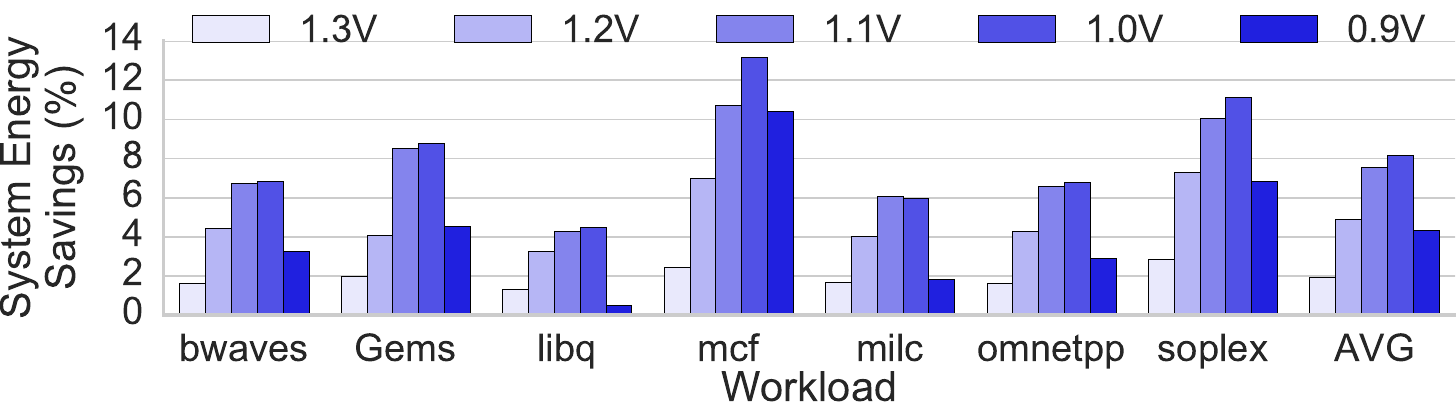}
        \vspace{-0.1in}
    }
    \caption{System performance loss and energy
    savings due to array voltage scaling for memory-intensive workloads.}
    \label{fig:vsweep_high_mpki}
\end{figure}


First, system performance loss \fixIII{increases} as we lower \varr, due to the increased
DRAM access latency. However, different workloads \fix{experience} a different
rate of performance loss, as they tolerate memory latency differently. Among the
\fix{memory-intensive} workloads, \emph{mcf} exhibits the lowest performance
degradation since it has the highest memory intensity and \fix{high memory-level
  parallelism}, leading to high queuing delays in the memory controller. The
queuing delays \fix{and memory-level parallelism} hide the longer DRAM access
latency \fix{more than in other workloads}. Other workloads lose more
performance because they are \fixIII{less able to tolerate/hide the} increased latency. Therefore,
workloads with very high memory intensity \fix{and memory-level parallelism can
  be less sensitive} to the increased memory latency.


Second, DRAM power savings increase with lower \varr since reducing the DRAM
array voltage decreases \emph{both} the dynamic and static power components of
DRAM. \fix{However, \emph{system} energy savings does \fixIII{\emph{not}}
  monotonically increase with lower \varr.} We find that using $V_{array}$=0.9V
\fix{provides} lower system energy savings than \fixIV{using} $V_{array}$=1.0V,
as the processor takes \emph{much longer} to run the applications \fixIV{at
  $V_{array}$=0.9V}. In this case, the increase in static DRAM and CPU energy
\fix{outweighs} the dynamic DRAM energy savings.


Third, \fix{reducing \varr leads to a system energy reduction only when} the
reduction in DRAM energy outweighs the increase \fixVI{in} CPU energy (due to the
longer execution time). For \varr=1.1V, the system energy reduces by an average
of 7.6\%. Therefore, we conclude that array voltage scaling is an effective
technique that improves system energy consumption, with a small performance
loss, for \fix{memory-intensive} workloads.



\paratitle{\fix{Non-Memory-Intensive} Workloads} \label{ssec:comp}
\tabref{non_mem_intensive_results} summarizes the \fixIV{system performance
  loss} and energy savings of 20 \fix{non-memory-intensive} workloads as \varr
varies from 1.30V to 0.90V, over the performance and energy consumption under a
nominal \varr of 1.35V. Compared to the \fix{memory-intensive} workloads,
\fix{non-memory-intensive} workloads \fixIV{obtain} smaller system energy
savings, as the system energy is dominated by the processor. Although the
workloads are more compute-intensive, lowering \varr \emph{does} reduce their
system energy consumption, by decreasing the energy consumption of DRAM. For
example, at 1.2V, array voltage scaling achieves an overall system energy
savings of 2.5\% with a performance loss of only 1.4\%.


\begin{table}[h]
  \centering
    \setlength{\tabcolsep}{.45em}
    \begin{tabular}{crrrrr}
        \toprule
        \textbf{\varr} & 1.3V & 1.2V & 1.1V & 1.0V & 0.9V  \\
        \midrule

        \textbf{System Performance Loss} (\%) & 0.5 & 1.4 & 3.5 & 7.1 & 14.2 \\
        \textbf{DRAM Power Savings} (\%) & 3.4 & 10.4 & 16.5 & 22.7 & 29.0 \\
        \textbf{System Energy Savings} (\%) & 0.8 & 2.5 & 3.5 & 4.0 & 2.9 \\

        \bottomrule
    \end{tabular}

  \caption{System performance loss and energy savings due to array voltage
    scaling for \fix{non-memory-intensive} workloads.}
  \label{tab:non_mem_intensive_results}
\end{table}


\subsection{Effect of Performance-Aware Voltage Control}
\label{ssec:pavc_eval}

In this section, we evaluate the effectiveness of \fix{our complete proposal for Voltron,
which incorporates} \fixIII{our} \emph{performance-aware voltage control} mechanism
\fixIII{to drive the array voltage scaling component intelligently}. The \fixIII{performance-aware voltage control}
mechanism selects the lowest voltage level that satisfies the performance loss
bound (provided by the user or system designer) based on our performance model
(see \ssecref{pavc}). We evaluate \voltron with a target performance loss of
5\%. \voltron executes the performance-aware voltage control mechanism once
every four million cycles.\footnote{\fix{We evaluate the sensitivity to the \fixIII{frequency} at
  which we execute the mechanism (i.e., the interval \fix{length} of Voltron) in
  \ssecref{interval}.}} We quantitatively
compare Voltron \fixIV{to} \textit{\memdvfs}, a dynamic DRAM frequency and voltage
scaling mechanism proposed by prior work~\cite{david-icac2011}, which we
describe in \ssecref{dvfs}. Similar to the configuration used in the prior work,
we enable \memdvfs to switch dynamically between three frequency steps: 1600,
1333, and 1066 MT/s, which employ \fix{supply voltages of} 1.35V, 1.3V, and 1.25V, respectively.

\figref{memdvfs} shows the system performance (WS) loss, DRAM power savings, and
system energy savings due to \memdvfs and \voltron, compared to a baseline DRAM
with \fix{a supply voltage of} 1.35V. We show one graph per metric, where each graph uses boxplots to show
the distribution among all workloads. \fix{In each graph, we categorize the workloads as}
\fixV{either non-memory-intensive or memory-intensive}.
Each box illustrates the quartiles of the population, and the
whiskers indicate the minimum and maximum values. The red dot indicates the
mean. We make four major observations.

\begin{figure}[!h]
    \centering
    \subcaptionbox{\label{fig:volt_vs_mem_perf}}[0.32\linewidth]
    {
      \includegraphics[scale=1.2]{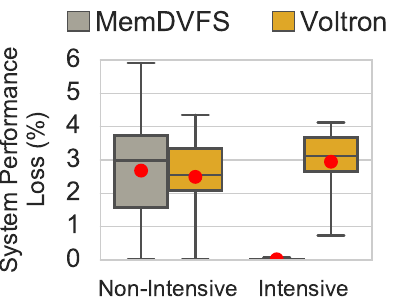}
    }
    \subcaptionbox{\label{fig:volt_vs_mem_dpow}}[0.32\linewidth]
    {
      \includegraphics[scale=1.2]{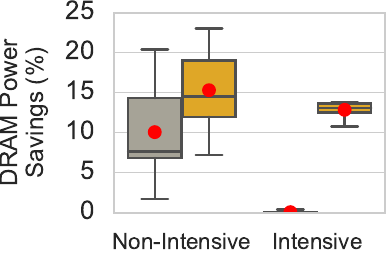}
    }
    \subcaptionbox{\label{fig:volt_vs_mem_syse}}[0.32\linewidth]
    {
      \includegraphics[scale=1.2]{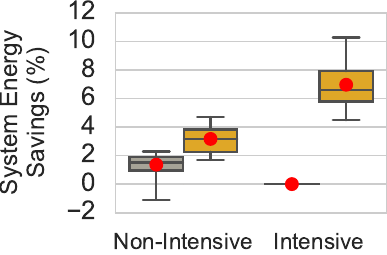}
    }
    \vspace{-0.15in}
    \caption{\fixIV{Performance and energy} comparison between \voltron and
      \memdvfs on \fixV{non-memory-intensive and memory-intensive workloads}.}
    \label{fig:memdvfs}
\end{figure}

First, \fix{as shown in \figref{volt_vs_mem_perf}}, \voltron consistently
selects a \varr value that satisfies the performance loss bound of 5\% across
all workloads. \voltron incurs \fix{ an average (maximum) performance loss of 2.5\% (4.4\%) and 2.9\% (4.1\%)}
for \fixV{non-memory-intensive and memory-intensive} workloads,
respectively. This demonstrates that our performance model enables \voltron to
select a low voltage value that saves energy while bounding performance loss
based on the user's requirement. We evaluate \voltron with a range of different
performance targets in \ssecref{vary_perf_target}.

Second, \memdvfs has almost zero effect on \fixIV{memory-intensive} workloads.
This is because \memdvfs avoids scaling DRAM frequency \fix{(and hence voltage)}
when an application's memory bandwidth \fixIV{utilization} is above a fixed
threshold. Reducing the frequency can result in a large performance loss since
the \fixIV{memory-intensive} workloads require high data throughput. As
\fix{memory-intensive} applications have high memory bandwidth consumption that
easily exceeds the \fixIV{fixed} threshold \fixIV{used by \memdvfs}, \memdvfs
\emph{cannot} perform frequency and voltage scaling during most of the execution
time. These results are consistent with the results reported in
\memdvfs\cite{david-icac2011}. In contrast, \voltron reduces system energy
\fix{(shown in \figref{volt_vs_mem_syse})} by 7.0\% on average for
\fix{memory-intensive} workloads, at the cost of 2.9\% system performance loss,
which is well within the specified performance loss target of 5\% \fix{(shown in
  \figref{volt_vs_mem_perf})}.

Third, both \memdvfs and \voltron reduce the average system energy consumption
for \fix{non-memory-intensive} workloads. \memdvfs reduces system energy by
dynamically scaling the frequency and voltage of DRAM, which lowers the DRAM
power consumption \fix{(as shown in \figref{volt_vs_mem_dpow})}. By reducing the
DRAM array voltage to a lower value than \memdvfs, \voltron is able to provide a
slightly higher DRAM power and system energy reduction \fix{for
  \fix{non-memory-intensive} workloads} \fixIV{than \memdvfs}.

Fourth, although \voltron reduces the system energy with a small performance
loss, the average system energy efficiency, in terms of \emph{performance per
  watt} (not shown in the figure), still improves by 3.3\% and 7.4\% for
\fixIV{\mbox{non-memory-intensive} and memory-intensive} workloads, respectively, over
the baseline. \fix{Thus, we} demonstrate that \voltron is an effective mechanism
that improves \fixIV{system} energy efficiency not only on
\fix{non-memory-intensive} applications, but also \fixIV{(especially)} on
\fix{memory-intensive} workloads where prior work was unable to do so.

\fix{To summarize, across \fixIV{non-memory-intensive and memory-intensive}
  workloads, Voltron reduces the average system energy consumption by 3.2\% and
  7.0\% while limiting average system performance loss to only 2.5\% and 2.9\%,
  respectively.  Voltron ensures that no workload loses performance by more than
  the specified target of 5\%. We conclude that Voltron is an effective DRAM
  \fixIV{and system} energy reduction mechanism that significantly outperforms
  prior memory DVFS mechanisms.}

\fix{
\subsection{\fixIII{System Energy Breakdown}}
\label{ssec:energy_breakdown}

To demonstrate the source of energy savings from Voltron, \figref{breakdown}
compares the system energy breakdown of Voltron to the baseline\fix{, which}
operates at the nominal voltage level \fix{of 1.35V}. The breakdown shows the
average CPU and DRAM energy consumption across workloads\fix{, which} are
categorized into \fixIV{non-memory-intensive and memory-intensive workloads}.
We make \fix{two observations from the figure}.

\figputHSL{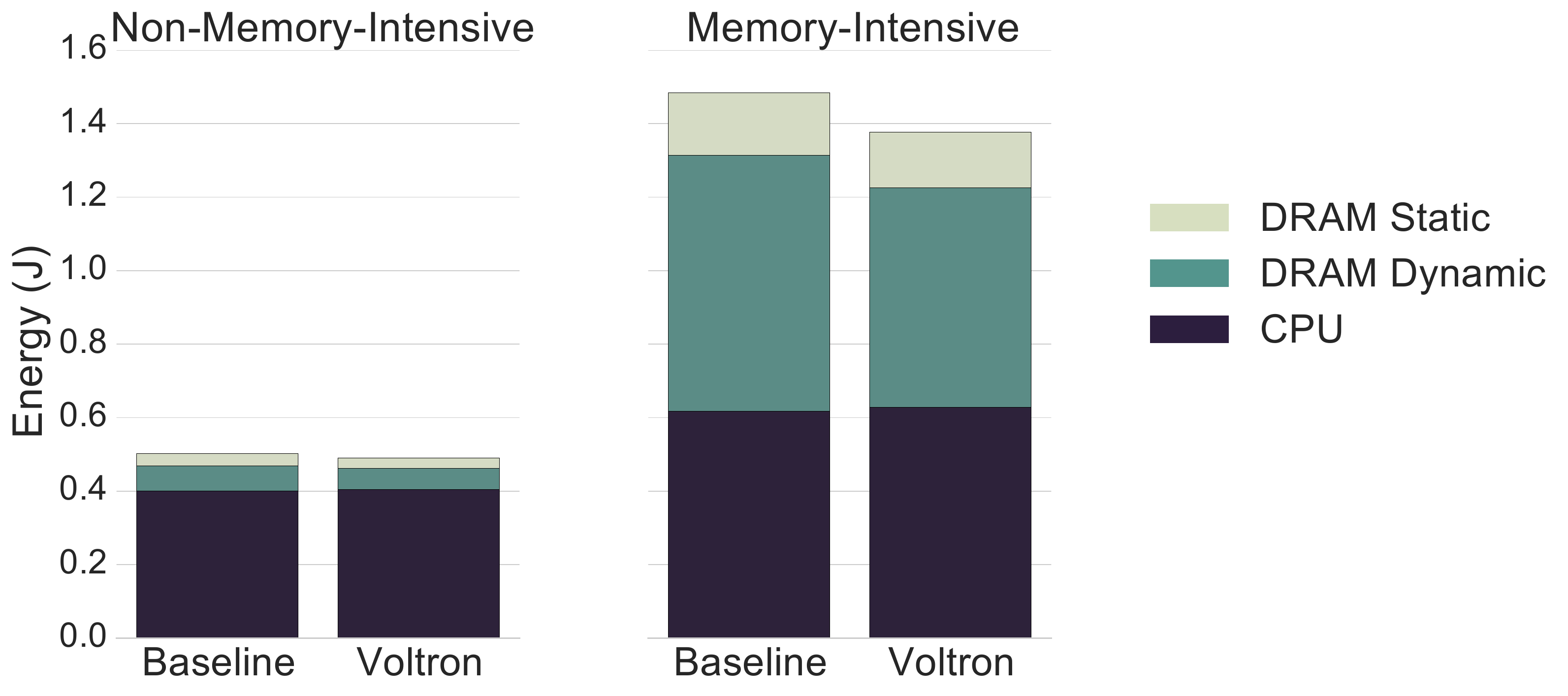}{0.31}{Breakdown of system energy consumption (lower is better).}{breakdown}

First, in the \fixIV{non-memory-intensive} workloads, \fix{the} CPU consumes an
average of 80\% of the total system energy when \fix{the} DRAM uses the nominal
voltage level. As a result, Voltron has less potential \fix{to reduce} the
overall system energy \fix{as} it \fix{reduces \emph{only}} the DRAM energy,
which \fix{makes up only} 20\% of the \fix{total} system energy. Second, DRAM
consumes an average of 53\% of the total system energy in the
\fix{memory-intensive} workloads.  \fixIII{As a result, Voltron has a larger
  room for potential improvement for memory-intensive workloads than for}
\fix{non-memory-intensive} workloads. Across the \fix{memory-intensive}
workloads, Voltron reduces the average dynamic and static DRAM energy by 14\%
and 11\%, respectively. However, Voltron increases the CPU energy consumption by
1.7\%, because \fix{the application} incurs a small system performance
degradation (due to \fix{the} increased \fix{memory} access latency), which is
within \fix{our 5\%} performance loss target (as shown in \ssecref{pavc_eval}).
\fixIV{We conclude that Voltron is effective in reducing DRAM energy, and it is
  an effective system energy reduction mechanism, especially when DRAM is a
  major consumer of energy in the system.} }

\ignore{
\subsection{Comparison to Memory DVFS}

In this section, we quantitatively compare Voltron against \textit{\memdvfs}, a
dynamic DRAM frequency and voltage scaling mechanism proposed by a prior
work~\cite{david-icac2011}, which we already described in \ssecref{dvfs}.
Similar to the configuration used in the prior work, we enable \memdvfs to
dynamically switch between three frequency steps: 1600, 1333, and 1066 MT/s,
which employ 1.35V, 1.3V, and 1.25V, respectively.

\figref{memdvfs} shows the system performance loss and system energy savings due
to \memdvfs and \voltron. Each metric (y-axis) is shown in a single plot that
uses boxplots to show the distribution among multiple workloads, which are
categorized into high and low memory intensity (x-axis). Each box demonstrates
the quartiles of the population and the whiskers indicate the minimum and
maximum value. The red dot on each box indicates the mean. We make several major
observations.

\begin{figure}[!h]
    \centering
    \subcaptionbox{\label{fig:volt_vs_mem_perf}}[0.32\linewidth]
    {
      \includegraphics[scale=1.2]{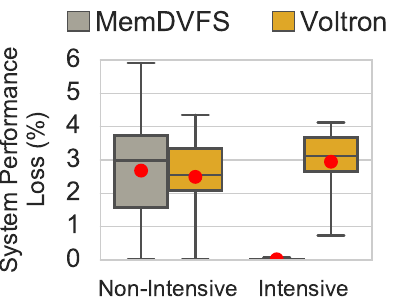}
    }
    \subcaptionbox{\label{fig:volt_vs_mem_dpow}}[0.32\linewidth]
    {
      \includegraphics[scale=1.2]{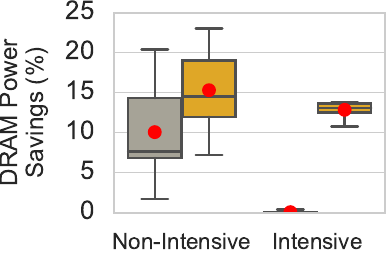}
    }
    \subcaptionbox{\label{fig:volt_vs_mem_syse}}[0.32\linewidth]
    {
      \includegraphics[scale=1.2]{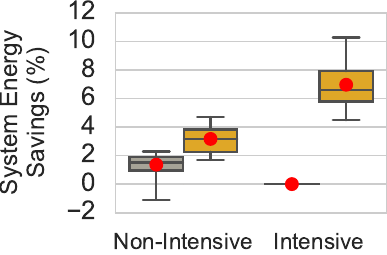}
    }
    \vspace{-0.15in}
    \caption{\fixIV{Performance and energy} comparison between \voltron and
      \memdvfs on \fixV{non-memory-intensive and memory-intensive workloads}.}
    \label{fig:memdvfs}
\end{figure}

First, for the \fix{memory-intensive} workloads, \memdvfs has almost zero effect on
system energy and performance. The reason is that \memdvfs avoids scaling DRAM
frequency when an application's memory bandwidth is above a fixed threshold
since reducing the frequency can result in a large performance loss due to lower
data throughput. As \fix{memory-intensive} applications have high memory bandwidth
consumption that is easily above the threshold, \memdvfs cannot perform
frequency and voltage scaling during most of the applications' runtime. These
results are consistent with that of reported in \memdvfs\cite{david-icac2011}. In
contrast, \voltron reduces system energy by 7.0\% on average for the \fix{memory-intensive}
workloads at the cost of 2.9\% system performance loss, which meets the
specified target of 5\%.

Second, both \memdvfs and \voltron reduce system energy consumption for low
memory-intensity workloads. By reducing the DRAM array voltage to a lower value
than \memdvfs, \voltron is able to provide slightly higher reduction. Third,
although \voltron reduces the system energy at a small performance loss, the
system energy efficiency, in terms of \emph{performance per watt}, still
improves by 7.4\% and 3.3\% for \fix{memory-intensive and non-memory-intensive} workloads,
respectively, on average over the baseline. We demonstrate that \voltron is a new
optimization technique that enables energy efficiency improvement not only on
non-memory-intensive applications, but also on memory-intensive workloads that
prior work cannot achieve.
}

\response{
\subsection{Effect of Exploiting \fixIV{Spatial} Locality \fixV{of Errors}}
\label{ssec:eval_el}
In \ssecref{spatial}, our experimental results show that errors due to reduced
voltage concentrate in certain regions, specifically in select DRAM banks for some
vendors' DIMMs. This implies that when we lower the voltage, only the banks with errors require a higher
access latency to read or write data correctly, whereas error-free banks can
be accessed reliably with the standard latency. Therefore, in this section, we
enhance our \voltron mechanism by exploiting the spatial locality of
\fixIV{errors caused by reduced-voltage operations}. The key idea is to dynamically change the access latency \fixIII{on a
  per-bank basis (i.e., based on the
DRAM banks being accessed)} to account for the reliability of each bank.
\fixIII{In other words, we would like to increase the latency only for banks
that would otherwise experience \emph{errors}, and do so just enough such that these banks operate
reliably.}

\fix{For our evaluation, we model the behavior based on a subset
  \fixIII{(three)} of Vendor C's DIMMs, which show that the number of banks with
  errors increases as we reduce the \fixIV{supply} voltage
  \fixIII{(\ssecref{spatial})}. We observe that these DIMMs start experiencing
  errors at 1.1V using the standard latency values. However, only one bank
  observes errors when we reduce the voltage level from 1.15V to 1.1V
  \fixIII{(i.e., 50mV reduction)}. We evaluate a \fixIV{conservative} model that
  increases the number of banks that need higher latency by one for every 50mV
  reduction from the nominal voltage of 1.35V. Note that this model is
  conservative, because we start increasing the latency when the voltage is
  reduced to 1.3V, which is much higher than the lowest voltage level (1.15V)
  for which we observe that DIMMs operate reliably without \fixIV{requiring} a
  latency increase. Based on this conservative model, we choose the banks
  \fixIV{whose latencies should} increase \emph{sequentially} starting from the
  first bank, while the remaining banks operate at the standard latency. For
  example, at 1.25V (100mV lower than the nominal voltage of 1.35V), \voltron
  needs to increase the latency for the first two out of the eight banks to
  ensure reliable \fixIII{operation}. }


\fix{\figref{voltron_bankloc} compares the system performance and energy
efficiency of our bank-error locality aware version of \voltron (denoted as
\emph{Voltron+BL}) to the previously-evaluated \voltron mechanism, which is not
aware of such locality.} \fix{By increasing the memory latency for only} a
subset of banks at each voltage step, Voltron+BL reduces the average performance
loss from 2.9\% to 1.8\% and \fix{increases} the average system energy savings
from 7.0\% to 7.3\% for \fix{memory-intensive} workloads, with similar
improvements for \fix{non-memory-intensive} workloads. We show that enhancing
\voltron by adding awareness of the spatial locality of errors can further
mitigate the latency penalty due to reduced voltage, \fixIV{even with \fixV{the}
  conservative bank error locality model we assume and evaluate in this
  example.} \fix{We believe that a mechanism that exploits \fixIV{spatial} error
  locality at a finer granularity could lead to even higher performance and
  energy savings, but we leave \fixIV{such an evaluation} to future work.}

\begin{figure}[!h]
    \centering
    \subcaptionbox*{}[0.49\linewidth]
    {
      \includegraphics[scale=1.2]{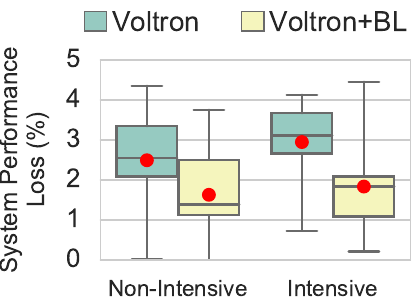}
    }
    \subcaptionbox*{}[0.49\linewidth]
    {
      \includegraphics[scale=1.2]{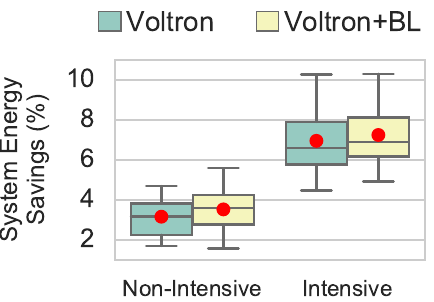}
    }
    \vspace{-0.25in}
    \caption{\fixIV{Performance and energy benefits} of exploiting bank-error locality in Voltron (denoted as
      Voltron+BL) on \fixIV{non-memory-intensive and memory-intensive workloads}.}
    \label{fig:voltron_bankloc}
\end{figure}


\subsection{Effect on Heterogeneous Workloads}
\label{ssec:eval_hetero}

So far, we have evaluated \voltron on \fixIII{\emph{homogeneous} multi-core} workloads, where
each workload consists of the same benchmark running on all cores. In this
section, we evaluate the effect of \voltron on \textit{heterogeneous} workloads,
where each workload consists of \emph{different} benchmarks running on each
core. We categorize the workloads based on the fraction of memory-intensive
benchmarks in the workload (0\%, 25\%, 50\%, 75\%, and 100\%). Each category
consists of 10~workloads, resulting in a total of 50~workloads across all
categories.


\fix{\figref{voltron_hetero} shows the system performance loss and energy
efficiency improvement (in terms of performance per watt) with \voltron and with
\memdvfs for heterogeneous workloads. The error bars indicate the 95\%
confidence interval \fixIII{across all workloads in the category}. We make two
observations from the figure.}
\fix{First, for each category of the heterogeneous workloads, \voltron is able
  to meet the 5\% performance loss target on average.
However, since \voltron is not designed to provide a \emph{hard} performance
guarantee for every single workload, \voltron exceeds the performance loss
target for 10 out of the 50~workloads, though it exceeds the target by only
0.76\% on average. Second,} the energy efficiency improvement due to
\voltron\fixIII{\ becomes larger} as the memory intensity of the workload
increases. This is because the fraction of system energy coming from memory
grows with higher memory intensity, \fixIII{due to the higher amount of} memory
traffic. Therefore, the memory energy reduction from \voltron has \fix{a
  greater} impact at the system level with more \fix{memory-intensive}
workloads. On the other hand, \memdvfs becomes \emph{less} effective with higher
memory intensity, as the memory bandwidth \fixIV{utilization more frequently}
exceeds the fixed threshold \fixIII{employed by \memdvfs}. Thus, \memdvfs has
\fix{a} smaller opportunity to scale the frequency and voltage. \fix{We conclude
  that \voltron is an effective mechanism that can adapt to different
  applications' characteristics to improve system energy efficiency.}

\begin{figure*}[!h]
    \centering
    \subcaptionbox*{}[0.49\linewidth]
    {
      \includegraphics[scale=1.0]{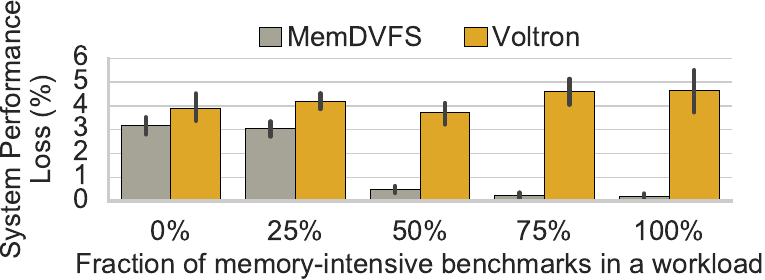}
    }
    \subcaptionbox*{}[0.49\linewidth]
    {
      \includegraphics[scale=1.0]{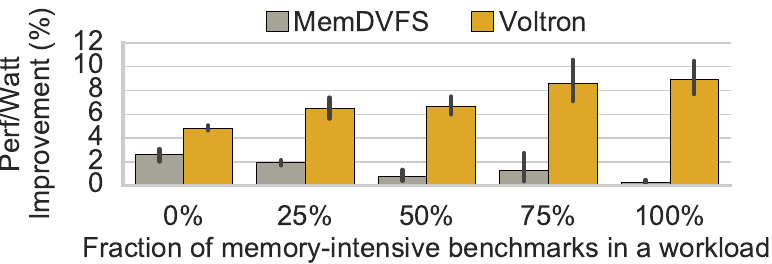}
    }
    \vspace{-0.25in}

    \caption{\response{System performance loss and energy efficiency improvement of Voltron
        and MemDVFS across 50 different heterogeneous workload mixes.}}
    \label{fig:voltron_hetero}
\end{figure*}


\subsection{Effect of Varying the Performance Target}
\label{ssec:vary_perf_target}

\figref{voltron_sweep} shows the performance loss and energy efficiency
\fixV{improvement due to} \voltron as we vary the system performance loss target for
\fixIV{both homogeneous and} heterogeneous workloads. For each target, we use a
boxplot to show the distribution across all workloads. In total, we evaluate
\voltron on 1001 combinations of workloads and performance targets: \fixIV{27
  homogeneous workloads $\times$ 13 targets + 50 heterogeneous workloads
  $\times$ 13 targets}. The first major observation is that \voltron's
performance-aware voltage control mechanism adapts to different performance
targets by dynamically selecting different voltage values at runtime. Across all
1001~runs, \voltron\fixIII{\ keeps performance within the performance loss}
target for 84.5\% of them. Even though \voltron cannot enforce a
\fixIII{\emph{hard}} performance guarantee for all workloads, \fixIII{it}
exceeds the target by only 0.68\% on average for those workloads \fixIII{where
  it does} not strictly meet the target.

\begin{figure}[!h]
    \centering
    \subcaptionbox*{}[\linewidth][c]
    {
      \includegraphics[scale=1.2]{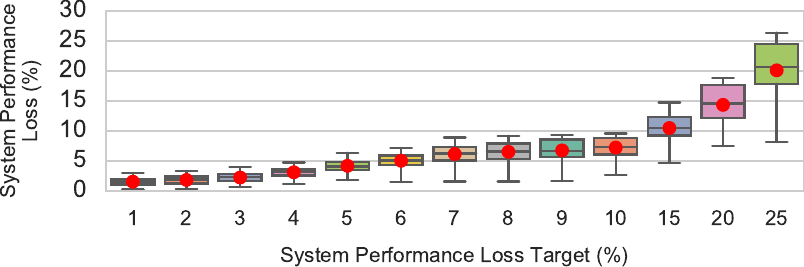}
      \vspace{-0.2in}
    }
    \subcaptionbox*{}[\linewidth][c]
    {
      \includegraphics[scale=1.2]{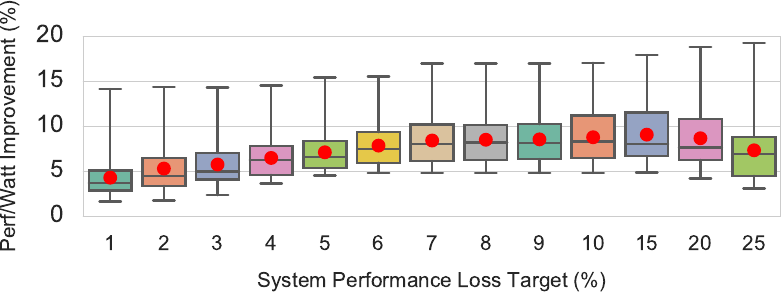}
    }
    \vspace{-0.2in}
    \caption{System performance loss and energy efficiency improvement of Voltron as the system performance loss target varies.}
    \label{fig:voltron_sweep}
\end{figure}

Second, system energy efficiency increases with higher performance loss targets,
but the gains plateau at around a target of 10\%. Beyond the 10\% target,
\voltron starts selecting smaller \varr values (e.g., 0.9V) that result in much
higher memory latency, which in turn increases both the CPU runtime and system
energy. \fix{In \ssecref{scaling_eval}, we observed that employing a \varr value
  less than 1.0V can result in smaller system energy savings than using
  \varr=1.0V.}

\fixIV{We conclude that,}
compared to prior work on memory DVFS, \voltron is a more
flexible mechanism, as it allows the \fixIV{users or system} designers to select
a performance and energy trade-off that best suits their target system or
applications.

\subsection{Sensitivity to the Profile Interval Length}
\label{ssec:interval}

\figref{interval} shows the average \fixV{system} energy efficiency improvement
due to \voltron with different profile interval lengths \fixIV{measured across
  27 homogeneous workloads}. As the \fixIV{profile} interval length increases
beyond two million cycles, we observe that the energy efficiency \fixIII{benefit
  of \voltron} \fixV{starts reducing}. This is because longer intervals prevent
\voltron from making \fixV{faster \varr} adjustments based on the
\fixV{collected new profile} information. Nonetheless, \voltron consistently
improves system energy efficiency \fixIV{for all evaluated profile interval
  lengths}.

\begin{figure}[h]
\begin{minipage}{\linewidth}
\begin{center}
\includegraphics[scale=1.1]{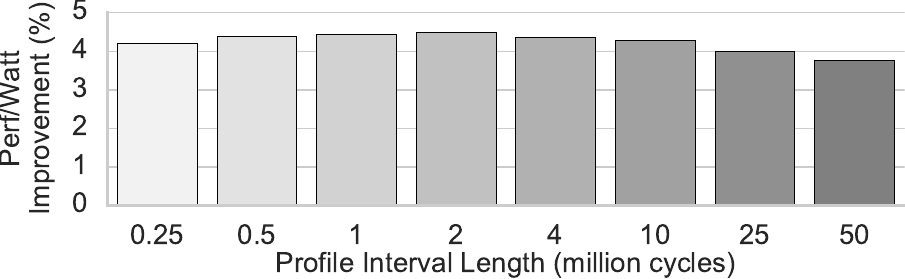}
\end{center}
\vspace{-0.1in}
\caption{Sensitivity of Voltron's \fixIV{system energy efficiency improvement} to profile interval length.\label{fig:interval}}
\end{minipage}
\end{figure}

} 


\section{Summary}

In this chapter, we provide the first experimental study that comprehensively
characterizes and analyzes the behavior of DRAM chips when the supply voltage is
reduced below its nominal value. We demonstrate, \fixIV{using 124 DDR3L DRAM
  chips}, that the \fix{DRAM} supply voltage can be reliably reduced to a
certain level, beyond which errors arise within the data. We then experimentally
demonstrate the relationship between the supply voltage and the latency of the
fundamental DRAM operations (activation, restoration, and precharge). \fixIV{We
  show that bit errors caused by reduced-voltage operation can be eliminated by
  increasing the latency of the three fundamental DRAM operations.} By
\fix{changing} the memory controller configuration to allow for the longer
latency of these operations, we can thus \emph{further} lower the supply voltage
without inducing errors in the data. We also \fixIV{experimentally} characterize
the relationship between reduced supply voltage and error locations, stored data
patterns, temperature, and data retention.


Based on these observations, we propose and evaluate \voltron, a low-cost energy
reduction mechanism that reduces DRAM energy \emph{without} \fix{affecting}
memory data throughput. \voltron reduces the supply voltage for \emph{only} the
DRAM array, while maintaining the nominal voltage for the peripheral circuitry
to continue operating the memory channel at a high frequency. \voltron uses a
new piecewise linear performance model to find the array supply voltage that
maximizes the system energy reduction within a given performance loss target.
\fix{Our experimental evaluations across a wide variety of workloads}
demonstrate that \voltron significantly reduces system energy \fix{consumption}
with only \fix{very} modest performance loss.

\ignore{
We conclude that it is very promising to understand and exploit reduced-voltage
operation in modern DRAM chips. We hope that the experimental characterization,
analysis, and optimization techniques presented in this chapter will enable the
development of other new mechanisms that can effectively \fixIII{exploit the
trade-offs between voltage, reliability, and latency in DRAM to improve system
performance, efficiency, and/or reliability.}
}

\begin{subappendices}

\section{FPGA Schematic of DRAM Power Pins}
\label{sec:pin_layout}

\fix{\figref{pin_layout} shows a schematic of the DRAM pins that our FPGA
board~\cite{ml605_schematic} connects to (see \secref{fpga} for our experimental
methodology). Since there are a large number of pins that are used for different
purposes (e.g., data address), we zoom in on the right side of the figure to
focus on the power pins that we adjust for our experiments in this chapter.}
\fixIII{Power pin numbering information can be found on the datasheets provided
  by all major vendors (e.g., \cite{micronDDR3L_2Gb, hynix-ddr3l,
    samsungddr3l_2Gb}).} In particular, we tune the VCC1V5 pin \fix{on the
  FPGA}, which is directly connected to all of the $V_{DD}$ and $V_{DDQ}$ pins
on the DIMM. The reference voltage VTTVREF is automatically adjusted by the DRAM
to half of VCC1V5.

\figputHS{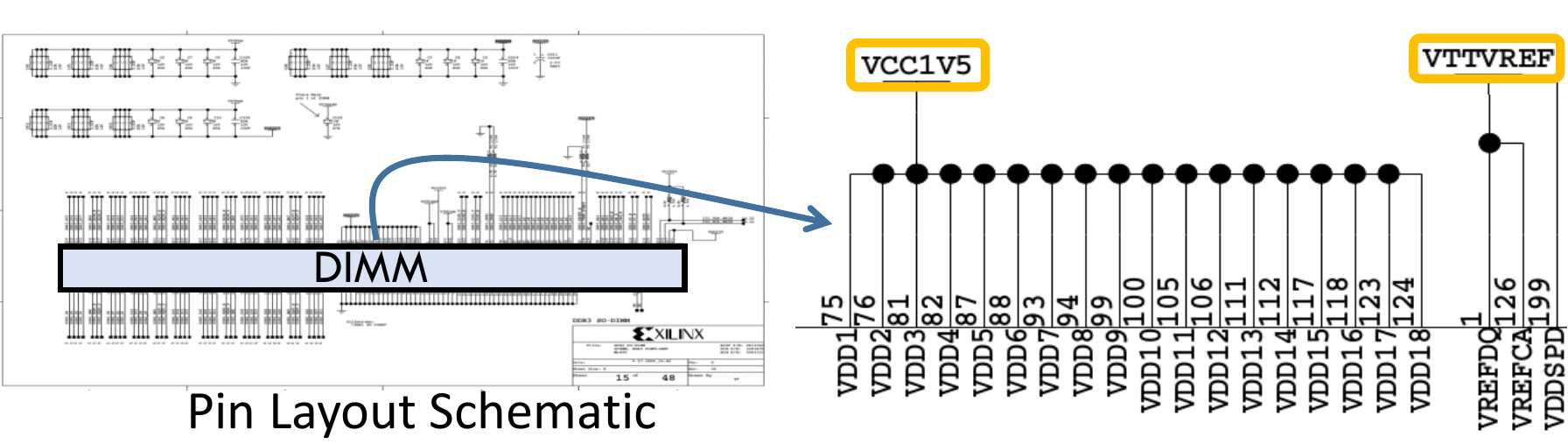}{0.7}{DRAM power pins \fix{controlled by the ML605 FPGA board}.}

\section{Effect of Data Pattern on Error Rate}
\label{sec:datapatt}

\fix{As
discussed in \ssecref{volt_sensitivity}, we do \emph{not} observe a significant
effect \fix{of} different stored data patterns \fix{on the DRAM error rate when we
  reduce the supply voltage.} \figref{datapatt_volt} shows the average bit error rate
(BER) of three different data patterns (\patt{aa}, \patt{cc}, and \patt{ff})
across different supply voltage levels for each vendor. Each data pattern
represents the byte value (shown in hex) that we fill into the DRAM. The error
bars indicate the 95\% confidence interval.  We make two observations from the
figure.



\figputHSL{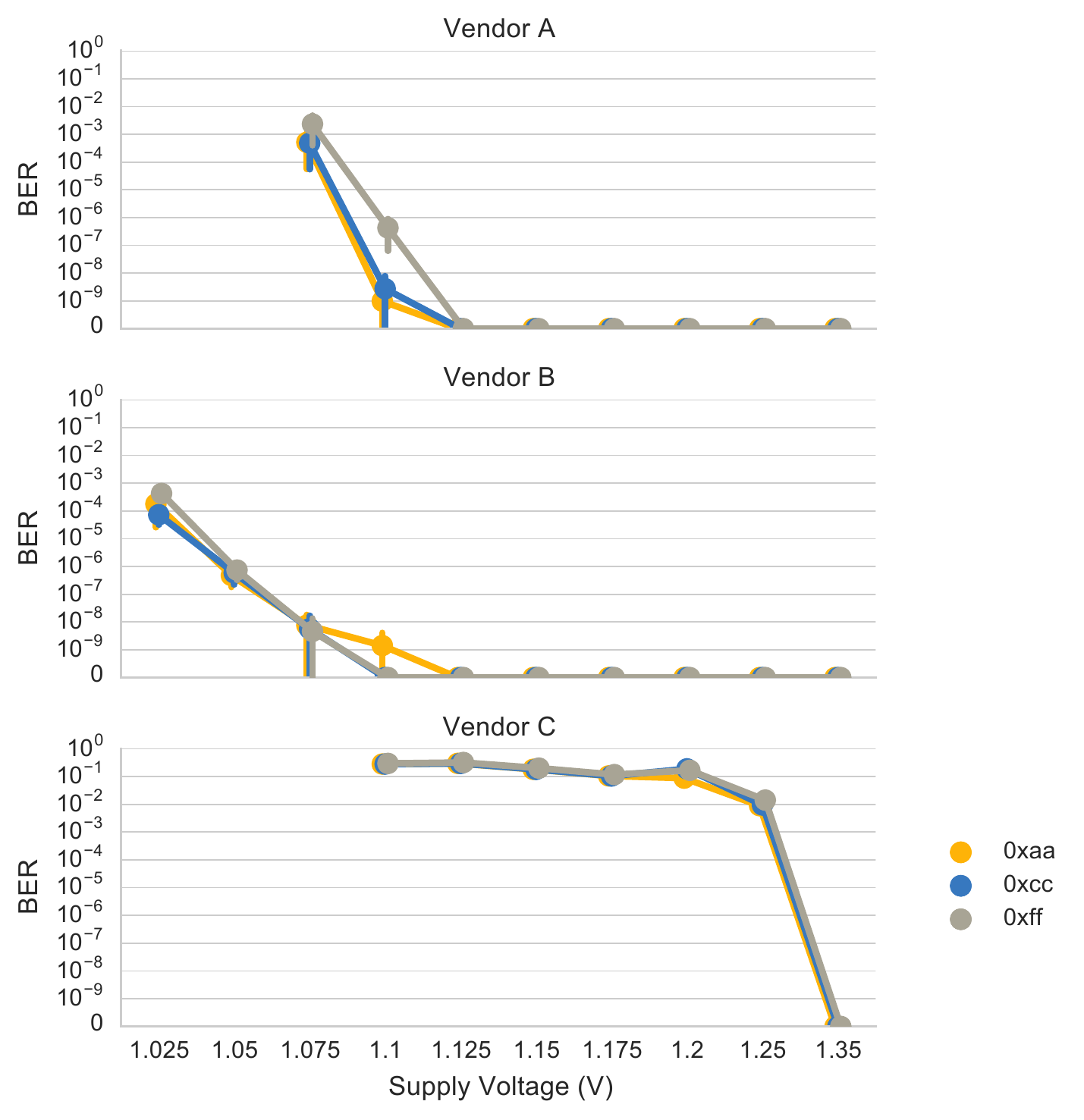}{0.72}{Effect of \fixIII{stored
  data pattern} on \fix{bit error rate (BER)} across different supply voltage
levels.}{datapatt_volt}

First, the BER increases as we reduce the supply
voltage for all three data patterns. We made a similar observation in
\ssecref{volt_sensitivity}, which shows that the fraction of errors increases as the
supply voltage drops. We explained our hypothesis on the cause of the errors, and
used both experiments and simulations to test the
hypothesis, in \ssecref{low_volt_latency}.

Second, we do \emph{not} observe a significant difference across the BER values from
the three different data patterns. We attempt to answer the following question:
\fix{Do} different data patterns induce BER values that are statistically different
from each other at each voltage level? To answer this, we conduct a one-way
ANOVA (analysis of variance) test across the measured BERs from all three data patterns at each supply
voltage level to calculate a \emph{p-value}. If the p-value is below 0.05, we
can claim that these three data patterns induce a statistically-significant
difference on the error rate. \tabref{pval} shows the calculated p-value at each
supply voltage level. At certain supply voltage levels, we do not have a p-value
listed (shown as --- or $\triangle$ in the table), \fixIII{either} because there
are no errors \fixIV{(indicated as ---)} or we cannot
reliably access data \fixIV{from the DIMMs} even if the access latency is
higher than the standard value \fixIV{(indicated as $\triangle$)}.


\begin{table}[h]
  \small
  \centering
    \begin{tabular}{crrr}
        \toprule
        Supply & \multicolumn{3}{c}{\bf Vendor} \\

         \cmidrule(lr){2-4}
        Voltage & \multicolumn{1}{c}{A} & \multicolumn{1}{c}{B} & \multicolumn{1}{c}{C} \\
        \midrule
        1.305  & --- & --- & --- \\
        1.250  & --- & --- & \textbf{0.000000} \\
        1.200   & --- & --- & 0.029947 \\
        1.175 & --- & --- & 0.856793 \\
        1.150  & --- & --- & 0.872205 \\
        1.125 & --- & 0.375906 & 0.897489 \\
        1.100 & \textbf{0.028592} & 0.375906 & \textbf{0.000000} \\
        1.075 & 0.103073 & 0.907960 & $\triangle$ \\
        1.050 & $\triangle$  & 0.651482 & $\triangle$ \\
        1.025 & $\triangle$  & \textbf{0.025167} & $\triangle$ \\
        \bottomrule
    \end{tabular}

  \caption{Calculated p-values from the BERs across three data patterns at each
    supply voltage level. A p-value less \fixIII{than} 0.05 indicates that the BER is
    statistically different across the three data patterns (indicated in bold). \fix{--- indicates
    that the BER is zero. \fixIV{$\triangle$ indicates that we cannot reliably
      access data from the DIMM}.}}
  \label{tab:pval}
\end{table}


Using the one-way ANOVA test, we find that using different data patterns does
\emph{not} have a statistically significant (i.e., p-value $\geq$ 0.05) effect
on the error rate at \emph{all} supply voltage levels. Significant effects
(i.e., p-value $<$ 0.05) occur at 1.100V for Vendor~A, at 1.025V for Vendor~B, and
at both 1.250V and 1.100V for Vendor~C. As a result, our study does \emph{not}
\fixIV{provide} enough evidence to conclude that using any of the three data
patterns (\patt{aa}, \patt{cc}, and \patt{ff}) induces higher or lower error
rates than the other two patterns at reduced voltage levels. }

%
%
%
%

\fix{
\section{SPICE Simulation Model}
\label{spice_model}

We perform circuit-level SPICE simulations to understand in detail how the DRAM cell
arrays operate at low supply voltage. We model a DRAM cell array in SPICE,
and simulate its behavior for different supply voltages. We have released our
SPICE model online~\cite{volt-github}.

\paratitle{DRAM Cell Array Model} We build a detailed cell array model, as shown
in \figref{spice_model}. In the cell array, the DRAM cells are organized as $512
x 512$ array, which is a common organization in modern DRAM
chips~\cite{vogelsang-micro2010}. Each column is vertical, and corresponds to
512~cells sharing a bitline that connects to a sense amplifier. Due to the
bitline wire and the cells that are connected to the bitline, there is parasitic
resistance and capacitance on each bitline. Each row consists of 512~cells
sharing the same wordline, which also has parasitic resistance and capacitance.
The amount of parasitic resistance and capacitance on the bitlines and wordlines
is a major factor that affects the latency of DRAM operations accessing a cell
array~\cite{lee-hpca2013,lee-sigmetrics2017}.


\figputHS{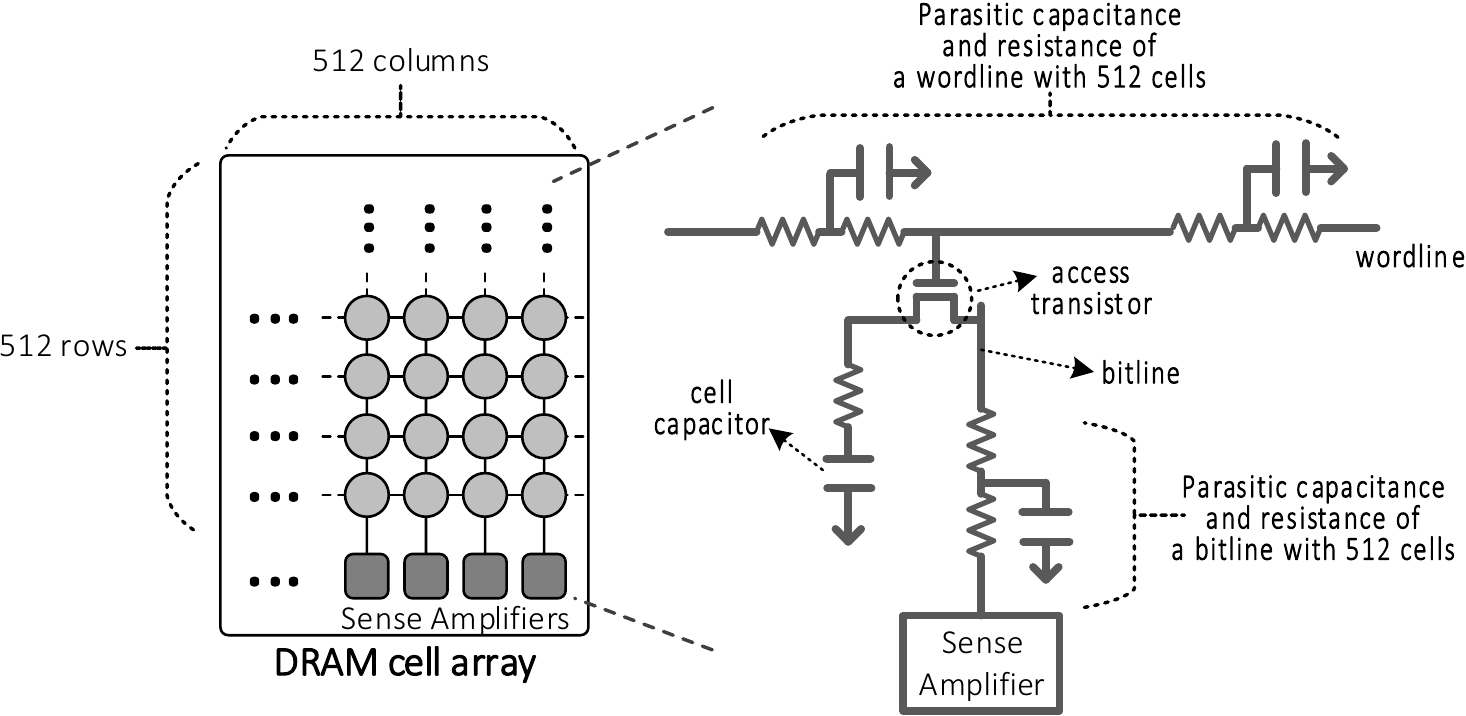}{1.0}{Our SPICE model \fixIV{schematic} of a DRAM cell array.}

\paratitle{Simulation Methodology} We use the LTspice~\cite{ltspice} SPICE simulator
to perform our simulations.
To find the access latency of the DRAM
operations under different supply voltages, we build a DRAM cell array using
technology parameters that we derive from a \SI{55}{\nano\meter} DRAM
model~\cite{vogelsang-micro2010} and from a \SI{45}{\nano\meter} process technology model~\cite{ptm,
  zhao-isqed2006}. By default, we assume that the cell capacitance is \SI{24}{\femto\farad} and
the bitline capacitance is \SI{144}{\femto\farad}~\cite{vogelsang-micro2010}. The nominal
\varr is 1.35V, and we perform simulations to obtain the latency of DRAM operations
at every 25mV step from 1.35V down to 0.9V.  The results of our SPICE simulations
are discussed in Section~\ref{ssec:volt_sensitivity} and \ref{ssec:low_volt_latency}.
}


\newpage

\section{Spatial Distribution of Errors}
\label{spatial}

In this section, we expand upon the spatial locality data presented in
\ssecref{spatial}. \fix{Figures~\ref{fig:full_loc_A}, \ref{fig:full_loc_B}, and
\ref{fig:full_loc_C} show the physical locations of errors that occur when the
supply voltage is reduced for a representative DIMM from Vendors~A, B, and C,
respectively.} At higher voltage levels, even if errors occur, they
tend to cluster in certain regions of a DIMM. However, as we reduce the supply
voltage further, the number of errors increases, and the errors start to spread
across the \fix{entire} DIMM.

\begin{figure}[!h]
    \centering
    \subcaptionbox{Supply voltage=1.075V.}[0.49\linewidth][l]
    {
        \includegraphics[width=0.49\linewidth]{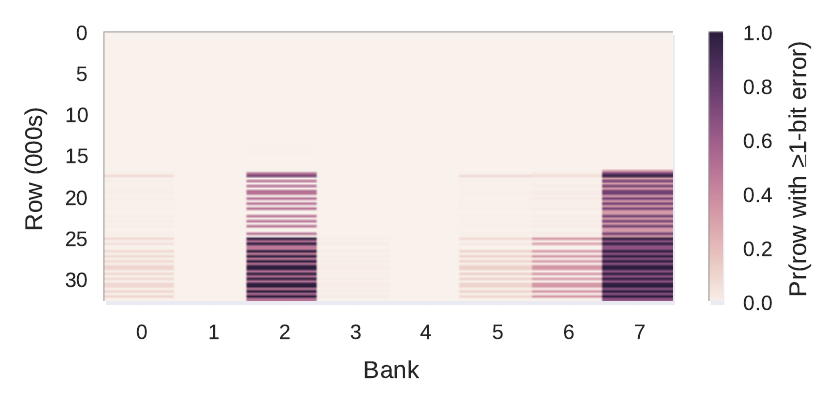}
    }
    \subcaptionbox{Supply voltage=1.1V.}[0.49\linewidth][r]
    {
        \includegraphics[width=0.49\linewidth]{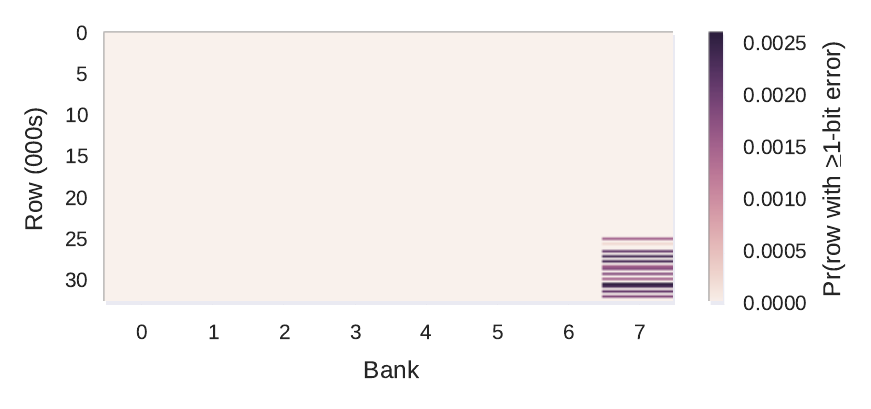}
    }

    \caption{Probability of error occurrence due to \fixIV{reduced-voltage operation} in a DIMM from \fix{Vendor~A}.}
    \label{fig:full_loc_A}
\end{figure}

\begin{figure}[!h]
    \centering
    \subcaptionbox{Supply voltage=1.025V.}[0.49\linewidth][l]
    {
        \includegraphics[width=0.47\linewidth]{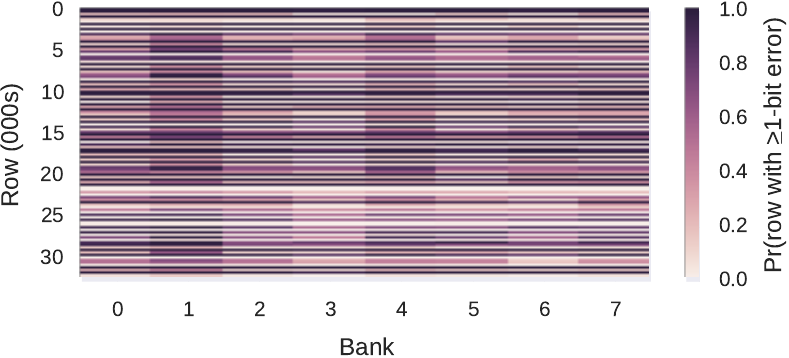}
    }
    \subcaptionbox{Supply voltage=1.05V.}[0.49\linewidth][r]
    {
        \includegraphics[width=0.47\linewidth]{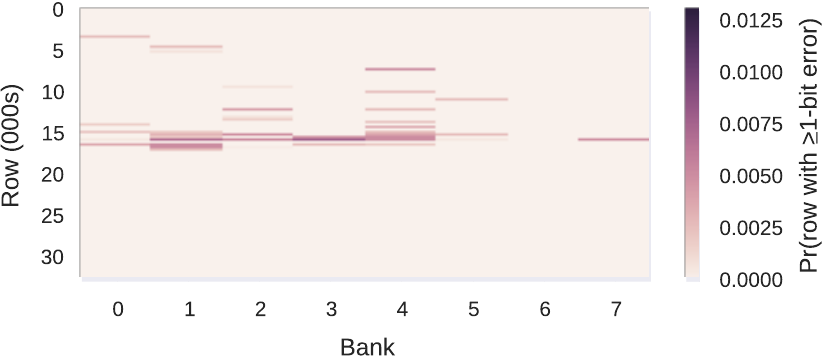}
    }

    \subcaptionbox{Supply voltage=1.1V.}[0.8\linewidth][c]
    {
        \includegraphics[width=0.49\linewidth]{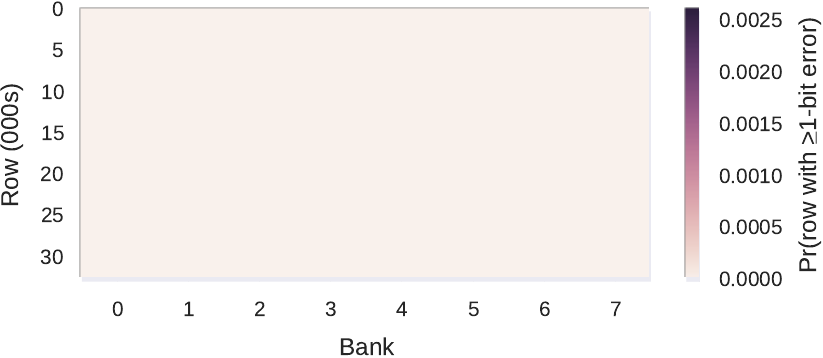}
    }
    \caption{Probability of error occurrence due to \fixIV{reduced-voltage operation} in a DIMM from \fix{Vendor~B}.}
    \label{fig:full_loc_B}
\end{figure}

\begin{figure}[!h]
    \centering
    \subcaptionbox{Supply voltage=1.1V.}[0.49\linewidth][l]
    {
        \includegraphics[width=0.47\linewidth]{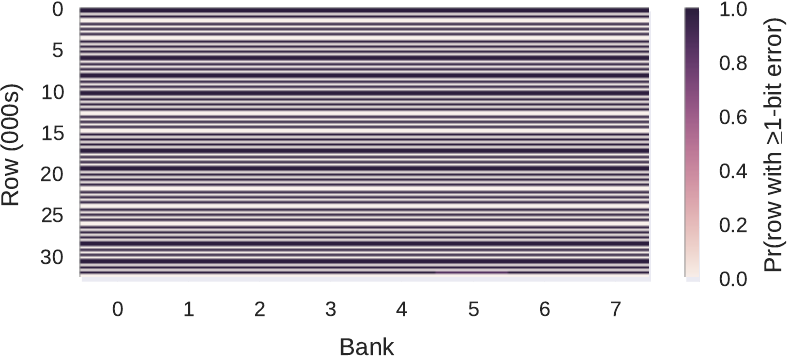}
    }
    \subcaptionbox{Supply voltage=1.15V.}[0.49\linewidth][l]
    {
        \includegraphics[width=0.47\linewidth]{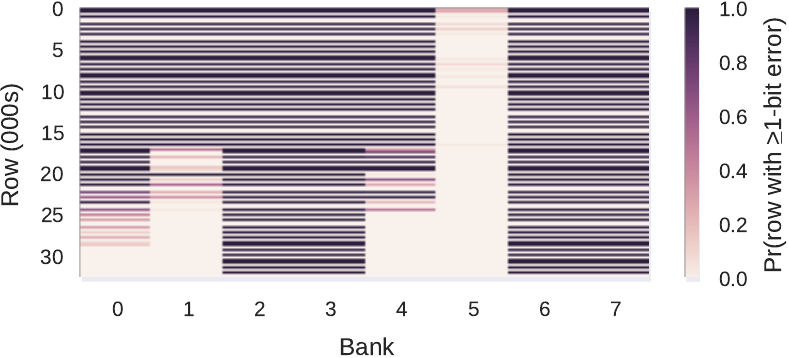}
    }
    %
    \subcaptionbox{Supply voltage=1.2V.}[0.49\linewidth][l]
    {
        \includegraphics[width=0.49\linewidth]{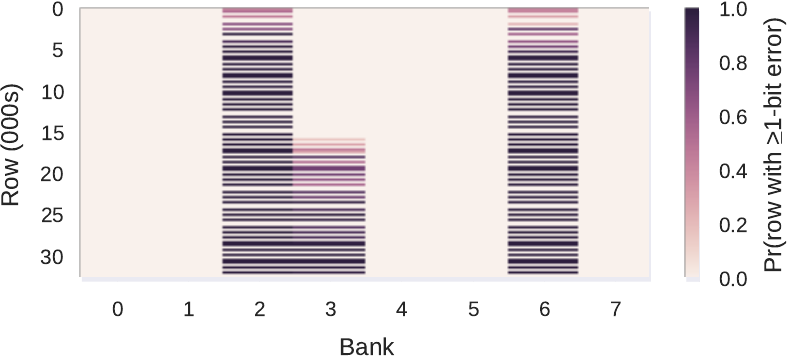}
    }
    \caption{Probability of error occurrence due to \fixIV{reduced-voltage operation} in a DIMM from \fix{Vendor~C}.}
    \label{fig:full_loc_C}
\end{figure}

\newpage
\
\newpage

\section{Full Information of Every Tested DIMM}
\label{sec:dimm_info}

\tabref{modules} lists the parameters of every DRAM module that we evaluate,
\fixIII{along with the \vmin we discovered for each module based on our
  experimental characterization (\ssecref{volt_sensitivity}).} \fix{We provide
  all results for all DIMMs in our GitHub repository~\cite{volt-github}.}
\definecolor{lightgray}{gray}{0.95}

\begin{table}[h]

\renewcommand{\arraystretch}{1.25}
\setlength{\tabcolsep}{1pt}
\centering
\scriptsize

\begin{tabular}{ccccccccccccc}

\toprule

\multirow{2}{*}[-2pt]{\em \scriptsize \centering Vendor} &
\multirow{2}{*}[-2pt]{\em \scriptsize \centering Module} &
                     {\em \scriptsize \centering Date$^{\ast}$} &
\multicolumn{4}{c}{\em \scriptsize \centering Timing$^{\dagger}$} &
\multicolumn{2}{c}{\em \scriptsize \centering Organization} &
\multicolumn{4}{c}{\em \scriptsize \centering Chip}
\\

\cmidrule(lr){3-3} \cmidrule(lr){4-7} \cmidrule(lr){8-9} \cmidrule(lr){10-13}

 &
 & {\em \scriptsize (yy-ww)}
 & {\em \scriptsize Freq~(MT/s)}
 & {\em \scriptsize tRCD (ns)}
 & {\em \scriptsize tRP (ns)}
 & {\em \scriptsize tRAS (ns)}
 & {\em \scriptsize Size (GB)$^{\ddagger}$}
 & {\em \scriptsize Chips$^{\star}$}
 & {\em \scriptsize Size (Gb)}
 & {\em \scriptsize Pins}
 & {\em \scriptsize Die Version$^{\S}$}
 & {\em \scriptsize \fix{V\textsubscript{min}} (V)$^{\circ}$}
 \\

\midrule


 & \module{A}{ 1}{} & 15-46 & 1600 & 13.75 & 13.75 & 35 & 2
& 4 & 4 & $\times$16  & $\mathcal{B}$ & 1.100 \\
 & \module{A}{ 2}{} & 15-47 & 1600 & 13.75 & 13.75 & 35 & 2
& 4 & 4 & $\times$16  & $\mathcal{B}$ & 1.125 \\
 & \module{A}{ 3}{} & 15-44 & 1600 & 13.75 & 13.75 & 35 & 2
& 4 & 4 & $\times$16  & $\mathcal{F}$ & 1.125 \\
 & \module{A}{ 4}{} & 16-01 & 1600 & 13.75 & 13.75 & 35 & 2
& 4 & 4 & $\times$16  & $\mathcal{F}$ & 1.125 \\
 & \module{A}{ 5}{} & 16-01 & 1600 & 13.75 & 13.75 & 35 & 2
& 4 & 4 & $\times$16  & $\mathcal{F}$ & 1.125 \\
 & \module{A}{ 6}{} & 16-10 & 1600 & 13.75 & 13.75 & 35 & 2
& 4 & 4 & $\times$16  & $\mathcal{F}$ & 1.125 \\
 & \module{A}{ 7}{} & 16-12 & 1600 & 13.75 & 13.75 & 35 & 2
& 4 & 4 & $\times$16  & $\mathcal{F}$ & 1.125 \\
 & \module{A}{ 8}{} & 16-09 & 1600 & 13.75 & 13.75 & 35 & 2
& 4 & 4 & $\times$16  & $\mathcal{F}$ & 1.125 \\
 & \module{A}{ 9}{} & 16-11 & 1600 & 13.75 & 13.75 & 35 & 2
& 4 & 4 & $\times$16  & $\mathcal{F}$ & 1.100 \\
\multirow{-10}{3em}{\centering {\small\em A} \\ {\tiny \quad \\} \centering
  Total of \\ \centering 10 DIMMs}
 & \module{A}{10}{} & 16-10 & 1600 & 13.75 & 13.75 & 35 & 2
& 4 & 4 & $\times$16  & $\mathcal{F}$ & 1.125 \\

\midrule

 & \module{B}{ 1}{} & 14-34 & 1600 & 13.75 & 13.75 & 35 & 2 &
4 & 4 & $\times$16  & $\mathcal{Q}$ & 1.100 \\
 & \module{B}{ 2}{} & 14-34 & 1600 & 13.75 & 13.75 & 35 & 2 &
4 & 4 & $\times$16  & $\mathcal{Q}$ & 1.150 \\
 & \module{B}{ 3}{} & 14-26 & 1600 & 13.75 & 13.75 & 35 & 2 &
4 & 4 & $\times$16  & $\mathcal{Q}$ & 1.100 \\
 & \module{B}{ 4}{} & 14-30 & 1600 & 13.75 & 13.75 & 35 & 2 &
4 & 4 & $\times$16  & $\mathcal{Q}$ & 1.100 \\
 & \module{B}{ 5}{} & 14-34 & 1600 & 13.75 & 13.75 & 35 & 2 &
4 & 4 & $\times$16  & $\mathcal{Q}$ & 1.125 \\
 & \module{B}{ 6}{} & 14-32 & 1600 & 13.75 & 13.75 & 35 & 2 &
4 & 4 & $\times$16  & $\mathcal{Q}$ & 1.125 \\
 & \module{B}{ 7}{} & 14-34 & 1600 & 13.75 & 13.75 & 35 & 2 &
4 & 4 & $\times$16  & $\mathcal{Q}$ & 1.100 \\
 & \module{B}{ 8}{} & 14-30 & 1600 & 13.75 & 13.75 & 35 & 2 &
4 & 4 & $\times$16  & $\mathcal{Q}$ & 1.125 \\
 & \module{B}{ 9}{} & 14-23 & 1600 & 13.75 & 13.75 & 35 & 2 &
4 & 4 & $\times$16  & $\mathcal{Q}$ & 1.125 \\
 & \module{B}{10}{} & 14-21 & 1600 & 13.75 & 13.75 & 35 & 2 &
4 & 4 & $\times$16  & $\mathcal{Q}$ & 1.125 \\
 & \module{B}{11}{} & 14-31 & 1600 & 13.75 & 13.75 & 35 & 2 &
4 & 4 & $\times$16  & $\mathcal{Q}$ & 1.100 \\
\multirow{-12}{3em}{\centering {\small\em B} \\ {\tiny \quad \\} \centering
  Total of \\ \centering 12 DIMMs}
 & \module{B}{12}{} & 15-08 & 1600 & 13.75 & 13.75 & 35  & 2 & 4 & 4 &
$\times$16  & $\mathcal{Q}$ & 1.100 \\

\midrule

 & \module{C}{ 1}{} & 15-33 & 1600 & 13.75 & 13.75 & 35  & 2 & 4 & 4 & $\times$16
& $\mathcal{A}$ & 1.300 \\
 & \module{C}{ 2}{} & 15-33 & 1600 & 13.75 & 13.75 & 35  & 2
& 4 & 4 & $\times$16 & $\mathcal{A}$ & 1.250 \\
 & \module{C}{ 3}{} & 15-33 & 1600 & 13.75 & 13.75 & 35  & 2
& 4 & 4 & $\times$16  & $\mathcal{A}$ & 1.150 \\
 & \module{C}{ 4}{} & 15-33 & 1600 & 13.75 & 13.75 & 35  & 2
& 4 & 4 & $\times$16  & $\mathcal{A}$ & 1.150 \\
 & \module{C}{ 5}{} & 15-33 & 1600 & 13.75 & 13.75 & 35  & 2
& 4 & 4 & $\times$16  & $\mathcal{C}$ & 1.300 \\
 & \module{C}{ 6}{} & 15-33 & 1600 & 13.75 & 13.75 & 35  & 2
& 4 & 4 & $\times$16  & $\mathcal{C}$ & 1.300 \\
 & \module{C}{ 7}{} & 15-33 & 1600 & 13.75 & 13.75 & 35  & 2
& 4 & 4 & $\times$16  & $\mathcal{C}$ & 1.300 \\
 & \module{C}{ 8}{} & 15-33 & 1600 & 13.75 & 13.75 & 35  & 2
& 4 & 4 & $\times$16  & $\mathcal{C}$ & 1.250 \\
\multirow{-9}{3em}{\centering {\small\em C} \\ {\tiny \quad \\} \centering
  Total of \\ \centering 9 DIMMs}
 & \module{C}{9}{} & 15-33 & 1600 & 13.75 & 13.75 & 35  & 2
& 4 & 4 & $\times$16  & $\mathcal{C}$ & 1.300 \\

\bottomrule
\end{tabular}

\begin{tabular}{l}
\small
{$\,$} \\

{$\ast$ The manufacturing date in the format of year-week
(yy-ww). For example, 15-01 indicates that the DIMM}\\
{was manufactured during the first week of 2015.} \\

{$\,$} \vspace{-7pt}\\

$\dagger$ The timing factors associated with each DIMM:\\
\hspace{0.1in}$Freq$: the channel frequency\\
\hspace{0.1in}$tRCD$: the minimum required latency for an \act to complete \\
\hspace{0.1in}$tRP$: the minimum required latency for a \pre to complete \\
\hspace{0.1in}$tRAS$: the minimum required latency for to restore the charge in an activated row of cells \\

{$\,$} \vspace{-7pt}\\

{$\ddagger$ The maximum DRAM module size supported by our testing platform is 2GB.} \\

{$\,$} \vspace{-7pt}\\

{$\star$ The number of DRAM chips mounted on each DRAM module. }\\

{$\,$} \vspace{-7pt}\\

{$\S$ The DRAM die versions that are marked on the chip package. } \\

{$\,$} \vspace{-7pt}\\

${\circ}$ The \emph{minimum voltage level} that allows error-free \changes{operation}, as
described in \ssecref{volt_sensitivity}.

{$\,$} \\

\end{tabular}

\caption{Characteristics of the evaluated DDR3L DIMMs.}
\vspace{-35pt}
\label{tab:modules}
\end{table}

\end{subappendices}

\chapter{Conclusions and Future Directions}
\label{chap:conclusions}

Over the past few decades, long DRAM access latency has been a critical
bottleneck in system performance. Increasing core counts and the emergence of
increasingly more data-intensive and latency-critical applications further
exacerbate the performance penalty of high memory latency. Therefore, providing
low-latency memory accesses is more critical now than ever before for achieving
high system performance. While certain specialized DRAM architectures provide
low memory latency, they come at a high cost (e.g., 39x higher than the common
DDRx DRAM chips, as described in \chapref{intro}) with low chip density. As a
result, the goal of this dissertation is to enable low-latency DRAM-based memory
systems at \emph{low cost}, with a solid understanding of the latency behavior
in DRAM based on experimental characterization on real DRAM chips.

To this end, we propose a series of mechanisms to reduce DRAM access latency at
\emph{low cost}. First, we propose Lost-Cost Inter-Linked Subarrays (LISA) to
enable \emph{low-latency, high-bandwidth} inter-subarray connectivity within
each bank at a very modest cost of 0.8\% DRAM area overhead. Using this new
inter-subarray connection, DRAM can perform inter-subarray data movement at 26x
the bandwidth of a modern 64-bit DDR4-2400 memory channel. We exploit LISA's
fast inter-subarray movement to propose three new architectural mechanisms that
reduce the latency of two frequently-used system API calls (i.e.,
\texttt{memcpy} and \texttt{memmove}) and the three fundamental DRAM operations,
i.e., activation, restoration, and precharge. We describe and evaluate three
such mechanisms in this dissertation: (1)~Rapid Inter-Subarray Copy
(\emph{\lisarcnol}), which copies data across subarrays at low latency and low
DRAM energy; (2)~Variable Latency (\emph{\lisavillanol}) DRAM, which reduces the
access latency of frequently-accessed data by caching it in fast subarrays; and
(3)~Linked Precharge (\emph{\lisaprenol}), which reduces the precharge latency
for a subarray by linking its precharge units with neighboring idle precharge
units. Our evaluations show that the three new mechanisms of LISA significantly
improve system performance and energy efficiency when used individually or
together, across a variety of workloads and system configurations.


Second, we mitigate the refresh interference, which incurs long memory latency,
by proposing two access-refresh parallelization mechanisms that enable
overlapping more accesses with refreshes inside DRAM.
These two refresh mechanisms are 1) \darp, a new per-bank refresh scheduling
policy that proactively schedules refreshes to banks that are idle or that are
draining writes and 2) \is, a refresh architecture that enables a bank to serve
memory requests in idle subarrays while other subarrays are being refreshed.
\darp introduces minor modifications to only the memory controller, and \sarp
incurs a very modest cost of 0.7\% DRAM area overhead. Our extensive evaluations
on a wide variety of systems and workloads show that these two mechanisms
significantly improve system performance and outperform state-of-the-art refresh
policies. These two techniques together achieve performance close to an
idealized system that does not require refresh.

Third, this dissertation provides the first experimental study that
comprehensively characterizes and analyzes the latency variation within modern
DRAM chips for three fundamental DRAM operations (activation, precharge, and
restoration). We experimentally demonstrate that significant variation is
present across DRAM cells within our tested DRAM chips. Based on our
experimental characterization, we propose a new mechanism, \mech, which exploits
the lower latencies of DRAM regions with faster cells by introducing
heterogeneous timing parameters into the memory controller. We demonstrate that
\mech can greatly reduce DRAM latency, leading to significant system performance
\newt{improvements} on a variety of workloads.

Finally, for the first time, we perform detailed experimental characterization
that studies the critical relationship between DRAM supply voltage and DRAM access
latency in modern DRAM chips. Our detailed characterization of real commodity
DRAM chips demonstrates that memory access latency reduces with increasing
supply voltage. Based on our characterization, we propose Voltron, a new
mechanism that improves system energy efficiency by dynamically adjusting the
DRAM supply voltage based on a performance model.



\cc{
\section{Summary of Latency Reduction}

In this section, we summarize the memory latency reduction due to mechanisms
proposed in this dissertation. In the particular systems that we evaluated in
this dissertation, a last-level cache (LLC) miss generates a DRAM request that
requires multiple fundamental DRAM operations (e.g., activation, restoration,
precharge). Our mechanisms focus on improving the latency of these fundamental
DRAM operations after an LLC miss. Note that our proposals can potentially be
applied to different memory technologies in the memory hierarchy, such as eDRAM
(which typically serves as an LLC), providing additional latency benefits.


Specifically, DRAM has five major timing parameters associated with the DRAM
operations that are used to access a cache line in a closed row: \trcd, \tras,
\tcl, \tbl, and \trp, which are shown in \figref{baseline_latency}. Since we
have already explained the details of these timing parameters in
\chapref{background}, we focus on summarizing the improvements on these timing
parameters due to our proposed techniques in this section.

\figputHS{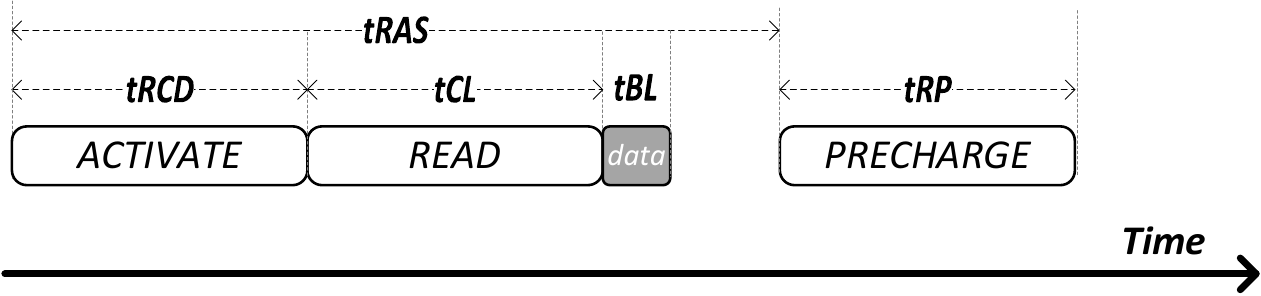}{1}{The major timing parameters required to access a
  cache line in DRAM.}

In this dissertation, our proposals reduce three of the five timing parameters:
\trcd, \tras, and \trp. These three timing parameters are crucial for systems
that generate a large number of random accesses (e.g., reading buffered network
packets) or data dependent accesses (e.g., pointer chasing). Since the read
timing parameter (\tcl) is a DRAM-internal timing that is determined by a clock
inside DRAM, our testing platform does not have the capability to characterize
its behavior. We leave the study on \tcl to future work. On the
other hand, \tbl is determined by the width and frequency of the DRAM channel,
which is not the focus of this dissertation. In addition to addressing the three
major DRAM timing parameters, our dissertation also reduces the bulk copy
latency and the refresh-induced latency. \tabref{summary} lists the quantitative
latency improvements due to each of our proposed mechanisms for the
high-density DDRx DRAM chips.

\begin{table}[h]
  \small
  \centering
    \setlength{\tabcolsep}{1.0em}
    \begin{tabular}{llrr}
        \toprule
        Mechanisms & Improved Latency Components & Latency (ns) & Improvement \\
        \midrule
        LISA-RISC (\symsref{lisaclone}) & Copy latency of 4KB data & 148.5 & 9.2x\\
        LISA-VILLA (\symsref{villa}) & tRCD/\tras /\trp & 7.5/13/8.5 & 1.8x/2.7x/1.5x\\
        LISA-LIP (\symsref{lisapre})   & \trp & 5 & 2.6x\\
        \midrule
        FLY-DRAM (\symsref{var_latency_mc})   & tRCD/\tras/\trp & 7.5/27/7.5 & 1.8x/1.3x/1.8x\\
        \midrule
        \multirow{2}{*}{DSARP (\symsref{mechanism})} & Avg. latency of read requests &
        \multirow{2}{*}{199/200/202} &
        \multirow{2}{*}{1.2x/1.3x/1.5x}\\
              & for 8/16/32Gb DRAM chips&  & \\

        \bottomrule
    \end{tabular}
  \caption{Summary of latency improvements due to our proposed mechanisms.}
  \label{tab:summary}
\end{table}

The three mechanisms built on top of LISA reduce various latency components.
First, Rapid Inter-Subarray Copy (\emph{\lisarcnol}) significantly reduces the
bulk copy latency between subarrays by 9.2x. Second, Variable Latency
(\emph{\lisavillanol}) DRAM reduces the access latency (i.e., \trcd, \tras, and
\trp) of frequently-accessed data by caching it in fast subarrays with shorter
bitlines. Third, LIP reduces the precharge latency of every subarray by 2.6x.
LIP connects two precharge units of adjacent subarrays together using \lisa to
accelerate the precharge operation. In total, the three LISA mechanisms together
incur a small DRAM chip area overhead of 2.4\%.

Flexible-Latency (FLY) DRAM reduces the three major timing parameters by
exploiting our experimental observation on latency variation within commodity
DDR3 DRAM chips. The key idea of \mech is to determine the shortest reliable
access latency of each DRAM region, and to use the memory controller to apply
that latency to the corresponding DRAM region at runtime. Overall, FLY-DRAM
reduces the latency of \trcd/\tras/\trp by 1.8x/1.3x/1.8x for accesses to those
DRAM regions without slow cells. FLY-DRAM does not require any modification to
the DRAM chips since it leverages the innate latency behavior that varies across
DRAM cells within the same DRAM chip.

To address the refresh-induced latency, DSARP mitigates refresh latency by
parallelizing refresh operations with memory accesses within the DRAM chip. As a
result, DSARP reduces the average latency of read requests across 100 different
8-core workloads by 1.2x/1.3x/1.5x for 8/16/32Gb DRAM chips. DSARP incurs a low
DRAM chip area overhead of 0.7\%.

We conclude that our dissertation enables significant latency improvements at
very low cost in high-density DRAM chips by augmenting DRAM architecture with
simple and low-cost features, and developing a better understanding of
manufactured DRAM chips.

}

\section{Future Research Directions}

This dissertation opens up several avenues of future research directions. In
this section, we describe several directions that can tackle other problems
related to memory systems based on the ideas and approaches proposed in this
dissertation.

\subsection{Enabling LISA to Perform 1-to-N Memory Copy or Move Operations}

A typical \texttt{memcpy} or \texttt{memmove} call only allows the data to be
copied from one source location to one destination location. To copy or move
data from one source location to multiple different destinations, repeated calls
are required. The problem is that such repeated calls incur long latency and
high bandwidth consumption. In \chapref{lisa}, we propose to use the LISA
substrate to accelerate \texttt{memcpy} and \texttt{memmove} in DRAM without the
intervention of CPU. One potential application that can be enabled by LISA is
performing \texttt{memcpy} or \texttt{memmove} from one source location to
\emph{multiple destinations} completely in DRAM without requiring multiple calls
of these operations.

By using LISA, we observe that moving data from the source subarray to the
destination subarray latches the source row's data in all the intermediate
subarrays' row buffer. As a result, activating these intermediate subarrays
would copy their row buffers' data into the specified row. By extending LISA
to perform multi-point (1-to-N) copy or move operations, we can significantly
increase system performance of several commonly-used system operations. For
example, forking multiple child processes can utilize 1-to-N copy operations to
copy those memory regions that are likely to be modified by the children.

\subsection{In-Memory Computation with LISA}

One important requirement of efficient in-memory computation is being able to
move data from its stored location to the computation units with very low
latency and energy. In \chapref{lisa}, we discussed the benefits of using LISA
to extend the data range of in-memory bitwise operations. We believe using the
LISA substrate can enable a new in-memory computation framework. The idea is to
add a small computation unit inside each or a subset of banks, and connect these
computation units to the neighboring subarrays which store the data. Doing so
allows the system to utilize LISA to move bulk data from the subarrays to the
computation under low latency with low area overhead.

Two potential types of computation units to add are bitwise shifters and
ripple-carry adders since simple integer addition and bitwise shifting between
two arrays of data are common operations in many applications. One key challenge
of adding computation units would be fitting each single unit that processes a
single bit within a pitch of DRAM array's column. For example, a single-bit
shifter requires 12 transistors which is much bigger than a sense amplifier (4
transistors). This implementation overhead can restrict the computation to
process data at the granularity of a row size. Nonetheless, this general
in-memory computation framework still has the potential to enable simple
filtering operations in memory to provide high system performance or energy
efficiency at low cost.

\subsection{Extending LISA to Non-Volatile Memory}

In this dissertation, we only focus on the DRAM technology. A class of emerging
memory technology is non-volatile memory (NVM), which has the capability of
retaining data without power supply. We believe that the LISA substrate can be
extended to NVM (e.g., STT-RAM) since the memory organization of NVM mostly
resembles that of DRAM. A potential application of LISA in NVM is an efficient file
copy operation that does not incur costly I/O data transfer.

\subsection{Data Prefetching with Variable Latency (VILLA) DRAM}

Data prefetching utilizes unused memory bandwidth to speculatively transfer
data from memory to caches. However, if memory bandwidth is heavily
utilized, prefetch requests can degrade system performance by interfering with
demand requests. Therefore, a prefetching scheme that does not exert
pressure on memory channels can potentially attain higher system performance.

In \secref{villa}, we described a new heterogeneous DRAM design, called
\emph{Variable Latency (VILLA) DRAM}, which introduces fast subarrays in each
DRAM bank. VILLA utilizes the LISA substrate to efficiently transfer row-size
data (8KB) from a slow subarray to a fast subarray without using the memory
channel. We believe a new prefetching scheme can be designed with the VILLA
cache by prefetching a whole row of data before demand requests occur. The
primary benefit is that prefetching to VILLA cache does not cause bandwidth
contention. Also, VILLA can increase the prefetch coverage since the prefetching
granularity is large, with hundreds of cache lines.

\subsection{Reducing Activation Latency with Error Detection Codes}

In \chapref{latvar}, we observed that activation errors (due to reduced
activation latency) are permanent by propagating back into the first accessed
column of data. If the errors were \emph{transient}, a new mechanism could be
devised to read data with aggressively-reduced activation latency and re-read
data when activation errors occur. Activation errors can be detected using error
detection codes. Therefore, this raises a key question: can we modify the DRAM
sensing circuit to make activation errors transient? Answering this question
requires a thorough understanding of the modern DRAM circuit to find out how
activation errors propagate back into DRAM cells.

\subsection{Characterizing Latency Behavior of DRAM Cells Over Time}

In \chapref{latvar}, we experimentally demonstrated that individual cells within
the same DRAM chip exhibit different latency behavior. However, we do not
examine the latency behavior of each cell over a controlled period of time,
except for the fact that we perform the tests for multiple rounds per DIMM. The
latency of a cell could potentially change over time, within a short period of
time (e.g., similar effect as Variable Retention Time) or long period of time
(e.g., aging and wearout). Therefore, a future direction is to experimentally
study the latency behavior of DRAM cells over time.

\subsection{Avoiding Worst-Case Data Patterns for Higher Reliability}

Our experimental characterization in \secref{act_lat_analysis} showed that
errors caused by reduced activation latency are dependent on the stored data
pattern. Reading bit 1 is significantly more reliable than bit 0 at reduced
activation latencies. To improve reliability of future DRAM, a future research
direction is to design new encoding scheme that will (1) increase the number of
bit 1 and (2) store the encoding metadata at low cost.

\ignore{
\subsection{LISA for SRAM cache resizing}
migrate dirty and frequently used clean cache line

http://ieeexplore.ieee.org/abstract/document/5763191/
http://dl.acm.org/citation.cfm?id=2695763
EnCache: Improving Cache Energy Efficiency
Using a Software-Controlled Profiling Cache
http://ieeexplore.ieee.org/document/6404160/

in-dram fast scrubbing with lisa
add soft-error correcting circuits at the end of each bank
: An On-Chip ECC Circuit for Correcting Soft Errors in DRAM’s with Trench Capacitors
mazumder-jssc1992
}

%

\section{Final Concluding Remarks}

In this dissertation, we highlighted problems that cause or affect long DRAM
latency and presented extensive experimental characterization on studying DRAM
latency behavior in commodity DRAM chips. Overall, we presented four new
techniques: 1) LISA, which is a versatile DRAM substrate that provides fast data
movement between subarrays to enable several low-latency mechanisms, 2) DSARP,
which overlaps accesses with refreshes to reduce refresh-induced latency, 3)
FLY-DRAM, which exploits our experimental characterization on latency variation
within a chip to reduce latency to access regions with faster DRAM cells, and 4)
Voltron, which exploits our experimental characterization on the critical
relationship between access latency and supply voltage to improve energy
efficiency. We conclude and hope the proposed low-latency architectural
mechanisms and the detailed experimental characterization on commodity DRAM
chips in this dissertation will pave the way for new research that can develop
new mechanisms to improve system performance, energy efficiency, or reliability
of future memory systems.

\chapter*{Other Works of the Author}

Throughout the course of my Ph.D. study, I have worked on several different
topics with many fellow graduate students from CMU and collaborators from other
institutions. In this chapter, I would like to acknowledge these works.

In the early years of my Ph.D., I worked on a number of projects on
\emph{networks-on-chip (NoCs)}. In collaboration with Rachata Ausavarungnirun,
Chris Fallin, and others, we have contributed to a new congestion control
algorithm (HAT~\cite{chang-sbacpad2012}), a new router architecture
(MinBD~\cite{fallin-nocs2012}), and a new hierarchical ring design
(HiRD~\cite{ausavarungnirun-sbacpad2014}). We show that these new techniques
can significantly improve the energy efficiency of NoCs.

Another topic that I have developed an interest and worked on was \emph{memory
scheduling policy} for heterogeneous processors that consist of conventional CPU
cores and other types of accelerators. In collaboration with Rachata
Ausavarungnirun and Lavanya Subramanian, we have developed a new memory
scheduler, SMS~\cite{ausavarungnirun-isca2012}, that improves system performance
and fairness of a CPU-GPU processor by reducing the application interference
between CPU and GPU. I have also contributed to a memory scheduler that targets
another type of heterogeneous processor that consists of conventional CPU cores
and hardware accelerators for image processing and recognition. In collaboration
with Hiroyuki Usui and Lavanya Subramanian, we have developed a memory
scheduler, DASH~\cite{usui-taco2016}, that enables the accelerators to meet
their deadlines while attaining high system performance.

In collaboration with Hasan Hassan, I have worked on developing a DRAM-testing
infrastructure, SoftMC~\cite{hassan-hpca2017}, that has facilitated my research
on DRAM characterization and other works. In collaboration with Donghyuk Lee, I
have contributed to another low DRAM latency architecture,
AL-DRAM~\cite{lee-hpca2015}, that adaptively adjusts latency of DRAM based on
the ambient temperature.

Finally, we have released the simulators used for these different works on GitHub.
The simulators that I contributed to are as follows: (1) \texttt{NoCulator} for
NoCs evaluation, (2) \texttt{Ramulator} (in C and C\#) for memory projects, and
(3) \texttt{SoftMC}, which is an FPGA-based memory controller design, for DRAM
characterization. The source code is available on GitHub at
\url{https://github.com/CMU-SAFARI}.

\small
\singlespacing
\bibliography{merge2}

\begin{thebibliography}{100}

\bibitem{ram-github}
Ramulator.
\newblock \url{https://github.com/CMU-SAFARI/ramulator}, 2015.

\bibitem{volt-github}
{DRAM Voltage Study}.
\newblock \url{https://github.com/CMU-SAFARI/DRAM-Voltage-Study}, 2017.

\bibitem{agarwal-asplos2015}
N.~Agarwal, D.~Nellans, M.~Stephenson, M.~O'Connor, and S.~W. Keckler.
\newblock {Page Placement Strategies for GPUs Within Heterogeneous Memory
  Systems}.
\newblock In {\em ASPLOS}, 2015.

\bibitem{agrawal-hpca2014}
A.~Agrawal, A.~Ansari, and J.~Torrellas.
\newblock {Mosaic: Exploiting the spatial locality of process variation to
  reduce refresh energy in on-chip eDRAM modules}.
\newblock In {\em HPCA}, 2014.

\bibitem{agrawal-hpca2013}
A.~Agrawal et~al.
\newblock {Refrint: Intelligent Refresh to Minimize Power in On-Chip
  Multiprocessor Cache Hierarchies}.
\newblock In {\em {HPCA}}, 2013.

\bibitem{agrawal-memsys2016}
A.~Agrawal, M.~O'Connor, E.~Bolotin, N.~Chatterjee, J.~Emer, and S.~Keckler.
\newblock {CLARA: Circular Linked-List Auto and Self Refresh Architecture}.
\newblock In {\em MEMSYS}, 2016.

\bibitem{ahn-isca2015}
J.~Ahn, S.~Hong, S.~Yoo, O.~Mutlu, and K.~Choi.
\newblock A scalable processing-in-memory accelerator for parallel graph
  processing.
\newblock In {\em ISCA}, 2015.

\bibitem{ahn-isca2015-2}
J.~Ahn, S.~Yoo, O.~Mutlu, and K.~Choi.
\newblock Pim-enabled instructions: A low-overhead, locality-aware
  processing-in-memory architecture.
\newblock In {\em ISCA}, 2015.

\bibitem{ahn-taco2012}
J.~H. Ahn, N.~P. Jouppi, C.~Kozyrakis, J.~Leverich, and R.~S. Schreiber.
\newblock {Improving System Energy Efficiency with Memory Rank Subsetting}.
\newblock {\em TACO}, 9(1):4:1--4:28, 2012.

\bibitem{ahn-cal2009}
J.~H. Ahn, J.~Leverich, R.~Schreiber, and N.~P. Jouppi.
\newblock {Multicore DIMM: an Energy Efficient Memory Module with Independently
  Controlled DRAMs}.
\newblock {\em CAL}, 2009.

\bibitem{ailamaki-vldb1999}
A.~Ailamaki, D.~J. DeWitt, M.~D. Hill, and D.~A. Wood.
\newblock {DBMSs on a Modern Processor: Where Does Time Go?}
\newblock In {\em VLDB}, 1999.

\bibitem{akin-isca2015}
B.~Akin, F.~Franchetti, and J.~C. Hoe.
\newblock {Data Reorganization in Memory Using 3D-stacked DRAM}.
\newblock In {\em ISCA}, 2015.

\bibitem{al-ars-ddecs2007}
Z.~Al-Ars, S.~Hamdioui, and G.~Gaydadjiev.
\newblock {Manifestation of Precharge Faults in High Speed DRAM Devices}.
\newblock 2007.

\bibitem{al-ars-vts2004}
Z.~Al-Ars, S.~Hamdioui, and A.~J. van~de Goor.
\newblock {Effects of bit line coupling on the faulty behavior of DRAMs}.
\newblock 2004.

\bibitem{alameldeen-hpca2007}
A.~R. Alameldeen and D.~A. Wood.
\newblock {Interactions Between Compression and Prefetching in Chip
  Multiprocessors}.
\newblock In {\em HPCA}, 2007.

\bibitem{appenzeller-sigcomm2004}
G.~Appenzeller, I.~Keslassy, and N.~McKeown.
\newblock {Sizing Router Buffers}.
\newblock In {\em SIGCOMM}, 2004.

\bibitem{arma9}
{ARM Ltd.}
\newblock {Cortex-A9 Processor}.
\newblock \url{https://www.arm.com/products/processors/cortex-a/cortex-a9.php}.

\bibitem{ausavarungnirun-isca2012}
R.~Ausavarungnirun, K.~K.-W. Chang, L.~Subramanian, G.~H. Loh, and O.~Mutlu.
\newblock Staged memory scheduling: Achieving high performance and scalability
  in heterogeneous systems.
\newblock In {\em ISCA}, 2012.

\bibitem{ausavarungnirun-sbacpad2014}
R.~Ausavarungnirun, C.~Fallin, X.~Yu, K.~K.~W. Chang, G.~Nazario, R.~Das, G.~H.
  Loh, and O.~Mutlu.
\newblock Design and evaluation of hierarchical rings with deflection routing.
\newblock In {\em SBAC-PAD}, 2014.

\bibitem{medic}
R.~Ausavarungnirun, S.~Ghose, O.~Kayıran, G.~H. Loh, C.~R. Das, M.~T.
  Kandemir, and O.~Mutlu.
\newblock {Exploiting Inter-Warp Heterogeneity to Improve GPGPU Performance}.
\newblock In {\em PACT}, 2015.

\bibitem{autran-tns2009}
J.~L. Autran, P.~Roche, S.~Sauze, G.~Gasiot, D.~Munteanu, P.~Loaiza,
  M.~Zampaolo, and J.~Borel.
\newblock {Altitude and Underground Real-Time SER Characterization of CMOS 65
  nm SRAM}.
\newblock {\em IEEE TNS}, 56(4):2258--2266, 2009.

\bibitem{awan-bdcloud2015}
A.~J. Awan, M.~Brorsson, V.~Vlassov, and E.~Ayguade.
\newblock {Performance Characterization of In-Memory Data Analytics on a Modern
  Cloud Server}.
\newblock In {\em BDCloud}, 2015.

\bibitem{awan-bdcloud2016}
A.~J. Awan, M.~Brorsson, V.~Vlassov, and E.~Ayguade.
\newblock {Micro-Architectural Characterization of Apache Spark on Batch and
  Stream Processing Workloads}.
\newblock In {\em BDCloud}, 2016.

\bibitem{babarinsa-2015}
O.~O. Babarinsa and S.~Idreos.
\newblock Jafar: Near-data processing for databases.
\newblock In {\em SIGMOD}, 2015.

\bibitem{baek-tc2014}
S.~Baek, S.~Cho, and R.~Melhem.
\newblock Refresh now and then.
\newblock {\em IEEE TC}, 63(12):3114--3126, 2014.

\bibitem{baer-1995}
J.-L. Baer and T.-F. Chen.
\newblock {Effective Hardware-Based Data Prefetching for High-Performance
  Processors}.
\newblock {\em IEEE TC}, 44(5):609--623, 1995.

\bibitem{bairavasundaram2008analysis}
L.~N. Bairavasundaram, A.~C. Arpaci-Dusseau, R.~H. Arpaci-Dusseau, G.~R.
  Goodson, and B.~Schroeder.
\newblock An analysis of data corruption in the storage stack.
\newblock {\em TOS}, 4(3):8, 2008.

\bibitem{bairavasundaram2007analysis}
L.~N. Bairavasundaram, G.~R. Goodson, S.~Pasupathy, and J.~Schindler.
\newblock {An analysis of latent sector errors in disk drives}.
\newblock In {\em SIGMETRICS}, 2007.

\bibitem{baker-dram}
R.~J. Baker.
\newblock {\em {CMOS Circuit Design, Layout, and Simulation}}.
\newblock Wiley-IEEE Press, 2010.

\bibitem{bauer-mckinsey2016}
H.~Bauer, S.~Burghardt, S.~Tandon, and F.~Thalmayr.
\newblock {{Memory: Are challenges ahead?}}, March 2016.

\bibitem{begum-iiswc2015}
R.~Begum, D.~Werner, M.~Hempstead, G.~Prasad, and G.~Challen.
\newblock {Energy-Performance Trade-offs on Energy-Constrained Devices with
  Multi-component DVFS}.
\newblock In {\em IISWC}, 2015.

\bibitem{bhati-ispled2013}
I.~Bhati, Z.~Chishti, and B.~Jacob.
\newblock {Coordinated refresh: Energy efficient techniques for DRAM refresh
  scheduling}.
\newblock In {\em ISLPED}, 2013.

\bibitem{bhati-isca2015}
I.~Bhati, Z.~Chishti, S.-L. Lu, and B.~Jacob.
\newblock Flexible auto-refresh: Enabling scalable and energy-efficient dram
  refresh reductions.
\newblock In {\em ISCA}, 2015.

\bibitem{bhattacharjee-isca2009}
A.~Bhattacharjee and M.~Martonosi.
\newblock {Thread Criticality Predictors for Dynamic Performance, Power, and
  Resource Management in Chip Multiprocessors}.
\newblock In {\em ISCA}, 2009.

\bibitem{blagodurov-usenix2011}
S.~Blagodurov, S.~Zhuralev, M.~Dashti, and A.~Fedorova.
\newblock {A Case for NUMA-Aware Contention Management on Multicore Systems}.
\newblock In {\em USENIX ATC}, 2011.

\bibitem{bloom-cacm70}
B.~H. Bloom.
\newblock {Space/Time Tradeoffs in Hash Coding with Allowable Errors}.
\newblock {\em CACM}, July 1970.

\bibitem{boncz-1999}
P.~A. Boncz, S.~Manegold, and M.~L. Kersten.
\newblock {Database Architecture Optimized for the New Bottleneck: Memory
  Access}.
\newblock In {\em VLDB}, 1999.

\bibitem{amirali-cal2016}
A.~Boroumand, S.~Ghose, B.~Lucia, K.~Hsieh, K.~Malladi, H.~Zheng, and O.~Mutlu.
\newblock Lazypim: An efficient cache coherence mechanism for
  processing-in-memory.
\newblock {\em CAL}, 2016.

\bibitem{spectre}
{Cadence Design Systems, Inc.}
\newblock {Spectre Circuit Simulator}.
\newblock
  \url{http://www.cadence.com/products/rf/spectre_circuit/pages/default.aspx}.

\bibitem{cai-ieee2017}
Y.~Cai, S.~Ghose, E.~F. Haratsch, Y.~Luo, and O.~Mutlu.
\newblock Error characterization, mitigation, and recovery in
  flash-memory-based solid-state drives.
\newblock {\em Proceedings of the IEEE}, 105(9):1666--1704, 2017.

\bibitem{cai.hpca17}
Y.~Cai, S.~Ghose, Y.~Luo, K.~Mai, O.~Mutlu, and E.~F. Haratsch.
\newblock {Vulnerabilities in MLC NAND Flash Memory Programming: Experimental
  Analysis, Exploits, and Mitigation Techniques}.
\newblock In {\em HPCA}, 2017.

\bibitem{cai-fccm2011}
Y.~Cai, E.~F. Haratsch, M.~McCartney, and K.~Mai.
\newblock {FPGA-Based Solid-State Drive Prototyping Platform}.
\newblock In {\em FCCM}, 2011.

\bibitem{cai.date12}
Y.~Cai, E.~F. Haratsch, O.~Mutlu, and K.~Mai.
\newblock {Error Patterns in MLC NAND Flash Memory: Measurement,
  Characterization, and Analysis}.
\newblock In {\em DATE}, 2012.

\bibitem{cai.date13}
Y.~Cai, E.~F. Haratsch, O.~Mutlu, and K.~Mai.
\newblock {Threshold Voltage Distribution in MLC NAND Flash Memory:
  Characterization, Analysis, and Modeling}.
\newblock In {\em DATE}, 2013.

\bibitem{cai.dsn15}
Y.~Cai, Y.~Luo, S.~Ghose, E.~F. Haratsch, K.~Mai, and O.~Mutlu.
\newblock {Read Disturb Errors in MLC NAND Flash Memory: Characterization and
  Mitigation}.
\newblock In {\em DSN}, 2015.

\bibitem{cai.hpca15}
Y.~Cai, Y.~Luo, E.~F. Haratsch, K.~Mai, and O.~Mutlu.
\newblock {Data Retention in MLC NAND Flash Memory: Characterization,
  Optimization, and Recovery}.
\newblock In {\em HPCA}, 2015.

\bibitem{cai.iccd13}
Y.~Cai, O.~Mutlu, E.~F. Haratsch, and K.~Mai.
\newblock {Program Interference in MLC NAND Flash Memory: Characterization,
  Modeling, and Mitigation}.
\newblock In {\em ICCD}, 2013.

\bibitem{cai.iccd12}
Y.~Cai, G.~Yalcin, O.~Mutlu, E.~F. Haratsch, A.~Cristal, O.~Unsal, and K.~Mai.
\newblock {Flash Correct and Refresh: Retention Aware Management for Increased
  Lifetime}.
\newblock In {\em ICCD}, 2012.

\bibitem{cai-itj2013}
Y.~Cai, G.~Yalcin, O.~Mutlu, E.~F. Haratsch, A.~Cristal, O.~Unsal, and K.~Mai.
\newblock {Error Analysis and Retention-Aware Error Management for NAND Flash
  Memory}.
\newblock In {\em ITJ}, 2013.

\bibitem{cai.sigmetrics14}
Y.~Cai, G.~Yalcin, O.~Mutlu, E.~F. Haratsch, O.~Unsal, A.~Cristal, and K.~Mai.
\newblock Neighbor-cell assisted error correction for mlc nand flash memories.
\newblock In {\em SIGMETRICS}, 2014.

\bibitem{cao-sigmetrics1995}
P.~Cao, E.~W. Felten, A.~R. Karlin, and K.~Li.
\newblock {A Study of Integrated Prefetching and Caching Strategies}.
\newblock In {\em SIGMETRICS}, 1995.

\bibitem{carter-hpca1999}
J.~Carter, W.~Hsieh, L.~Stoller, M.~Swanson, L.~Zhang, E.~Brunvand, A.~Davis,
  C.-C. Kuo, R.~Kuramkote, M.~Parker, L.~Schaelicke, and T.~Tateyama.
\newblock {Impulse: building a smarter memory controller}.
\newblock In {\em HPCA}, 1999.

\bibitem{kanad_dram_book}
K.~Chakraborty and P.~Mazumder.
\newblock {\em {Fault-Tolerance and Reliability Techniques for High-Density
  Random-Access Memories}}.
\newblock Prentice Hall, 2002.

\bibitem{chandra-asplos1994}
R.~Chandra, S.~Devine, B.~Verghese, A.~Gupta, and M.~Rosenblum.
\newblock {Scheduling and Page Migration for Multiprocessor Compute Servers}.
\newblock In {\em ASPLOS}, 1994.

\bibitem{chandrasekar-date2014}
K.~Chandrasekar, S.~Goossens, C.~Weis, M.~Koedam, B.~Akesson, N.~Wehn, and
  K.~Goossens.
\newblock {Exploiting Expendable Process-Margins in DRAMs for Run-Time
  Performance Optimization}.
\newblock In {\em DATE}, 2014.

\bibitem{drampower}
K.~Chandrasekar, C.~Weis, Y.~Li, S.~Goossens, M.~Jung, O.~Naji, B.~Akesson,
  N.~Wehn, and K.~Goossens.
\newblock {DRAMPower: Open-source DRAM Power \& Energy Estimation Tool}.
\newblock \url{http://www.drampower.info}.

\bibitem{chang-sigmetrics2016}
K.~K. Chang, A.~Kashyap, H.~Hassan, S.~Ghose, K.~Hsieh, D.~Lee, T.~Li,
  G.~Pekhimenko, S.~Khan, and O.~Mutlu.
\newblock {Understanding Latency Variation in Modern DRAM Chips: Experimental
  Characterization, Analysis, and Optimization}.
\newblock In {\em SIGMETRICS}, 2016.

\bibitem{chang-hpca2014}
K.~K. Chang, D.~Lee, Z.~Chishti, A.~Alameldeen, C.~Wilkerson, Y.~Kim, and
  O.~Mutlu.
\newblock {Improving DRAM Performance by Parallelizing Refreshes with
  Accesses}.
\newblock In {\em HPCA}, 2014.

\bibitem{chang-tr2016}
K.~K. Chang, P.~J. Nair, D.~Lee, S.~Ghose, M.~K. Qureshi, and O.~Mutlu.
\newblock {Low-Cost Inter-Linked Subarrays (LISA): A New DRAM Substrate with
  Higher Connectivity}.
\newblock Technical report, Carnegie Mellon Univ., SAFARI Research Group, 2016.

\bibitem{chang-hpca2016}
K.~K. Chang, P.~J. Nair, D.~Lee, S.~Ghose, M.~K. Qureshi, and O.~Mutlu.
\newblock {Low-Cost Inter-Linked Subarrays (LISA): Enabling Fast Inter-Subarray
  Data Movement in DRAM}.
\newblock In {\em HPCA}, 2016.

\bibitem{chang-sbacpad2012}
K.~K.-W. Chang, R.~Ausavarungnirun, C.~Fallin, and O.~Mutlu.
\newblock {HAT: Heterogeneous Adaptive Throttling for On-Chip Networks}.
\newblock In {\em SBAC-PAD}, 2012.

\bibitem{chatterjee-hpca2012}
N.~Chatterjee, N.~Muralimanohar, R.~Balasubramonian, A.~Davis, and N.~P.
  Jouppi.
\newblock Staged reads: Mitigating the impact of {DRAM} writes on {DRAM} reads.
\newblock In {\em HPCA}, 2012.

\bibitem{chatterjee-micro2012}
N.~Chatterjee, M.~Shevgoor, R.~Balasubramonian, A.~Davis, Z.~Fang, R.~Illikkal,
  and R.~Iyer.
\newblock Leveraging heterogeneity in dram main memories to accelerate critical
  word access.
\newblock In {\em MICRO}, 2012.

\bibitem{clapp-2015}
R.~Clapp, M.~Dimitrov, K.~Kumar, V.~Viswanathan, and T.~Willhalm.
\newblock Quantifying the performance impact of memory latency and bandwidth
  for big data workloads.
\newblock In {\em IISWC}, 2015.

\bibitem{safari-github}
{CMU SAFARI Research Group}.
\newblock \url{https://github.com/CMU-SAFARI}.

\bibitem{cooksey-asplos2002}
R.~Cooksey, S.~Jourdan, and D.~Grunwald.
\newblock {A Stateless, Content-directed Data Prefetching Mechanism}.
\newblock In {\em ASPLOS}, 2002.

\bibitem{cooper-socc2010}
B.~F. Cooper, A.~Silberstein, E.~Tam, R.~Ramakrishnan, and R.~Sears.
\newblock {Benchmarking Cloud Serving Systems with YCSB}.
\newblock In {\em SOCC}, 2010.

\bibitem{dahlgren-1995}
F.~Dahlgren, M.~Dubois, and P.~Stenstr\"{o}m.
\newblock Sequential hardware prefetching in shared-memory multiprocessors.
\newblock {\em IEEE TPDS}, 6(7):733--746, 1995.

\bibitem{das-hpca2013}
R.~Das, R.~Ausavarungnirun, O.~Mutlu, A.~Kumar, and M.~Azimi.
\newblock {Application-to-core mapping policies to reduce memory system
  interference in multi-core systems}.
\newblock In {\em HPCA}, 2013.

\bibitem{das-micro2009}
R.~Das, O.~Mutlu, T.~Moscibroda, and C.~Das.
\newblock Application-aware prioritization mechanisms for on-chip networks.
\newblock In {\em MICRO}, 2009.

\bibitem{das-isca2010}
R.~Das, O.~Mutlu, T.~Moscibroda, and C.~R. Das.
\newblock {A{\'e}Rgia: Exploiting Packet Latency Slack in On-chip Networks}.
\newblock In {\em ISCA}, 2010.

\bibitem{david-icac2011}
H.~David, C.~Fallin, E.~Gorbatov, U.~R. Hanebutte, and O.~Mutlu.
\newblock {Memory Power Management via Dynamic Voltage/Frequency Scaling}.
\newblock In {\em ICAC}, 2011.

\bibitem{deng-micro2012}
Q.~Deng, D.~Meisner, A.~Bhattacharjee, T.~F. Wenisch, and R.~Bianchini.
\newblock {CoScale: Coordinating CPU and Memory System DVFS in Server Systems}.
\newblock In {\em MICRO}, 2012.

\bibitem{deng-islped2012}
Q.~Deng, D.~Meisner, A.~Bhattacharjee, T.~F. Wenisch, and R.~Bianchini.
\newblock {MultiScale: Memory System DVFS with Multiple Memory Controllers}.
\newblock In {\em ISLPED}, 2012.

\bibitem{deng-asplos2011}
Q.~Deng, D.~Meisner, L.~Ramos, T.~F. Wenisch, and R.~Bianchini.
\newblock {MemScale: Active Low-power Modes for Main Memory}.
\newblock In {\em ASPLOS}, 2011.

\bibitem{dennard-scaling}
R.~H. Dennard, F.~H. Gaensslen, V.~L. Rideout, E.~Bassous, and A.~R. LeBlanc.
\newblock {Design of ion-implanted MOSFET's with very small physical
  dimensions}.
\newblock {\em IEEE JSSC}, 9(5):256--268, 1974.

\bibitem{digikey}
{Digi-Key}.
\newblock \url{http://www.digikey.com}.

\bibitem{draper-ics2002}
J.~Draper, J.~Chame, M.~Hall, C.~Steele, T.~Barrett, J.~LaCoss, J.~Granacki,
  J.~Shin, C.~Chen, C.~W. Kang, I.~Kim, and G.~Daglikoca.
\newblock {The Architecture of the DIVA Processing-in-memory Chip}.
\newblock In {\em ICS}, 2002.

\bibitem{dubois-isca2013}
K.~Du~Bois, S.~Eyerman, J.~B. Sartor, and L.~Eeckhout.
\newblock {Criticality Stacks: Identifying Critical Threads in Parallel
  Programs Using Synchronization Behavior}.
\newblock In {\em ISCA}, 2013.

\bibitem{ebrahimi-asplos2010}
E.~Ebrahimi, C.~J. Lee, O.~Mutlu, and Y.~N. Patt.
\newblock {Fairness via Source Throttling: A Configurable and High-performance
  Fairness Substrate for Multi-core Memory Systems}.
\newblock In {\em ASPLOS}, 2010.

\bibitem{ebrahimi-isca2011}
E.~Ebrahimi, C.~J. Lee, O.~Mutlu, and Y.~N. Patt.
\newblock Prefetch-aware shared resource management for multi-core systems.
\newblock In {\em ISCA}, 2011.

\bibitem{ebrahimi-micro2011}
E.~Ebrahimi, R.~Miftakhutdinov, C.~Fallin, C.~J. Lee, J.~A. Joao, O.~Mutlu, and
  Y.~N. Patt.
\newblock Parallel application memory scheduling.
\newblock In {\em MICRO}, 2011.

\bibitem{1669154}
E.~Ebrahimi, O.~Mutlu, C.~J. Lee, and Y.~N. Patt.
\newblock {Coordinated control of multiple prefetchers in multi-core systems}.
\newblock In {\em MICRO}, 2009.

\bibitem{ebrahimi-hpca2009}
E.~Ebrahimi, O.~Mutlu, and Y.~N. Patt.
\newblock Techniques for bandwidth-efficient prefetching of linked data
  structures in hybrid prefetching systems.
\newblock In {\em HPCA}, 2009.

\bibitem{el-sayed-sigmetrics2012}
N.~El-Sayed, I.~A. Stefanovici, G.~Amvrosiadis, A.~A. Hwang, and B.~Schroeder.
\newblock {Temperature Management in Data Centers: Why Some (Might) Like It
  Hot}.
\newblock In {\em SIGMETRICS}, 2012.

\bibitem{eyerman-ieeemicro2008}
S.~Eyerman and L.~Eeckhout.
\newblock {{System-Level Performance Metrics for Multiprogram Workloads}}.
\newblock {\em IEEE Micro}, 2008.

\bibitem{fallin-hpca2011}
C.~Fallin, C.~Craik, and O.~Mutlu.
\newblock {CHIPPER: A Low-complexity Bufferless Deflection Router}.
\newblock In {\em HPCA}, 2011.

\bibitem{fallin-nocs2012}
C.~Fallin, G.~Nazario, X.~Yu, K.~Chang, R.~Ausavarungnirun, and O.~Mutlu.
\newblock {MinBD: Minimally-Buffered Deflection Routing for Energy-Efficient
  Interconnect}.
\newblock In {\em NOCS}, 2012.

\bibitem{7056040}
A.~Farmahini-Farahani, J.~H. Ahn, K.~Morrow, and N.~S. Kim.
\newblock Nda: Near-dram acceleration architecture leveraging commodity dram
  devices and standard memory modules.
\newblock In {\em HPCA}, 2015.

\bibitem{fraguela-2003}
B.~B. Fraguela, J.~Renau, P.~Feautrier, D.~Padua, and J.~Torrellas.
\newblock {Programming the FlexRAM Parallel Intelligent Memory System}.
\newblock In {\em PPoPP}, 2003.

\bibitem{aya-2017}
A.~Fukami, S.~Ghose, Y.~Luo, Y.~Cai, and O.~Mutlu.
\newblock Improving the reliability of chip-off forensic analysis of nand flash
  memory devices.
\newblock {\em DFRWS EU}, 20(S):S1--S11, 2017.

\bibitem{7429299}
M.~Gao, G.~Ayers, and C.~Kozyrakis.
\newblock Practical near-data processing for in-memory analytics frameworks.
\newblock In {\em PACT}, 2015.

\bibitem{7446059}
M.~Gao and C.~Kozyrakis.
\newblock {HRL: Efficient and flexible reconfigurable logic for near-data
  processing}.
\newblock In {\em HPCA}, 2016.

\bibitem{ghose-isca2013}
S.~Ghose, H.~Lee, and J.~F. Mart{\'\i}nez.
\newblock {Improving Memory Scheduling via Processor-Side Load Criticality
  Information}.
\newblock In {\em ISCA}, 2013.

\bibitem{glew-asploswac98}
A.~Glew.
\newblock {MLP Yes! ILP No! Memory Level Parallelism, or, Why I No Longer Worry
  About IPC}.
\newblock In {\em Proc.~of the ASPLOS Wild and Crazy Ideas Session}, San Jose,
  CA, October 1997.

\bibitem{375174}
M.~Gokhale, B.~Holmes, and K.~Iobst.
\newblock {Processing in memory: the Terasys massively parallel PIM array}.
\newblock {\em Computer}, 28(4):23--31, 1995.

\bibitem{chromebook}
Google.
\newblock {Chromebook}.
\newblock \url{https://www.google.com/chromebook/}.

\bibitem{grot-hpca2009}
B.~Grot, J.~Hestness, S.~W. Keckler, and O.~Mutlu.
\newblock {Express Cube Topologies for on-Chip Interconnects}.
\newblock In {\em HPCA}, 2009.

\bibitem{grot-isca2011}
B.~Grot, J.~Hestness, S.~W. Keckler, and O.~Mutlu.
\newblock {Kilo-NOC: A Heterogeneous Network-on-chip Architecture for
  Scalability and Service Guarantees}.
\newblock In {\em ISCA}, 2011.

\bibitem{grot-micro2009}
B.~Grot, S.~W. Keckler, and O.~Mutlu.
\newblock {Preemptive Virtual Clock: A flexible, efficient, and cost-effective
  QOS scheme for networks-on-chip}.
\newblock In {\em MICRO}, 2009.

\bibitem{gschwind-cf2006}
M.~Gschwind.
\newblock {Chip Multiprocessing and the Cell Broadband Engine}.
\newblock In {\em CF}, 2006.

\bibitem{gummaraju-pact2007}
J.~Gummaraju, M.~Erez, J.~Coburn, M.~Rosenblum, and W.~J. Dally.
\newblock {Architectural Support for the Stream Execution Model on
  General-Purpose Processors}.
\newblock In {\em PACT}, 2007.

\bibitem{guo-wondp14}
Q.~Guo, N.~Alachiotis, B.~Akin, F.~Sadi, G.~Xu, T.-M. Low, L.~Pileggi, J.~C.
  Hoe, and F.~Franchetti.
\newblock {3D-Stacked Memory-Side Acceleration: Accelerator and System Design}.
\newblock In {\em WONDP}, 2014.

\bibitem{halper-hpca2016}
M.~Halpern, Y.~Zhu, and V.~J. Reddi.
\newblock {Mobile CPU's rise to power: Quantifying the impact of generational
  mobile CPU design trends on performance, energy, and user satisfaction}.
\newblock In {\em HPCA}, 2016.

\bibitem{hart-compcon1994}
C.~A. Hart.
\newblock {{CDRAM} in a Unified Memory Architecture}.
\newblock In {\em Intl.~Computer Conference}, 1994.

\bibitem{hashemi-isca2016}
M.~Hashemi, Khubaib, E.~Ebrahimi, O.~Mutlu, and Y.~N. Patt.
\newblock {Accelerating Dependent Cache Misses with an Enhanced Memory
  Controller}.
\newblock In {\em ISCA}, 2016.

\bibitem{hashemi-micro2016}
M.~Hashemi, O.~Mutlu, and Y.~N. Patt.
\newblock Continuous runahead: Transparent hardware acceleration for memory
  intensive workloads.
\newblock In {\em MICRO}, 2016.

\bibitem{hassan-hpca2016}
H.~Hassan et~al.
\newblock {ChargeCache: Reducing DRAM Latency by Exploiting Row Access
  Locality}.
\newblock In {\em HPCA}, 2016.

\bibitem{hassan-hpca2017}
H.~Hassan et~al.
\newblock {SoftMC: A Flexible and Practical Open-Source Infrastructure for
  Enabling Experimental DRAM Studies}.
\newblock In {\em HPCA}, 2017.

\bibitem{hermsmeyer-bell2009}
C.~Hermsmeyer, H.~Song, R.~Schlenk, R.~Gemelli, and S.~Bunse.
\newblock {Towards 100G Packet Processing: Challenges and Technologies}.
\newblock {\em Bell Lab. Tech. J.}, 14(2):57--79, 2009.

\bibitem{herrero-tc2013}
E.~Herrero, J.~Gonzalez, R.~Canal, and D.~Tullsen.
\newblock Thread row buffers: Improving memory performance isolation and
  throughput in multiprogrammed environments.
\newblock {\em IEEE TC}, 62(9):1879--1892, 2013.

\bibitem{hidaka-ieeemicro90}
H.~Hidaka, Y.~Matsuda, M.~Asakura, and K.~Fujishima.
\newblock {The Cache DRAM Architecture}.
\newblock {\em IEEE Micro}, 1990.

\bibitem{hoseinzadeh-taco2015}
M.~Hoseinzadeh, M.~Arjomand, and H.~Sarbazi-Azad.
\newblock {SPCM: The Striped Phase Change Memory}.
\newblock {\em TACO}, 12(4):38:1--38:25, 2015.

\bibitem{randombench}
{HPC Challenge}.
\newblock {RandomAccess}.
\newblock \mbox{http://icl.cs.utk.edu/hpcc}.

\bibitem{hsieh-isca2016}
K.~Hsieh, E.~Ebrahim, G.~Kim, N.~Chatterjee, M.~O'Connor, N.~Vijaykumar,
  O.~Mutlu, and S.~W. Keckler.
\newblock {Transparent Offloading and Mapping (TOM): Enabling
  Programmer-Transparent Near-Data Processing in GPU Systems}.
\newblock In {\em ISCA}, 2016.

\bibitem{hsieh-iccd2016}
K.~Hsieh, S.~Khan, N.~Vijaykumar, K.~K. Chang, A.~Boroumand, S.~Ghose, and
  O.~Mutlu.
\newblock {Accelerating pointer chasing in 3D-stacked memory: Challenges,
  mechanisms, evaluation}.
\newblock In {\em ICCD}, 2016.

\bibitem{hsu-isca1993}
W.-C. Hsu and J.~E. Smith.
\newblock {Performance of Cached {DRAM} Organizations in Vector
  Supercomputers}.
\newblock In {\em ISCA}, 1993.

\bibitem{hur-micro2004}
I.~Hur and C.~Lin.
\newblock Adaptive history-based memory schedulers.
\newblock In {\em MICRO}, 2004.

\bibitem{hur-micro2006}
I.~Hur and C.~Lin.
\newblock Memory prefetching using adaptive stream detection.
\newblock In {\em MICRO}, 2006.

\bibitem{hwang-asplos2012}
A.~A. Hwang, I.~A. Stefanovici, and B.~Schroeder.
\newblock {Cosmic Rays Don't Strike Twice: Understanding the Nature of DRAM
  Errors and the Implications for System Design}.
\newblock In {\em ASPLOS}, 2012.

\bibitem{scikit}
{INRIA}.
\newblock scikit-learn.
\newblock \url{http://scikit-learn.org/stable/index.html}.

\bibitem{ti-usb}
T.~Instrument.
\newblock {USB Interface Adapter EVM}.
\newblock \url{http://www.ti.com/tool/usb-to-gpio}.

\bibitem{intelioat}
{Intel Corp.}
\newblock {Intel\textregistered I/O Acceleration Technology}.
\newblock
  \url{http://www.intel.com/content/www/us/en/wireless-network/accel-technology.html}.

\bibitem{intel-optmanual2012}
{Intel Corp.}
\newblock {Intel 64 and IA-32 Architectures Optimization Reference Manual},
  2012.

\bibitem{ipek-isca2008}
E.~Ipek, O.~Mutlu, J.~F. Mart\'{\i}nez, and R.~Caruana.
\newblock Self-optimizing memory controllers: A reinforcement learning
  approach.
\newblock In {\em ISCA}, 2008.

\bibitem{isen-micro2009}
C.~Isen and L.~John.
\newblock {ESKIMO - energy savings using semantic knowledge of inconsequential
  memory occupancy for DRAM subsystem}.
\newblock In {\em MICRO}, 2009.

\bibitem{ishii-msc2012}
Y.~Ishii, K.~Hosokawa, M.~Inaba, and K.~Hiraki.
\newblock High performance memory access scheduling using compute-phase
  prediction and writeback-refresh overlap.
\newblock In {\em {JILP Memory Scheduling Championship}}, 2012.

\bibitem{itoh-vlsi}
K.~Itoh.
\newblock {\em {VLSI Memory Chip Design}}.
\newblock Springer, 2001.

\bibitem{itoh-ultra-low-memories}
K.~Itoh, M.~Horiguchi, and H.~Tanaka.
\newblock {\em {Ultra-Low Voltage Nano-Scale Memories}}.
\newblock Springer, 2007.

\bibitem{itrs-dram}
{ITRS}.
\newblock International technology roadmap for semiconductors executive
  summary.
\newblock
  \url{http://www.itrs.net/Links/2011ITRS/2011Chapters/2011ExecSum.pdf}, 2011.

\bibitem{itrs-fep-2013}
{ITRS}.
\newblock \url{http://www.itrs.net/ITRS 1999-2014 Mtgs, Presentations &
  Links/2013ITRS/2013Tables/FEP_2013Tables.xlsx}, 2013.

\bibitem{itrs-interconnect-2013}
{ITRS}.
\newblock \url{http://www.itrs.net/ITRS 1999-2014 Mtgs, Presentations &
  Links/2013ITRS/2013Tables/Interconnect_2013Tables.xlsx}, 2013.

\bibitem{jacobsen-rts2015}
M.~Jacobsen, D.~Richmond, M.~Hogains, and R.~Kastner.
\newblock {RIFFA 2.1: A Reusable Integration Framework for FPGA Accelerators}.
\newblock {\em RTS}, 2015.

\bibitem{jedec-ddr2}
{JEDEC}.
\newblock {DDR2 SDRAM Standard}, 2009.

\bibitem{jedec-ddr3}
{JEDEC}.
\newblock {DDR3 SDRAM Standard}, 2010.

\bibitem{jedec-spd}
{JEDEC}.
\newblock {Standard No. 21-C. Annex K: Serial Presence Detect (SPD) for DDR3
  SDRAM Modules}, 2011.

\bibitem{jedec-ddr4}
{JEDEC}.
\newblock {DDR4 SDRAM Standard}, 2012.

\bibitem{jedec-lpddr3}
{JEDEC}.
\newblock {Low Power Double Data Rate 3 (LPDDR3)}, 2012.

\bibitem{jedec-ddr3l}
{JEDEC}.
\newblock {Addendum No.1 to JESD79-3 - 1.35V DDR3L-800, DDR3L-1066, DDR3L-1333,
  DDR3L-1600, and DDR3L-1866}, 2013.

\bibitem{jedec-lpddr4}
{JEDEC}.
\newblock {Low Power Double Data Rate 4 (LPDDR4)}, 2014.

\bibitem{jiang-hpca2012}
L.~Jiang, B.~Zhao, Y.~Zhang, J.~Yang, and B.~R. Childers.
\newblock {Improving Write Operations in MLC Phase Change Memory}.
\newblock In {\em HPCA}, 2012.

\bibitem{jiang-hpca2010}
X.~Jiang, N.~Madan, L.~Zhao, M.~Upton, R.~Iyer, S.~Makineni, D.~Newell,
  Y.~Solihin, and R.~Balasubramonian.
\newblock {{CHOP}: Adaptive Filter-Based {DRAM} Caching for {CMP} Server
  Platforms}.
\newblock In {\em HPCA}, 2010.

\bibitem{jiang-pact2009}
X.~Jiang, Y.~Solihin, L.~Zhao, and R.~Iyer.
\newblock {Architecture Support for Improving Bulk Memory Copying and
  Initialization Performance}.
\newblock In {\em PACT}, 2009.

\bibitem{joao-asplos2012}
J.~A. Joao, M.~A. Suleman, O.~Mutlu, and Y.~N. Patt.
\newblock {Bottleneck Identification and Scheduling in Multithreaded
  Applications}.
\newblock In {\em ASPLOS}, 2012.

\bibitem{joao-isca2013}
J.~A. Joao, M.~A. Suleman, O.~Mutlu, and Y.~N. Patt.
\newblock {Utility-based Acceleration of Multithreaded Applications on
  Asymmetric CMPs}.
\newblock In {\em ISCA}, 2013.

\bibitem{osp-isca13}
A.~Jog, O.~Kay{\i}ran, A.~K. Mishra, M.~T. Kandemir, O.~Mutlu, R.~Iyer, and
  C.~R. Das.
\newblock {Orchestrated Scheduling and Prefetching for GPGPUs}.
\newblock In {\em ISCA}, 2013.

\bibitem{owl-asplos13}
A.~Jog, O.~Kay{\i}ran, N.~C. Nachiappan, A.~K. Mishra, M.~T. Kandemir,
  O.~Mutlu, R.~Iyer, and C.~R. Das.
\newblock {OWL: Cooperative Thread Array Aware Scheduling Techniques for
  Improving GPGPU Performance}.
\newblock In {\em ASPLOS}, 2013.

\bibitem{jog-sigmetrics2016}
A.~Jog, O.~Kayiran, A.~Pattnaik, M.~T. Kandemir, O.~Mutlu, R.~Iyer, and C.~R.
  Das.
\newblock {Exploiting Core Criticality for Enhanced GPU Performance}.
\newblock In {\em SIGMETRICS}, 2016.

\bibitem{joseph-isca1997}
D.~Joseph and D.~Grunwald.
\newblock {Prefetching Using Markov Predictors}.
\newblock In {\em ISCA}, 1997.

\bibitem{jouppi-isca90}
N.~P. Jouppi.
\newblock {Improving Direct-Mapped Cache Performance by the Addition of a Small
  Fully-Associative Cache and Prefetch Buffers}.
\newblock In {\em ISCA}, 1990.

\bibitem{jung-patmos2016}
M.~Jung, D.~M. Mathew, {\'E}.~F. Zulian, C.~Weis, and N.~Wehn.
\newblock {A New Bank Sensitive DRAMPower Model for Efficient Design Space
  Exploration}.
\newblock In {\em PATMOS}, 2016.

\bibitem{jung-memsys2016}
M.~Jung, C.~C. Rheinl\"{a}nder, C.~Weis, and N.~Wehn.
\newblock {Reverse Engineering of DRAMs: Row Hammer with Crosshair}.
\newblock In {\em MEMSYS}, 2016.

\bibitem{kahle-ibmjrd2005}
J.~A. Kahle, M.~N. Day, H.~P. Hofstee, C.~R. Johns, T.~R. Maeurer, and
  D.~Shippy.
\newblock {Introduction to the Cell Multiprocessor}.
\newblock {\em IBM JRD}, 2005.

\bibitem{kanev-isca2015}
S.~Kanev, J.~P. Darago, K.~Hazelwood, P.~Ranganathan, T.~Moseley, G.-Y. Wei,
  and D.~Brooks.
\newblock {Profiling a Warehouse-Scale Computer}.
\newblock In {\em ISCA}, 2015.

\bibitem{kang14}
U.~Kang, H.-S. Yu, C.~Park, H.~Zheng, J.~Halbert, K.~Bains, S.~Jang, and
  J.~Choi.
\newblock {Co-Architecting Controllers and {DRAM} to Enhance {DRAM} Process
  Scaling}.
\newblock In {\em {The Memory Forum}}, 2014.

\bibitem{808425}
Y.~Kang, W.~Huang, S.-M. Yoo, D.~Keen, Z.~Ge, V.~Lam, P.~Pattnaik, and
  J.~Torrellas.
\newblock {FlexRAM: toward an advanced intelligent memory system}.
\newblock In {\em ICCD}, 1999.

\bibitem{kedem-1997}
G.~Kedem and R.~P. Koganti.
\newblock {{WCDRAM}: A Fully Associative Integrated Cached-{DRAM} with Wide
  Cache Lines}.
\newblock {\em CS-1997-03, Duke}, 1997.

\bibitem{keeth-dram-tutorial}
B.~Keeth and R.~J. Baker.
\newblock {\em DRAM Circuit Design: A Tutorial}.
\newblock Wiley, 2001.

\bibitem{khan-dsn2016}
S.~Khan et~al.
\newblock {PARBOR: An Efficient System-Level Technique to Detect Data Dependent
  Failures in DRAM}.
\newblock In {\em DSN}, 2016.

\bibitem{khan-sigmetrics2014}
S.~Khan, D.~Lee, Y.~Kim, A.~R. Alameldeen, C.~Wilkerson, and O.~Mutlu.
\newblock {The Efficacy of Error Mitigation Techniques for DRAM Retention
  Failures: A Comparative Experimental Study}.
\newblock In {\em SIGMETRICS}, 2014.

\bibitem{khan-cal2016}
S.~Khan, C.~Wilkerson, D.~Lee, A.~R. Alameldeen, and O.~Mutlu.
\newblock {A Case for Memory Content-Based Detection and Mitigation of
  Data-Dependent Failures in DRAM}.
\newblock {\em CAL}, 2016.

\bibitem{kim-rtas2014}
H.~Kim, D.~de~Niz, B.~Andersson, M.~Klein, O.~Mutlu, and R.~Rajkumar.
\newblock Bounding memory interference delay in cots-based multi-core systems.
\newblock In {\em RTAS}, 2014.

\bibitem{kim-rts2016}
H.~Kim, D.~de~Niz, B.~Andersson, M.~Klein, O.~Mutlu, and R.~Rajkumar.
\newblock {Bounding and Reducing Memory Interference Delay in COTS-Based
  Multi-Core Systems}.
\newblock {\em RTS}, {52}({3}):{356--395}, 2016.

\bibitem{kim-asic2001}
J.~Kim and M.~C. Papaefthymiou.
\newblock Block-based multi-period refresh for energy efficient dynamic memory.
\newblock In {\em ASIC}, 2001.

\bibitem{kim-iedm2005}
K.~Kim.
\newblock {Technology for Sub-50nm DRAM and NAND Flash Manufacturing}.
\newblock {\em IEDM}, pages 323--326, December 2005.

\bibitem{kim-edl2009}
K.~Kim and J.~Lee.
\newblock {A New Investigation of Data Retention Time in Truly Nanoscaled
  DRAMs}.
\newblock {\em EDL}, 30(8):846--848, 2009.

\bibitem{kim-thesis}
Y.~Kim.
\newblock {\em {Architectural Techniques to Enhance DRAM Scaling}}.
\newblock PhD thesis, Carnegie Mellon University, 2015.

\bibitem{kim-isca2014}
Y.~Kim, R.~Daly, J.~Kim, C.~Fallin, J.~H. Lee, D.~Lee, C.~Wilkerson, K.~Lai,
  and O.~Mutlu.
\newblock {Flipping Bits in Memory Without Accessing Them: An Experimental
  Study of DRAM Disturbance Errors}.
\newblock In {\em ISCA}, 2014.

\bibitem{ramulator}
Y.~Kim et~al.
\newblock {Ramulator}.
\newblock \url{https://github.com/CMU-SAFARI/ramulator}.

\bibitem{kim-hpca2010}
Y.~Kim, D.~Han, O.~Mutlu, and M.~Harchol-Balter.
\newblock {{ATLAS}: A Scalable and High-Performance Scheduling Algorithm for
  Multiple Memory Controllers}.
\newblock In {\em HPCA}, 2010.

\bibitem{kim-micro2010}
Y.~Kim, M.~Papamichael, O.~Mutlu, and M.~Harchol-Balter.
\newblock {Thread Cluster Memory Scheduling: Exploiting Differences in Memory
  Access Behavior}.
\newblock In {\em MICRO}, 2010.

\bibitem{kim-isca2012}
Y.~Kim, V.~Seshadri, D.~Lee, J.~Liu, and O.~Mutlu.
\newblock {A Case for Exploiting Subarray-Level Parallelism (SALP) in DRAM}.
\newblock In {\em ISCA}, 2012.

\bibitem{kim-cal2015}
Y.~Kim, W.~Yang, and O.~Mutlu.
\newblock {Ramulator: A Fast and Extensible DRAM Simulator}.
\newblock {\em CAL}, 2015.

\bibitem{kleveland-ieeemicro2013}
B.~Kleveland, M.~J. Miller, R.~B. David, J.~Patel, R.~Chopra, D.~K. Sikdar,
  J.~Kumala, S.~D. Vamvakos, M.~Morrison, M.~Liu, and J.~Balachandran.
\newblock {An Intelligent RAM with Serial I/Os}.
\newblock {\em IEEE Micro}, 2013.

\bibitem{4115697}
P.~M. Kogge.
\newblock {EXECUBE-A New Architecture for Scaleable MPPs}.
\newblock In {\em ICPP}, 1994.

\bibitem{kongetira-ieeemicro2005}
P.~Kongetira, Kathirgamar, and K.~Olukotun.
\newblock {Niagara: A 32-Way Multithreaded Sparc Processor}.
\newblock {\em IEEE Micro}, 25(2):21--29, March--April 2005.

\bibitem{koob-transvlsi2011}
J.~C. Koob, S.~A. Ung, B.~F. Cockburn, and D.~G. Elliott.
\newblock {Design and Characterization of a Multilevel DRAM}.
\newblock {\em IEEE Transactions on VLSI}, 19(9):1583--1596, Sept 2011.

\bibitem{kroft-isca81}
D.~Kroft.
\newblock {Lockup-Free Instruction Fetch/Prefetch Cache Organization}.
\newblock In {\em ISCA}, 1981.

\bibitem{ku-ispass2013}
E.~Kultursay, M.~Kandemir, A.~Sivasubramaniam, and O.~Mutlu.
\newblock {Evaluating STT-RAM as an energy-efficient main memory alternative}.
\newblock In {\em ISPASS}, 2013.

\bibitem{lai-isca2001}
A.-C. Lai, C.~Fide, and B.~Falsafi.
\newblock {Dead-block Prediction \& Dead-block Correlating Prefetchers}.
\newblock In {\em ISCA}, 2001.

\bibitem{lee-isca2009}
B.~C. Lee, E.~Ipek, O.~Mutlu, and D.~Burger.
\newblock {Architecting Phase Change Memory as a Scalable DRAM Alternative}.
\newblock In {\em ISCA}, 2009.

\bibitem{lee-cacm2010}
B.~C. Lee, E.~Ipek, O.~Mutlu, and D.~Burger.
\newblock {Phase Change Memory Architecture and the Quest for Scalability}.
\newblock {\em CACM}, 53(7):99--106, 2010.

\bibitem{lee-ieeemicro2010}
B.~C. Lee, P.~Zhou, J.~Yang, Y.~Zhang, B.~Zhao, E.~Ipek, O.~Mutlu, and
  D.~Burger.
\newblock {Phase-Change Technology and the Future of Main Memory}.
\newblock {\em {IEEE Micro}}, 30(1):143--143, 2010.

\bibitem{lee-micro2008}
C.~J. Lee, O.~Mutlu, V.~Narasiman, and Y.~N. Patt.
\newblock {Prefetch-Aware DRAM Controllers}.
\newblock In {\em MICRO}, 2008.

\bibitem{lee-tc2011}
C.~J. Lee, O.~Mutlu, V.~Narasiman, and Y.~N. Patt.
\newblock {Prefetch-Aware Memory Controllers}.
\newblock {\em IEEE TC}, 60(10):1406--1430, 2011.

\bibitem{lee-tech2010}
C.~J. Lee, V.~Narasiman, E.~Ebrahimi, O.~Mutlu, and Y.~N. Patt.
\newblock {{DRAM-Aware} Last-Level Cache Writeback: Reducing Write-Caused
  Interference in Memory Systems}.
\newblock Technical report, 2010.

\bibitem{lee-micro2009}
C.~J. Lee, V.~Narasiman, O.~Mutlu, and Y.~N. Patt.
\newblock {Improving Memory Bank-Level Parallelism in the Presence of
  Prefetching}.
\newblock In {\em MICRO}, 2009.

\bibitem{lee-thesis2016}
D.~Lee.
\newblock {\em {Reducing DRAM Latency at Low Cost by Exploiting
  Heterogeneity}}.
\newblock PhD thesis, Carnegie Mellon University, 2016.

\bibitem{lee-thesis}
D.~Lee.
\newblock {Reducing DRAM Latency at Low Cost by Exploiting Heterogeneity}.
\newblock In {\em arXiv:1604.08041v1}, 2016.

\bibitem{lee-sigmetrics2017}
D.~Lee, S.~Khan, L.~Subramanian, S.~Ghose, R.~Ausavarungnirun, G.~Pekhimenko,
  V.~Seshadri, and O.~Mutlu.
\newblock {Design-Induced Latency Variation in Modern DRAM Chips:
  Characterization, Analysis, and Latency Reduction Mechanisms}.
\newblock In {\em SIGMETRICS}, 2017.

\bibitem{lee-arxiv2016}
D.~Lee, S.~M. Khan, L.~Subramanian, R.~Ausavarungnirun, G.~Pekhimenko,
  V.~Seshadri, S.~Ghose, and O.~Mutlu.
\newblock Reducing {DRAM} latency by exploiting design-induced latency
  variation in modern {DRAM} chips.
\newblock In {\em arXiv:1610.09604v1}, 2016.

\bibitem{lee-hpca2015}
D.~Lee, Y.~Kim, G.~Pekhimenko, S.~Khan, V.~Seshadri, K.~Chang, and O.~Mutlu.
\newblock {Adaptive-Latency DRAM: Optimizing DRAM Timing for the Common-Case}.
\newblock In {\em HPCA}, 2015.

\bibitem{lee-hpca2013}
D.~Lee, Y.~Kim, V.~Seshadri, J.~Liu, L.~Subramanian, and O.~Mutlu.
\newblock {Tiered-Latency {DRAM}: A Low Latency and Low Cost {DRAM}
  Architecture}.
\newblock In {\em HPCA}, 2013.

\bibitem{lee-taco2016}
D.~Lee, G.~Pekhimenko, S.~M. Khan, S.~Ghose, and O.~Mutlu.
\newblock {Simultaneous Multi Layer Access: A High Bandwidth and Low Cost
  3D-Stacked Memory Interface}.
\newblock {\em TACO}, 2016.

\bibitem{lee-pact2015}
D.~Lee, L.~Subramanian, R.~Ausavarungnirun, J.~Choi, and O.~Mutlu.
\newblock {Decoupled Direct Memory Access: Isolating CPU and IO Traffic by
  Leveraging a Dual-Data-Port DRAM}.
\newblock In {\em PACT}, 2015.

\bibitem{lee-2006}
E.~A. Lee.
\newblock The problem with threads.
\newblock {\em Computer}, 39(5):33--42, 2006.

\bibitem{lee-pact2010}
M.~M. Lee, J.~Kim, D.~Abts, M.~Marty, and J.~W. Lee.
\newblock {Approximating Age-based Arbitration in On-chip Networks}.
\newblock In {\em PACT}, 2010.

\bibitem{mcpat:micro}
S.~Li, J.~H. Ahn, R.~D. Strong, J.~B. Brockman, D.~M. Tullsen, and N.~P.
  Jouppi.
\newblock {McPAT: An Integrated Power, Area, and Timing Modeling Framework for
  Multicore and Manycore Architectures}.
\newblock In {\em MICRO}, 2009.

\bibitem{li-usenixatc2010}
X.~Li, M.~C. Huang, K.~Shen, and L.~Chu.
\newblock {A Realistic Evaluation of Memory Hardware Errors and Software System
  Susceptibility}.
\newblock In {\em USENIX ATC}, 2010.

\bibitem{li11}
Y.~Li, H.~Schneider, F.~Schnabel, R.~Thewes, and D.~Schmitt-Landsiedel.
\newblock {DRAM Yield Analysis and Optimization by a Statistical Design
  Approach}.
\newblock In {\em IEEE TCSI}, 2011.

\bibitem{li2016maxpb}
Z.~Li, F.~Wang, D.~Feng, Y.~Hua, J.~Liu, and W.~Tong.
\newblock {MaxPB: Accelerating PCM write by maximizing the power budget
  utilization}.
\newblock {\em TACO}, 13(4):46, 2016.

\bibitem{lim-isscc2012}
K.-N. Lim, W.-J. Jang, H.-S. Won, K.-Y. Lee, H.~Kim, D.-W. Kim, M.-H. Cho,
  S.-L. Kim, J.-H. Kang, K.-W. Park, and B.-T. Jeong.
\newblock {A 1.2V 23nm 6F2 4Gb DDR3 SDRAM with Local-Bitline Sense Amplifier,
  Hybrid LIO Sense Amplifier and Dummy-Less Array Architecture}.
\newblock In {\em ISSCC}, 2012.

\bibitem{kyunam-isscc2012}
K.-N. Lim, W.-J. Jang, H.-S. Won, K.-Y. Lee, H.~Kim, D.-W. Kim, M.-H. Cho,
  S.-L. Kim, J.-H. Kang, K.-W. Park, and B.-T. Jeong.
\newblock {A 1.2V 23nm 6F2 4Gb DDR3 SDRAM With Local-Bitline Sense Amplifier,
  Hybrid LIO Sense Amplifier and Dummy-Less Array Architecture}.
\newblock In {\em ISSCC}, 2012.

\bibitem{lin-iccd2012}
C.~H. Lin, D.~Y. Shen, Y.~J. Chen, C.~L. Yang, and M.~Wang.
\newblock Secret: Selective error correction for refresh energy reduction in
  drams.
\newblock In {\em ICCD}, 2012.

\bibitem{lin-hpca2008}
J.~Lin, Q.~Lu, X.~Ding, Z.~Zhang, and P.~Sadayappan.
\newblock {Gaining Insights into Multicore Cache Partitioning: Bridging the Gap
  between Simulation and Real Systems}.
\newblock In {\em HPCA}, 2008.

\bibitem{lindholm-2008}
E.~Lindholm, J.~Nickolls, S.~Oberman, and J.~Montrym.
\newblock {NVIDIA Tesla: A Unified Graphics and Computing Architecture}.
\newblock {\em IEEE Micro}, 28(2):39--55, 2008.

\bibitem{ltspice}
{Linear Technology Corp.}
\newblock {LTspice IV}.
\newblock \url{http://www.linear.com/LTspice}.

\bibitem{lipasti-micro1996}
M.~H. Lipasti and J.~P. Shen.
\newblock Exceeding the dataflow limit via value prediction.
\newblock In {\em MICRO}, 1996.

\bibitem{lipasti-asplos1996}
M.~H. Lipasti, C.~B. Wilkerson, and J.~P. Shen.
\newblock {Value Locality and Load Value Prediction}.
\newblock In {\em ASPLOS}, 1996.

\bibitem{liu-isca2013}
J.~Liu, B.~Jaiyen, Y.~Kim, C.~Wilkerson, and O.~Mutlu.
\newblock {An Experimental Study of Data Retention Behavior in Modern {DRAM}
  Devices: Implications for Retention Time Profiling Mechanisms}.
\newblock In {\em ISCA}, 2013.

\bibitem{liu-isca2012}
J.~Liu, B.~Jaiyen, R.~Veras, and O.~Mutlu.
\newblock {{RAIDR}: Retention-Aware Intelligent {DRAM} Refresh}.
\newblock In {\em ISCA}, 2012.

\bibitem{liu-pact2012}
L.~Liu, Z.~Cui, M.~Xing, Y.~Bao, M.~Chen, and C.~Wu.
\newblock {A Software Memory Partition Approach for Eliminating Bank-level
  Interference in Multicore Systems}.
\newblock In {\em PACT}, 2012.

\bibitem{liu2014going}
L.~Liu, Y.~Li, Z.~Cui, Y.~Bao, M.~Chen, and C.~Wu.
\newblock {Going vertical in memory management: Handling multiplicity by
  multi-policy}.
\newblock In {\em ISCA}, 2014.

\bibitem{lu-micro2015}
S.-L. Lu, Y.-C. Lin, and C.-L. Yang.
\newblock {Improving DRAM Latency with Dynamic Asymmetric Subarray}.
\newblock In {\em MICRO}, 2015.

\bibitem{luk-pldi2005}
C.-K. Luk, R.~Cohn, R.~Muth, H.~Patil, A.~Klauser, G.~Lowney, S.~Wallace, V.~J.
  Reddi, and K.~Hazelwood.
\newblock {Pin: Building Customized Program Analysis Tools with Dynamic
  Instrumentation}.
\newblock In {\em PLDI}, 2005.

\bibitem{luo-ispass2001}
K.~Luo, J.~Gummaraju, and M.~Franklin.
\newblock Balancing throughput and fairness in {SMT} processors.
\newblock In {\em ISPASS}, 2001.

\bibitem{luo.jsac16}
Y.~Luo, S.~Ghose, Y.~Cai, E.~F. Haratsch, and O.~Mutlu.
\newblock Enabling accurate and practical online flash channel modeling for
  modern mlc nand flash memory.
\newblock {\em JSAC}, 2016.

\bibitem{luo-dsn2014}
Y.~Luo, S.~Govindan, B.~Sharma, M.~Santaniello, J.~Meza, A.~Kansal, J.~Liu,
  B.~Khessib, K.~Vaid, and O.~Mutlu.
\newblock {Characterizing Application Memory Error Vulnerability to Optimize
  Datacenter Cost via Heterogeneous-Reliability Memory}.
\newblock In {\em DSN}, 2014.

\bibitem{mai-isca2000}
K.~Mai, T.~Paaske, N.~Jayasena, R.~Ho, W.~J. Dally, and M.~Horowitz.
\newblock Smart memories: a modular reconfigurable architecture.
\newblock In {\em ISCA}, 2000.

\bibitem{maiz-iedm2003}
J.~Maiz, S.~Hareland, K.~Zhang, and P.~Armstrong.
\newblock {Characterization of multi-bit soft error events in advanced SRAMs}.
\newblock In {\em IEDM}, 2003.

\bibitem{mao-2013}
Y.~Mao, C.~Cutler, and R.~Morris.
\newblock {Optimizing RAM-latency Dominated Applications}.
\newblock In {\em APSys}, 2013.

\bibitem{marathe-ppopp2006}
J.~Marathe and F.~Mueller.
\newblock {Hardware Profile-Guided Automatic Page Placement for ccNUMA
  Systems}.
\newblock In {\em PPoPP}, 2006.

\bibitem{massobrio1993semiconductor}
G.~Massobrio and P.~Antognetti.
\newblock {\em {Semiconductor Device Modeling with SPICE}}.
\newblock McGraw-Hill, 1993.

\bibitem{mathew-rapido2017}
D.~M. Mathew, E.~F. Zulian, S.~Kannoth, M.~Jung, C.~Weis, and N.~Wehn.
\newblock {A Bank-Wise DRAM Power Model for System Simulations}.
\newblock In {\em RAPIDO}, 2017.

\bibitem{stream}
J.~D. McCalpin.
\newblock {STREAM Benchmark}.

\bibitem{meza-cal2012}
J.~Meza, J.~Chang, H.~Yoon, O.~Mutlu, and P.~Ranganathan.
\newblock {Enabling Efficient and Scalable Hybrid Memories Using
  Fine-Granularity DRAM Cache Management}.
\newblock {\em CAL}, 2012.

\bibitem{meza-weed2013}
J.~Meza, Y.~Luo, S.~Khan, J.~Zhao, Y.~Xie, and O.~Mutlu.
\newblock {A Case for Efficient Hardware/Software Cooperative Management of
  Storage and Memory}.
\newblock In {\em WEED}, 2013.

\bibitem{meza-sigmetrics2015}
J.~Meza, Q.~Wu, S.~Kumar, and O.~Mutlu.
\newblock {A Large-Scale Study of Flash Memory Failures in the Field}.
\newblock In {\em SIGMETRICS}, 2015.

\bibitem{meza-dsn2015}
J.~Meza, Q.~Wu, S.~Kumar, and O.~Mutlu.
\newblock {Revisiting Memory Errors in Large-Scale Production Data Centers:
  Analysis and Modeling of New Trends from the Field}.
\newblock In {\em DSN}, 2015.

\bibitem{micron-tr}
{Micron Technology}.
\newblock {Calculating Memory System Power for DDR3}, 2007.

\bibitem{micronLPDDR2_2Gb}
{Micron Technology}.
\newblock {2Gb: x16, x32 Mobile LPDDR2 SDRAM S4}, 2010.

\bibitem{micronDDR3_8Gb}
{Micron Technology}.
\newblock {8Gb: x4, x8 1.5V TwinDie DDR3 SDRAM}, 2011.

\bibitem{micronSDR_128Mb}
{Micron Technology, Inc.}
\newblock {128Mb: x4, x8, x16 Automotive SDRAM}, 1999.

\bibitem{micronDDR3_4Gb}
{Micron Technology, Inc.}
\newblock {4Gb: x4, x8, x16 DDR3 SDRAM}, 2011.

\bibitem{micron-rldram3}
{Micron Technology, Inc.}
\newblock {576Mb: x18, x36 RLDRAM3}, 2011.

\bibitem{micronDDR3L_2Gb}
{Micron Technology, Inc.}
\newblock {2Gb: x4, x8, x16 DDR3L SDRAM}, 2015.

\bibitem{moore-law}
G.~E. Moore.
\newblock {Cramming More Components Onto Integrated Circuits}.
\newblock {\em Proceedings of the IEEE}, 86(1):82--85, 1998.

\bibitem{moscibroda-usenix2007}
T.~Moscibroda and O.~Mutlu.
\newblock {Memory Performance Attacks: Denial of Memory Service in Multi-core
  Systems}.
\newblock In {\em USENIX Security Symposium}, 2007.

\bibitem{moscibroda-podc2008}
T.~Moscibroda and O.~Mutlu.
\newblock {Distributed Order Scheduling and its Application to Multi-Core DRAM
  Controllers}.
\newblock In {\em PODC}, 2008.

\bibitem{moscibroda-isca2009}
T.~Moscibroda and O.~Mutlu.
\newblock {A Case for Bufferless Routing in On-chip Networks}.
\newblock In {\em ISCA}, 2009.

\bibitem{mukundan-isca2013}
J.~Mukundan, H.~Hunter, K.-h. Kim, J.~Stuecheli, and J.~F. Mart\'{\i}nez.
\newblock {Understanding and mitigating refresh overheads in high-density DDR4
  DRAM systems}.
\newblock In {\em ISCA}, 2013.

\bibitem{muralidhara-micro2011}
S.~P. Muralidhara, L.~Subramanian, O.~Mutlu, M.~Kandemir, and T.~Moscibroda.
\newblock {Reducing Memory Interference in Multicore Systems via
  Application-aware Memory Channel Partitioning}.
\newblock In {\em MICRO}, 2011.

\bibitem{mutlu-imw2013}
O.~Mutlu.
\newblock {Memory Scaling: A Systems Architecture Perspective}.
\newblock {\em IMW}, 2013.

\bibitem{mutlu-micro2005}
O.~Mutlu, H.~Kim, and Y.~N. Patt.
\newblock {Address-value delta (AVD) prediction: increasing the effectiveness
  of runahead execution by exploiting regular memory allocation patterns}.
\newblock In {\em MICRO}, 2005.

\bibitem{mutlu-isca2005}
O.~Mutlu, H.~Kim, and Y.~N. Patt.
\newblock {Techniques for Efficient Processing in Runahead Execution Engines}.
\newblock In {\em ISCA}, 2005.

\bibitem{mutlu-ieeetc2006}
O.~Mutlu, H.~Kim, and Y.~N. Patt.
\newblock {Address-Value Delta (AVD) Prediction: A Hardware Technique for
  Efficiently Parallelizing Dependent Cache Misses}.
\newblock {\em IEEE TC}, 55(12):1491--1508, 2006.

\bibitem{mutlu-ieeemicro2006}
O.~Mutlu, H.~Kim, and Y.~N. Patt.
\newblock {Efficient Runahead Execution: Power-Efficient Memory Latency
  Tolerance}.
\newblock {\em IEEE Micro}, 2006.

\bibitem{mutlu-micro2007}
O.~Mutlu and T.~Moscibroda.
\newblock {Stall-Time Fair Memory Access Scheduling for Chip Multiprocessors}.
\newblock In {\em MICRO}, 2007.

\bibitem{mutlu-isca2008}
O.~Mutlu and T.~Moscibroda.
\newblock {Parallelism-Aware Batch Scheduling: Enhancing Both Performance and
  Fairness of Shared {DRAM} Systems}.
\newblock In {\em ISCA}, 2008.

\bibitem{mutlu-hpca2003}
O.~Mutlu, J.~Stark, C.~Wilkerson, and Y.~N. Patt.
\newblock {Runahead Execution: An Alternative to Very Large Instruction Windows
  for Out-of-Order Processors}.
\newblock In {\em HPCA}, 2003.

\bibitem{mutlu-ieeemicro2003}
O.~Mutlu, J.~Stark, C.~Wilkerson, and Y.~N. Patt.
\newblock {Runahead execution: An effective alternative to large instruction
  windows}.
\newblock {\em IEEE Micro}, 23(6):20--25, 2003.

\bibitem{superfri}
O.~Mutlu and L.~Subramanian.
\newblock {Research Problems and Opportunities in Memory Systems}.
\newblock {\em SUPERFRI}, 2015.

\bibitem{nagel-spice}
L.~W. Nagel and D.~Pederson.
\newblock {SPICE (Simulation Program with Integrated Circuit Emphasis)}.
\newblock Technical Report UCB/ERL M382, EECS Department, University of
  California, Berkeley, 1973.

\bibitem{nair-hpca2013}
P.~Nair, C.-C. Chou, and M.~K. Qureshi.
\newblock A case for refresh pausing in {DRAM} memory systems.
\newblock In {\em HPCA}, 2013.

\bibitem{nair-hpca2015}
P.~J. Nair, C.~Chou, B.~Rajendran, and M.~K. Qureshi.
\newblock {Reducing read latency of phase change memory via early read and
  Turbo Read}.
\newblock In {\em HPCA}, 2015.

\bibitem{nair-isca2013}
P.~J. Nair, D.-H. Kim, and M.~K. Qureshi.
\newblock {ArchShield: Architectural Framework for Assisting DRAM Scaling by
  Tolerating High Error Rates}.
\newblock In {\em ISCA}, 2013.

\bibitem{nair-micro2014}
P.~J. Nair, D.~A. Roberts, and M.~K. Qureshi.
\newblock {Citadel: Efficiently Protecting Stacked Memory from Large
  Granularity Failures}.
\newblock In {\em MICRO}, 2014.

\bibitem{largewarps}
V.~Narasiman, M.~Shebanow, C.~J. Lee, R.~Miftakhutdinov, O.~Mutlu, and Y.~N.
  Patt.
\newblock {Improving GPU Performance via Large Warps and Two-Level Warp
  Scheduling}.
\newblock In {\em MICRO}, 2011.

\bibitem{narayanan.systor16}
I.~Narayanan, D.~Wang, M.~Jeon, B.~Sharma, L.~Caulfield, A.~Sivasubramaniam,
  B.~Cutler, J.~Liu, B.~Khessib, and K.~Vaid.
\newblock {SSD Failures in Datacenters: What? When? And Why?}
\newblock In {\em SYSTOR}, 2016.

\bibitem{nassif-isscc2000}
S.~Nassif.
\newblock {Delay Variability: Sources, Impacts and Trends}.
\newblock In {\em ISSCC}, 2000.

\bibitem{nesbit-micro2006}
K.~J. Nesbit, N.~Aggarwal, J.~Laudon, and J.~E. Smith.
\newblock Fair queuing memory systems.
\newblock In {\em MICRO}, 2006.

\bibitem{nesbit-pact2004}
K.~J. Nesbit, A.~S. Dhodapkar, and J.~E. Smith.
\newblock {AC/DC: An Adaptive Data Cache Prefetcher}.
\newblock In {\em PACT}, 2004.

\bibitem{ncsu-freepdk45}
{North Carolina State Univ.}
\newblock {FreePDK45}.
\newblock \url{http://www.eda.ncsu.edu/wiki/FreePDK}.

\bibitem{shield}
NVIDIA.
\newblock {SHIELD Tablet}.
\newblock \url{https://www.nvidia.com/en-us/shield/tablet/}.

\bibitem{o-isca2014}
S.~O, Y.~H. Son, N.~S. Kim, and J.~H. Ahn.
\newblock {Row-Buffer Decoupling: A Case for Low-Latency DRAM
  Microarchitecture}.
\newblock In {\em ISCA}, 2014.

\bibitem{ohsawa-islped1998}
T.~Ohsawa, K.~Kai, and K.~Murakami.
\newblock {Optimizing the DRAM Refresh Count for Merged DRAM/Logic LSIs}.
\newblock In {\em ISLPED}, 1998.

\bibitem{pvt_book}
M.~Onabajo and J.~Silva-Martinez.
\newblock {\em Analog Circuit Design for Process Variation-Resilient
  Systems-on-a-Chip}.
\newblock Springer, 2012.

\bibitem{694774}
M.~Oskin, F.~T. Chong, and T.~Sherwood.
\newblock Active pages: a computation model for intelligent memory.
\newblock In {\em ISCA}, 1998.

\bibitem{ousterhout-usenix1990}
J.~K. Ousterhout.
\newblock {Why Aren't Operating Systems Getting Faster as Fast as Hardware?}
\newblock In {\em USENIX Summer Conf.}, 1990.

\bibitem{ouyang-asplos2014}
J.~Ouyang, S.~Lin, S.~Jiang, Z.~Hou, Y.~Wang, and Y.~Wang.
\newblock {SDF: Software-defined Flash for Web-scale Internet Storage Systems}.
\newblock In {\em ASPLOS}, 2014.

\bibitem{parnell.globecom14}
T.~Parnell, N.~Papandreou, T.~Mittelholzer, and H.~Pozidis.
\newblock {Modelling of the Threshold Voltage Distributions of Sub-20nm NAND
  Flash Memory}.
\newblock In {\em GLOBECOM}, 2014.

\bibitem{patel-isca2017}
M.~Patel, J.~Kim, and O.~Mutlu.
\newblock {The Reach Profiler (REAPER): Enabling the Mitigation of DRAM
  Retention Failures via Profiling at Aggressive Conditions}.
\newblock In {\em ISCA}, 2017.

\bibitem{patil-micro2004}
H.~Patil, R.~Cohn, M.~Charney, R.~Kapoor, A.~Sun, and A.~Karunanidhi.
\newblock {Pinpointing Representative Portions of Large Intel Itanium Programs
  with Dynamic Instrumentation}.
\newblock In {\em MICRO}, 2004.

\bibitem{patt-micro1985}
Y.~N. Patt, W.~M. Hwu, and M.~Shebanow.
\newblock Hps, a new microarchitecture: Rationale and introduction.
\newblock In {\em Proceedings of the 18th Annual Workshop on Microprogramming},
  MICRO 18, 1985.

\bibitem{592312}
D.~Patterson, T.~Anderson, N.~Cardwell, R.~Fromm, K.~Keeton, C.~Kozyrakis,
  R.~Thomas, and K.~Yelick.
\newblock A case for intelligent ram.
\newblock {\em IEEE Micro}, 17(2):34--44, 1997.

\bibitem{pattnaik-pact2016}
A.~Pattnaik, X.~Tang, A.~Jog, O.~Kayiran, A.~K. Mishra, M.~T. Kandemir,
  O.~Mutlu, and C.~R. Das.
\newblock {Scheduling Techniques for GPU Architectures with
  Processing-In-Memory Capabilities}.
\newblock In {\em PACT}, 2016.

\bibitem{paul-isca2015}
I.~Paul, W.~Huang, M.~Arora, and S.~Yalamanchili.
\newblock {Harmonia: Balancing Compute and Memory Power in High-performance
  GPUs}.
\newblock In {\em ISCA}, 2015.

\bibitem{phadke-date2011}
S.~Phadke and S.~Narayanasamy.
\newblock {MLP aware heterogeneous memory system}.
\newblock In {\em DATE}, 2011.

\bibitem{pinheiro2007failure}
E.~Pinheiro, W.-D. Weber, and L.~A. Barroso.
\newblock {Failure Trends in a Large Disk Drive Population}.
\newblock In {\em FAST}, 2007.

\bibitem{1369204}
A.~Pirovano, A.~Redaelli, F.~Pellizzer, F.~Ottogalli, M.~Tosi, D.~Ielmini,
  A.~L. Lacaita, and R.~Bez.
\newblock {Reliability study of phase-change nonvolatile memories}.
\newblock {\em IEEE T-DMR}, 4(3):422--427, 2004.

\bibitem{dynograph}
J.~Poovey et~al.
\newblock {DynoGraph}.
\newblock \url{https://github.com/sirpoovey/DynoGraph}.

\bibitem{ptm}
PTM.
\newblock {{Predictive technology model}}.

\bibitem{6844483}
S.~H. Pugsley, J.~Jestes, H.~Zhang, R.~Balasubramonian, V.~Srinivasan,
  A.~Buyuktosunoglu, A.~Davis, and F.~Li.
\newblock {NDC: Analyzing the impact of 3D-stacked memory+logic devices on
  MapReduce workloads}.
\newblock In {\em ISPASS}, 2014.

\bibitem{qureshi-isca2010}
M.~K. Qureshi, M.~M. Franceschini, L.~A. Lastras-Monta\~{n}o, and J.~P.
  Karidis.
\newblock {Morphable Memory System: A Robust Architecture for Exploiting
  Multi-level Phase Change Memories}.
\newblock In {\em ISCA}, 2010.

\bibitem{qureshi-isca2007}
M.~K. Qureshi, A.~Jaleel, Y.~N. Patt, S.~C.~S. Jr., and J.~Emer.
\newblock {Adaptive Insertion Policies for High-Performance Caching}.
\newblock In {\em ISCA}, 2007.

\bibitem{qureshi-micro2009}
M.~K. Qureshi, J.~Karidis, M.~Franceschini, V.~Srinivasan, L.~Lastras, and
  B.~Abali.
\newblock {Enhancing Lifetime and Security of PCM-based Main Memory with
  Start-gap Wear Leveling}.
\newblock In {\em MICRO}, 2009.

\bibitem{qureshi-dsn2015}
M.~K. Qureshi, D.~H. Kim, S.~Khan, P.~J. Nair, and O.~Mutlu.
\newblock {AVATAR: A Variable-Retention-Time (VRT) Aware Refresh for DRAM
  Systems}.
\newblock In {\em DSN}, 2015.

\bibitem{qureshi-isca2009}
M.~K. Qureshi, V.~Srinivasan, and J.~A. Rivers.
\newblock {Scalable High Performance Main Memory System Using Phase-change
  Memory Technology}.
\newblock In {\em ISCA}, 2009.

\bibitem{radaelli2005investigation}
D.~Radaelli, H.~Puchner, S.~Wong, and S.~Daniel.
\newblock {Investigation of multi-bit upsets in a 150 nm technology SRAM
  device}.
\newblock {\em IEEE TNS}, 52(6):2433--2437, 2005.

\bibitem{rafique-pact2007}
N.~Rafique, W.~T. Lim, and M.~Thottethodi.
\newblock {Effective Management of DRAM Bandwidth in Multicore Processors}.
\newblock In {\em PACT}, 2007.

\bibitem{rambus_powermodel}
{Rambus}.
\newblock {DRAM Power Model}, 2010.

\bibitem{ramos-ics2011}
L.~E. Ramos, E.~Gorbatov, and R.~Bianchini.
\newblock {Page Placement in Hybrid Memory Systems}.
\newblock In {\em ICS}, 2011.

\bibitem{rixner-isca2000}
S.~Rixner, W.~Dally, U.~Kapasi, P.~Mattson, and J.~Owens.
\newblock {Memory Access Scheduling}.
\newblock In {\em ISCA}, 2000.

\bibitem{rosenblum-sosp1995}
M.~Rosenblum, E.~Bugnion, S.~A. Herrod, E.~Witchel, and A.~Gupta.
\newblock {The Impact of Architectural Trends on Operating System Performance}.
\newblock In {\em SOSP}, 1995.

\bibitem{rosenfeld-cal2011}
P.~Rosenfeld, E.~Cooper-Balis, and B.~Jacob.
\newblock {DRAMSim2: A cycle accurate memory system simulator}.
\newblock {\em CAL}, 2011.

\bibitem{samsungddr3l_2Gb}
{Samsung Electronics Co., Ltd.}
\newblock {2Gb D-die DDR3L SDRAM}, 2011.

\bibitem{sato-vlsic1998}
Y.~Sato, T.~Suzuki, T.~Aikawa, S.~Fujioka, W.~Fujieda, H.~Kobayashi, H.~Ikeda,
  T.~Nagasawa, A.~Funyu, Y.~Fuji, K.~Kawasaki, M.~Yamazaki, and M.~Taguchi.
\newblock {Fast cycle RAM (FCRAM): A 20-ns Random Row Access, Pipe-Lined
  Operating DRAM}.
\newblock In {\em VLSIC}, 1998.

\bibitem{sazeides-micro1997}
Y.~Sazeides and J.~E. Smith.
\newblock {The Predictability of Data Values}.
\newblock In {\em MICRO}, 1997.

\bibitem{schroeder-fast07}
B.~Schroeder and G.~A. Gibson.
\newblock {Disk Failures in the Real World: What Does an MTTF of 1,000,000
  Hours Mean to You?}
\newblock In {\em FAST}, 2007.

\bibitem{schroeder2016flash}
B.~Schroeder, R.~Lagisetty, and A.~Merchant.
\newblock {Flash Reliability in Production: The Expected and the Unexpected}.
\newblock In {\em FAST}, 2016.

\bibitem{schroeder-sigmetrics2009}
B.~Schroeder, E.~Pinheiro, and W.-D. Weber.
\newblock {DRAM Errors in the Wild: A Large-Scale Field Study}.
\newblock In {\em SIGMETRICS}, 2009.

\bibitem{seo-patent}
S.-Y. Seo.
\newblock {Methods of Copying a Page in a Memory Device and Methods of Managing
  Pages in a Memory System}.
\newblock U.S. Patent Application 20140185395, 2014.

\bibitem{seshadri-thesis2016}
V.~Seshadri.
\newblock {\em {Simple DRAM and Virtual Memory Abstractions to Enable Highly
  Efficient Memory Systems}}.
\newblock PhD thesis, Carnegie Mellon University, 2016.

\bibitem{seshadri-isca2014}
V.~Seshadri, A.~Bhowmick, O.~Mutlu, P.~Gibbons, M.~Kozuch, and T.~Mowry.
\newblock {The Dirty-Block Index}.
\newblock In {\em ISCA}, 2014.

\bibitem{seshadri-pact2012}
V.~Seshadri et~al.
\newblock {The Evicted-Address Filter: A Unified Mechanism to Address Both
  Cache Pollution and Thrashing}.
\newblock In {\em PACT}, 2012.

\bibitem{seshadri-cal2015}
V.~Seshadri, K.~Hsieh, A.~Boroumand, D.~Lee, M.~Kozuch, O.~Mutlu, P.~Gibbons,
  and T.~Mowry.
\newblock {Fast Bulk Bitwise AND and OR in DRAM}.
\newblock {\em CAL}, 2015.

\bibitem{seshadri-micro2013}
V.~Seshadri, Y.~Kim, C.~Fallin, D.~Lee, R.~Ausavarungnirun, G.~Pekhimenko,
  Y.~Luo, O.~Mutlu, P.~B. Gibbons, M.~A. Kozuch, and T.~C. Mowry.
\newblock {RowClone: Fast and Energy-Efficient In-{DRAM} Bulk Data Copy and
  Initialization}.
\newblock In {\em MICRO}, 2013.

\bibitem{seshadri-arxiv2016}
V.~Seshadri, D.~Lee, T.~Mullins, H.~Hassan, A.~Boroumand, J.~Kim, M.~A. Kozuch,
  O.~Mutlu, P.~B. Gibbons, and T.~C. Mowry.
\newblock Buddy-ram: Improving the performance and efficiency of bulk bitwise
  operations using {DRAM}.
\newblock In {\em arXiv:1611.09988v1}, 2016.

\bibitem{seshadri-micro2015}
V.~Seshadri, T.~Mullins, A.~Boroumand, O.~Mutlu, P.~B. Gibbons, M.~A. Kozuch,
  and T.~C. Mowry.
\newblock {Gather-Scatter DRAM: In-DRAM Address Translation to Improve the
  Spatial Locality of Non-Unit Strided Accesses}.
\newblock In {\em MICRO}, 2015.

\bibitem{seshadri-taco2015}
V.~Seshadri, S.~Yedkar, H.~Xin, O.~Mutlu, P.~B. Gibbons, M.~A. Kozuch, and
  T.~C. Mowry.
\newblock {Mitigating Prefetcher-Caused Pollution Using Informed Caching
  Policies for Prefetched Blocks}.
\newblock {\em TACO}, 11(4):51:1--51:22, 2015.

\bibitem{shafiee-hpca2014}
A.~Shafiee, M.~Taassori, R.~Balasubramonian, and A.~Davis.
\newblock {MemZip: Exploring Unconventional Benefits from Memory Compression}.
\newblock In {\em HPCA}, 2014.

\bibitem{shao-hpca2007}
J.~Shao and B.~T. Davis.
\newblock A burst scheduling access reordering mechanism.
\newblock In {\em HPCA}, 2007.

\bibitem{sharifi-micro2012}
A.~Sharifi, E.~Kultursay, M.~Kandemir, and C.~R. Das.
\newblock {Addressing End-to-End Memory Access Latency in NoC-Based
  Multicores}.
\newblock In {\em MICRO}, 2012.

\bibitem{shevgoor-micro2013}
M.~Shevgoor, J.-S. Kim, N.~Chatterjee, R.~Balasubramonian, A.~Davis, and A.~N.
  Udipi.
\newblock Quantifying the relationship between the power delivery network and
  architectural policies in a {3D}-stacked memory device.
\newblock In {\em MICRO}, 2013.

\bibitem{shin-hpca2014}
W.~Shin, J.~Yang, J.~Choi, and L.-S. Kim.
\newblock {NUAT: A Non-Uniform Access Time Memory Controller}.
\newblock In {\em HPCA}, 2014.

\bibitem{hynix-ddr3l}
{SK Hynix}.
\newblock {DDR3L SDRAM Unbuffered SODIMMs Based on 4Gb A-die}, 2014.

\bibitem{smith-mimd}
B.~J. Smith.
\newblock {A pipelined, shared resource MIMD computer}.
\newblock In {\em ICPP}, 1978.

\bibitem{smith-spie1981}
B.~J. {Smith}.
\newblock {Architecture and applications of the HEP multiprocessor computer
  system}.
\newblock In {\em SPIE}, 1981.

\bibitem{snavely-asplos2000}
A.~Snavely and D.~Tullsen.
\newblock {Symbiotic Jobscheduling for a Simultaneous Multithreading
  Processor}.
\newblock In {\em ASPLOS}, 2000.

\bibitem{son-isca2013}
Y.~H. Son, S.~O, Y.~Ro, J.~W. Lee, and J.~H. Ahn.
\newblock {Reducing Memory Access Latency with Asymmetric DRAM Bank
  Organizations}.
\newblock In {\em ISCA}, 2013.

\bibitem{sridharan-asplos2015}
V.~Sridharan, N.~DeBardeleben, S.~Blanchard, K.~B. Ferreira, J.~Stearley,
  J.~Shalf, and S.~Gurumurthi.
\newblock Memory errors in modern systems: The good, the bad, and the ugly.
\newblock In {\em ASPLOS}, 2015.

\bibitem{sridharan-sc2012}
V.~Sridharan and D.~Liberty.
\newblock {A Study of DRAM Failures in the Field}.
\newblock In {\em SC}, 2012.

\bibitem{srinath-hpca2007}
S.~Srinath, O.~Mutlu, H.~Kim, and Y.~N. Patt.
\newblock Feedback directed prefetching: Improving the performance and
  bandwidth-efficiency of hardware prefetchers.
\newblock In {\em HPCA}, 2007.

\bibitem{spec2006}
{Standard Performance Evaluation Corp.}
\newblock {SPEC CPU2006 Benchmarks}.
\newblock \mbox{http://www.spec.org/cpu2006}.

\bibitem{stone-1970}
H.~S. Stone.
\newblock {A Logic-in-Memory Computer}.
\newblock {\em IEEE Transactions on Computers}, C-19(1):73--78, 1970.

\bibitem{stuecheli-isca2010}
J.~Stuecheli, D.~Kaseridis, D.~Daly, H.~C. Hunter, and L.~K. John.
\newblock {The virtual write queue: Coordinating DRAM and last-level cache
  policies}.
\newblock In {\em ISCA}, 2010.

\bibitem{stuecheli-micro2010}
J.~Stuecheli, D.~Kaseridis, H.~Hunter, and L.~John.
\newblock {Elastic refresh: Techniques to mitigate refresh penalties in high
  density memory}.
\newblock In {\em MICRO}, 2010.

\bibitem{subramanian-thesis2015}
L.~Subramanian.
\newblock {\em {Providing High and Controllable Performance in Multicore
  Systems Through Shared Resource Management}}.
\newblock PhD thesis, Carnegie Mellon University, 2015.

\bibitem{subramanian-iccd2014}
L.~Subramanian, D.~Lee, V.~Seshadri, H.~Rastogi, and O.~Mutlu.
\newblock {The Blacklisting Memory Scheduler: Achieving High Performance and
  Fairness at Low Cost}.
\newblock In {\em ICCD}, 2014.

\bibitem{subramanian-tpds2016}
L.~Subramanian, D.~Lee, V.~Seshadri, H.~Rastogi, and O.~Mutlu.
\newblock {BLISS: Balancing Performance, Fairness and Complexity in Memory
  Access Scheduling}.
\newblock In {\em IEEE TPDS}, 2016.

\bibitem{subramanian-micro2015}
L.~Subramanian, V.~Seshadri, A.~Ghosh, S.~Khan, and O.~Mutlu.
\newblock {The Application Slowdown Model: Quantifying and Controlling the
  Impact of Inter-application Interference at Shared Caches and Main Memory}.
\newblock In {\em MICRO}, 2015.

\bibitem{subramanian-hpca2013}
L.~Subramanian, V.~Seshadri, Y.~Kim, B.~Jaiyen, and O.~Mutlu.
\newblock Mise: Providing performance predictability and improving fairness in
  shared main memory systems.
\newblock In {\em HPCA}, 2013.

\bibitem{sudan-asplos2010}
K.~Sudan, N.~Chatterjee, D.~Nellans, M.~Awasthi, R.~Balasubramonian, and
  A.~Davis.
\newblock {Micro-Pages: Increasing DRAM Efficiency with Locality-Aware Data
  Placement}.
\newblock In {\em ASPLOS}, 2010.

\bibitem{suleman-isca2010}
M.~A. Suleman, O.~Mutlu, J.~A. Joao, Khubaib, and Y.~N. Patt.
\newblock {Data Marshaling for Multi-core Architectures}.
\newblock In {\em ISCA}, 2010.

\bibitem{suleman-asplos2009}
M.~A. Suleman, O.~Mutlu, M.~K. Qureshi, and Y.~N. Patt.
\newblock {Accelerating Critical Section Execution with Asymmetric Multi-core
  Architectures}.
\newblock In {\em ASPLOS}, 2009.

\bibitem{sundriyal2016}
V.~Sundriyal and M.~Sosonkina.
\newblock {Joint frequency scaling of processor and DRAM}.
\newblock {\em The Journal of Supercomputing}, 2016.

\bibitem{sura-2015}
Z.~Sura, A.~Jacob, T.~Chen, B.~Rosenburg, O.~Sallenave, C.~Bertolli, S.~Antao,
  J.~Brunheroto, Y.~Park, K.~O'Brien, and R.~Nair.
\newblock Data access optimization in a processing-in-memory system.
\newblock In {\em CF}, 2015.

\bibitem{takahashi-jsscc2001}
T.~Takahashi, T.~Sekiguchi, R.~Takemura, S.~Narui, H.~Fujisawa, S.~Miyatake,
  M.~Morino, K.~Arai, S.~Yamada, S.~Shukuri, M.~Nakamura, Y.~Tadaki,
  K.~Kajigaya, K.~Kimura, and B.~Kiyoo~Itoh.
\newblock {A Multigigabit DRAM Technology with 6F2 Open-Bitline Cell,
  Distributed Overdriven Sensing, and Stacked-Flash Fuse}.
\newblock {\em IEEE JSSC}, 2001.

\bibitem{tavva-taco2014}
V.~K. Tavva, R.~Kasha, and M.~Mutyam.
\newblock {EFGR: An Enhanced Fine Granularity Refresh Feature for
  High-Performance DDR4 DRAM Devices}.
\newblock {\em TACO}, 11(3), 2014.

\bibitem{itrs2-2015}
{The Intl.~Technology Roadmap for Semiconductors}.
\newblock {\mbox{http://www.itrs2.net}}.
\newblock 2015.

\bibitem{thornton-1964}
J.~E. Thornton.
\newblock {Parallel operation in the Control Data 6600}.
\newblock In {\em Fall Joint Computer Conference}, 1964.

\bibitem{thwaites-pact2014}
B.~Thwaites, G.~Pekhimenko, H.~Esmaeilzadeh, A.~Yazdanbakhsh, O.~Mutlu,
  J.~Park, G.~Mururu, and T.~Mowry.
\newblock {Rollback-free Value Prediction with Approximate Loads}.
\newblock In {\em PACT}, 2014.

\bibitem{tipton2006multiple}
A.~D. Tipton, J.~A. Pellish, R.~A. Reed, R.~D. Schrimpf, R.~A. Weller, M.~H.
  Mendenhall, B.~Sierawski, A.~K. Sutton, R.~M. Diestelhorst, G.~Espinel,
  et~al.
\newblock {Multiple-bit upset in 130 nm CMOS technology}.
\newblock {\em IEEE TNS}, 53(6):3259--3264, 2006.

\bibitem{toal-ahs2007}
C.~Toal, D.~Burns, K.~McLaughlin, S.~Sezer, and S.~O'Kane.
\newblock {An RLDRAM II Implementation of a 10Gbps Shared Packet Buffer for
  Network Processing}.
\newblock In {\em AHS}, 2007.

\bibitem{toh-vlsic2010}
S.~O. Toh, Z.~Guo, and B.~Nikolić.
\newblock {Dynamic SRAM stability characterization in 45nm CMOS}.
\newblock In {\em VLSIC}, 2010.

\bibitem{tomasulo-ibmj67}
R.~M. Tomasulo.
\newblock {An Efficient Algorithm for Exploiting Multiple Arithmetic Units}.
\newblock {\em IBM Journal}, pages 25--33, January 1967.

\bibitem{tosaka2004comprehensive}
Y.~Tosaka, H.~Ehara, M.~Igeta, T.~Uemura, H.~Oka, N.~Matsuoka, and K.~Hatanaka.
\newblock {Comprehensive study of soft errors in advanced CMOS circuits with
  90/130 nm technology}.
\newblock In {\em IEDM}, 2004.

\bibitem{tpc}
{Transaction Performance Processing Council}.
\newblock {TPC Benchmarks}.
\newblock \url{http://www.tpc.org/}.

\bibitem{tullsen-isca1995}
D.~M. Tullsen, S.~J. Eggers, and H.~M. Levy.
\newblock {Simultaneous multithreading: Maximizing on-chip parallelism}.
\newblock In {\em ISCA}, 1995.

\bibitem{udipi-isca2010}
A.~N. Udipi, N.~Muralimanohar, N.~Chatterjee, R.~Balasubramonian, A.~Davis, and
  N.~P. Jouppi.
\newblock {Rethinking DRAM Design and Organization for Energy-Constrained
  Multi-Cores}.
\newblock In {\em ISCA}, 2010.

\bibitem{umuroglu-fpl2015}
Y.~Umuroglu, D.~Morrison, and M.~Jahre.
\newblock {Hybrid breadth-first search on a single-chip FPGA-CPU heterogeneous
  platform}.
\newblock In {\em FPL}, 2015.

\bibitem{usui-taco2016}
H.~Usui, L.~Subramanian, K.~K.-W. Chang, and O.~Mutlu.
\newblock {DASH: Deadline-Aware High-Performance Memory Scheduler for
  Heterogeneous Systems with Hardware Accelerators}.
\newblock {\em TACO}, 12(4):65:1--65:28, 2016.

\bibitem{venkatesan-hpca2006}
R.~Venkatesan, S.~Herr, and E.~Rotenberg.
\newblock {Retention-aware placement in DRAM (RAPID): Software methods for
  quasi-non-volatile DRAM}.
\newblock In {\em HPCA}, 2006.

\bibitem{verghese-asplos1996}
B.~Verghese, S.~Devine, A.~Gupta, and M.~Rosenblum.
\newblock {Operating System Support for Improving Data Locality on CC-NUMA
  Compute Servers}.
\newblock In {\em ASPLOS}, 1996.

\bibitem{caba}
N.~Vijaykumar, G.~Pekhimenko, A.~Jog, A.~Bhowmick, R.~Ausavarungnirun, C.~Das,
  M.~Kandemir, T.~C. Mowry, and O.~Mutlu.
\newblock {A Case for Core-Assisted Bottleneck Acceleration in {GPUs}: Enabling
  Flexible Data Compression with Assist Warps}.
\newblock In {\em ISCA}, 2015.

\bibitem{vishwanath-conext2011}
A.~Vishwanath, V.~Sivaraman, Z.~Zhao, C.~Russell, and M.~Thottan.
\newblock {Adapting Router Buffers for Energy Efficiency}.
\newblock In {\em {CoNEXT}}, 2011.

\bibitem{vogelsang-micro2010}
T.~Vogelsang.
\newblock {Understanding the Energy Consumption of Dynamic Random Access
  Memories}.
\newblock In {\em MICRO}, 2010.

\bibitem{wang-micro1997}
K.~Wang and M.~Franklin.
\newblock {Highly Accurate Data Value Prediction Using Hybrid Predictors}.
\newblock In {\em MICRO}, 1997.

\bibitem{ware-iccd2006}
F.~A. Ware and C.~Hampel.
\newblock {Improving Power and Data Efficiency with Threaded Memory Modules}.
\newblock In {\em ICCD}, 2006.

\bibitem{wong-fpt2006}
S.~Wong, F.~Duarte, and S.~Vassiliadis.
\newblock {A Hardware Cache memcpy Accelerator}.
\newblock In {\em FPT}, 2006.

\bibitem{xi-2015}
S.~L. Xi, O.~Babarinsa, M.~Athanassoulis, and S.~Idreos.
\newblock {Beyond the Wall: Near-Data Processing for Databases}.
\newblock In {\em DaMoN}, 2015.

\bibitem{xiang-ics2017}
X.~Xiang, W.~Shi, S.~Ghose, L.~Peng, O.~Mutlu, and N.-F. Tzeng.
\newblock {Carpool: A Bufferless On-chip Network Supporting Adaptive Multicast
  and Hotspot Alleviation}.
\newblock In {\em ICS}, 2017.

\bibitem{xilinx-ml605}
{Xilinx}.
\newblock {ML605 Hardware User Guide}, Oct. 2012.

\bibitem{ml605_schematic}
{Xilinx, Inc.}
\newblock {Xilinx XTP052 -- ML605 Schematics (Rev D)}.
\newblock
  \url{https://www.xilinx.com/support/documentation/boards_and_kits/xtp052_ml605_schematics.pdf}.

\bibitem{xu-iiswc2014}
Q.~Xu, H.~Jeon, and M.~Annavaram.
\newblock {Graph processing on GPUs: Where are the bottlenecks?}
\newblock In {\em IISWC}, 2014.

\bibitem{yasin-iiswc2014}
A.~Yasin, Y.~Ben-Asher, and A.~Mendelson.
\newblock {Deep-dive analysis of the data analytics workload in CloudSuite}.
\newblock In {\em IISWC}, 2014.

\bibitem{yazdanbakhsh-taco2016}
A.~Yazdanbakhsh, G.~Pekhimenko, B.~Thwaites, H.~Esmaeilzadeh, O.~Mutlu, and
  T.~C. Mowry.
\newblock {RFVP: Rollback-Free Value Prediction with Safe-to-Approximate
  Loads}.
\newblock {\em TACO}, 12(4):62:1--62:26, 2016.

\bibitem{yoon-iccd2012}
H.~Yoon, J.~Meza, R.~Ausavarungnirun, R.~Harding, and O.~Mutlu.
\newblock {Row Buffer Locality Aware Caching Policies for Hybrid Memories}.
\newblock In {\em ICCD}, 2012.

\bibitem{yoon-taco2014}
H.~Yoon, J.~Meza, N.~Muralimanohar, N.~P. Jouppi, and O.~Mutlu.
\newblock {Efficient Data Mapping and Buffering Techniques for Multilevel Cell
  Phase-Change Memories}.
\newblock {\em TACO}, 11(4):40:1--40:25, 2014.

\bibitem{yu-micro2017}
X.~Yu et~al.
\newblock {Banshee: Bandwidth-Efficient DRAM Caching via Software/Hardware
  Cooperation}.
\newblock In {\em MICRO}, 2017.

\bibitem{zabinski-csm2013}
P.~J. Zabinski, B.~K. Gilbert, and E.~S. Daniel.
\newblock {Coming Challenges with Terabit-per-Second Data Communication}.
\newblock {\em IEEE CSM}, 13(3):10--20, 2013.

\bibitem{zhang-2014}
D.~Zhang, N.~Jayasena, A.~Lyashevsky, J.~L. Greathouse, L.~Xu, and
  M.~Ignatowski.
\newblock {TOP-PIM: Throughput-oriented Programmable Processing in Memory}.
\newblock In {\em HPDC}, 2014.

\bibitem{zhang-ieee2001}
L.~Zhang, Z.~Fang, M.~Parker, B.~K. Mathew, L.~Schaelicke, J.~B. Carter, W.~C.
  Hsieh, and S.~A. McKee.
\newblock {The Impulse Memory Controller}.
\newblock {\em IEEE TC}, 50(11):1117--1132, 2001.

\bibitem{zhang-micro2009}
W.~Zhang and T.~Li.
\newblock {Characterizing and Mitigating the Impact of Process Variations on
  Phase Change Based Memory Systems}.
\newblock In {\em MICRO}, 2009.

\bibitem{6378664}
Z.~Zhang, W.~Xiao, N.~Park, and D.~J. Lilja.
\newblock Memory module-level testing and error behaviors for phase change
  memory.
\newblock In {\em ICCD}, 2012.

\bibitem{zhao-micro2014}
J.~Zhao, O.~Mutlu, and Y.~Xie.
\newblock Firm: Fair and high-performance memory control for persistent memory
  systems.
\newblock In {\em MICRO}, 2014.

\bibitem{zhao-iccd2005}
L.~Zhao, R.~Iyer, S.~Makineni, L.~Bhuyan, and D.~Newell.
\newblock {Hardware Support for Bulk Data Movement in Server Platforms}.
\newblock In {\em ICCD}, 2005.

\bibitem{zhao-isqed2006}
W.~Zhao and Y.~Cao.
\newblock New generation of predictive technology model for sub-45nm design
  exploration.
\newblock In {\em ISQED}, 2006.

\bibitem{zheng-micro2008}
H.~Zheng, J.~Lin, Z.~Zhang, E.~Gorbatov, H.~David, and Z.~Zhu.
\newblock {Mini-rank: Adaptive DRAM Architecture for Improving Memory Power
  Efficiency}.
\newblock In {\em MICRO}, 2008.

\bibitem{zheng-icpp2008}
H.~Zheng, J.~Lin, Z.~Zhang, and Z.~Zhu.
\newblock Memory access scheduling schemes for systems with multi-core
  processors.
\newblock In {\em ICPP}, 2008.

\bibitem{zhuravlev-asplos2010}
S.~Zhuravlev, S.~Blagodurov, and A.~Fedorova.
\newblock Addressing shared resource contention in multicore processors via
  scheduling.
\newblock In {\em ASPLOS}, 2010.

\bibitem{zuravleff-patent}
W.~Zuravleff and T.~Robinson.
\newblock {Controller for a Synchronous DRAM That Maximizes Throughput by
  Allowing Memory Requests and Commands to Be Issued Out of Order}.
\newblock U.S. Patent 5630096, 1997.

\end{thebibliography}

\end{document}